\documentclass[aps,prfluids,print,groupedaddress]{revtex4-2}

\usepackage{amsthm,bm} 
\usepackage{natbib}

\usepackage{epsfig,wrapfig}
\usepackage{amstext,amsfonts,amsmath,amscd, amssymb}
\usepackage{latexsym,amsthm}
\usepackage{mathrsfs}
\usepackage[small,it]{caption}
\usepackage{subfigure}
\usepackage[colorlinks=true,linkcolor=blue,urlcolor=blue,citecolor=blue]{hyperref}
\usepackage{epstopdf}
\usepackage{color}
\usepackage{mdwlist}
\usepackage{wrapfig}
\usepackage{todonotes}
\usepackage{comment}
\usepackage{float}
\usepackage{tikz}
\usepackage[percent]{overpic}
\usepackage[page]{appendix}
\usepackage{listings}
\usepackage{amsmath,amsfonts,amsthm,amssymb,bm} 
\usepackage{siunitx}
\usepackage{paralist}
\usepackage{xcolor}
\newcommand\crule[3][black]{\textcolor{#1}{\rule{#2}{#3}}}
\def\pd{\partial}
\def\la{\langle}
\def\ra{\rangle}
\def\lpm{\{}
\def\rpm{\}}
\def\lppm{[}
\def\rppm{]}

\def\v2{\overline{v^2}}
\def\v2f{\overline{v^2}f}

\def\bar{\overbar}

\def\cM{\mathcal{M}}

\def\cL{\mathcal{L}}
\def\be{\mathbf{e}}
\def\cP{\mathcal{P}}
\def\cC{\mathcal{C}}
\def\cR{\mathcal{R}}

\definecolor{dy}{rgb}{0.6,0.6,0.1}
\definecolor{dg}{rgb}{0,0.6,0}

\definecolor{green1}{RGB}{221,242,151}
\definecolor{green2}{RGB}{30,150,30}
\definecolor{purple1}{RGB}{149,87,164}
\definecolor{blue1}{RGB}{42,144,158}
\definecolor{dblue}{RGB}{11,48,84}
\definecolor{lblue}{RGB}{150,150,250}
\definecolor{mygray}{gray}{0.6}

\graphicspath{{./figs/}}

\def\pd{\partial}
\def\la{\langle}
\def\ra{\rangle}

\newcommand{\overbar}[1]{\mkern 1.5mu\overline{\mkern-1.5mu#1\mkern-1.5mu}\mkern 1.5mu}

\def\respa#1{\textcolor{blue}{#1}}
\def\respb#1{\textcolor{blue}{#1}}

\newcommand{\commentout}[1]{}

\newcommand{\RN}[1]{%
  \textup{\uppercase\expandafter{\romannumeral#1}}%
}

\renewcommand{\S}{Sec.~}

\graphicspath{{./figures/}}

\begin{document}

\title{Active model split hybrid RANS/LES}
\author{Sigfried W. Haering}
%\affiliation{The Oden Institute for Computational Engineering and
%  Science,\\ The University of Texas at Austin, Austin, Texas, 78712 USA.}
\altaffiliation[Present Address: ]{Sandia National Laboratories, Livermore, CA 94551, USA.}
\author{Todd A. Oliver}
\affiliation{The Oden Institute for Computational Engineering and
  Science,\\ The University of Texas at Austin, Austin, Texas, 78712 USA.}
\author{Robert D. Moser}
\affiliation{Department of Mechanical Engineering, The Oden Institute for Computational Engineering and
  Science \\The University of Texas at Austin, Austin, Texas, 78712 USA.}

\date{\today}

\begin{abstract}
Reliably predictive simulation of complex flows requires a level of
model sophistication and robustness exceeding the capabilities of
current Reynolds-averaged Navier-Stokes (RANS) models.  The necessary 
capability can often be provided by well-resolved large eddy simulation 
(LES), but, for many flows of interest, such simulations are too 
computationally intensive to be performed routinely.  In principle, 
hybrid RANS/LES (HRL) models capable of transitioning through 
arbitrary levels of modeled and resolved turbulence would ameliorate 
both RANS deficiencies and LES expense.  However, these HRL 
approaches have led to a host of unique complications, in addition to 
those already present in RANS and LES.  This work proposes a 
modeling approach aimed at overcoming such challenges.  The 
approach presented here relies on splitting the turbulence model into 
three distinct components: two responsible for the standard subgrid 
model roles of either providing the unresolved stress or dissipation 
and a third which reduces the model length scale by creating resolved 
turbulence.  This formulation renders blending functions unnecessary 
in HRL.  Further, the split-model approach both reduces the 
physics-approximation burden on simple eddy-viscosity-based models 
and provides convenient flexibility in model selection.  In regions where 
the resolution is adequate to support additional turbulence, fluctuations 
are generated at the smallest locally resolved scales of motion.  This active 
forcing drives the system towards a balance between RANS and 
grid-resolved LES for any combination of resolution and flow while the 
split-model formulation prevents local disruption to the total stress.  
The model is demonstrated on fully-developed, incompressible channel 
flow \cite{LeeMoser2015} and the periodic hill \cite{breu:2009}, in which 
it is shown to produce accurate results and avoid common HRL 
shortcomings, such as model stress depletion.
\end{abstract}

\maketitle

\section{Introduction}
\label{sec:intro}

% ADD: FASEL 2002 (FSM)
% PATIL 2012, zonal, two layer, synthetic inflow\
% mockett and thiele 2012: phill w/des
% choauat 2013 : phill pitm
% schmidt 2014 : phill hybrid
% Forcing: tabor 2010, huang 2010, laraufie 2011, yu 2014, davidson 2006
% grey area: probst 2017

It has long been recognized that current Reynolds-averaged Navier-Stokes
models (RANS) are inadequate for the prediction of complex turbulent
flows. On the other hand, sufficiently well-resolved large-eddy
simulation (LES) models provide an accurate representation of
turbulence in many circumstances, but are commonly too expensive to
apply in practice. Turbulence modeling methods that allow for a
flexible balance between resolving some of the turbulent fluctuations, as in LES,
and modeling their effect, as in RANS, offer an attractive compromise
between these approaches. In particular, by resolving turbulence
fluctuations only where RANS models are deficient, significant
improvements in mean flow predictions are possible with minimal
sacrifice of computational efficiency. For instance, RANS models are
well-known to be inadequate in regions of flow separation,
reattachment, or three-dimensionality in the
mean~\cite{wilcox:2007,cara:1996,edwards:1996,celic:2006,rodi:1997}.  
Enabling LES to be active in such regions will
effectively avoid the deficiencies of the RANS models, while also
avoiding the cost of using LES resolution where RANS models are
sufficient.

Due to the combination of limited computational resources and the need for
high fidelity simulations, methods that partially resolve turbulent
fluctuations can be expected to remain a necessity for engineering
applications until the turn of the next century \cite{spal:2000}.  In
response, a myriad of such turbulence modeling techniques have been
proposed and developed with varying degrees of success.  The
techniques fall into two main categories: wall-modeled large eddy
simulation
(WMLES) \cite{bose:2018,lars:2016,piom:2002,park:2014,yang:2015},
which focuses specifically on relieving the computational expense of
resolving the near-wall layer in an LES, and hybrid RANS/LES (HRL) \cite{froh:2008,ment:2010,spez:1996,spal:1997,xiao:2017,bech:2009,quem:2002,lync:2008,baur:2003,hamb:2011,giri:2006,brun:2013,fan:2004},
which addresses the more general objective of employing LES only where
RANS is deficient.  Despite extensive effort in both categories, a truly robust and
predictive approach remains elusive.

\begin{comment}
In the absence of revolutions in either computational capabilities or
the accuracy and generality of Reynolds-averaged Navier-Stokes
(RANS) based flow simulations, flexible turbulence modeling methods
allowing for variable levels of modeled and resolved turbulence will
remain a necessity until the turn of the century \cite{spal:2000}.  Such 
scale-resolving methods are attractive for two reasons.  First, they 
promise to improve upon the mean flow predictions of existing RANS 
models, which are well-known to be inadequate in the presence of 
complex flow features, such as separation, reattachment, or three-
dimensionality in the mean~\cite{wilcox:2001}.  Second, they promise 
to provide a computationally feasible tool for simulating large-scale 
turbulent fluctuations, which are important in many contexts, including, 
for instance, predictions of unsteady fluid-structure interactions or 
turbulent combustion instabilities.  Because of these needs, a myriad 
of scale-resolving turbulence modeling techniques have been 
proposed and developed \cite{??} with the techniques falling into two 
main categories: wall-modeled large eddy simulation (WMLES) 
\cite{dear:19XX} and hybrid RANS/LES (HRL) \cite{spez:1998,spal:1996}.
\end{comment}

There is a great deal of overlap between WMLES and HRL.  For
instance, a HRL in which the RANS model is only active near the
wall \cite{spal:1997} can be considered a WMLES.  To avoid ambiguity 
between the two general categories, the taxonomy of Larsson 
\emph{et.al.} is observed \cite{lars:2016}.  In WMLES, LES is 
active in the entire simulated domain, but with resolution inadequate
to resolve the near-wall layer.  Wall-models provide the wall
stress, and the LES field provides velocity information for the wall
model.  In contrast, in HRL, there are distinct regions in which RANS
and LES models are active. The near-wall layer in HRL will generally
be treated with RANS, so wall-normal resolution must be sufficient to
capture the layer allowing direct computation of the wall
stress. While WMLES does not require this wall-normal resolution of the viscous
wall layer, it does require wall-parallel
resolution as dictated by the LES in the outer layer.  Like RANS, HRL
has no explicit wall-parallel resolution requirements in the near-wall
layer, other than the need to resolve the geometry and the mean flow. In effect, the
resolution used in an HRL determines the turbulence scales that can be resolved,
if any.  

It is this difference in near-wall resolution requirements that
primarily defines the strengths of each method.  A RANS model is often
sufficiently accurate throughout a boundary layer, generally failing
only when representing the interactions with large detached
structures. At
least in principle then, HRL requires the use of LES resolution
only in the vicinity of flow features of interest (e.g. separations).
WMLES, on the other hand, requires such resolution throughout the
domain. Nonetheless, WMLES may be more efficient than HRL techniques
in turbulent flows that are so complex that RANS is inadequate
everywhere except the thin viscous wall layer. In this case both
techniques would require LES resolution almost everywhere yet, HRL
techniques would also require resolution of the mean velocity in the
viscous near-wall layer, while WMLES would not.
This of course assumes that the wall model used in a WMLES is both predictive and does
not require a RANS-quality near-wall grid for complex
flows.  Fortunately, recent advances in LES wall
modeling \cite{bose:2014,loza:2017,park:2014,yang:2015} appear to be
on a path to developing the robust and truly predictive wall models
required for this purpose. Despite this, HRL techniques will continue
to be preferable in many technologically relevant flows, such as most
external aerodynamic applications, precisely because in these flows
RANS models are adequate to represent boundary layers over most of a
body surface.

Because of this promise of tractable high fidelity simulation in many
important flow scenarios, HRL has been of great interest
to turbulence modelers since its introduction by Spalart \cite{spal:1997} and Speziale \cite{spez:1996}. 
Largely due to the
formal similarities in the RANS and filtered Navier-Stokes equations,
most HRL methods are based on the attractive, yet perhaps misleading,
prospect of simply blending between eddy-viscosity-based RANS and LES
models in a manner amenable to implementation in existing CFD code
structures.  An excellent general review of HRL methods is presented by
Fr\"{o}lich and von Terzi \cite{froh:2008}.  More focused reviews of
detached eddy simulation (DES)\cite{spal:1997} and its variants \cite{spal:2006,yan:2005,trav:2000,riou:2009,deck:2005}, the most ubiquitous HRL approach, are
provided by Spalart \cite{spal:2009} and, for the particular
combination of DES with wall functions, by Gritskevich \emph{et.al.}
\cite{grits:2017}.  In general, HRL methods operate in one of three
ways: 1) by reducing the RANS model stress with some specified function
\cite{spez:1996} to allow resolved fluctuations to develop, 2) by carrying distinct RANS and LES models with some
blending of the respective modeled stress terms \cite{fan:2004}, or 3)
by carrying a single model which internally transitions between RANS
and LES-like modes of operation (\emph{e.g.} DES).  For all
methods, the model transition between RANS and LES modes may be
in response to the local model or flow parameters and a measure of the
grid size (``unified'' methods) \cite{spal:2006,Girimaji2005,Chaouat2005} or specified \emph{ab-initio}
(``interfaced'' methods) \cite{deck:2005,quem:2002,shur:2014}.  Transition from RANS to LES in unified
methods tends to rely on \emph{ad-hoc} functions requiring tuning for
specific flows (e.g. model reduction factor in the ``flow simulation
methodology'' (FSM) approach \cite{spez:1996, fasel:2002} or delaying
function in DDES\cite{spal:2006}).  Because of the need for this tuning, amongst other issues, the
resulting methods may not be predictive, in general. 

%\todo{I added quotes here b/c FSM is such a ridiculously generic name I didn't realize it was a name and thought it was a typo}

%RDM: The following isn't really the issue is it?
%Finally, all these methods are based on the basic
%Boussinesq assumption with the modeled stress being approximated by
%some eddy viscosity operating on the resolved strain rate tensor.
%This form is not sufficiently general, as discussed in~\S\ref{sec:scalar_eddy}.
%\todo{Repetition here... ok?}

%Additional problems with blending in unified methods are discussed in~\S\ref{sec:blending}. 
% \todo{added schumann 1975, still need to add leveque 1997, and moin 1982 }
Since the publication of the aforementioned review articles, there has
been further development of HRL techniques.  Of particular interest to
the developments presented here are the two-velocity hybrid RANS/LES
(TVHRL) \cite{urib:2010}, the Reynolds-stress-constrained subgrid scale (RSC-SGS) model \cite{chen:2012}, the dynamic hybrid RANS/LES
(DHRL) \cite{bhus:2012,walt:2013} and the
dual-mesh hybrid RANS/LES (DMHRL) \cite{xiao:2017} approaches.  This family of methods build on the stress decomposition of Schumann \cite{schu:1975} into locally isotropic and inhomogeneous portions where distinct eddy viscosity models act on the mean and fluctuating strains.   In Schumann's original work, a transport-based subgrid model was used for the fluctuating portion while a wall-sensitive velocity-difference model was used for the mean.  The main advantage of this structure was to allow for a more standard Reynolds stress closures to take over near walls were grid scales become large in comparison to some characteristic mixing length scale.  Thus, this may actually be considered the first hybrid method.  Until recently, mean and fluctuating stress decomposition methods have only seen pure-LES application \cite{moin1982,leveque2007}.

The TVHRL \cite{urib:2010} method blends distinct RANS and LES models (category 2 above),
and builds on the innovation of \cite{schu:1975} to have each model act on a
different portion of the resolved strain rate tensor.  In particular,
the RANS eddy viscosity acts only on the mean strain rate while the
LES eddy viscosity acts only on the resolved fluctuating strain rate.
Additionally, all quantities entering the RANS model are mean values
while a standard Smagorinsky model acts on the fluctuating portion of
the strain.  This spitting of the model was designed to allow the
resolved turbulent stress to develop independently from the RANS
viscosity.  The authors further noted that the portion of the model
acting on the mean is intended to provide the entirety of the subgrid
contribution to the mean stress while the fluctuating portion is to
contribute only to the rate of transfer of energy from the resolved
turbulent motions to the unresolved scales. This assumes, however,
that the fluctuating eddy viscosity is uncorrelated with the resolved
strain rate magnitude, which is not the case for the
Smagorinsky model as used in that work.  Time averaging over many eddy turnovers was used
to establish the mean velocity.  The model still made use of
an \emph{ad-hoc} hyperbolic tangent lengthscale comparison blending function to switch between RANS and LES
models, which can result in drastic model shifts depending on the local
ratio of model-to-grid length scales.  Nonetheless, as discussed
further in \S\ref{sec:motivateMS} and \S\ref{sec:ms}, the two velocity model splitting
approach is significant and may indeed be a \emph{necessity} for any
LES based on eddy viscosity models (EVM), when a
non-negligible portion of the mean turbulent stress must be provided
by the model.

The RSC-SGS model \cite{chen:2012} eliminated the ad-hoc blending used 
in \cite{urib:2010} by directly reduced the mean-strain contribution, as 
calculated with either the SA model \cite{spal:1994} or the van Driest 
mixing-length approximation, to the total stress by subtracting an averaged 
resolved stress.  The dynamic Smagorinsky model (DSM) \cite{lilly:1992} 
was used on the fluctuating strain.  However, the fluctuating portion of the 
model will again contribute to the mean Reynolds stress due the 
correlation between the Smagorinsky model and resolved strain.  
Without this contribution, the approach will yield the unaltered RANS stress 
in expectation.  The method also relied on specification of the wall-normal 
location where the model transitioned from the ``constrained'' mixed model 
to a standard DSM.  Thus, from the hybrid perspective, this would be 
considered a zonal approach.  

The DHRL method \cite{bhus:2012,walt:2013} extends  the two-velocity 
approach by blending between RANS, acting only on the mean, and an 
implicit SGS model (monotonically integrated LES 
\cite{grinstein2002recent}) using the ratio of resolved turbulent production 
to the difference between the production resulting from use of only the 
RANS model and the production from use of only the SGS model.  Thus, 
the state of the resolved field is considered in the model blending.  Further, 
both models are active throughout the entire domain.  The DMHRL 
\cite{xiao:2017} approach represents the extreme of interfaced methods 
by coupling distinct LES and RANS simulations on separate grids.  
Additional forcing terms are added to each set of governing equations 
which enforce consistency of the two simulations in expectation.  In the 
LES simulation, regions are designated as either RANS or LES regions 
based on a measure of the local resolution.
%\todo{SWH: they dont actually say how they determine this}.
In designated LES regions, the 
difference between the RANS simulation velocity and time-average LES 
simulation velocity is used along with a time scale
%\todo{SWH: they also dont define this}
to construct an artificial forcing acceleration that is added to the 
RANS simulation.  In regions designated as under-resolved in the LES 
simulation, the difference between the total mean stress, as determined by 
the RANS model and the subgrid scale model plus the resolved fluctuations, 
is added to the LES simulation to effectively enforce the RANS stress.
%\todo{TAO: The previous sentence is confusing.  I don't understand what 
%consistency is enforced.  SWH: is this better?  the method is really messy...}

%However, there must be
%resolved turbulence present as initiating from a RANS solution will
%result in no production of resolved fluctuations while starting
%from an arbitrarily imposed field will result in the transient
%solution being erroneously considered resolved production. \todo{check if additional averaging was used in rans part}
%\todo{RDM: What does this sentence mean? Why is it relevant?}

Despite extensive modeling developments, hybrid models based on
blending between RANS and LES have not led to generally reliable HRL
methods capable of traversing through arbitrary levels of resolved
turbulence.  There are two key reasons for this shortcoming.  First,
HRL models often rely on passive generation of resolved turbulence in
the presence of reduced modeled stress.  This approach necessarily
leads to inconsistency between the modeled, resolved, and total
turbulent stresses, as exhibited in regions of
modeled-stress-depletion.  This issue is described further
in \S\ref{sec:consistency}.  Second, most HRL approaches rely on
scalar eddy viscosity models to simultaneously represent both the
unresolved portion of the mean turbulent stress as well as the
transfer of energy from the resolved fluctuations to the subgrid, with
the notable exception of the TVHRL method and related approaches
described previously. However, a single eddy viscosity is insufficient
for this task, as discussed in~\S\ref{sec:scalar_eddy}.  Additional
problems arise due to ad-hoc blending approaches, discussed
in \S\ref{sec:blending}, and inappropriate application of transport
equations formulated to govern mean turbulence properties in RANS
closure models to fluctuating quantities, as described
in \S\ref{sec:transport}.

Here, we introduce an HRL method that directly addresses these common
issues, which we will refer to as the ``active model split'' (AMS)
method.
\begin{comment}
%RDM too much forshadowing
The method defines separate models for the unresolved mean stress (similar to
common RANS models) and for the resolved to subgrid energy transfer
(similar to common LES models) which act simultaneously throughout the
domain, similar to the TVHRL method \cite{urib:2010}.  A
simple gradient-diffusion argument is presented which both justifies
the model-split approach and obviates \emph{ad hoc} blending between
RANS and LES models.  An active forcing mechanism
(\S\ref{sec:forcing}) generates resolved turbulent fluctuations which
the model-split approach directly responds to by reducing the modeled
contribution to the total stress.  The coupling of these two
techniques allows the modeled, resolved, and total turbulence to
maintain consistency.  The method requires approximations of the
expected values of several quantities including mean velocities; here,
a weighted causal time average is used. A dual-mesh method like that
employed in DMHRL might also be appropriate, but is not pursued here.
\end{comment}
The remainder of the paper is organized as follows.
Section~\ref{sec:motivation} describes in more detail the common HRL
deficiencies that AMS is designed to address.
Section~\ref{sec:gencon} discusses general modeling issues that arise
for partial turbulence resolving models, and details of AMS are
described in Section~\ref{sec:ms}.
%including the advantages of
%model-splitting and culminating with a gradient-diffusion argument
%that provides an appropriate mean stress scaling.
%Then, \S\ref{sec:ms} gives the details of the approach as it is
%currently implemented, including the hybridization via
%model-splitting and the active forcing formulation.
Results of applying AMS in channel flow and a periodic hill case
\cite{breu:2009} are evaluated in Section~\ref{sec:results} and
in Section~\ref{sec:conclusion} conclusions and opportunities for
further developments are discussed. 

\section{Hybrid Modeling Issues}
\label{sec:motivation}

In this section, we recall several common issues with HRL and propose
strategies to either correct or circumvent them.  As this work is
primarily focused on a framework for predictive HRL, those issues
inherent to either RANS or LES are not considered here to allow
focus on those unique to hybrid methods.  However, as discussed
in \S\ref{sec:gencon}, formulating a robust HRL has led to modeling
concepts applicable to LES in general.  The particular hybrid modeling
issues addressed here are those
of
\begin{inparaenum}[(1)]
\item maintaining consistency between resolved and modeled turbulence,
\item over-reliance on simple eddy viscosity models,
\item ad-hoc RANS to LES blending, and
\item misuse of the RANS transport models.
\end{inparaenum}
Many of these issues were first identified in \cite{haer:2019} with
the exception of the first, which has been reported on extensively
elsewhere \cite{shur:2008,riou:2009,grits:2017,piom:2003}.

\subsection{Consistency of Resolved and Modeled Turbulence}
\label{sec:consistency}
  
  \begin{comment}
  
  \begin{itemize}

    \item {Consistency: Modeled stress depletion (MSD) arises from an imbalance between
      resolved and modeled turbulence
      \begin{itemize} 
      \item{Resolved turbulence necessary to balance reduction in model}
      \item{Reliance on self-generation of turbulence necessarily leads to MSD}
      \end{itemize}
      }
      
    \item{Active energy transfer from modeled to resolved turbulence 
    \begin{itemize}
    \item{Exchange synthetic zonal inflow velocity condition for body forcing}
    \item{Still must ``make-up'' turbulence}
    %\item{Eliminate zonal inflow synthetic turbulence requirements}
    \end{itemize}
    }
      
   \end{itemize}
   
   \end{comment}
   
   \begin{comment}
As discussed in 
\S\ref{??}, well-resolved LES circumvents this issue by resolving the majority of the  
stress so that inconsistencies are negligible.  In a hybrid simulation, there will \emph{necessarily} 
be regions where this is not the case.  
\end{comment}

For any scale resolving turbulence model to be predictive, the
resolved and modeled contributions to the mean turbulent stress (the
Reynolds stress) must be consistent. That is, they should sum to the
correct total stress.  As more turbulence is resolved, the resolved
contribution to the Reynolds stress should increase and the modeled
Reynolds stress should decrease by the same amount, until ultimately all of
the Reynolds stress is being carried by resolved turbulence
fluctuations as in a DNS.

In typical HRL methods, the ability of the simulation to resolve
turbulence fluctuations in some region is signaled by a reduction in
the modeled Reynolds stress.  In response, assuming the mean shear is
large enough, natural flow instabilities will then lead to resolved
fluctuations.
%These methods then rely on natural flow
%instabilities to generate such fluctuations in those regions.
By construction, such HRL methods \emph{must} exhibit regions of
modeled-stress-depletion (MSD) \cite{spal:2006,shur:2008,spal:2009} as the instabilities cannot
develop without the total stress being depleted first.
%That is, assuming the mean shear is large enough to be unstable.
In fact, this is the best case scenario, as it is possible that
turbulent regions of flow exist where reducing the model stress to zero will
not result in the development of resolved fluctuations.  Simulations by a variety of researchers with disparate
models have shown that some type of active forcing to generate
resolved fluctuations is needed in simulations of mixed levels of resolved turbulence
\cite{ment:2010,piom:2003,prob:2017}.  It appears that explicitly introducing
resolved fluctuations, at a rate consistent with the reduction of
modeled stress is the only way to prevent regions of MSD.  Otherwise,
common consequences of MSD such as log-layer mismatch, reduced body
forces, premature flow separation, and delayed flow reattachment will
persist in HRL.

Existing forcing methods typically address this issue by generating
fluctuations only at a prescribed LES inlet
\cite{shur:2014,Spalart2017,tabor:2010,patil:2012}.  In addition to
requiring specification of the distinct LES region and the entire
spectrum of locally resolved synthetic turbulent scales and
intensities, the LES inlet location must be sufficiently upstream of
the flow features of interest so that the artificial inflow condition
can ``heal'' to a realistic state.  An alternative is to use spatially
distributed body forcing to introduce synthetic turbulence slowly so
that the fluctuations maintain a realistic turbulent state through the
entire forcing region.  In this way, none of the simulated domain and
associated computational cost is sacrificed to the healing process,
which should result in an accurate solution in the entire simulated
domain. Further, the simulation domain size can be reduced or
turbulence-resolved regions limited, since no healing regions are needed,

%\todo{TAO: I like the next three paragraphs, but I'm not sure where they belong...}
% 
While it is clear that some forcing is required, the method and form
it should take is not.  The problem of introducing turbulent
structures in a hybrid simulation requires these structures to be
artificially constructed.  While one can make principled estimates of
the appropriate strength of the body forcing based on the smallest
resolved scales and the length and time scales of the modeled
turbulence, the structure of the forcing must be prescribed.

Formally, body forcing can be considered a model for the commutator of
the filter operator, denoted $\overline{\cdot}$, defining the resolved scales of turbulence and
the substantial derivative in the momentum equation,
$\mathcal{C}_e=\overbar{D u_i}/Dt - D \overbar{u}_i/Dt$, \respb{where here
the filter operator is the projection of the turbulent field onto the
discrete representation of the numerical solution (a so-called
implicit filter, see \cite{Moser2020})}.  
%\todo{add standard formal filter definition}.
From this
perspective, the only way to avoid ``making up'' fluctuations would be
to have the full DNS solution as a function of time, which would, of
course, obviate the LES. However, formulating the forcing to be as
representative of the commutator as possible should result in more
realistic forced fluctuations which can be introduced more rapidly
while maintaining a realistic resolved turbulent state.  \respb{Improving the
prescribed forcing structure will allow the use of shorter regions of
LES resolution upstream of flow features of interest, such as flow
separation.}

There is also evidence that even unrealistic forcing structures may be
sufficient to reduce inconsistencies between modeled and resolved
turbulence.  For instance, in channel flow, Piomelli \emph{et.al.}
\cite{piom:2003} found that introducing a region of broad spectrum,
tuned stochastic forcing in a thin region along the RANS/LES interface
region in an HRL eliminated log-layer mismatch in fully developed
channel flow.  While this forcing is problem dependent and difficult
to implement in general, this numerical test showed that forcing with
random fluctuations in regions of transition from RANS to LES can address
problems related to MSD.  The notion of adding user-specified regional
body forcing was further explored by Menter \emph{et.al.}  in the
SAS-F model \cite{ment:2010}.  In their work, a broad spectrum of
velocity fluctuations were produced using the random flow generator method (RFG)
\cite{Batt:2004} and used to construct an acceleration term for each discrete
timestep.  The unique aspect of their construction is that the basic
SAS model form allows for the model to detect the added fluctuations
and naturally respond by lowering the internal model length scale and
corresponding eddy viscosity.  Even so, channel flow results still
exhibited log layer mismatch.

%\todo{TAO: Next paragraph seems out of place}
In sum, inconsistency in the resolved and modeled turbulence arise from 
combinations of reliance on passive self-generation of resolved turbulence 
in the presence of reduced modeled stress and the overarching hybridization 
strategy that only manipulates the modeled turbulence.  In addition to 
representing the unresolved contribution to the turbulent stress, a hybrid 
method should ensure accurate representation of the total turbulence stress.  
We propose that \emph{body forcing is a necessary component of an HRL
framework} and that \emph{modeled stress must only be reduced in response to
the presence of locally resolved fluctuations}.  As presented in 
\S\ref{sec:forcing}, a body forcing method is formulated with no explicit time
step dependance and a structure that only introduces fluctuations at the 
scale of the smallest locally resolved turbulence, thereby reducing the 
potential corruption of existing, and presumably realistic, resolved turbulence.  
Further, no user specification of forcing-regions is necessary.

\begin{comment}
based on the variable length scale TG field and subgrid velocity 
scale rather than relying on generation of a wide spectrum of fluctuations using the 
Random Flow Generator method \cite{krai:1969,batt:2004}.  By restricting the forcing 
structure to be only at the scale of the smallest locally resolved turbulence, existing 
larger, and presumably realistic, structures are not corrupted.
\end{comment}

\subsection{Simple Eddy Viscosities}
\label{sec:scalar_eddy}

\begin{comment}

  \begin{itemize}
      \item {Overloading of $\nu_t$ in LES: required to model both stress and dissipation
      \begin{itemize}
      \item{Only required to provide momentum transfer in RANS}
      \item{In a \emph{well-resolved} LES, only required to provided energy transfer to unresolved scales}
      \item{Can't do both [Jim\'{e}nez and Moser, 2000]}
      \end{itemize}
      }
      
    \item{Eliminate ambiguous hybrid blending and activation parameters
    \begin{itemize}
    \item{Divorce physics of the flow from the hybridization}
    \item{Both mean and fluctuating portions of model always active}
    \item{Resolution tensor to obviate scalar grid measures}
    \end{itemize}
    }

  \end{itemize}
  
\end{comment}
  
The second issue stems from a limitation of simple eddy viscosity
models that has been known in the context of LES for at least two
decades \cite{Jimenez2000}, but has been ignored in hybrid
model formulations.  HRL requires the model to both make a
non-negligible contribution to the unresolved mean turbulence stress
(Reynolds stress) and represent the transfer of energy from the resolved
scales of motion to the modeled subgrid scales.  In LES, simple
eddy viscosities only represent the latter while
significantly under-predicting the former.  Jimen\'{e}z and Moser
\cite{Jimenez2000} demonstrated that the dynamic Smagorinsky model produces
subgrid stress values that are nearly five times smaller than the true
filtered stress for fully-developed channel flow.  In these cases, the LES is
``well-resolved'' in that the contribution of the subgrid turbulence
to the mean turbulent shear stress is small, so that the large
relative errors in the mean subgrid shear stress have minimal impact on the
mean velocity.  The ability of subgrid models to accurately predict the dissipation but not the momentum transport 
has also been observed in experimental measurements of a turbulent plume \cite{bast:1998}.
The canonical SGS model test case of homogeneous isotropic
turbulence (HIT) further demonstrates this point as the divergence of
the mean stress is zero by homogeneity and any contribution to local stress
from the model is irrelevant.  In either of the above cases, the function of the
subgrid ``stress'' model is only to model the transfer of energy from
the resolved to unresolved scales, with no concern for the mean stress.
On the other hand, even for flows where an eddy viscosity model
provides a good approximation of the total mean stress for RANS,
there is no reason to expect the same model to provide the correct
energy transfer for any resolved fluctuations.

The difficulty arises because the
statistical characteristics of the mean subgrid stress and the mean
energy transfer are different. The latter is governed by the correlation
coefficient between the subgrid stress and the resolved strain rate
tensor, as opposed to the former which is just the mean of the
stress. In filtered turbulence, the correlation coefficient is small
(of order 20\% or less \cite{Jimenez2000,liu:1994}), but with an eddy viscosity model, the
correlation coefficient is one or nearly so. To get the energy
transfer rate right, it is necessary to reduce the eddy viscosity compared
to that needed to get the magnitude, and the mean, of the subgrid stresses right.
Therefore, \emph{a simple eddy viscosity model is not equipped to model both
stress and energy transfer to the subgrid scales}, as commonly demanded in HRL.  It seems natural then to
divorce these two model functions in HRL.  That is, rather than use a
single SGS model, introduce a model primarily responsible for the
mean stress, operating only on the mean gradients, and a model primarily
responsible for energy transfer, operating only on resolved fluctuating
gradients.  This ``model-split'' approach is presented in \S\ref{sec:ms}.

\subsection{Model Blending}
\label{sec:blending}

\begin{comment}

  \begin{itemize}

    \item{Eliminate ambiguous hybrid blending and activation parameters
    \begin{itemize}
    \item{Divorce physics of the flow from the hybridization}
    \item{Both mean and fluctuating portions of model always active}
    \item{Resolution tensor to obviate scalar grid measures}
    \end{itemize}
    }

  \end{itemize}
  
\end{comment}
  
Third, the methods used to reduce the model stress in HRL often
involve \emph{ad-hoc} blending functions and simplified
resolution indicators. Typically, a blending function is
specified to internally modify a model's behavior, transition between two separate
models, or directly reduce the modeled Reynolds stress. In all cases,
it has a direct effect on the physics of the flow being simulated, and
unintended consequences may arise.

In some methods \cite{Girimaji2005,brun:2013}, the
complexity of determining the correct blending
%level of resolved turbulence
provided the grid, domain, model, and flow is placed entirely on the user with
\emph{ab intitio} specification of the level of resolved turbulence.
The optimal balance of resolved and modeled turbulence will generally
be a function of space and time making explicit specification by the
user difficult.

To avoid this explicit specification, it is more common to construct
the blending function based on a resolution indicator that is intended
to determine when and where LES fluctuations can be resolved based on
the local grid and flow.  However, commonly used indicators use
simplified measures of the generally anisotropic resolution and
resolved turbulence scales, if they make use of information from the
resolved fluctuations at all.  For instance, the most commonly used
grid length scale, the cube-root of the cell volume
\cite{yan:2005,dear:1973}, tends to the smallest cell length scale as
the cell aspect ratio increases.
Therefore, it implicitly assumes structures of the smallest grid scale
are resolved in every direction, which over-estimates the effective
resolution of the grid.  From this perspective, grid measures based on
the cell diagonal are preferable. Certainly, with highly anisotropic
resolutions common to near-wall meshes, different definitions will
provide different results for any particular combination of grid and
flow.
%Length scale definition is a general LES issue but is amplified
%by its use in HRL for model transition.
In determining the turbulence length scale, the state of resolved
field is often neglected in favor of turbulence length scales derived
from the model.  This practice implicitly assumes isotropy of the
resolved scales, and consistency between the resolved and unresolved
turbulence. \respb{However, the resolved turbulence is generally not
  isotropic both because the large scales are generally not isotropic
  and because anisotropic resolution will necessarily yield
  anisotropy. Anisotropy of both the resolution and the resolved
  turbulence should clearly be accounted for in a resolution
  indicator.}
%
%The coarser the LES, the more likely these assumptions
%will be violated.

Ideally, \emph{HRL methods should be constructed that eliminate blending
entirely}.  We do so here with the aforementioned model-split approach,
which yields two separate models, both active everywhere in space and
time so that no transition functions need be specified.  Additionally,
ambiguous grid measures are eliminated by introducing a resolution
tensor that represents the anisotropy of the resolution. Finally, the
indicator metric depends on both modeled and resolved quantities
everywhere in the flow.

\subsection{Transport Models}
\label{sec:transport}

\begin{comment}

  \begin{itemize}
    \item {Misuse of RANS in LES: Transport models designed for use with expected quantities
      \begin{itemize}
      \item{Models for source terms are often non-linear}
      \item{Use with fluctuating terms yields drastically different behavior}
      \end{itemize}
      }
      
    \item{Use models only for what they are intended
    \begin{itemize}
    \item{RANS transport a function of, and acts only on, the mean}
    \item{Subgrid ``stress'' models used to dissipate resolved turbulence}
    \end{itemize}
    }
    
    \end{itemize}
    
\end{comment}
    
The final issue is the use of transport equations formulated to govern
mean turbulence properties for RANS models.  These models can
dramatically misbehave when used with resolved fluctuations.  This
problem is specific to single-model unified HRL approaches based on
RANS models, a class including many common methods
(DES\cite{spal:2009}, PANS \cite{giri:2006},
PITM\cite{brun:2013}). However, the fundamental issue has largely been
ignored or unrecognized because its symptoms have been ameliorated
with numerical dissipation, coefficient tuning for a particular
numerical method and code, or the averaging of selected terms.

In these HRL methods it is often tacitly assumed that nonlinear source
terms in the turbulence model transport equations, which have been
constructed and calibrated to operate on mean quantities in a RANS
model, will perform similarly when applied to fluctuating quantities.
Of course, these terms are themselves models for actual terms in transport
equations, and in general, their use with fluctuating quantities will
result in spurious correlations with non-negligible effects on the
mean behavior.  For instance, consider the model for the destruction
of dissipation in the $\varepsilon$ transport equation
\cite{mans:1989}
\begin{equation}
2\overbar{\partial_j\partial_k{}u_i\partial_j\partial_k{}u_i}\approx{}C_{\varepsilon{}2}\frac{\varepsilon^2}{k},
\end{equation}
which is used in all $k$-$\varepsilon$ based RANS models.  The model has little physical justification 
but after coefficient tuning, it is functional for RANS in
conjunction with models of other unclosed terms.  The impact of
applying this model to fluctuations can be seen by expanding $1/k$ in a Taylor series about $1/\la{}k\ra$ and retaining up to 
quadratic terms in $k'$. The result is
  \begin{equation}
 \bigg{\la}\frac{\varepsilon^2}{k}\bigg{\ra} \approx  \frac{ \langle{} \varepsilon \rangle{}^2}{\langle k \rangle}
\left( 
1
+
\frac{ \langle{} \varepsilon'^2 \rangle{} }{\langle \varepsilon \rangle^2}
+
\frac{ \langle k'^2 \rangle }{\langle k \rangle^2}
-
\frac{ 2 \langle \varepsilon' k' \rangle }{\langle \varepsilon \rangle \langle k \rangle}
-
\frac{ \langle{} \varepsilon'^2 k' \rangle }{\langle \varepsilon \rangle^2 \langle k \rangle}
+
\frac{ 2 \langle{} \varepsilon' k'^2 \rangle{} }{\langle \varepsilon \rangle \langle k \rangle^2}
+
\frac{ \langle{} \varepsilon'^2 k'^2 \rangle{} }{\langle \varepsilon \rangle^2 \langle k \rangle^2}
\right).
\end{equation}
%RDM: Following seems unnecessary. Preserving in case I'm wrong 
%While the behavior of mixed terms is not obvious, variances are
%clearly positive definite for any level of fluctuations.
Thus, the
mean destruction of dissipation becomes a direct, and complicated,
function of the amount of resolved turbulence.  Even the standard RANS
production term, $\mathcal{P}_k=2\nu_t\la{}S\ra^2$, is a model and
when used with fluctuating quantities will exhibit spurious
correlations, depending on the particular RANS model used to determine
$\nu_t$.
%
% TAO: following a bit redundant with below, and seems out of place here
%To avoid this issue, all nonlinear terms in model transport
%equations should either have coefficients that depend on the level of
%resolved turbulence or operate only on mean quantities.

This difficulty would appear to be inconsistent with the fact that
transport-based $k$-SGS models have been used successfully for some time
\cite{dear:1973}. However, in those applications, the
model length scale is taken as the grid scale, which results in
production and dissipation of kinetic energy that scale like $k^{1/2}$
and $k^{3/2}$, respectively. The stronger dependence of dissipation on
$k$ effectively damps the $k$ fluctuations, minimizing the effects of
this problem by effectively adjusting the mean of each modeled term so
that the length scale will remain as prescribed.  In a hybrid context,
this approach is not possible due to the lack of an intrinsic model
length scale, unless the local simulation is fully LES or RANS.  Among
other issues, such an approach implies production scaling
with $k^{1/2}$ which is entirely incorrect when the SGS length scale
is between the integral and grid length scales.

To fully address the problem of using RANS transport models in HRL
without resorting to problem- and numerics-specific coefficient tuning
for all non-linear model terms, such models should be used to
represent mean quantities as a function of only mean quantities, as
they were designed to do.  This is not to say that they have no place in
HRL, nor that they cannot function as subgrid models, only that
\emph{what goes in to, and what comes out of, RANS-based transport models must be expected
values}.

\subsection{An Alternate Hybrid Modeling Formulation} 
\label{sec:method}

Motivated by the HRL difficulties discussed
in \S\ref{sec:consistency}-\ref{sec:transport}, an alternate approach
has been developed. The proposed HRL formulation includes: 1) a
generalizations of the resolution adequacy indicator used in
DES \cite{spal:1997}, 2) different models for use in RANS and LES
similar to TVHRL and DHRL though without blending, and 3) forcing to
generate resolved fluctuations as in SAS-F \cite{ment:2010} though with a different structure.  Referred
to as the active model-split (AMS) formulation, the critical features
are: 1) the use of distinct turbulence models to act on different
portions of the resolved velocity gradient tensor, and 2) directly
incorporating forcing into the model formulation so that the modeled
stress can be reduced only in response to the local level of resolved
turbulence.  Resolved turbulence is continuously generated where
indicated by a resolution adequacy parameter which accounts for the
anisotropy in the resolution and resolved velocity field along with
the model quantities.  Before presenting the new formulation
in \S\ref{sec:ms}, we present a simple argument which motivates a
significant change to LES in general and directly leads to a merger of
RANS and LES modes of operation in HRL.

%\section{General LES Considerations}
\section{General Considerations for Turbulence-Resolving Methods}
\label{sec:gencon}

To begin, we argue that the model-split approach offers an attractive
paradigm for constructing general turbulence resolving methods.
Toward this end, consider the filtered Navier-Stokes equations:
\begin{equation*}
\partial_t \overbar{u}_i + \partial_j \left( \overbar{u}_i \overbar{u}_j \right)
=
- \partial_i \overbar{p} + \partial_j \tau^{sgs}_{ij} + \partial_j \overbar{\tau}^{visc}_{ij},
\end{equation*}
where $\overbar{u}_i$ denotes the filtered velocity, $\overbar{p}$ is
the filtered pressure, $\overbar{\tau}^{visc}_{ij}$ is the filtered
viscous stress, and $\tau^{sgs}_{ij} = -(\overbar{u_i u_j} -
\overbar{u}_i \overbar{u}_j)$ represents the subgrid stress.  It is
common to represent the deviatoric part of the subgrid stress with an
eddy viscosity based model:
\begin{equation}
\tau^{sgs}_{ij} + \frac{2}{3} k_{sgs} \delta_{ij}
\approx
2 \nu_{sgs} \overbar{S}_{ij},
\label{eqn:eddyViscModel}
\end{equation}
where $2 k_{sgs}$ is the trace of $- \tau^{sgs}_{ij}$,
$\overbar{S}_{ij}$ is the filtered strain rate tensor, $\nu_{sgs}$
is the eddy viscosity, \respb{and the filter operation has been assumed to commute with differentiation}.  Of course, this model form cannot be correct
in general, simply because it relates two tensors through a scalar.
Further, it has been shown by \emph{a priori} tests~\cite{liu:1994,Jimenez2000} that common
eddy-viscosity-based subgrid models (e.g., Smagorinsky) do not lead to
high correlation between the modeled and actual subgrid stesses.

Fortunately, accurate prediction of the mean flow does not require
that~\eqref{eqn:eddyViscModel} hold in an instantaneous sense.
Instead, it is only necessary to match important statistics involving
the subgrid scales.  At a minimum, for general flows, it is necessary
that two conditions are satisfied.  First, the mean of the subgrid
contribution to the total mean stress anisotropy must be
well-represented, to ensure that errors are not introduced into the
mean momentum balance.  This implies that
\begin{equation}
\langle \tau^{sgs}_{ij} + \frac{2}{3} k_{sgs} \delta_{ij} \rangle
\approx
\langle 2 \nu_{sgs} \overbar{S}_{ij} \rangle.
\label{eqn:mean_tausgs}
\end{equation}
Second, the contribution of the subgrid stress to the mean rate of transfer of
energy from the resolved to the unresolved scales must also be
well-represented, such that the resolved fluctuations have the correct energy, or
% Mathematically, this requirement implies that
\begin{equation}
\langle \tau^{sgs}_{ij} \partial_j \bar{u}_i \rangle
\approx
\langle 2 \nu_{sgs} \overbar{S}_{ij} \partial_j \bar{u}_i \rangle.
\label{eqn:mean_res2un}
\end{equation}
Other statistics may also be important.  For example, one may also wish
to correctly capture the anisotropy of the transfer from the resolved to unresolved
turbulence.  However, it is sufficient for the present purposes to
consider just~\eqref{eqn:mean_tausgs} and~\eqref{eqn:mean_res2un},
because they are enough to reveal both the deficiencies of common
approaches and one of the major benefits of model-splitting.

If a model existed such that~\eqref{eqn:eddyViscModel} were
instantaneously true or nearly so, then~\eqref{eqn:mean_tausgs}
and~\eqref{eqn:mean_res2un} would be satisfied automatically.
However, as discussed in \S\ref{sec:scalar_eddy}, it is known that
existing RANS and LES models do not do a good job at simultaneously
predicting both the mean subgrid stress and the energy transfer rate.
Indeed, they are generally not even designed with this goal in mind.
RANS models are built entirely to capture~\eqref{eqn:mean_tausgs}.  In
RANS, since the only ``resolved'' component is the mean, an accurately
predicted mean stress is sufficient to accurately predict the transfer
from resolved to unresolved, which is just the usual production of TKE.
However, once the state is fluctuating, there is no longer any reason
to expect a RANS-like eddy viscosity to give a good prediction of the
energy transfer, even if it is adequate for the mean subgrid stress.

Alternatively, LES models are commonly built for and tested on either
homogenous flows, where the mean subgrid stress does not vary in space
and is therefore dynamically insignificant, or for well-resolved
simulations, where the mean subgrid stress is
negligible.  Thus, common LES models are calibrated to capture the
energy transfer rate, which leads to poor predictions of the mean
subgrid stress, as discussed in \S\ref{sec:scalar_eddy}.  Advanced
subgrid models \cite{nico:1999,vrem:2004,roze:2015,lilly:1992},
developed for transitional and wall-bounded flows, are constructed so
the eddy viscosity vanishes where both the resolved and modeled stress
become dynamically insignificant.  Again, the model construction does
not speak to the general scenario of both the modeled and total stress
being dynamically significant.  Successful application of such models
to well-resolved LES has no bearing on HRL.

\subsection{Model Splitting}
\label{sec:motivateMS}

%In general these two models need not be
%based on eddy viscosities, but since such models are common, we
%examine this case.  

In general turbulence resolving simulations, one cannot
expect to have good LES resolution everywhere, so at least in
some regions of the flow, both~\eqref{eqn:mean_tausgs}
and~\eqref{eqn:mean_res2un} are important.  This is especially true in a HRL as the absence of such a region would mean the simulation is just a regular well-resolved LES.  Further, in the context of
HRL methods specifically, there is no reason to expect an \emph{ad ho}c blending of
models designed to work in the limits of RANS or well-resolved LES
will succeed in simultaneously capturing~\eqref{eqn:mean_tausgs}
and~\eqref{eqn:mean_res2un} in the intermediate regime.  
Instead, recognizing these requirements motivates splitting the model
into two parts: one aimed at~\eqref{eqn:mean_tausgs} and the second
at~\eqref{eqn:mean_res2un}.  In general, these two models need not be
based on eddy viscosities.  For instance, upwinding of the convective term 
is known to introduce numerical diffusivity \cite{broo:1982} and could conceivably be used as the 
energy transfer portion. But, since one of the aims of this work is to develop 
an HRL which works with existing RANS models, we
proceed in the vein of eddy viscosity models.  In this situation, the goal is to develop two
eddy-viscosity-based models, one denoted $\nu_s$ to represent the mean
subgrid stress,
\begin{equation}
\langle \tau^{sgs}_{ij} + \frac{2}{3} k_{sgs} \delta_{ij} \rangle
\approx
2 \nu_s \langle \overbar{S}_{ij} \rangle,
\label{eqn:ms_mean_tausgs}
\end{equation}
and a second denoted $\nu_e$ to capture the energy transfer from the resolved fluctuations to unresolved scales
\begin{equation}
\langle \tau^{sgs}_{ij} \partial_j {u}^>_i \rangle
\approx
\langle 2 \nu_{e} S^>_{ij} \partial_j {u}^>_i \rangle.
\label{eqn:ms_mean_res2un}
\end{equation}
%. 2 \nu_s \langle \overbar{S}_{ij} \partial_j  {u}^>_i \rangle + 
where the greater than superscript, $(\cdot)^>$, denotes the resolved
fluctuation of the quantity about its mean.  Note that the component
of production of subgrid energy due to the mean, which is given by
$2 \nu_s \langle \overbar{S}_{ij}\ra \partial_j \la{u}_i \rangle$, is
not part of~\eqref{eqn:ms_mean_res2un}, but if
(\ref{eqn:ms_mean_tausgs}) is satisfied, the corresponding modeled production will
also be automatically be correct.
%Also, note we have excluded the modeled contribution to the total production in (\ref{eqn:ms_mean_res2un}), or $2 \nu_s \langle \overbar{S}_{ij}\ra \partial_j  \la{u}_i \rangle$.  If (\ref{eqn:ms_mean_tausgs}) is satisfied, the modeled production will also be automatically satisfied.

The above relations~\eqref{eqn:ms_mean_tausgs}
and~\eqref{eqn:ms_mean_res2un} suggest the following instantaneous
model:
\begin{equation}
\tau^{sgs}_{ij} + \frac{2}{3} k_{sgs} \delta_{ij}
\approx
2 \nu_s \langle \overbar{S}_{ij} \rangle + 
2 \nu_{e} S^>_{ij}.
\label{eqn:genericMS}
\end{equation}
In general, due to correlations between the eddy viscosities and the
velocity fluctuations, this form would result in a the contribution
from $\nu_s$ to the energy transfer,
$2 \langle \nu_s \la\overbar{S}_{ij}\ra \partial_j {u}^>_i \rangle$,
and a contribution from $\nu_e$ to the stress, $2 \la \nu_e
{S}_{ij}^>\rangle$.  To eliminate these contributions, we require that
both $\nu_s$ and $\nu_e$ not fluctuate, \emph{i.e.} they must
only depend on mean quantities or be explicitly averaged after
construction.
%Were this not true, the correlations between $\nu_e$
%and the strain would contribute to (\ref{eqn:ms_mean_tausgs}) and
%correlations between $\nu_s$ and the velocity gradient would
%contribute to (\ref{eqn:ms_mean_res2un}).
While not disqualifying,
such correlations would complicate the roles, and hence the
construction, of the two desired models.

As with~\eqref{eqn:eddyViscModel}, there is no reason to believe
that~\eqref{eqn:genericMS} represents a good approximation of the
instantaneous subgrid stress, but it has two substantial advantages.
First, for flows with a one-dimensional mean, it gives sufficient
flexibility to simultaneously satisfy the two 
requirements of predicting the mean subgrid stress
and the mean energy transfer rate.  For complex three-dimensional flows, at a minimum, the mean production ($\la{}u^\prime_iu^\prime_j\ra\la{}S_{ij}\ra$) and subgrid energy transfer constraints can be simultaneously satisfied.  Second, since $\nu_s$ is
responsible for the mean subgrid stress and $\nu_e$ is responsible for
the portion of the energy transfer not associated with the mean, these
components map naturally to existing RANS and LES models, offering a
formulation in which the models can function together simultaneously rather
than requiring \emph{ad hoc} blending.

\subsection{Mean Stress Scaling}
\label{sec:genLES}

The model-split form provides a natural segregation of the model in
a partial turbulence-resolving simulation into ``RANS-like'' and
``LES-like'' components that can function simultaneously.  However, it
does not give any insight into how these components should scale with
the level of resolved turbulence, which is important to the performance
of any practical model.  
To deduce this scaling, we apply basic eddy viscosity arguments to 
a decomposition of the subgrid stress tensor. 

Begin by decomposing the velocity field into
its mean and fluctuating parts: $u_i = \la u_i \ra + u^{\prime}_i$.
Further, apply the standard triple decomposition: $u_i = \la
u_i \ra + u^>_i + u^<_i$, where $u^>_i$ denotes the resolved
fluctuation, and $u^<_i$ denotes the subgrid fluctuation, so
$u^{\prime}_i = u^>_i + u^<_i$.  In this notation, the filtered field
is given by $\bar{u}_i = \la u_i \ra + u^>_i$.  In the following, it
is assumed that $\la{}u^>_i\ra = 0$, with the result that $\la \bar{u}_i \ra = \la
u_i \ra$.  While this does not hold for general filters applied to
general fields, one can always construct a filter that
is consistent with this assumption by applying any standard filter to just the
fluctuating field and then adding the mean.  Applying this expansion 
to the true subgrid stress term, we have
\def\brk{\mathcal{B}}
       \begin{align}
         -\tau_{ij}^{sgs}=\brk(u_i,u_j) &= \bar{u_i^>u_j^<} + \bar{u_i^<u_j^>} + \bar{u_i^<u_j^<}\nonumber\\
         &+\brk(\la{}u_i\ra,\la{}u_j\ra) + \brk(\la{}u_i\ra{},u_j') + \brk(u_i',\la{}u_j\ra) + \brk(u_i^>,u_j^>)
       \label{tau2}
       \end{align}
%       \begin{align}
%         \tau(u_i,u_j) &= \bar{u_i^>u_j^<} + \bar{u_i^<u_j^>} + \bar{u_i^<u_j^<}\nonumber\\
%         &+{\tau(\la{}u_i\ra,\la{}u_j\ra)} +  {\tau(\la{}u_i\ra{},u_j^>)} + {\tau(\la{}u_i\ra{},u_j^<)} \nonumber\\ 
%         &+ {\tau(u_i^>,\la{}u_j\ra)}+  {\tau(u_i^>,u_j^>)} + {\tau(u_i^<,\la{}u_j\ra)}  \nonumber
%       \label{tau2}
%       \end{align}
where $\brk$ is a generic ``breaking operator'' given by
$\brk({a,c})=\overbar{ac}-\bar{a}\bar{c}$.  The first three terms are
those most commonly considered as constituting the subgrid stress
tensor. The four terms on the second line written in terms of
$\brk$ arise because of non-idealities that may occur when the mean is
not well resolved in the LES and due to the non-linear interaction of
the resolved fluctuations. The standard Leonard stress is formed from the combination 
of $\brk(\la{}u_i\ra,\la{}u_j\ra)$ and $\brk(u_i^>,u_j^>)$ while the ``cross'' stress is formed from $\bar{u_i^>u_j^<} + \bar{u_i^<u_j^>} $ and $\brk(\la{}u_i\ra{},u_j') + \brk(u_i',\la{}u_j\ra)$ \cite{leon:1974}.
Stress contributions left in $\brk$-form are akin to aliasing errors in
the numerical discretization of nonlinear terms, and are often
neglected or explicitly discarded with dealiased numerical approximations. The term $\brk(u_i^>,u_j^>)$ may be of modeling interest to implicitly filtered LES.  However, we currently neglect all terms on the second line of (\ref{tau2}) to obtain
%All terms represented 
%with a $\tau$ on the r.h.s. are essentially aliasing errors of nonlinear products 
%of fluctuating quantities and the mean or aliasing of the resolved fluctuations in ${\tau(u_i^>,u_j^>)}$ which does actual%ly contain any subgrid $u$.  Ignoring such errors, we simply have
%
\begin{equation}
- \tau^{sgs}_{ij} 
\approx
\bar{u^>_i u^<_j} + \bar{u^<_i u^>_j} + \bar{u^<_i u^<_j}.
\label{eqn:tau_sgs}
\end{equation}

To ensure that a turbulence-resolving simulation produces the correct
mean flow, it is necessary that the expected value of the
filtered Navier-Stokes equations including the modeled subgrid stress, is
consistent with the RANS equations.  In terms of the total convective
flux, this requires 
\begin{equation*}
\la u_i \ra \la u_j \ra + \la u^{\prime}_i u^{\prime}_j \ra
=
\la \bar{u}_i \bar{u}_j \ra - \la \tau_{ij}^{sgs} \ra,
\end{equation*}
which, using the triple decomposition and properties of the filter
introduced above, implies that
\begin{equation*}
\la u^{\prime}_i u^{\prime}_j \ra
=
\la u^>_i u^>_j \ra - \la \tau^{sgs}_{ij} \ra.
\end{equation*}
A turbulence-resolving model is an improvement over RANS due to the
direct representation of the resolved Reynolds stress, $(u^>_iu^>_j)$, but we are
left with the difficulty of modeling the expected value of the subgrid
stress $\la \tau^{sgs}_{ij} \ra$, that is
\begin{equation*}
- \la \tau^{sgs}_{ij} \ra
=
\la u^>_i u^<_j \ra + \la u^<_i u^>_j \ra + \la u^<_i u^<_j \ra.
\label{eqn:tau}
\end{equation*}
In the model-split paradigm, the mean subgrid stress is modeled
separately, which,
assuming an eddy-viscosity form for the model is
\begin{equation}
\la u^>_i u^<_j \ra + \la u^<_i u^>_j \ra + \la u^<_i u^<_j \ra
\approx
- 2 \nu_s \la \bar{S}_{ij} \ra + \tfrac{2}{3} k_{sgs} \delta_{ij}.
\label{eqn:tau_decomp_model}
\end{equation}
The objective is then to model $\nu_s$, and we do so by considering
the theoretical basis for a linear eddy viscosity model.

Gradient-diffusion models represent the expected convective transport
of some conserved fluctuating quantity, $\phi$, by
velocity fluctuations through a linear approximation of their covariance \cite{tayl:1921,batc:1949,schw:2016},
       \begin{equation}
       \la{}u_i^\prime{}\phi^\prime\ra \approx -C\la u_i^\prime{}\chi_j\ra{}\pd_j\la{\phi}\ra=-C\la u_i^\prime{}u_j^\prime{}\ra{}T_\phi\pd_j\la{\phi}\ra,
       \label{uphi1}
       \end{equation}
where $\chi_j$ is the stochastic displacement of a fluid particle when
$\phi$ fluctuations become decorrelated with themselves. This
decorrelation occurs over a time $T_\phi$, which is defined such that
$\la{}\phi^\prime(0)\phi^\prime(T_\phi)\ra/\la\phi^{\prime2}\ra \ll1$.
Alternatively, $T_\phi$ may be taken as the the integral timescale of
the Lagrangian two-point correlation between the $i^{\rm th}$ and
$j^{\rm th}$ velocity components and ensemble averaged over the
Lagrangian particles \cite{schw:2016}.  However, such a timescale does
not ensure the validity of modeling the fluctuating convective
transport of some quantity through its gradient. It seems more natural
for the timescale to be determined from the conserved quantity
%so that the
%``gradient'' portion of the gradient-diffusion approximation will be
%applicable.\todo{swh: okay?}
leading to the substitution $\la
u_i'\chi_j\ra = \la u'_iu'_j\ra T_\phi$ in (\ref{uphi1}). Assuming
isotropy then yields,
       \begin{equation}
       \la{}u_i^\prime{}\phi^\prime\ra = -C\tfrac{2}{3}\la{}k_c\ra\delta_{ij}T_\phi \pd_j\la{\phi}\ra
       \label{uphi2}
       \end{equation}
which is a standard eddy viscosity model for the transport of
$\phi$, with the eddy
viscosity given by $\nu_\phi = Ck_cT_\phi$. Here the subscript ``c''
indicates that $k_c$ is the kinetic energy associated with the convecting
velocity fluctuations and the subscript ``$\phi$'' that $T_\phi$ is
the decorrelation time for the fluctuations of the conserved quantity $\phi$.
In LES, we essentially extend this approximation to 
filtered quantities and write
       \begin{equation}
       \bar{u_i^\prime{}\phi^\prime} = -C\tfrac{2}{3}\la{}k\ra{}T_\phi \pd_i\bar{\phi}.
       \label{uphi3}
       \end{equation}
However, $\bar{u_i^\prime{}\phi^\prime}$ is now itself a fluctuating
quantity, and there is no reason to expect that the fluctuation of the
model on the right hand side of (\ref{uphi3}), which was formulated to
represent the expected value, will match those of
$\bar{u_i^\prime{}\phi^\prime}$. Therefore, a subgrid model of the
form (\ref{uphi3}) takes on an implicit dual role. The equality in
(\ref{uphi3}) is satisfied only in expectation
(i.e. $\la\bar{u_i^\prime{}\phi^\prime}\ra =
-C\la\tfrac{2}{3}\la{}k_c\ra{}T_\phi \pd_i\bar{\phi}\ra$), while the
model fluctuations, if correlated with fluctuations in the gradient of
$\phi$, are responsible for the dissipation $\la\phi^{\prime 2}\ra$ when $\phi$ is the momentum.
We will proceed with the approximation of (\ref{uphi3}) and return to
its expected value in due course.
%
\begin{comment}
\begin{equation}
\la u'_k \phi' \ra
\approx
- C k_c^{1/2} L_c \partial_k\la \phi \ra
=
- C k_c T_c \partial_k\la \phi \ra
\label{levm}
\end{equation}
where isotropy in $\la u'_iu'_j\ra=2k_c\delta_{ij}$ has been assumed,
and $L_c$ is the expected distance the fluid particle travels before
$u'_k$ and $\phi'$ become de-correlated.  This correlation length
scale is proportional to the r.m.s. of the fluctuating velocity and
the correlation timescale of $u'_k$ and $\phi'$ as $L_c=k_c^{1/2}T_c$.
Thus, the transport is modeled in terms of an eddy viscosity
$\nu_t \propto k_c T_c$ and the mean gradient of the conserved
quantity.  The takeaway here is that the scaling of the eddy viscosity
depends on the choice of $k_c$ and $T_c$.  
\end{comment}

The purpose of this seemingly banal recounting is to highlight a
subtlety that is often overlooked when constructing subgrid stress
models. In LES modeling, it is important to consider how the
fluctuations being modeled are defined; that is, whether they are the
standard fluctuations about the mean, as in the Reynolds decomposition
(Reynolds fluctuations), or subgrid fluctuations defined relative to
the resolved turbulence in the LES. When the fluctuations of $u_k$ and
$\phi$ are defined in the same way (i.e. both Reynolds or both
subgrid), there are natural choices for $k_c$, $T_\phi$, and the
gradient.  For instance, if we were interested in the product of the
total fluctuations, \emph{e.g.}  $\bar{u^\prime_i\phi^\prime}$, the
approximation is proportional to
$k_{tot}T_{tot}\partial_i\la{}\phi\ra$ where the ``tot'' subscript
indicates values derived from the total fluctuations.  On the other
hand, were we concerned with the product of the subgrid fluctuations,
\emph{e.g.} $\bar{u^<_i\phi^<}$, the approximation is proportional to
$k_{sgs}T_{sgs}\partial_i\bar{\phi}$ where the ``sgs'' subscript
indicates quantities derived from subgrid scale fluctuations.
Alternatively, for ``mixed'' terms, such as $u'_i \phi^<$, the three
quantities forming the model, $k_c$, $T_\phi$, and the gradient of
$\phi$, must also be mixed. In particular, $k_c$ is associated with
the definition of the convecting velocity fluctuations, $T_\phi$ is
related to the definition of fluctuations in the transported quantity
($\phi$), and the gradient is taken of the complement of the $\phi$
fluctuations.

For mixed scale terms, we therefore have
\begin{equation}
\bar{u'_i \phi^<}\approx{}-C\tfrac{2}{3}\la{}k_{tot}\ra{}T_{sgs} \pd_i\bar{\phi}=-C\tfrac{2}{3}\la{}k_{tot}\ra{}T_{sgs} (\pd_i\la{\phi}\ra+\pd_i{\phi^>})
\end{equation}
and
\begin{equation}
\bar{u^<_i \phi^\prime}\approx{}-C\tfrac{2}{3}\la{}k_{sgs}\ra{}T_{tot} \pd_i\la{\phi}\ra.
\end{equation}

Let the level of resolved turbulence be measured by $\alpha=k_{sgs}/k_{tot}$, 
assume $T_{tot}=k_{tot}/\varepsilon$ and
$T_{sgs}=k_{sgs}/\varepsilon=\alpha T_{tot}$, and define a total eddy viscosity as 
$\nu_{tot}=Ck_{tot}T_{tot}$. Then the possible combinations of filtered
products would be approximated as
\begin{equation}
\bar{u'_i \phi^<}\approx{}-\alpha\nu_{tot}
(\pd_i\la{\phi}\ra+\pd_i{\phi^>}),
\label{upphil}
\end{equation}
\begin{equation}
  \bar{u^<_i \phi^\prime}\approx{}-\alpha\nu_{tot} \pd_i\la{\phi}\ra,
  \label{ulphip}
\end{equation}
and
\begin{equation}
\bar{u^<_i \phi^<}\approx{}-\alpha^2\nu_{tot}
(\pd_i\la{\phi}\ra+\pd_i{\phi^>}).
\label{ulphil}
\end{equation}
To apply these scaling arguments, we rearrange
the unresolved stress contributions in~\eqref{eqn:tau_sgs} as
\begin{equation}
- \tau^{sgs}_{ij}
=
\bar{u'_i u^<_j} + \bar{u^<_i u'_j}-  \bar{u^<_i u^<_j},
\label{tau_ur}
\end{equation}
The first two terms are transposes of each other so their sum is
symmetric, as is of course the third term. However, a straight-forward
application (\ref{upphil}--\ref{ulphil}) with $\phi=u_j$
does not yield a symmetric result for $\tau^{sgs}$. Instead, we
take the forms (\ref{upphil}--\ref{ulphil}) as guidance and symmetrize the
velocity gradients that appear,
\begin{align}
  \bar{u'_i u^<_j} + \bar{u^<_i u'_j}-\tfrac{2}{3}\bar{u'_k u^<_k}\delta_{ij}&\approx -2\alpha\nu_{tot}(2\la
  S_{ij}\ra + S_{ij}^>) \\
  \bar{u^<_i u^<_j}-\tfrac{1}{3}\bar{u^<_k u^<_k}\delta_{ij}&\approx -2\alpha^2\nu_{tot}(\la S_{ij}\ra +
  S_{ij}^>).
\end{align}
Gathering like terms we obtain 
\begin{equation}
\tau^{sgs}_{ij} -\tfrac{1}{3}\tau^{sgs}_{kk}\delta_{ij} = 2\nu_{s}\la{}S_{ij}\ra + 2\nu_{e}S^>_{ij},
\label{tau_ur2}
\end{equation}
where $\nu_s=\alpha(2-\alpha)\nu_{tot}$ and
$\nu_e=\alpha(1-\alpha)\nu_{tot}$.  Thus, from applying simple eddy
viscosity arguments to the total subgrid stress tensor, we see that
the model-split form arises naturally.  Further, this analysis indicates 
the contribution to the subgrid stress from the mean and
fluctuating gradients do not scale in the same way, \emph{i.e.} the
form $\tau^{sgs}_{ij}-\tfrac{1}{3}\tau^{sgs}_{kk}\delta_{ij}=2\nu_t\bar{S}_{ij}$ is not generally valid.
%
%\todo{RDM: in (6-8), these are $\nu_e$ and $\nu_s$ (lower case  subscripts), make consistent}
In expectation, (\ref{tau_ur2}) is simply
\begin{equation}
\la \tau^{sgs}_{ij} -\tfrac{1}{3}\tau^{sgs}_{kk}\delta_{ij}\ra = 2\nu_{s}\la{}S_{ij}\ra,
\label{tau_ur3}
\end{equation}
and the contribution of the model (\ref{eqn:mean_tausgs}) to the
Reynolds stress depends only on the mean.  On the other hand, the energy transfer provided by the model 
(\ref{eqn:mean_res2un}) arises entirely from the correlation of $\la{}S^>_{ij}S^>_{ij}\ra$, 
scaled by $2\nu_{e}$.

It is quite apparent why this mixed scaling issue has been overlooked.  For RANS, it is not 
an issue as all scales are the same.  Subgrid stress models were patterned 
off RANS and adopted identical forms using length scales from the filter width, 
timescales from the inverse of the filtered velocity gradient, and the filtered 
velocity gradient itself.  Therefore, such models are only directly relevant to the 
subgrid-subgrid term of (\ref{eqn:tau}), \emph{i.e.} $u^<_i u^<_j$.  Further, for filters 
and projections which minimize the $L^2$ error of  $||u^>-u^\prime||$, the 
mixed terms are zero as $u^>_i$ and $u^<_i$ live in orthogonal subspaces of the full space of Navier-Stokes 
solutions.  Many numerical method do not satisfy this requirement.  For instance, the 
second-order finite volume method used in this work (more details in \S\ref{sec:results}) most closely 
approximates box filtering in physical space. With such common methods, the decomposed velocity fluctuations 
are not uncorrelated and the above arguments are relevant.  This raises another practical 
issue.  The above arguments made use of $\alpha=\la{}u^<_iu^<_i\ra/\la{}u^\prime_ju^\prime_j\ra$.
In a simulation, we only have access to the information of the resolved field and an 
approximation of $\la{}u^\prime_ju^\prime_j\ra$ provided by the RANS model.  Therefore, 
we directly have an approximation of only $\beta = 1-\la{}u^>_iu^>_i\ra/\la{}u^\prime_ju^\prime_j\ra$ or $\tau_{kk}/\la{}u^\prime_ju^\prime_j\ra$.  
When the mixed terms are non-zero, $\alpha$ and $\beta$ are not equivalent.  We must then find a relationship 
between $\alpha$ and $\beta$ for the particular numerics of the code in use.  
\begin{figure}[th!]
\begin{center}
\subfigure[ Relation of subgrid energy definitions]
{\includegraphics[width=0.46\linewidth]{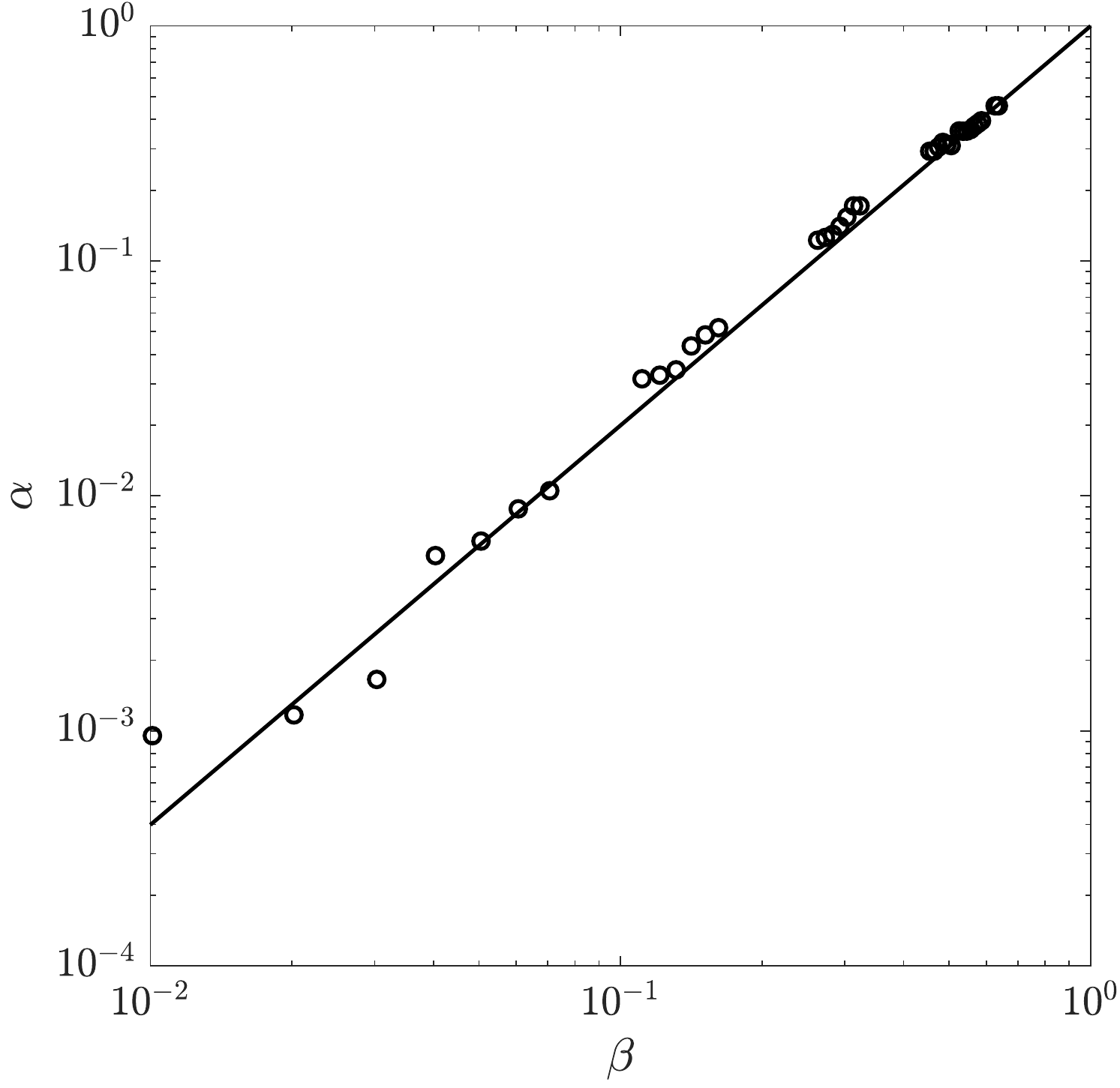}
\label{two_alpha}}
\subfigure[Subgrid shear stress scaling]{
\includegraphics[width=0.47\linewidth]{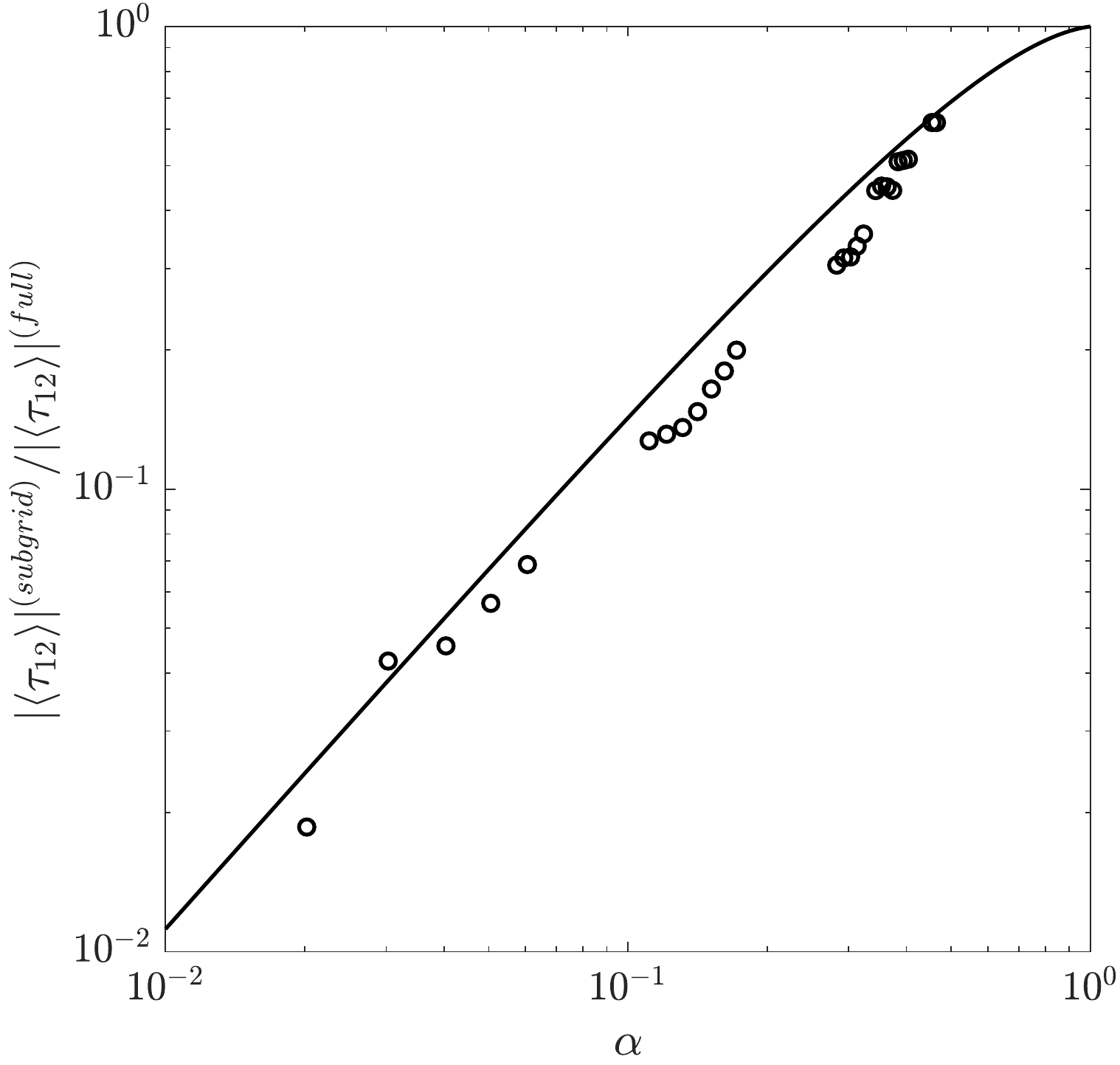}
\label{total_alpha}}
\end{center}
\caption{Subgrid information determined by applying box filters to
  channel flow DNS data at $Re_\tau=1000$\cite{LeeMoser2015}, (a)
  relationship between
  $\alpha=\la{}u^<_iu^<_i\ra/\la{}u^\prime_ju^\prime_j\ra$ and
  $\beta=1-\la{}u^>_iu^>_i\ra/\la{}u^\prime_ju^\prime_j\ra$ overlaid
  with $\alpha=\beta^{1.7}$ and (b) ratio of total subgrid shear
  stress (including mixed terms) and total shear stress, overlaid with
  $\alpha(2-\alpha)$.}
\label{fig:filtered_chan}
\end{figure}

\respa{In isotropic turbulence at infinite Reynolds number, the
  Kolmogorov inertial range expression for the two-point correlation
  leads to $\alpha\propto\beta$, with the proportionality constant
  depending on the details of the filter. However, we need to use this
  relationship for wall-bounded shear flows, and so here we estimate
  $\alpha$ as a function of $\beta$ by applying box filters of varying
  size to channel flow fields at $Re_\tau=1000$ \cite{LeeMoser2015},
  excluding the region $y^+<300$ since it is wall and viscous
  dominated. Averages were computed over the homogeneous spatial
  directions and six snapshots in time separated by two channel flow
  throughs. The values of $\alpha$ and $\beta$ determined at various
  $y$ locations and for six different filters sizes were aggregated to
  determine a single relationship between $\alpha$ and $\beta$. In
  Fig. \ref{fig:filtered_chan}, to reduce statistical noise, the data
  were binned by $\alpha$ value to obtain average $\beta$'s.
%
%A total of six fields are used, each separated by two channel flow
%throughs, along with six different filter stencils.  Filtered boxes
%with data below $y^+=300$ are discarded to eliminate viscous effects.
%In this way, a total of 1.5K data points are collected and binned in
%Fig. \ref{fig:filtered_chan}.
  We find a simple relation of $\alpha=\beta^{1.7}$ as shown in Figure
  \ref{two_alpha}. Clearly, given the difference between the results
  for isotropic turbulence and channel flow, the relationship between
  $\alpha$ and $\beta$ depends on characteristics of the turbulence.
  It will be useful to generalize this relationship to reflect that,
  but for now, we have used the relationship inferred from the channel
  flow DNS.}%\todo{RDM: is this OK}\todo{swh: looks good to me}

Of interest to the general eddy viscosity arguments presented here, we can also
examine the total subgrid stress scaling.  Figure \ref{total_alpha}
shows this relation for the box filtered channel shear stress data.
The $\alpha(2-\alpha)$ scaling determined from the arguments above is
consistent with the data and is used in the models discussed here,
though $\alpha^{9/8}(2-\alpha)$ fits the data slightly better.

\section{Active Model-Split HRL}
\label{sec:ms}

%\textcolor{blue}{
%{\color{blue}
The hybridization strategy proposed here is motivated by deficiencies
of previous methods, as described in~\S\ref{sec:motivation}.  To
overcome these problems, as described in~\S\ref{sec:gencon}, the total
model stress is split into two separately modeled components, and an
appropriate scaling of the mixed stress terms is incorporated.  Further, a
stirring force is used to introduce resolved fluctuations where
they are capable of being resolved.  In this ``active model-split'' (AMS)
hybridization, the resolved momentum equation takes the following
form:
\begin{equation*}
D_t\overbar{u}_i
= 
-\frac{1}{\rho}\partial_i{}\bar{p}
+
\nu\partial_j\partial_j\overbar{u}_i
+
\partial_j \left( \tau^s_{ij} + \tau^e_{ij} \right) +F_i,
\label{ms}
\end{equation*}
where $\overbar{u}_i$ is the resolved velocity, $\overbar{p}$ is the
resolved pressure, $\tau_{ij}^s$ is the ``mean'' subgrid stress,
$\tau_{ij}^e$ is the ``fluctuating'' subgrid stress which results in
energy transfer from the resolved to unresolved scales, and $F_i$ is
the forcing.  The forcing term subsumes all filter-width commutation errors.The models for $\tau^s_{ij}$ and $\tau^e_{ij}$ will be formulated
in terms of ``pseudo-mean'' quantities (thus the quotes above),
which are described in the following subsection before the detailed
description of the hybrid formulation.

%and simply note $\la{}\la{}\cdot\ra\ra\ne{}\la{}\cdot\ra$ for now. 

\begin{comment}
nd turbulent
velocity fluctuations ($u'_i$) are divided into resolved ($u^>_i$) and
discarded ($u^<_i$) components; \emph{i.e.}, the total resolved velocity field
is $\bar{u}_i = \la{}u\ra_i+u^>_i$.  The resolved momentum flux is
$(u^>_i\la{}u\ra_j+\la{}u\ra_iu^>_j+u^>_iu^>_j)$ so that the average turbulent stress to be
modeled is $\tau_{ij}^{SGRS}\approx\la{}u'_iu'_j\ra-\la{}u^>_iu^>_j\ra$.

However, when coupled with existing RANS models, scaling the Reynolds
stress anisotropy by $\alpha^2$ was found to perform worse than
$\alpha$.  

This behavior is attributed to the fact that the scalar
quantities, e.g., $k$ and $\epsilon$, in RANS models are only loosely
representative of the corresponding true quantities. RANS models
function by a series of error-canceling and coefficient tuning to
arrive at nearly the correct mean stress provided the mean
gradients; i.e., they get the ``correct'' eddy viscosity and not
the correct individual scalar contributions to the eddy viscosity.
Thus, it is possible that the theoretically correct scaling for any
given flow and level of resolved turbulence will not perform well in a
hybrid simulation built upon existing inadequate models.  Because of
this, the $\alpha$ scaling is used here, but improving, or removing
the necessity for, this scaling will be the subject of future works.

\end{comment}

\subsection{The Pseudo-Mean} \label{sec:pseudoMean}
The arguments presented in \S\ref{sec:gencon} have tacitly assumed
knowledge of the expected value for all required quantities. In
particular, $\la{}u_i\ra$ is used to define $u^>_i$ and then $\alpha$ and $\beta$
through the resolved TKE as $2k_{res}=\la u^>_iu^>_i\ra$.  Additional
averaged terms will appear in~\S\ref{sec:stress}
and~\S\ref{sec:forcing}.  Yet,
this information is not readily available in LES or HRL, necessitating
a method for obtaining mean quantities.  The natural
assumption would be that the method for approximating the mean should
be as close to the true expected value as possible, perhaps even
warranting solution of a separate set of RANS mean velocity equations
as in \cite{xiao:2017}. However, it was determined emperically that using the
true mean in the AMS model formulation yields poor simulation
results. The difficulty arises because of the use of an eddy viscosity
model to represent the mixed term $\overline{u_i^<u'_j}$
(\ref{tau_ur2}), which represents the transport of the unresolved
fluctuations, $u^<_i$, by the total fluctuations, $u'_j$. The total
fluctuations include resolved fluctuations with much larger length and
time scales than the unresolved scales. Interactions between these
very large resolved scales and the unresolved scales are poorly
described by dissipative processes, as the dominant effect is the
large-scale advection of the unresolved scales. That is, these slowly
varying and relatively energetic fluctuations behave more like an
unsteady mean in the context of an eddy viscosity model (EVM). For this reason, using a
pseudo-mean that filters out all but the largest and slowest scales of
motion when defining the resolved fluctuations leads to a more
accurate representation of the mean stress and a more consistent
definition of $\alpha$ for the purpose of stress modeling.

\begin{comment}
However, we have empirically 
determined that the ``mean'' operator used to decompose $\bar{u}_i$ must 
actually be a low-pass filter which permits low frequency fluctuations.  That is, 
if $\la{}\cdot\ra$ is our psuedo-mean and $\{\cdot\}$ the true mean, there are 
some fluctuations, $\hat{u}_i=\la{}u\ra_i - \{u\}_i$, which must be excluded 
from calculation of the $\alpha$ used to scale the stress as in \S\ref{}.  
While counterintuitive at first, this finding is consistent with the splitting and scaling
arguments being based on directly accounting for the mixed stress terms, particularly 
(\ref{tau_ur2}).  Indeed, it is (\ref{tau_ur2}) which provides the entirety of stress when 
using eddy viscosities which do not fluctuate.  This term represents the transport 
of the unresolved fluctuations, $u^<_i$, by the total fluctuations, $u'_i$.  The 
action of the low-frequency portion of $u'_i$, \emph{i.e.} $\hat{u}_i$, behaves 
much more like pure advection of a passive scalar than gradient diffusion.  As such, 
reducing the stress through their inclusion in $\alpha$ with eddy viscosity models 
is not appropriate.  That is, these slowly varying and relatively energetic fluctuations 
behave more like an unsteady mean in the context of LEVM.  Simply allowing these 
resolved fluctuations to act solely through the convection term has been found 
to yield optimal results.
\end{comment}

Anticipating use of the AMS formulation in complex geometries in which
the turbulence is inhomogeneous, time averaging is used to define the
pseudo-mean, rather than spatial averaging. For stationary flows \respb{and
flows with slowly evolving means} a
causal time average with an exponentially decaying kernel should be
sufficient, and this approach is used here. In this case, the
pseudo-mean $\{\phi\}$ of a quantity $\phi$ evolves according to
\begin{equation}
d_t\lpm\phi\rpm = \frac{1}{T_{ave}}\big{(}\bar{\phi}-\lpm{}\phi\rpm{}\big{)},
\label{ave}
\end{equation}
%\todo{RDM: it seems that we need a seperate nomenclature for pseudo-mean and pseudo-fluctuations. Maybe $\{u\}$ and $u^\succ$ and $u''=u^\succ+u^<$}
% swh: probably, but lets skip for this paper?
where the time constant is proportional to the large-scale turbulent
time scale, or $T_{ave} = C_\phi{}k_{tot} / \varepsilon$.  For all
averaged quantities $C_\phi=1$, with the exception of $\mathcal{P}_k$,
the production of turbulent kinetic energy, due to special
circumstances as described in \S\ref{sec:stress}. The performance of the model
appears somewhat insensitive to the precise value of $C_\phi$, but
no attempt has been made to optimize its value or to determine whether
it needs to vary depending on flow characteristics. Using (\ref{ave}) with $C_\phi$ of
order one, eliminates all but the lowest frequency fluctuations from
the definition of $\lpm\phi\rpm$.

%\subsection{Subgrid (Reynolds) Stress Model}
\subsection{Subgrid Mean Stress Model}
\label{sec:stress}
Following the argument presented in \S\ref{sec:genLES}, the deviatoric subgrid contribution to
the unresolved Reynolds stress tensor is assumed to scale with 
$\tau_{ij}^{s}\approx\alpha(2-\alpha)\tau^{R}_{ij}$
where $\alpha$ is computed from the psuedo-mean resolved turbulent
kinetic energy as
$\alpha=k_{sgs}/k_{tot}=\beta^{1.7}$, $\beta=(1-k_{res}/k_{tot})$, $k_{res}=\la{}u^>_iu^>_i\ra/2$, and 
 $\tau_{ij}^{R}$ denotes the
Reynolds stress determined by a RANS closure evaluated using only
pseudo-mean quantities, to be consistent with the arguments outlined
in \S\ref{sec:transport}.  
%Note that the extra averaging on the production 
%in the RANS closure models does not mean the RANS scalars are relative 
%to the longer average.  That is, the fluctuating terms in the RANS production 
%are relative to the psuedo-mean, the longer average on $\mathcal{P}_k$ si
%RDM: seems that this should be part of 3.2
\begin{comment}
Such a scaling is a
simplification of the actual behavior of the subgrid stress as a
function of resolved kinetic energy.  The true
scaling depends on the proximity of the resolution length scale to the
integral and Kolmogorov length scale.  For example, the similarity analysis
of Lumley \cite{luml:1967}, as discussed in \cite{Jimenez2000},
indicates that for weak simple shear the off-diagonal components of the
subgrid Reynolds stress scale like $\alpha^2$ for resolution
in the inertial range. Despite this discrepancy, the simplification of
scaling the entire subgrid stress with this function of $\alpha$
performs well (see \S\ref{}).
\end{comment}

Due to the presence of $\alpha$ in the subgrid stress model described
above, it is necessary that $k_{tot}$ be known. This can be obtained
from the $k$-equation of any two-equation RANS model, provided the
resulting $k$ is intended to be representative of
$k_{tot}$.  For example ``$k$'' in the SST model \cite{ment:1994} does not
represent the turbulent kinetic energy in attached boundary layers, but rather is more akin to
$\tfrac{3}{2}\overbar{v^2}$ where $\overbar{v^2}$ is the wall-normal
component of the Reynolds stress tensor.  Use of this ``$k$'' would be inconsistent with the current evaluation of the resolved TKE and the AMS approach would perform poorly near walls.  Recall that the velocity
fluctuations $u'_i=u_i-\lpm u_i\rpm$ is defined in terms of
$\lpm\cdot\rpm$, which is essentially a low-pass temporal filter
rather than the expected value.  Therefore, $u_i^\prime$ excludes 
the low frequency fluctuations $u_i^\gg=\lpm u_i\rpm-\la u_i\ra$. 
For consistency in the definition of $\alpha$, $k_{tot}$ is defined as
$\la{}u'_iu'_i\ra/2$, \emph{i.e.} excluding $u_i^\gg$ contributions. In determining a model equation for $k_{tot}$,
we assume that cross terms that would be zero if $\lpm\cdot\rpm$ were the
true expected value are negligible and arrive at the usual 
%\todo{SWH: dont follow why we are bringing this up, its not really
%relevant for the equation for the expected value of $k$ based on
%$\lpm{}\cdot\rpm$.  The $k_{tot}$ we calculate is just a subfilter
%$k$ (not the subgrid $k$, of course) when all of  $u_i^\gg$ is
%resolved (and something between this and RANS when it is not)}.
$k$ equation, with the exception of the convection and
production terms ($\cC_k$ and $\cP_k$ respectively), which are
\begin{equation}
\cC_k=\la u_i\ra \frac{\partial k_{tot}}{\partial x_i}\qquad
\cP_k=-\left\la u'_iu_j'\lpm S_{ij}\rpm\right\ra
\label{eq:kProducion}
\end{equation}
Thus, the modeled $k$-equation from a RANS model can be used, with the
exception of the production and convection terms. In the production
term, we use the model for $\tau^s_{ij}$ to account for the unresolved
portion of $u_i'u_j'$ and approximate the expected value by the
pseudo mean (\ref{ave}) with $C_\phi=4$ (signified as
$\lppm\cdot\rppm$). This yields 
\begin{equation}
\mathcal{P}_k=\lppm (\tau^s_{ij} - u^>_iu^>_j)\lpm S_{ij}\rpm\rppm,
\label{Pk}
\end{equation}
For simplicity $\la u_i\ra$ in the convection term is approximated as
$\lpm{}u_i\rpm$, though approximating it as $\lppm u_i\rppm$ would be
more consistent. This simplification did not appear to make a
significant difference.

\subsection{Energy Transfer Model} \label{sec:energyXfer}
As discussed in \S\ref{sec:motivateMS}, when $\tau_{ij}^{s}$ is not fluctuating, 
the transfer of energy from resolved to unresolved turbulence is entirely due
to $\tau_{ij}^{e}$.  Traditional LES SGS models primarily function to model this 
transfer of energy with the majority of the Reynolds stress assumed to 
be resolved \cite{Jimenez2000}.  Accordingly, one can use typical LES SGS 
model forms for $\tau_{ij}^{e}$ in addition to the scaling argument-based model 
presented in \S\ref{sec:genLES}.  In principle, any standard SGS model can be 
applied.  However, for typical eddy viscosity model forms (e.g. Smagorinsky 
and variants), $\tau_{ij}^{e}$ will contribute to the mean stress due to 
non-vanishing correlations between fluctuations in the model viscosity and the 
velocity gradient tensor. In this case, these contributions to the mean stress
must be accounted for in the subgrid Reynolds stress formulation described 
in \S\ref{sec:genLES}.

On the other hand, if the eddy viscosity does not fluctuate or if its
fluctuations are uncorrelated with the fluctuating velocity gradient,
then $\la \tau_{ij}^{e} \ra = 0$ so that $\tau^{s}_{ij}$ and $\tau^{e}_{ij}$ are
the mean and fluctuating parts of the model stress. The AMS
formulation can thus be considered a technique to segregate the mean and
fluctuating model stresses so they can be modeled differently,
consistent with their differing roles of representing mean momentum
transport and the transfer of energy from resolved to unresolved turbulence.

\begin{comment}
, perfectly delineating the roles of the
$\tau^s_{ij}$ and $\tau^e_{ij}$ as presented in \S\ref{}.  If the
$\tau_{ij}^e$ model is selected so that the SGET eddy viscosity is
uncorrelated with the fluctuating gradient, the AMS approach can be
considered as a method of segregating the mean and fluctuating
portions of the modeled stress and isolating their eddy-viscosity
dependencies, \emph{i.e.} the total model is decomposed into the mean
deviatoric stress ($\tau_{ij}^{R}=\la{}\tau-\frac{2}{3}k_{sgs}{I}\ra$)
and the fluctuating portion ($\tau_{ij}^e=\tau - \tau_{ij}^{R}$).
\end{comment}

Here, we use the tensor eddy viscosity M43 model, which is formulated
to account for anisotropy of the LES resolution \cite{haer:2019b}. Its
coefficients were determined for consistency with second-order finite
volume numerics as described in Appendix A of \cite{haer:2019b}, with
the resulting coefficients provided in Appendix B here.  Details of
the model formulation are provided in \cite{haer:2019b} but a
modification is required here.  As derived, M43 is an algebraic
subgrid model which dissipates at the anisotropic resolution
scales. It is formulated in terms of the mean dissipation rate, which
in \cite{haer:2019b} was prescribed \emph{a priori} from knowledge of
the forcing energy injection rate in simple forced homogeneous
isotropic turbulence.  Here, the mean dissipation is determined from
the dissipation equation in the RANS model. In this case, the M43
model becomes a two-equation ($k$ and $\varepsilon$) tensor eddy
viscosity model.  Where a non-negligible portion of the total dissipation 
is resolved, \emph{i.e.} where $\varepsilon\gg{}\nu\la{}\pd_ju^>_i\pd_ju^>_i\ra$ 
is not true, the dissipation used in the M43 model should be reduced 
by the resolved dissipation.  This modification would correctly observe 
DNS limits.  However, when resolutions are relatively coarse, as is the case 
with the simulations performed here, the resolved dissipation may be 
neglected.

\subsection{Active Forcing}
\label{sec:forcing}

%{\color{blue}
As discussed in \S\ref{sec:consistency}, resolved turbulence can be passively
generated only by creating regions of modeled-stress depletion.  Thus,
relying on natural self-generation of turbulence fluctuations
inherently requires reducing the fidelity of the model. When the
resolved flow includes turbulent fluctuations the energy cascade will
naturally populate additional finer scales of motion as they can be
resolved, if the modeled dissipation is also reduced.  However, the
near-RANS regions will not self-generate finer turbulent scales
without explicitly disrupting the mean stress. A high fidelity hybrid
formulation therefore requires a mechanism for generating turbulent
fluctuations, and here we propose a fluctuating body force (active
forcing) driven by a resolution metric.

An active forcing formulation requires three ingredients: 1)
identification of regions where more turbulence can be resolved, 2)
determination of the rate at which resolved fluctuations should be
introduced, and 3) specification of the structure of the generated
velocity fluctuations.  Where finer velocity fluctuations can be
resolved, they should be introduced with the length scale of the
smallest resolved turbulence so as not to disrupt existing larger, and
presumably accurate, structures.  Conversely, in under-resolved
regions, $\tau_{ij}^{e}$ will naturally dissipate resolved energy.
However, due to grid and flow inhomogeneity, $\tau_{ij}^{e}$ may not
affect this transfer fast enough.  Here, the forcing scheme is only
formulated to add energy to regions that can resolve more turbulence.
A forcing formulation capable of actively removing energy in
under-resolved regions would also be useful.  Currently, a
modification to $\tau_{ij}^{e}$, discussed in Appendix \ref{app:M43},
specifically~\eqref{M43_hack}, partially alleviates the issue of
under-resolution.  Each of the three necessary forcing
components is presented next.

%In such instances, forcing that more strongly damps
%the resolved turbulence would be appropriate.  However, in this work, the
%forcing scheme is only designed to generate turbulent fluctuations
%when that is needed.  
%which is a common requirement of hybrid models.  

\subsubsection{Resolution evaluation}
Identifying regions where forcing is needed requires evaluating the
grid's ability to resolve more, or any, of the local turbulent
fluctuations. Historically, scalar measures of grid resolution (e.g.,
cell diagonal or volume cube-root) have been compared to a scalar
turbulent length scale to make this assessment.  With anisotropic grids
and/or turbulence, such measures are incomplete indicators and
generally insufficient.  Instead, the evaluation should be based on
the locally least-resolved orientation to ensure adequate resolution
in all directions. Naturally, this measure must depend on both the
grid and the flow state.

First, the resolution capacity of the grid is described by the
resolution tensor, $\mathcal{M}_{ij}$, \cite{syng:1949} which
characterizes resolution anisotropy.  The eigenvalues,
$\lambda_i^\mathcal{M}$, of $\mathcal{M}_{ij}$ are the resolution
scale in $i^{\rm th}$ principal directions, while its eigenvectors define those
directions. Common grid measures are invariants or
eigenvalues of $\mathcal{M}_{ij}$ (\emph{e.g.}
$\delta_{vol}=(det(\mathcal{M}_{ij}))^{1/3}$). Defining a
resolution metric based on $\mathcal{M}_{ij}$ provides a tensorially
consistent mechanism to account for resolution anisotropy.

%\todo{swh: note to self, make sure all the signs are consistent with modeled and true taus}
%\todo{swh: looks like I started editing a section in progress, commenting the in-progress part for noe}

Second, the size and anisotropy of the largest unresolved turbulence scales,
which are also the smallest resolved turbulence scales, can be
characterized by the production of unresolved turbulence and
unresolved kinetic energy. This is analogous to the expression of the
integral scale as proportional to $k^{3/2}/\varepsilon$ in stationary, homogeneous, isotropic
turbulence, since then the dissipation, $\varepsilon$, is also the rate of
production. To represent the
anisotropy of the unresolved length scale, we adopt the ansatz that
the tensor representing the inverse unresolved length scale $\cL^{-1}$ is given by
\begin{align}
\cL^{-1}_{ij}&=\cP^{sgs}_{ijkl} \cR^{-3/2}_{kl}\label{LPsgs}\\
\noalign{where}
\cP^{sgs}_{ijkl} &= \tfrac{1}{2}\big{(}\tau_{ik} \partial_j\bar{u}_l + \tau_{jl} \partial_i\bar{u}_k\big{)}\label{Psgs}\\
\cR_{ij}&= \frac12\lpm u^<_iu^<_j\rpm\\
\tau_{ij}&=\tau^{s}_{ij}+\tau^{e}_{ij}+\tfrac{2}{3}\beta{}k_{tot}\delta_{ij}\label{tau_sgs}
\end{align}
In (\ref{tau_sgs}), the non-deviatoric part, $\frac23\beta
k_{tot}\delta_{ij}$,
is included because it affects the scale anisotropy
of the production, despite the fact that it does not
contribute to the production of kinetic energy, and is therefore often
ignored. Note that in (\ref{LPsgs}), the indices of $\cR^{-3/2}$ are
contracted with the velocity component indices of $\cP$, so that only
length scale anisotropy is represented in $\cL$.

The ratio $\rho(\be)$ of the grid resolution to the smallest resolved
scale in the direction defined by the unit vector $\be$ is then given
by
\begin{equation}
\rho(\be)=\cL^{-1}_{ij}\cM_{jk}e_ie_k.
\end{equation}
Clearly adequate representation of the resolved fluctuations requires
that the maximum of this ratio over all directions be less than some
constant, here taken to be one. Also, if the maximum ratio is
significantly less than one, then smaller scale fluctuations can be
resolved, indicating that forcing can be applied to generate such
fluctuations.

However, the anisotropy of the
unresolved fluctuation covariance $\cR_{ij}$ is commonly not
available. In many RANS models, such  anisotropy is only accounted for near
the wall, often through the use of wall functions. As discussed by
Durbin \cite{durb:1995}, such wall functions represent the effects of near wall
anisotropy, and in the context of the $\overline{v^2}$--$f$ model can
be seen to scale with the ratio of the wall normal velocity variance
($\overline{v^2}$) to the turbulent kinetic energy. Regardless of the
two-equation RANS model being used, we can thus take the
$\overline{v^2}$--$f$ eddy viscosity relation as an expression for
$\overline{v^2}$ and determine the anisotropy measure $\zeta$ as:
\begin{equation}
\zeta=\frac{3\overline{v^2}}{2k}=C_\zeta\frac{\nu_t}{kT}
\end{equation}
where $C_\zeta=7.5$, as determined from the coefficients of the
$\overline{v^2}$--$f$ model, and $T=k/\varepsilon$ is a
turbulent time scale. Since $\overline{v^2}$ is an estimate of the
minimum eigenvalue of the Reynolds stress, and because the unresolved
velocity covariance is expected to be less anisotropic than the
Reynolds stress, the following inequality should hold:
\begin{equation}
\max_{\be}\rho(\be)<r_\cM=C_r(\zeta\beta{}k_{tot})^{-3/2}\max(\cP^{sgs}_{ilkk}\cM_{lj})
\label{rM}
\end{equation}
where the maximum on the right hand side refers to the maximum
eigenvalue of its tensor argument.  The coefficient $C_r$ is related 
to how many grid spacings are required to resolve the structures  
produced at the model cutoff.  With the exception of the study
described in \S\ref{sec:cr_sens} 
we take this coefficient to be unity.  The quantity $r_\cM$ is easily
determined, and (\ref{rM}) implies that using it as a resolution
indicator is conservative in the sense that it will indicate that
scales resolvable on the grid are larger than it can actually support.
This is the resolution indicator used here. In particular, when $\langle
r_\cM\rangle<1$, the grid is locally capable of resolving smaller
scale fluctuations. When the grid resolution is isotropic and the
turbulence is isotropic and in equilibrium, $\langle r_\cM\rangle$
reduces to a length scale comparison similar to that used by DES \cite{spal:1997}.
Thus, it can be viewed as an anisotropic generalization of the DES
resolution metric.

\subsubsection{Forcing structure}
In regions where $\la{}r_{\mathcal{M}}\ra < 1$, the goal of forcing is
to gradually introduce resolved fluctuation, thereby allowing the
forced resolved structures to evolve into actual turbulence without
corrupting the mean.  Uniform forcing over some region where
$\la{}r_{\mathcal{M}}\ra < 1$ is clearly not desirable, as it would
necessarily result in forcing the mean.  Instead, we must prescribe
the spatial structure of the acceleration field.  Ideally, this
artificial structure would retain turbulence characteristics, such as
spatial correlation and intensity, of the largest of the unresolved
turbulence.  As more turbulence is added to the resolved field, the
characteristics of the largest of the unresolved scales necessarily
changes.  The construction of this field as used here is \emph{ad
hoc}. \respa{It is presented here as an example, which was implemented to
assess the utility of the greater AMS HRL framework. Improvements are certainly possible and desirable.}

An artificial turbulence-like vortex field is defined based on the structure of a Taylor-Green (TG) 
vortex field with variable length scale:
\begin{align}
  h_1(x,t) &= A\cos(a_1 x_1^p) \sin(a_2 x_2^p) \sin(a_3 x_3^p), \nonumber \\
  h_2(x,t) &= B\sin(a_1 x_1^p) \cos(a_2 x_2^p) \sin(a_3 x_3^p),  \label{TG} \\
  h_3(x,t) &= C\sin(a_1 x_1^p) \sin(a_2 x_2^p) \cos(a_3 x_3^p), \nonumber
\end{align}
where the magnitudes are somewhat arbitrary and selected such that
$h_ih_i\le{}1$ with $A=1$, $B=-1/3$, and $C=-2/3$.  The desired vortex sizes 
should closely mimic the local length scale at the implicit cutoff and be 
resolvable everywhere in the simulation domain.  Thus, we select 
the vortex scale to be $\ell=min(N_LL_{sgs},\delta_{wall})$, where 
$L_{sgs} = (\beta{}k_{tot})^{3/2} / \varepsilon$, $\delta_{wall}$ 
is the distance to the wall, and $N_L$ is a constant empirically set to $8$. Applying 
$\ell$ to (\ref{TG}), we have
\begin{equation}
a_i=\begin{cases}
\frac{\pi}{D_{(i)}}\textsf{nint}\left(\frac{D_{(i)}}{\min(\ell,D_{(i)})}\right)& \mbox{for $i$ direction periodic}\\
\pi/\ell& \mbox{otherwise}
\end{cases}
\end{equation}
where $D_{(i)}$ is the domain size in periodic directions, ``nint''
is the nearest integer, and subcripts in parenthesis indicate no summation is implied. This form is designed to ensure periodicity of
the $h$'s in periodic directions.
The TG vortex coordinates, $x_i^p$, 
are specified to translate with the mean flow as
\begin{equation}
 x_i^p(x,t) = x_i - \lpm u_i \rpm{}t
\end{equation}
%\todo{RDM: changed sign on $u_it$ here}
to provide a degree of temporal correlation in addition to the desired 
length scale.  With the prescribed structure in hand, we must now determine 
the appropriate magnitude for the forcing acceleration.

\subsubsection{Forcing magnitude} %\todo{swh: ehh, this is awkward}.
As previously discussed, the magnitude of the forcing acceleration should 
be based on the largest of the unresolved fluctuations.  Additionally, the rate 
of injection of manufactured fluctuations should be specified by 
the eddy turnover time of the smallest resolved structures so that the added 
fluctuations can be ``healed'' into realistic turbulence, as they are added, by 
the resolved turbulence.  Making use of 
near-wall anisotropy, the target forcing is then defined as
\begin{equation}
F_{tar} = C_F\frac{\sqrt{\zeta{}k_{tot}}}{\sqrt{\beta}T},
\label{Ftar}
\end{equation}
where $C_F$ is currently an empirical constant set to 8.  Note the $\sqrt{\beta}$ in the denominator of (\ref{Ftar}) results from both the velocity and timescale corresponding to the subgrid scales.  With the target 
acceleration magnitude (\ref{Ftar}) and structure (\ref{TG}) in hand, additional 
modifications are necessary to respect RANS and DNS limits.  The RANS 
limit will be naturally detected by $\la r_\mathcal{M} \ra$ while 
identifying near-DNS conditions requires an approximation of the local Kolmogorov length 
scale.  With the information provided by the RANS model, this limit is approximated as 
\begin{equation}
\beta_{kol}\approx\frac{3}{2}\frac{(\nu\varepsilon)^{1/2}}{k_{tot}}. 
\label{bkol}
\end{equation}
Limiters are incorporated in a scaling coefficient, $\eta$, as
\begin{equation}
%\eta = \min(F_r-D_{lim}-R_{lim},0)
\eta = F_r-D_{lim}
\end{equation}
where $D_{lim}$ enforces the DNS limit and $F_r$ responds to the
resolution evaluation and can thus be considered both a grid-resolved
LES and RANS limiter.  In the vicinity of either of these three
limits, the scaling of $h_i$ is attenuated to prevent sharp
transitions between full and no forcing.  Gradual attenuation is
prescribed in the following \emph{ad hoc} functional forms.

Overall forcing behavior is controlled through $F_r$ as
\begin{equation}
F_r = -\tanh(1 -  \min(\la r_\mathcal{M} \ra,1)^{-1/2}),
\end{equation}
When $\la r_\mathcal{M} \ra<1$, forcing is activated with $F_r\to1$.  In the vicinity 
of $\la r_\mathcal{M} \ra\approx1$, $F_r$ smoothly goes to zero.  Finally, in RANS 
and under-resolved regions, $\la r_\mathcal{M} \ra>1$ and so $F_r=0$.  The DNS limit 
is enforced with
\begin{equation}
D_{lim}=F_r(\tanh(10\hat{\beta})+1) ,
\label{Dlim}
\end{equation}
\begin{equation}
\hat{\beta} = (1-\beta)/(1-\beta_{kol})-1.
\end{equation}
%and finally
%\begin{equation}
%R_{lim}=\alpha\max(F_r,0),
%\end{equation}
%is used to prevent forcing in RANS limits.

\respa{Specifically, as DNS resolution is approached, $\beta\to\beta_{kol}$,
$\hat{\beta} \to 0$, and $D_{lim}\to{}F_r$.  Thus, $\eta$ also goes to
zero, and the forcing effectively ``shuts off'' in the limit of DNS
resolution.  Alternatively, when $\beta\gg{}\beta_{kol}$,
$D_{lim} \approx 0$ and the limiter does not affect the forcing.  Note
that (\ref{Ftar}) is poorly behaved in the limit of $\beta\to0$.  To
avoid this, one could replace $\beta$ in~\eqref{Ftar} with
$\max(\beta, \beta_{kol})$.}
%This modification would serve to only
%prevent undefined code behavior and would not affect the results as
%the DNS limit will set the forcing to zero.}
%\todo{TAO: Do we need the final three sentences of above revision?  Seems like an irrelevant
%detail, and it isn't so obvious to me it is true (i.e., one needs to
%define the limiting process carefully, right?).}
%\respa{The DNS limit is respected by the ``switch'' portion in parenthesis in (\ref{Dlim}). When $\beta\gg{}\beta_{kol}$, $D_{lim}\to0$ and the limiter does not affect the forcing. In the limit of $\beta\to\beta_{kol}$, the switch goes to unity and $D_{lim}\to{}F_r$, so that $\eta$ also goes to zero. This effectively ``shuts off'' the forcing in the limit of DNS resolution.  Note that (\ref{Ftar}) is poorly behaved in the limit of $\beta\to0$ and may be clipped with (\ref{bkol}).  This modification would serve to only prevent undefined code behavior and would not affect the results as the DNS limit will set the forcing to zero.}
In the RANS limit, the structure of $h_i$ may be applied everywhere in the 
domain to affect turbulence generation.  However, with existing
turbulent fluctuations, applying a generic acceleration structure,
such as $h_i$, will on average result in both the addition and removal
of energy.  So, the magnitude of $h_i$ must be further modified. This
issue is addressed by testing the resolved production due to $h_i$ as
\begin{equation}
\mathcal{P}^{test}_F = h_iu^>_i,
\label{Ptest}
\end{equation}
which will be used to clip (\ref{TG}).  With these considerations, the forcing 
acceleration vector is 
\begin{equation}
F_i =
\left\{ \begin{array}{ll}
F_{tar}\eta{}h_i & \textrm{ if }\mathcal{P}^{test}_F\ge0\\
0 & \textrm{ otherwise}.
\end{array} \right.
\label{Fi_actual}
\end{equation}
%and zero when $\mathcal{P}^{test}_F$ is negative.

The specification of $F_i$ along with the two model terms
(\S\ref{sec:stress} and~\S\ref{sec:energyXfer}) and the pseudo-mean
(\S\ref{sec:pseudoMean}) closes the AMS approach.  While there are
many advantages of the AMS hybrid formulation, the model-split form
introduces new challenges as well.  Before presenting several tests of
the new approach, we first briefly discuss these issues.

\subsection{Discussion of AMS Formulation}
\label{sec:discussion}

%Being such a drastic deviations from existing HRL, specific implications 
%of the AMS 

\respb{The components of the AMS hybrid RANS/LES approach include:
a RANS mean stress model, a subgrid scale dissipation model, under-utilized
resolution evaluation, active fluctuation forcing, and computation of
the resolved turbulent stress with a causal time average.  The
resolved mean stress is used throughout AMS, most notably to scale the RANS
stress to the appropriate subgrid scale stress.  All of these
components are dynamically and continuously coupled in a single
numerical simulation, on a single grid.}
%continually updated and calculated on a single mesh
%during a simulation and affect each another.}
\respa{There are two critical features of the AMS formulation that are
necessary for good model performance. First, is the explicit
separation of the mean stress and dissipation roles of the subgrid
scale turbulence model. This splitting allows the eddy-viscosity models
to be tailored to each role, enabling the overall AMS formulation to
remain valid regardless of resolution.} \respb{It is essentially an
LES that \emph{by construction} respects both RANS and DNS limits. In
particular, in the limit of DNS, all the model terms ($\tau^s$,
$\tau^e$ and $F$) go to zero, leaving only the numerical approximation
of the Navier-Stokes equations.}
\begin{comment}
 
The essence of 
the method is a combination of 1) the explicit separation of the two
of the roles of the subgrid scale turbulence model, and 2) the model
responding to what turbulence \emph{is} resolved as opposed to what
turbulence one \emph{wants} to be resolved.  The splitting of the
model into portions which explicitly account for either the unresolved
stress or the unresolved dissipation allows for eddy-viscosity models
to be tailored to each role, enabling the overall AMS formulation to
remain valid at levels of resolution where either or both roles are
not negligible.}  \respb{The resulting composite model can be thought
of as basically an LES which admits both RANS and DNS limits \emph{by
construction} and remains valid for levels of resolution where the
majority of the turbulent stress is not resolved.}
\end{comment}

\respa{The second critical feature is that the model hybridization
responds only to the turbulence fluctuations that \emph{are} resolved,
rather than the fluctuations that \emph{could be} resolved. This
ensures that the modeled mean stress is always consistent with the
resolved mean stress, which is
important to avoid common problems in HRL, such a model stress
depletion. Many HRL techniques promote the development of fluctuations where they
can be resolved by reducing the modeled stress and thereby promoting
instabilities, but while also disrupting the mean. In AMS, by contrast, resolved
fluctuations are introduced explicitly through forcing when
needed. The details of the forcing as described in
Section~\ref{sec:forcing} are necessarily \emph{ad hoc}, as is the
promotion of instabilities used in other approaches, since fluctuation
information must be
created. Using explicit forcing, however, provides an opportunity to
design the forcing that will allow fluctuations to be introduced as
rapidly as possible while ensuring that they realistically represent
their contributions to the mean stress, as discussed below.}

The model-split hybridization approach has numerous advantages over
traditional HRL methods.  First, following the argument
in \S\ref{sec:ms}, blending, which is responsible for many of the
difficulties in HRL, is obviated.
%RDM: redundant with above
%and the hybrid system
%naturally respects both RANS and DNS limits.
Second, it uses the RANS model as designed.  The turbulence model
state variables represent mean features of the turbulence, and the
governing PDEs depend only on mean, or psuedo-mean, quantities.  Thus,
pathological behaviors of the RANS transport models due to fluctuating
state variables are avoided.  Third, the RANS eddy viscosity appearing
in $\tau_{ij}^{s}$ makes no contribution to the dissipation of
resolved fluctuations, allowing turbulence at all resolvable scales to
form naturally without being overly dissipated.  \respb{The AMS
formulation will also naturally be inactive in laminar flow regions.
The RANS model that is part of the AMS formulation governs this
behavior. When the RANS model predicts zero or negligible TKE as in a
laminar region, the forcing will be zero and the eddy viscosity will
also go to zero.}  Finally, nearly any combination of base mean and
fluctuating models can be used.  Because of this flexibility, advanced
models are easily incorporated and used to treat complex flow and
domain features such as the effects of resolution anisotropy with the
M43 model.  Since stress anisotropies are mostly carried by the
largest of turbulent structures, it is reasonable to expect that even
modestly resolving turbulent kinetic energy will result in significant
improvement over RANS to the representation of total stress
anisotropy. Nonetheless, the AMS approach does present the opportunity
to include treatments for stress anisotropy such as explicit algebraic
Reynolds stress models (EARSM) \cite{mars:2009}.

%\todo{TAO: Ordering of issues seems off to me. SWH: rearranged, feel free to change order or add as you see fit.  Would be good to preempt any potential reviewer criticism here.}
%
However, there are several issues demanding future attention beyond
the scope of this paper.  Perhaps the most glaring is the specifics of
the psuedo-mean and precisely what large scale fluctuations should be
excluded from the $\alpha$ scaling in the model.  This concept, that
some of the interactions with large-scale turbulence should not be
modeled with gradient diffusion, is new to both HRL and LES in
general.  The degree of exclusion of low frequency turbulence from the
model is effectively controlled by the coefficient used in the
psuedo-mean.  Another way to approach the modeled Reynolds stress
would be to adopt the form of \cite{chen:2012} and simply subtract the
deviatoric portion of the mean resolved stress from the RANS model
stress.  This method would sidestep some of the ambiguity in RANS
scalar terms but, as formulated, leads to no improvement over basic
RANS behavior in expectation.  Nonetheless, this should be
explored.  \respb{Further, the use of the causal time average results
in a lagged response to large-scale unsteadiness and is strictly only
valid for stationary flows.  However, similar to the assumptions of
unsteady RANS, if the unsteady timescales are much larger than the
largest turbulence timescales and the averaging time, the averaging
will produce an appropriate pseudo-mean for use in the AMS
formulation.}

The second main issue is the rudimentary form of the forcing.  As 
discussed in \S\ref{sec:forcing} the prescribed structure of the synthetic 
forcing is necessarily \emph{ad hoc}.  The particular method outlined here has 
potential drawbacks.  Foremost, the resulting $F_i$ is not
divergence free.  This is a result of TG fields only being
divergence-free when constructed using a uniform overall scaling and
vortex length scale and without clipping.  In an incompressible solver, the dilatational
portion is projected out, which alters the structure of the effective
forcing.  The result is that the forcing can remove resolved
turbulence at some points in space and time.  However, on average, the
approach does result in net production of resolved fluctuations where desired.  In a
compressible solver, the method may result in spurious acoustic
sources.  Further, it should be possible to force more strongly, \emph{i.e.}
increase the coefficient $C_F$ (currently set to 8), if the forcing
structure produced fluctuations that were more consistent with
unresolved turbulence in equilibrium with the current  
$\overbar{u}_i$.
\respa{The forcing
design described in Section~\ref{sec:forcing} is certainly not the
best that can be done in this regard.}
%
%by making the
%forcing more representative of some acceleration which would produce a
%real state of turbulence with length scales between $L^n_{sgs}$ and
%$L^{n+1}_{sgs}$ over some $\Delta{}t$ for the particular
%$\overbar{u}_i$.  That is, if the forcing were more representative of
%the commutation error $\overbar{D u_i}/Dt - D \overbar{u}_i/Dt$,
%generation of realistic resolved content could be affected more
%efficiently.  There is no reason to expect the augmented TG field to
%be representative of these effects.
Improving forcing structure so that forcing strength can be increased
would allow the size of the LES-resolved region upstream of an area of
interest to be minimized.
%By improving on this structure,
%shorter lengths of LES resolution upstream to areas of interest, such
%as flow separation, will be necessary.
%
% BOB version:  However, while the forcing used here is \emph{ad hoc}, AMS results are not sensitive to the \emph{ad hoc} characteristics of the forcing.
%SWH version: The AMS approach responds to the presence of resolved turbulence while 
%being agnostic regarding how those fluctuations are introduced. 
% Compromise:  
However, while the forcing used here is \emph{ad hoc}, the AMS 
approach is not sensitive to the \emph{ad hoc} characteristics of 
the forcing.  That is, \emph{AMS responds to the presence of resolved 
turbulence and not to how the turbulence is
introduced}. \respa{As a result, any new or improved forcing technique
can easily be substituted into AMS.  Indeed, we regard the current forcing as a ``place holder'' to close the overall AMS modeling framework.}

\begin{comment}
RDM: I don't think we need all this.
\respa{ 
While some forcing is essential to AMS for this reason, the precise
structure of the forcing only affects the rate at which resolved
turbulence can be introduced during a simulation and the precise
structure of the forcing is \emph{not} essential to AMS.  The specific
type of forcing can be interchanged without any alterations to the AMS
method formulation. }

\respa{Another issue related to forcing is the use of a single local length scale as opposed to forcing a broad range of structures from the large-scale down to the grid scale at the onset of forcing. Broad spectrum forcing may lead to faster transition from RANS to grid-resolved LES.  Gradual addition of resolved turbulence is due to the fact that we are fabricating turbulent fluctuations at the local LES-scale.  Such manufactured turbulence must be ``healed'' by the resolved field to achieve some physically meaning state (in the sense of momentum transport and dissipation).  The use of a single length scale, which gradually decreases, was used here with the expectation that this approach would minimally affect the resolved large scales.  However, this is likely overly conservative.  Examining how many wavenumbers can be affected without disrupting the resolved turbulence will be the subject of future work and is beyond the scope of this introductory paper.}
\end{comment}

% transport of resolved turbulence
The RANS models used for $\tau^s$ are improved relative to their use
in a purely RANS computation through the 
inclusion of the resolved stress in the production term.
%and, presumably, an improved 
%mean convective velocity.
However, for the sake of simplicity, we have not analogously included
an explicitly computed resolved contribution $\pd_j\la
u^>_iu^>_iu_j^>\ra$ to the turbulent transport of $k$ (\emph{i.e.}
$\pd_j\la u^\prime_iu^\prime_iu_j^\prime\ra$), relying instead on the
standard eddy viscosity model to represent the entire term. This
simplification can lead to local inconsistency between the resolved
and modeled TKE, which we have observed as small regions of negative
$\beta$ where the total TKE is small. Such negative values were clipped to the
Kolmogorov microscale value of $\beta$, given by $\beta_{kol}$ (\ref{bkol}).  
%\todo{RDM: what does this mean?  SWH: fixed?}  
The results of Section~\ref{sec:results} suggest that, at least for
the flows considered, this treatment is adequate. Yet we are
essentially discarding useful information by not computing $\pd_j\la
u^>_iu^>_iu_j^>\ra$ explicitly, and so we expect that doing so would
improve the veracity and robustness of the model.
\begin{comment} 
this treatment is incomplete in that it 
does not consider the resolved turbulent transport in favor of the basic 
eddy viscosity model, \emph{i.e.} the resolved portion, $\pd_j\la u^>_iu^>_iu_j^>\ra$, 
of $\pd_j\la u^\prime_iu^\prime_iu_j^\prime\ra$ in the $k$-equation is not explicitly 
represented by $\pd_j(\nu_t\pd_j{}k)$.  This simplification can lead to local 
inconsistency between the resolved turbulence levels and the modeled total TKE.  
In fact, we have observed that small regions of negative $\beta$ can result in regions 
of small total TKE.  Here, we have simply clipped these value at 
the Kolmogorov $\beta$.  As we will show in the next section, this method 
seems to be acceptable for the relatively simple cases examined.  However, 
we are essentially discarding useful information which we are already paying 
for in terms of increased computational expense.  Thus, the 
expected value of the resolved triple-correlation should be included in the 
RANS models with an appropriate reduction in the modeled term.
\end{comment}

Finally, while we have touted model-flexibility as a benefit, we must
also caution that different RANS models may require different stress
scalings and a different resolution-adequacy coefficient.  This is an
unfortunate side-effect of how RANS models function through a series
of error cancellations to arrive at a reasonable eddy viscosity for a
given mean flow.  That is, the scalar terms which adopt the names of
``turbulent kinetic energy'', ``turbulent dissipation'', \emph{etc.}
are not, strictly speaking, models for those terms.  In particular the
$k$ from one model may be significantly different from the $k$ from
another.

In the model-split hybridization approach, this model form can be used for
$\tau^{SGET}$ with the necessary mean dissipation provided by the RANS
model.  This choice is used in the example in~\S\ref{sec:results}.

%\textcolor{red}{\emph{Move this bit elsewhere...}}
\subsection{Similarities with Existing Models}
Though the proposed hybrid structure is unique in total with particularly novel components, it does share 
qualities with existing methods and, in the case of DES, expanded on existing concepts.  The anisotropic resolution adequacy 
parameter collapses to the lengthscale comparison driving all DES methods under 
the conditions of isotropic resolution and equilibrium/isotropic turbulence.  
The DHRL method \cite{bhus:2012} similarly divides the total modeled 
stress into RANS-based and LES-based models but retains dependance 
on the total resolved velocity field for each contribution while introducing 
blending between each term based on their resulting production.   The 
dual-mesh hybrid methods of \cite{jenn:2017} explicitly solves a RANS 
and LES set of equations on disparate grids and also uses time-averaging 
to enforce consistency between the resolved and modeled fields by 
modification of the RANS stress with the average of the resolved Reynolds 
stress present in the LES equations.  Such an approach does provide 
and improved mean but it is unclear how it used to improve the turbulence 
modeling beyond the changes to mean terms.
}%\color
\end{comment}

\section{AMS Performance Tests}
\label{sec:results}
The proposed AMS hybrid formulation is evaluated using an
implementation in a branch of the  
finite volume incompressible Navier Stokes solver CDP v2.4 \cite{ham:2004,you:2008},
developed at the Stanford Center for Turbulence Research.  The solver
is $2^{nd}$ order accurate in time and space 
with no upwinding used for convective fluxes in the momentum equation and time 
advancement performed with Crank-Nicolson.  The convective term is linearized 
with an Adams-Bashforth prediction of the convective velocity.  Two base RANS 
models are considered: Chien's $k$-$\varepsilon$ \cite{chie:1982} and the 
``code-friendly'' version \cite{lien:2001} of the $\overbar{v^2}$-$f$ RANS 
model~\cite{durb:1995}.  Details of these models are provided in appendix~\ref{app:rans}.
The energy transfer model is the M43 model 
\cite{haer:2019b}, which is specified in appendix~\ref{app:M43}.  

\respb{The results presented here include AMS simulations with varying
resolutions. However, these should not be confused with convergence
studies in the usual sense of numerical analysis. Indeed, even the
finest resolutions reported here are very coarse by LES
standards, with significant contributions of unresolved scales to the
mean stresses, and there are always RANS regions near the walls. This
is the resolution regime in which hybrid methods are intended to
operate and standard SGS models are known to fail. Within this regime,
we consider a sequence of AMS models, with different resolutions, to
investigate sensitivities to the resolution. Models with finer
resolution will resolve more of the turbulent fluctuations, relying
less on RANS, and so one would generally expect them to yield more
accurate simulations. However, this may not always be true for all
quantities. We are most concerned that the mean velocity be accurately
predicted, since its prediction is the most common objective for HRL
in practice, with the expectation that HRL will improve on RANS
predictions. For a HRL formulation to be useful it is important that
its results not exhibit extreme sensitivities to the resolution, and
that relevant features of the solution, particularly the mean,
improve or at least not degrade as resolution increases.
It is in this context that we examine AMS model results at different
resolutions in the following subsections.}

\subsection{Periodic Channel}
\label{sec:fdc}

%\subsubsection{Fully-developed channel} \label{sec:channel_cdp}
To demonstrate the potential of the proposed hybrid modeling formulation, fully-developed,
incompressible, turbulent channel flow at $Re_\tau\approx5200$ is
simulated.  DNS data is available for this case~\cite{LeeMoser2015}
allowing detailed evaluation of the results.  It is common for RANS models 
to perform very well for channel flow.  Therefore, successful hybrid 
simulations for this case would simply not degrade RANS mean velocity 
profiles.  This may seem like a rather modest goal,  however,
the relatively coarse levels of LES resolution used here place the simulations 
firmly in a regime where existing SGS models will fail.

Results are presented in the following three subsections with focus on 1) 
hybrid steady-state with varying resolutions, 2) hybrid state evolution in 
time, and 3) hybrid state evolution in space.  The term ``hybrid state'' is 
used to indicate the level of resolved turbulence which, in general, will 
not be at a grid-resolved LES levels and will vary in both time and space.  
We emphasize that evaluation of HRL in the transition from RANS to 
the statistically stationary hybrid state, is absolutely critical for an HRL.  While all but one of the 
flow scenarios considered in this paper employ periodic boundary conditions, in practice, a HRL 
will be required to transition in space from a RANS, even if only at inlet
boundaries, to a hybrid LES-RANS state.  If this transition behavior 
is corrupted by the HRL formulation, the downstream solution may be 
corrupted.

\subsubsection{Stationary channel with varying resolution}
\label{sec:stationary}
\begin{figure}[th!]
\begin{center}
%
%\vspace{-0.5cm}
\begin{overpic}
[trim={5cm 51.25cm 5cm 51cm},clip=true,width=1.0\linewidth]{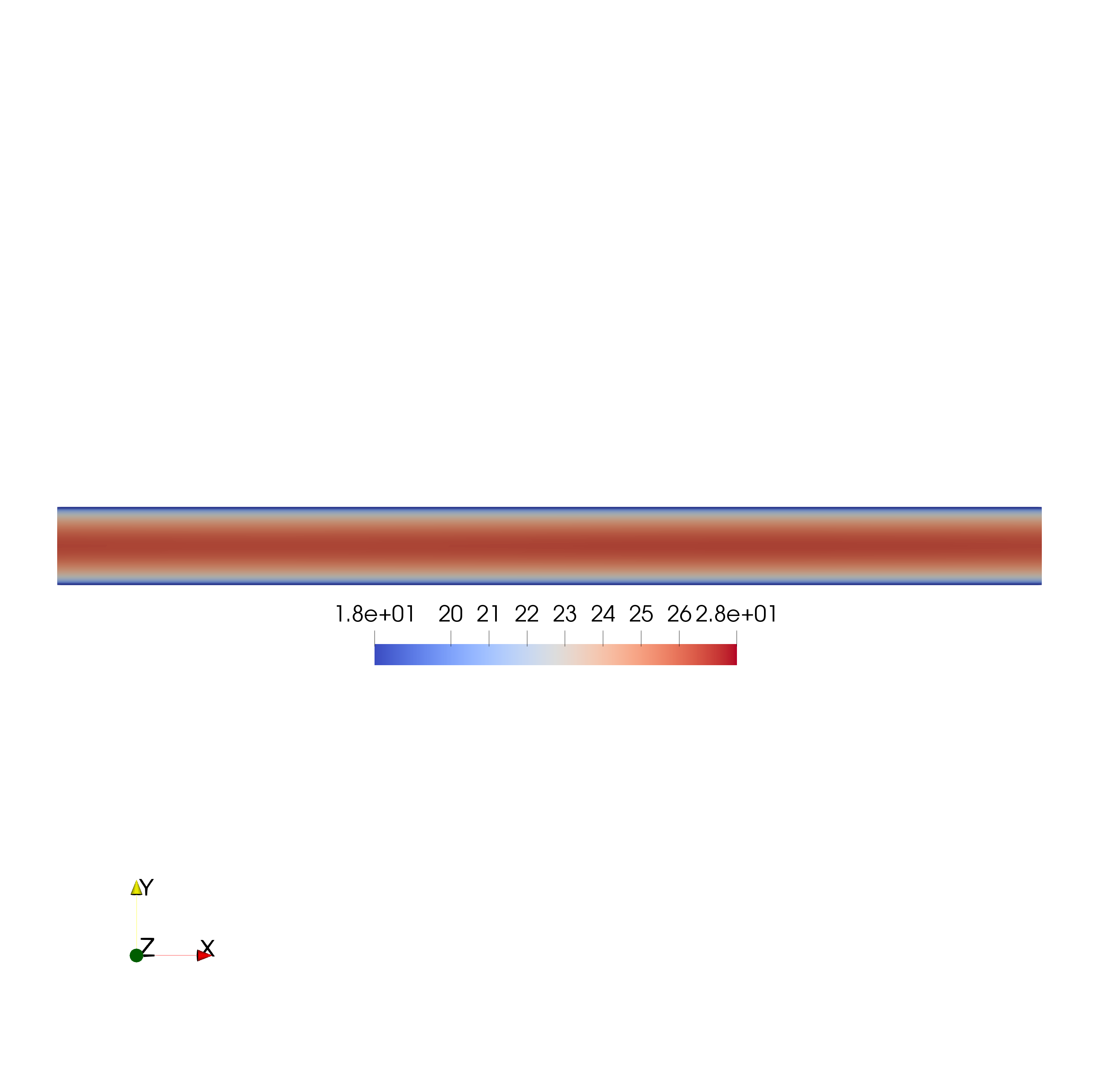}
%\put(7.0,8.0){\color{white}\rule{1.3pt}{2pt}}
\put(1.5,1.5){\color{white}\rotatebox{0}{RANS}}
%\put(-8.0,18.0){\color{black}\rotatebox{0}{$t_f$}}
\end{overpic}
%
%\vspace{-0.5cm}
\begin{overpic}
[trim={5cm 51.255cm 5cm 51.25cm},clip=true,width=1.0\linewidth]{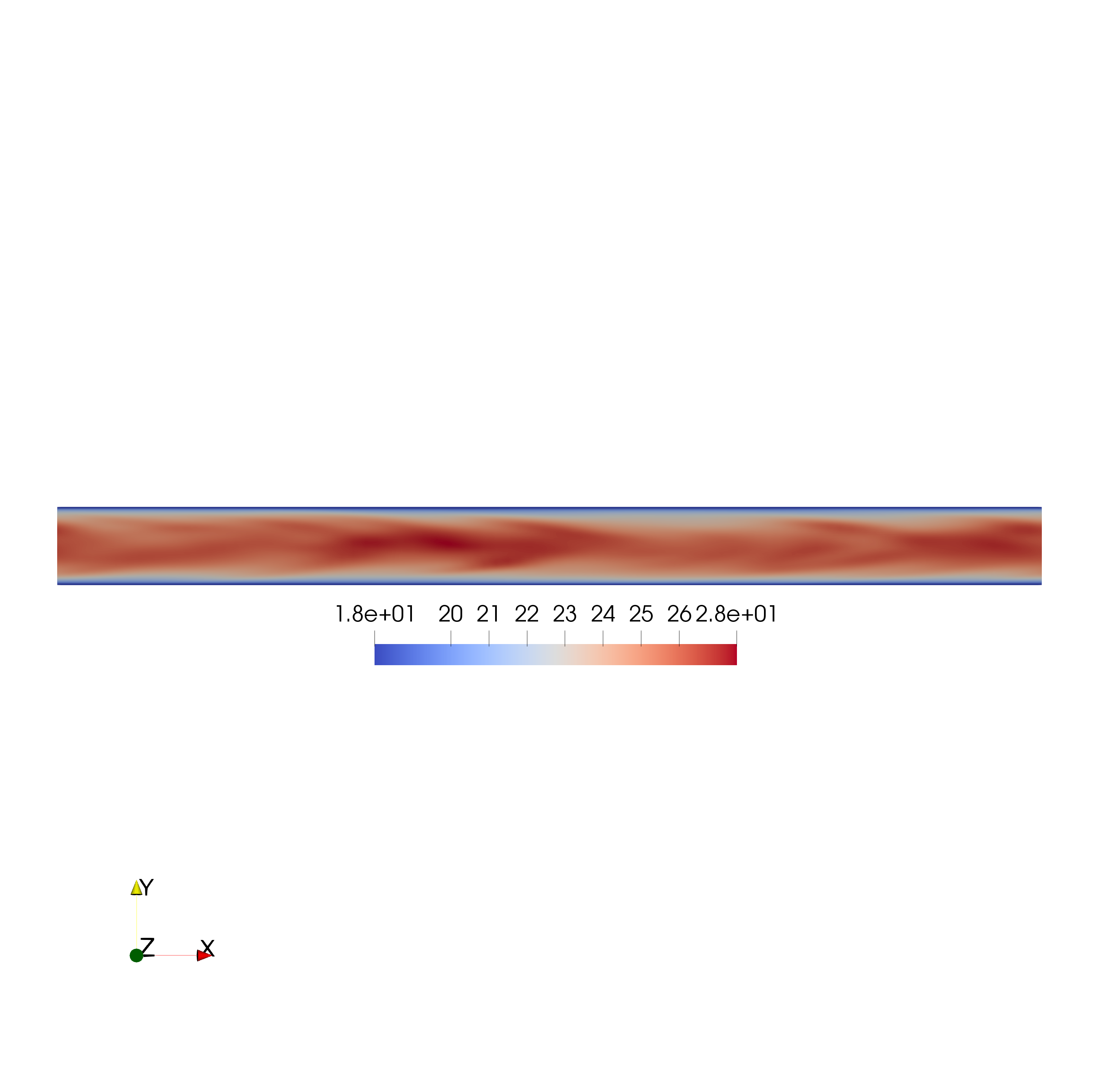}
%\put(7.0,8.0){\color{white}\rule{1.3pt}{2pt}}
\put(1.5,1.5){\color{white}\rotatebox{0}{Extra Coarse}}
%\put(-8.0,18.0){\color{black}\rotatebox{0}{$t_f$}}
\end{overpic}
%
%\vspace{-0.5cm}
\begin{overpic}
[trim={5cm 51.25cm 5cm 51.25cm},clip=true,width=1.0\linewidth]{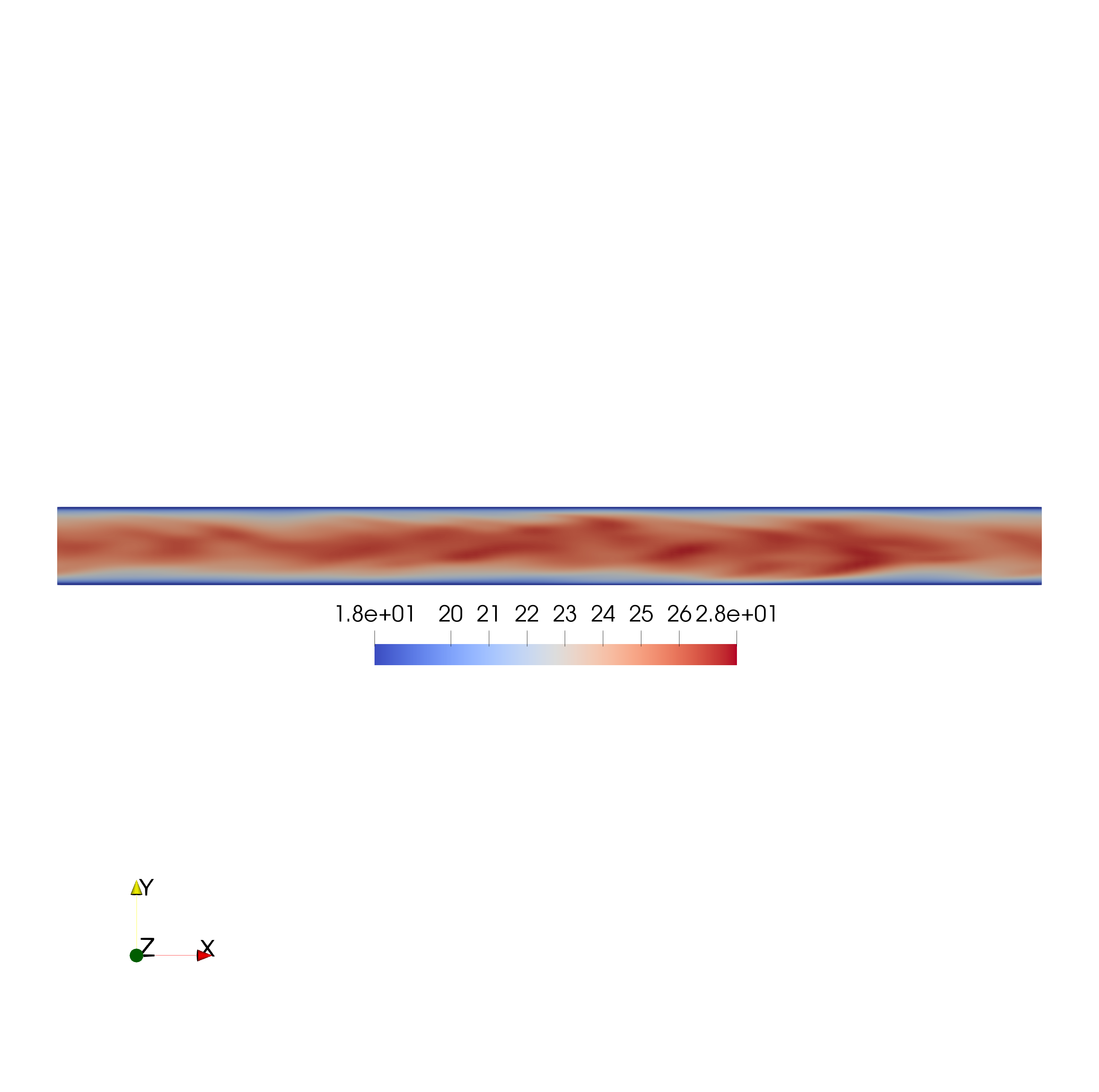}
%\put(7.0,8.0){\color{white}\rule{4.2pt}{2pt}}
\put(1.5,1.5){\color{white}\rotatebox{0}{Coarse}}
\end{overpic}
%
%\vspace{-0.5cm}
\begin{overpic}
[trim={5cm 51.25cm 5cm 51.25cm},clip=true,width=1.0\linewidth]{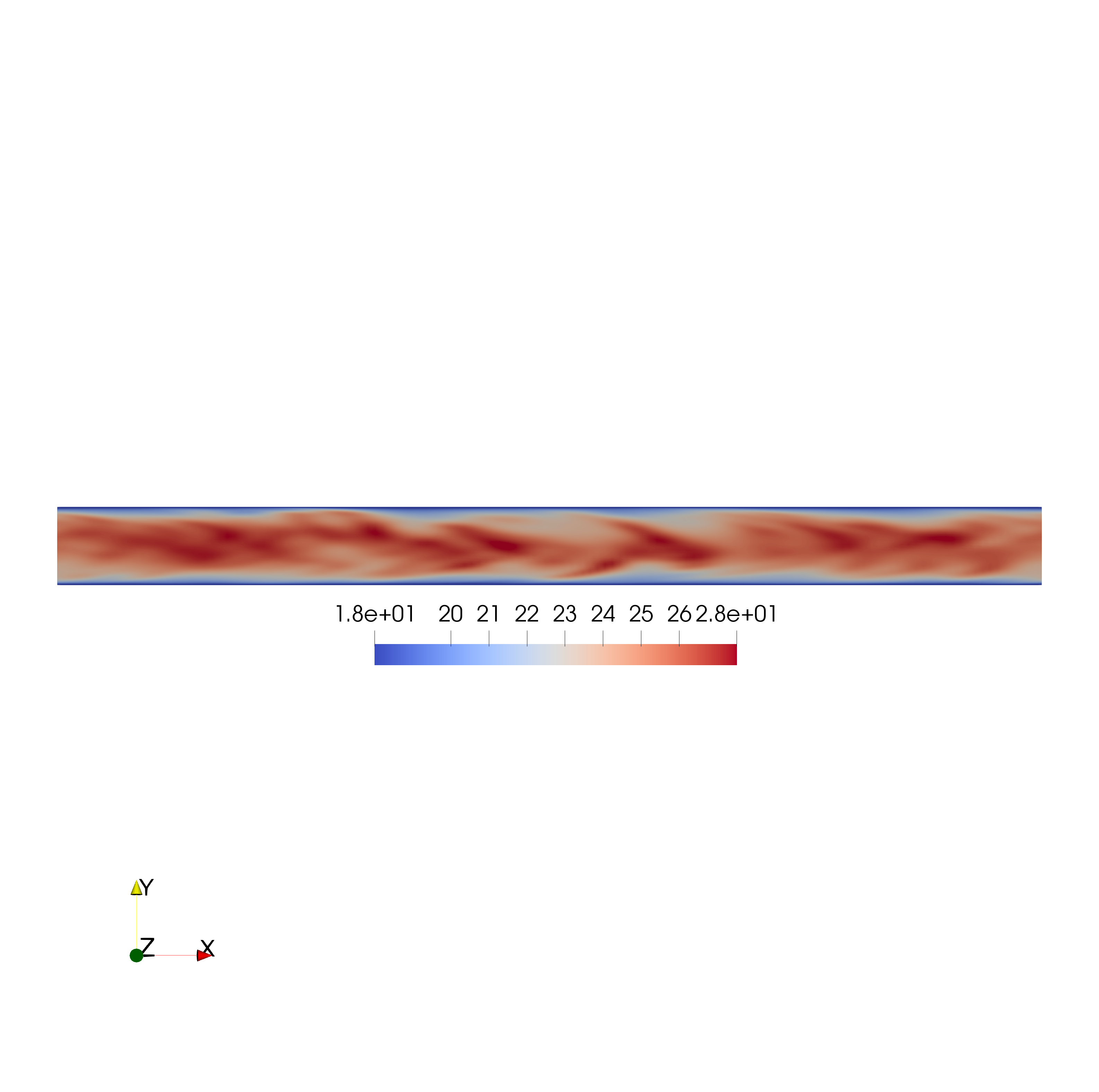}
%\put(7.0,8.0){\color{white}\rule{7.7pt}{2pt}}
\put(1.5,1.5){\color{white}\rotatebox{0}{Medium}}
\end{overpic}
\vspace{-0.5cm}
\begin{overpic}
[trim={5cm 45.0cm 5cm 51.25cm},clip=true,width=1.0\linewidth]{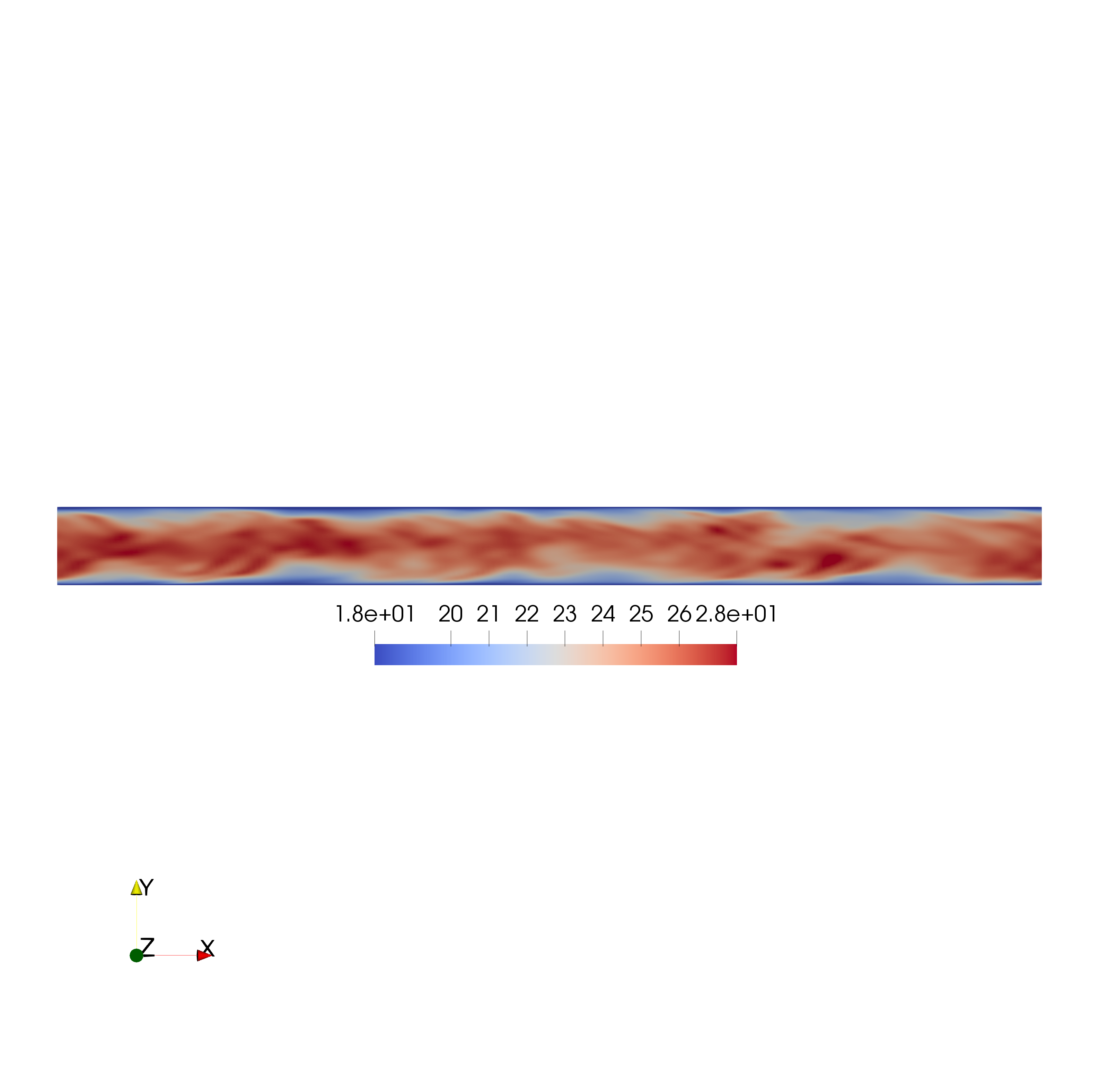}
%\put(7.0,8.0){\color{white}\rule{15.7pt}{2pt}}
\put(1.5,7.5){\color{white}\rotatebox{0}{Fine}}
\put(28.0,2.0){\crule[white]{8cm}{0.5cm}}
\put(31.4,2.5){\color{black}\rotatebox{0}{18}}
\put(39.0,2.5){\color{black}\rotatebox{0}{20}}
\put(46.6,2.5){\color{black}\rotatebox{0}{22}}
\put(54.2,2.5){\color{black}\rotatebox{0}{24}}
\put(61.8,2.5){\color{black}\rotatebox{0}{26}}
\put(67.5,2.5){\color{black}\rotatebox{0}{28}}
\end{overpic} % 18 to 28
\end{center}
\caption{Instantaneous streamwise velocity in an $x$--$y$ plane in fully developed channel
flow at $Re_\tau\approx5200$ using the Chien $k$-$\varepsilon$ RANS
model and the AMS hybrid formulation, with resolutions described in Table 1.\label{fig:ux_cont}}
\end{figure}

The domain for the hybrid simulation is $8\pi \delta \times
3\pi\delta \times 2\delta$, where $\delta$ is the channel
half-width; \emph{i.e.} identical to the DNS domain
in \cite{LeeMoser2015}.  The base RANS model used is Chien's
$k$-$\varepsilon$ \cite{chie:1982} in which $u_\tau$ is
specified \emph{a priori} for use in the wall functions based on DNS.
A spatially uniform streamwise body force (mean pressure gradient) was
applied throughout the channel, which varied in time to maintain a
constant bulk velocity.  Multiple grid resolutions are considered as
shown in Table \ref{tab:grids}.  For all resolutions, the wall-normal
grid spacing is fixed with $N_y=110$,
$\Delta_y^+(\mbox{wall})\approx1$, and
$\Delta_y^+(\mbox{center})\approx345$ while the spanwise and
streamwise spacing ranges from approximately $700-1500$ wall units.
\begin{table}
\centering
\begin{tabular}{ c c c c c c }
\hline
$Case$  & $N_x$ & $N_z$ & $\Delta_x^+$ & $\Delta_z^+$ & $Reduction$ \\
\hline
\hline
Fine & 186 & 70 & 701 & 698 & 84K \\
Medium & 134  & 50 & 973 & 978 & 164K \\
Coarse & 102  & 39  & 1278 & 1253 &  276K \\
Extra Coarse &  84 & 32 & 1552 & 1528 & 408K \\
\hline
\end{tabular}
\caption{Resolutions used in $Re_\tau\approx5200$, 
$8 \pi\delta \times 3\pi\delta \times 2\delta$ periodic channel AMS
simulations reported here.  For all simulations $N_y=110$ with
$\Delta_y^+(\mbox{wall})\approx1$ and
$\Delta_y^+(\mbox{center})\approx345$.  Grid size reductions, in the
last column are reported as the ratio of the
DNS \cite{LeeMoser2015} grid size to that of the AMS simulations.\label{tab:grids}}
\end{table}
\respb{Note that the finest resolution considered here is still some
70 times too coarse to approach the DNS limit for even
$Re_\tau=180$ \cite{LeeMoser2015}.}
%RDM: don't need this
%This is, of course, due to the spectral-distribution of energy in
%turbulent flows and serves to highlight the computational expense of
%demonstrating DNS limits in a hybrid RANS/LES.}
Results are obtained using $\tau^s_{ij}$ scaling of $\alpha (2-\alpha)$ 
as described in \S\ref{sec:genLES}.  
The hybrid simulation is initialized from a steady-state RANS solution.  
Snapshots of the resolved streamwise velocity fields are shown in Fig. 
\ref{fig:ux_cont} to illustrate the difference in resolved turbulence for each 
resolution case.
\begin{figure}[th!]
\begin{center}
\subfigure[Mean $\bar{u}^+_x$ and $\beta$]
{\includegraphics[trim={1.9cm 0cm 2cm 0cm},clip=true,width=0.45\linewidth]{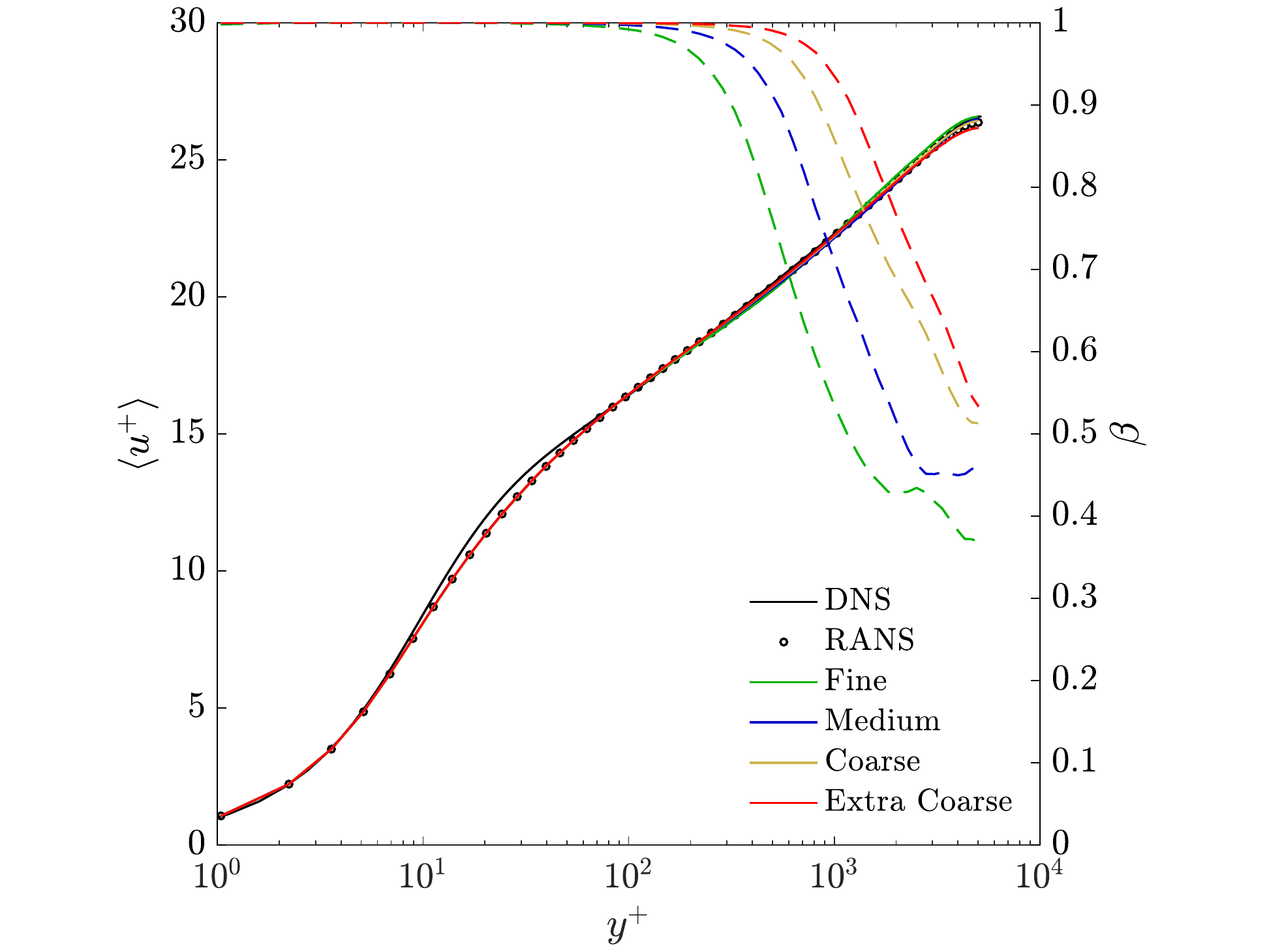}}
\label{mux}
\subfigure[Turbulent kinetic energy $k^+$]{
\includegraphics[trim={1.9cm 0cm 2cm 0cm},clip=true,width=0.45\linewidth]{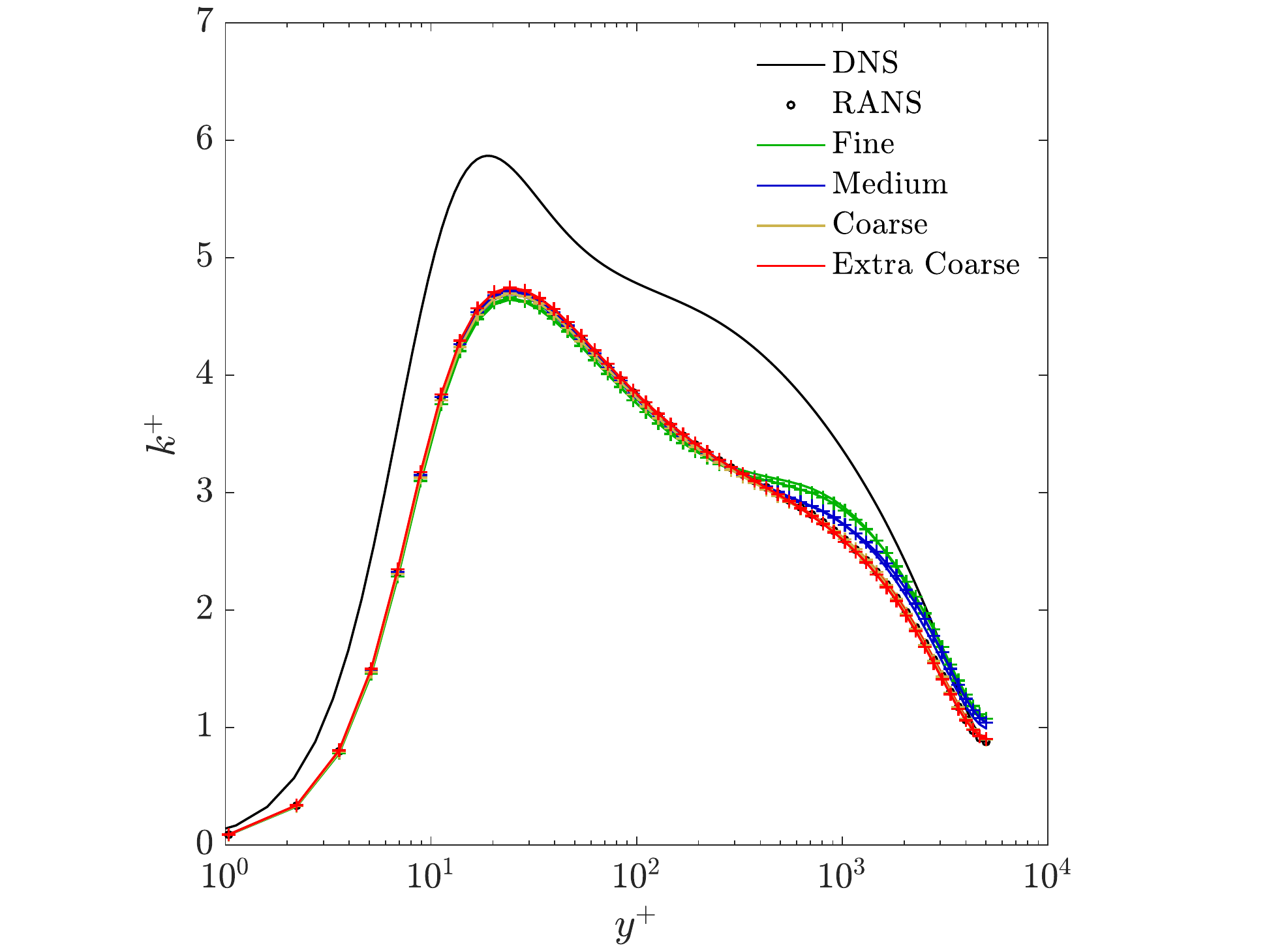}
\label{mtke}}
\end{center}
\caption{
%\todo[inline]{Vertical axis labels: $\la\bar{u}^+_x\ra$ ??, $k^+$}
Mean streamwise velocity in wall units (a) for fully
developed channel flow at $Re_\tau\approx5200$ along with the fraction
of unresolved turbulence $\beta$ (dashed), and (b) turbulent
kinetic energy $k^+$ of the statistically steady solution.  In (b),
lines marked with + symbols
are $k$ obtained directly from the RANS model while the
unmarked lines are the time-averaged resolved turbulence plus $\alpha$
times the RANS $k^+$.  Simulations have been run for approximately 80
flow-throughs.\label{fig:ux_res}}
\end{figure}
Statistically stationary mean velocity and resolution level ($\beta$)
profiles are shown in Fig. \ref{fig:ux_res}a.  As expected, the basic
$k$-$\varepsilon$ RANS model does a good job of reproducing the mean
velocity profile with only a slight under-prediction in the buffer
layer.  For all resolutions levels, the AMS hybrid formulation also
adequately predicts the mean flow with a very small deviation from the
RANS and DNS solution in the center of the channel.  As qualitatively
illustrated in Fig. \ref{fig:ux_cont}, and quantitatively shown by
$\beta$ in Fig. \ref{fig:ux_res}, the AMS formulation allows the
hybrid state to evolve to an appropriate level of resolved turbulence,
in the sense that the mean is correctly predicted, for the different
grids in each simulation.  With each increase in resolution in
Table 1, turbulence is resolved further into the log-layer while
introducing no log-layer mismatch.  Naturally, the most resolved
turbulence occurs in the fine case with $\beta$ as low as 0.4 at the
channel center.  For typical LES, $\beta\lesssim0.1$ is expected over
the entire domain for the model to perform well \cite{Jimenez2000}.
The model-splitting formulation presented here retains good
performance well past this threshold, enabling true ``coarse'' LES.
The nearly identical mean profiles for such varied levels of resolved
turbulence is an indication of the robustness of the formulation.

The response of the underlying RANS model to the modified production
(Eq. \ref{Pk}) is shown Fig. \ref{fig:ux_res}b.  Here, $k$ determined
from the RANS transport equation in the hybrid simulations remains
virtually unchanged from that in the pure RANS model, despite the
modified production terms and the use of the pseudo-mean velocity in
the RANS equations.  This confirms that the formulation is successful
in preserving the RANS solution characteristics in the presence of
resolved fluctuations.  Ostensibly, it may seem that the RANS $k$
should move towards the DNS value with increasing resolution; however,
RANS models function through a series or error cancelling.  That is,
they are tuned to produce an appropriate eddy viscosity and not the
correct individual scalar quantities whose name they bare ($k$,
$\varepsilon$, etc.).  Thus, changes to the RANS $k$ might actually
break the underlying RANS behavior.  Both the RANS $k$ and the
turbulent kinetic energy including the resolved fluctuations
($\tfrac{1}{2}\la{}u^>_iu^>_i\ra+\beta{}k$) remain virtually
unchanged.  With increasing resolution, the two measures of TKE do
increase towards the DNS values \respb{for $y^+>300$ due to increased
resolved kinetic energy and resolved production}.  The fact that these
two definitions of TKE are nearly indistinguishable indicates
consistency between the resolved a modeled turbulence.
\respb{However, resolved TKE must ultimately approach the DNS as
$\alpha\to{}0$, that is as all the fluctuations are
resolved. Therefore, the modeled TKE and resolved TKE may separate
with reducing $\alpha$, which begins to be visible in the finest
resolution case in Fig \ref{fig:ux_res}b.  Such a separation is due to deficiencies in 
the underlying RANS model.}

By changing the spanwise and streamwise grid spacing, we are also
varying the resolution anisotropy, from 2:1 cell aspect ratios at the
center of the channel in the fine case to 20:1 at $y^+=400$ in the
extra coarse case.  The converged steady-state mean profiles for these
different degrees of resolution anisotropy indicate that the
resolution adequacy parameter proposed in \S\ref{sec:forcing}
successfully quantifies the turbulence that an anisotropic grid is
capable of resolving in the presence of inhomogeneous mean shear.
Further, the use of the M43 model for the energy transfer portion of
the model, and its scaling with the pseudo-mean $\varepsilon$ obtained
from the RANS transport equations, appears to be valid.

\subsubsection{Channel with temporally evolving hybrid-state}
The results in the previous section are for stationary hybrid states.  In this 
section, we examine the temporal development of the hybrid state for the 
case of the ``coarse'' resolution.  The forcing method presented in 
\S\ref{sec:forcing} was formulated with the goal that that hybrid 
solutions remain valid through transition from RANS to an arbitrary LES 
state.  For stationary flows, the transient from an initial field to the final 
solution is of no consequence.  However, for unsteady flows, transition 
through varying turbulence states in time is of prime concern and may 
continue throughout the duration of a simulation.  The case examined 
here represents an extreme example of this situation, \emph{i.e.} transition 
from no resolved turbulence to a grid-resolved LES.  Such a drastic transition 
would not actually occur in most unsteady simulations. The opposite 
transition from less to more modeled turbulence is also of interest,
but is not considered here.

\begin{figure}[th!]
\begin{center}
% ROW 1
\begin{overpic}
%[trim={5cm 38.5cm 75cm 38.5cm},clip=true,width=0.3\linewidth]{figures/rM_tdev10.png}
[trim={1cm 18cm 55cm 18cm},clip=true,width=0.35\linewidth]{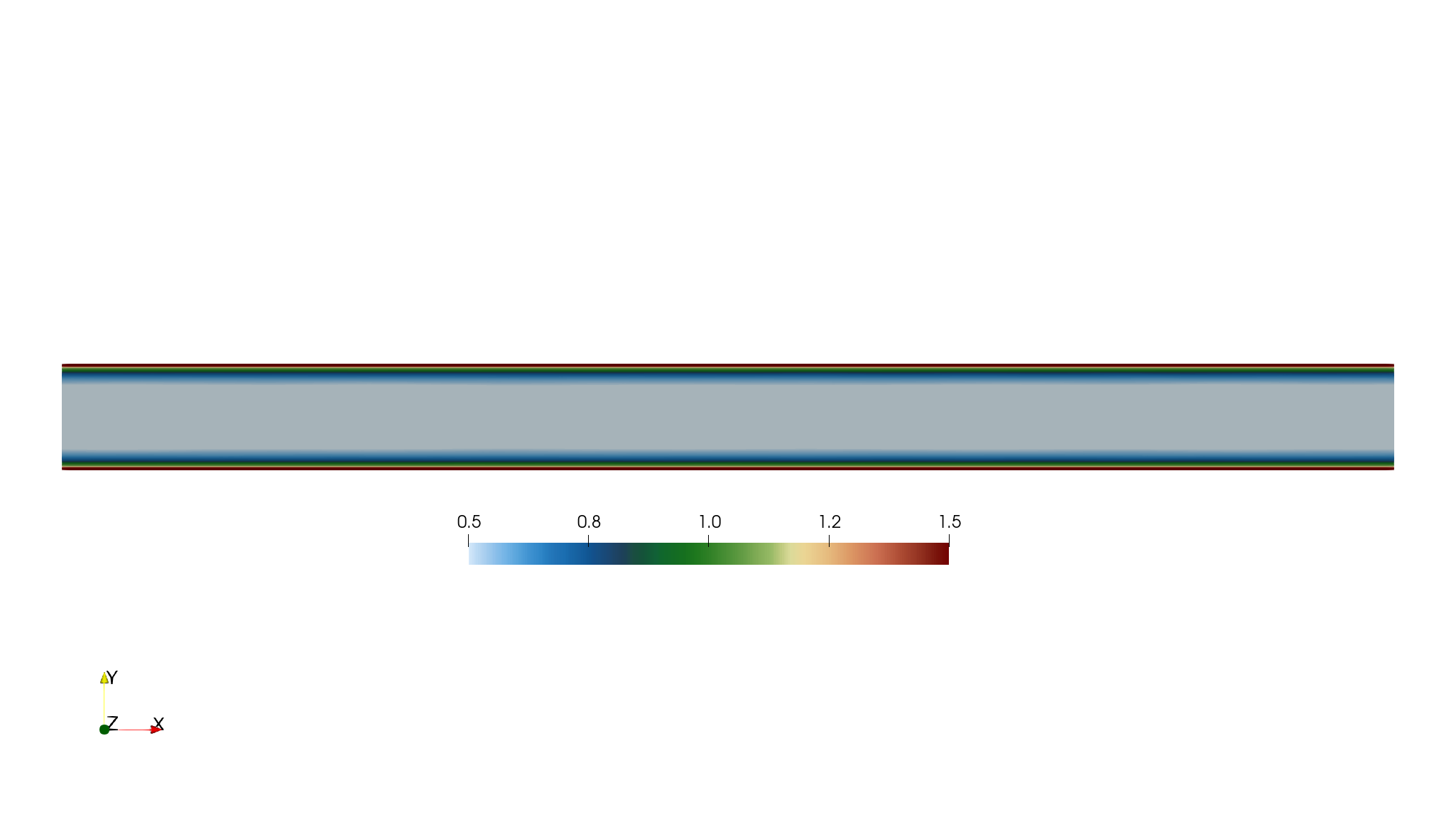}
\put(1.0,12.0){\color{black}\rotatebox{90}{$0.0^+$}}
\put(-1.0,30.0){\color{black}\rotatebox{0}{$t/t_f$}}
\end{overpic}\hspace{-0.7cm}
\includegraphics[trim={1cm 18cm 55cm 18cm},clip=true,width=0.35\linewidth]{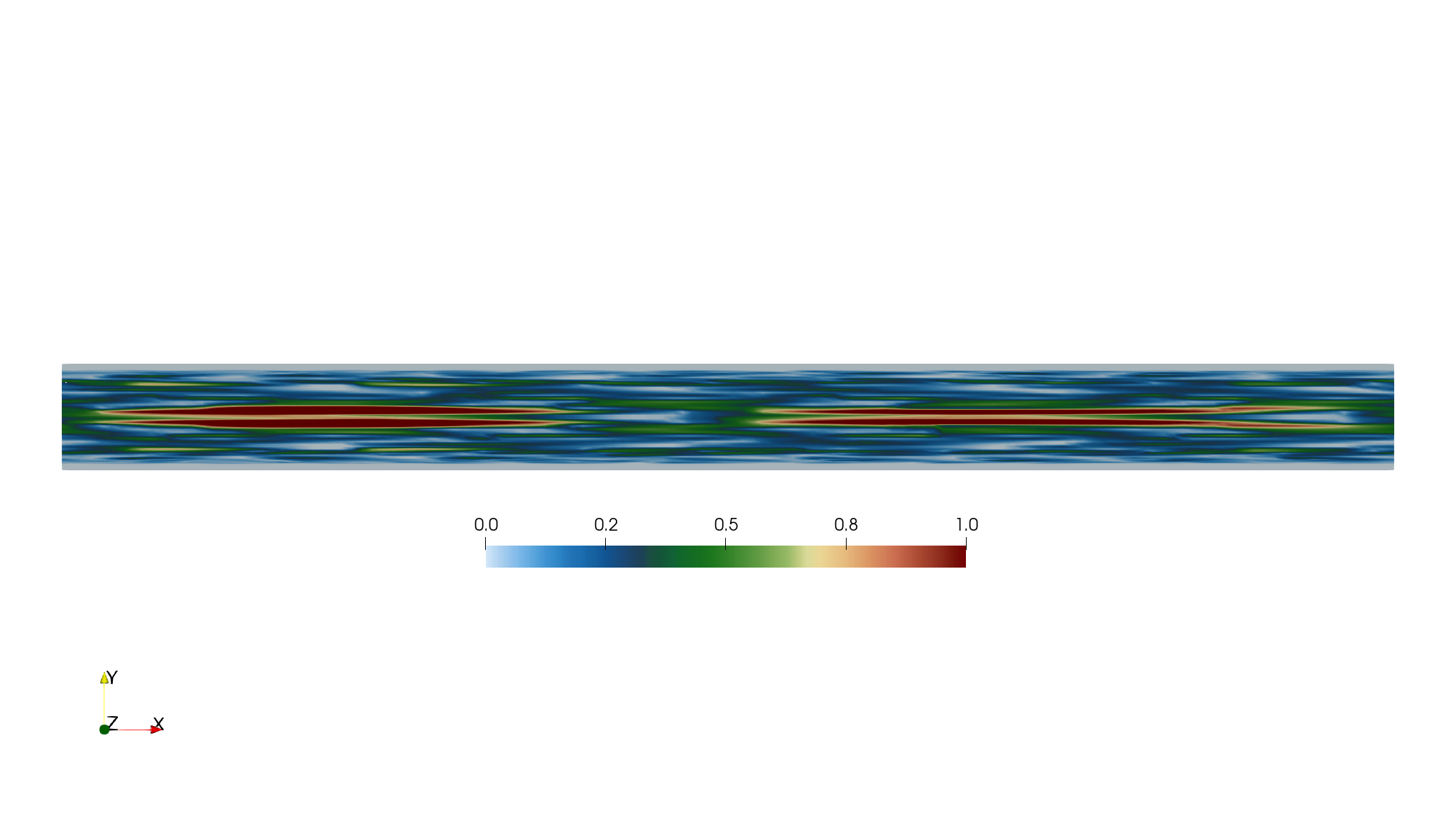}\hspace{-0.7cm}
\includegraphics[trim={1cm 18cm 55cm 18cm},clip=true,width=0.35\linewidth]{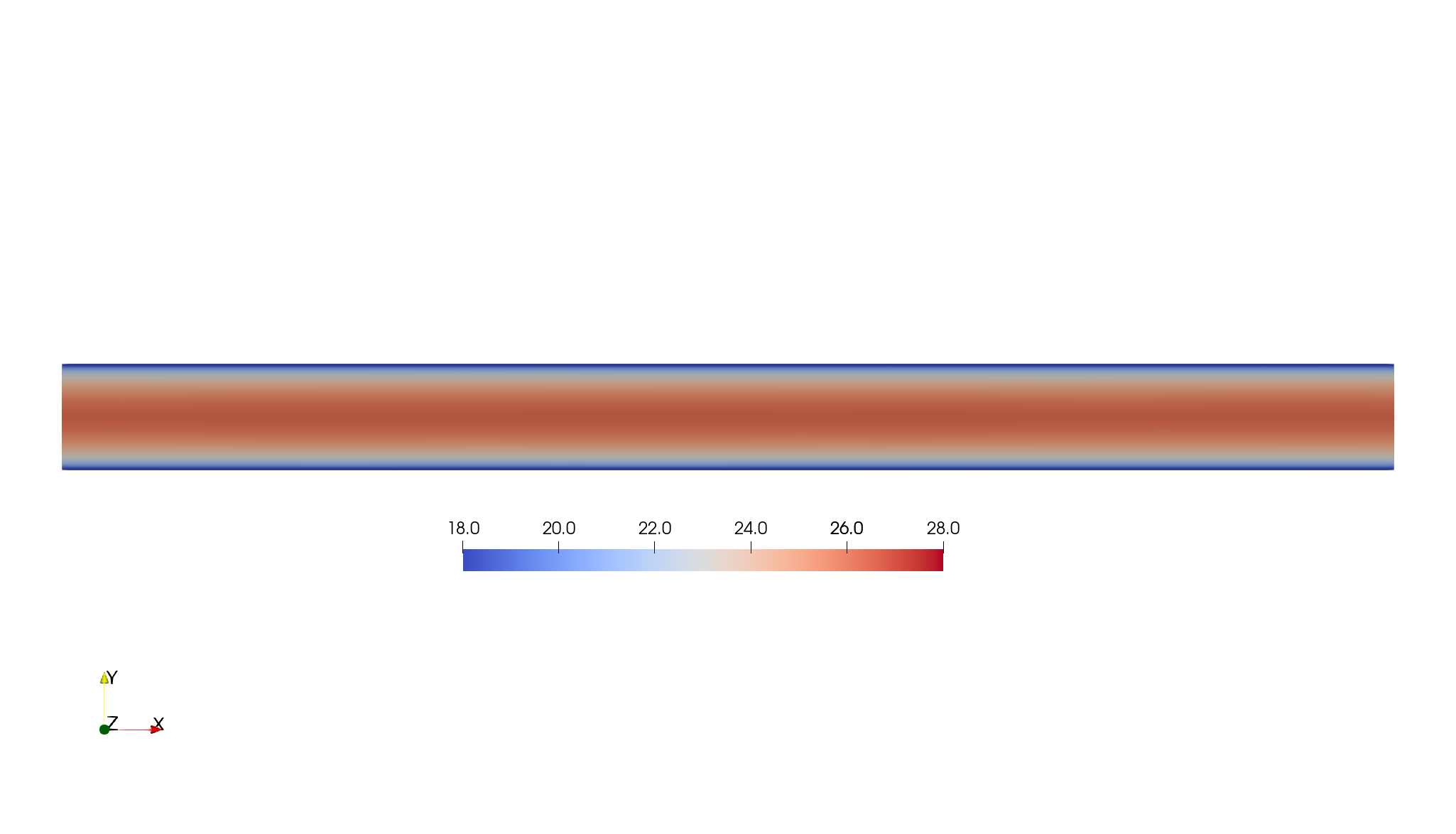}
%
% ROW 2
\begin{overpic}
[trim={1cm 18cm 55cm 18cm},clip=true,width=0.35\linewidth]{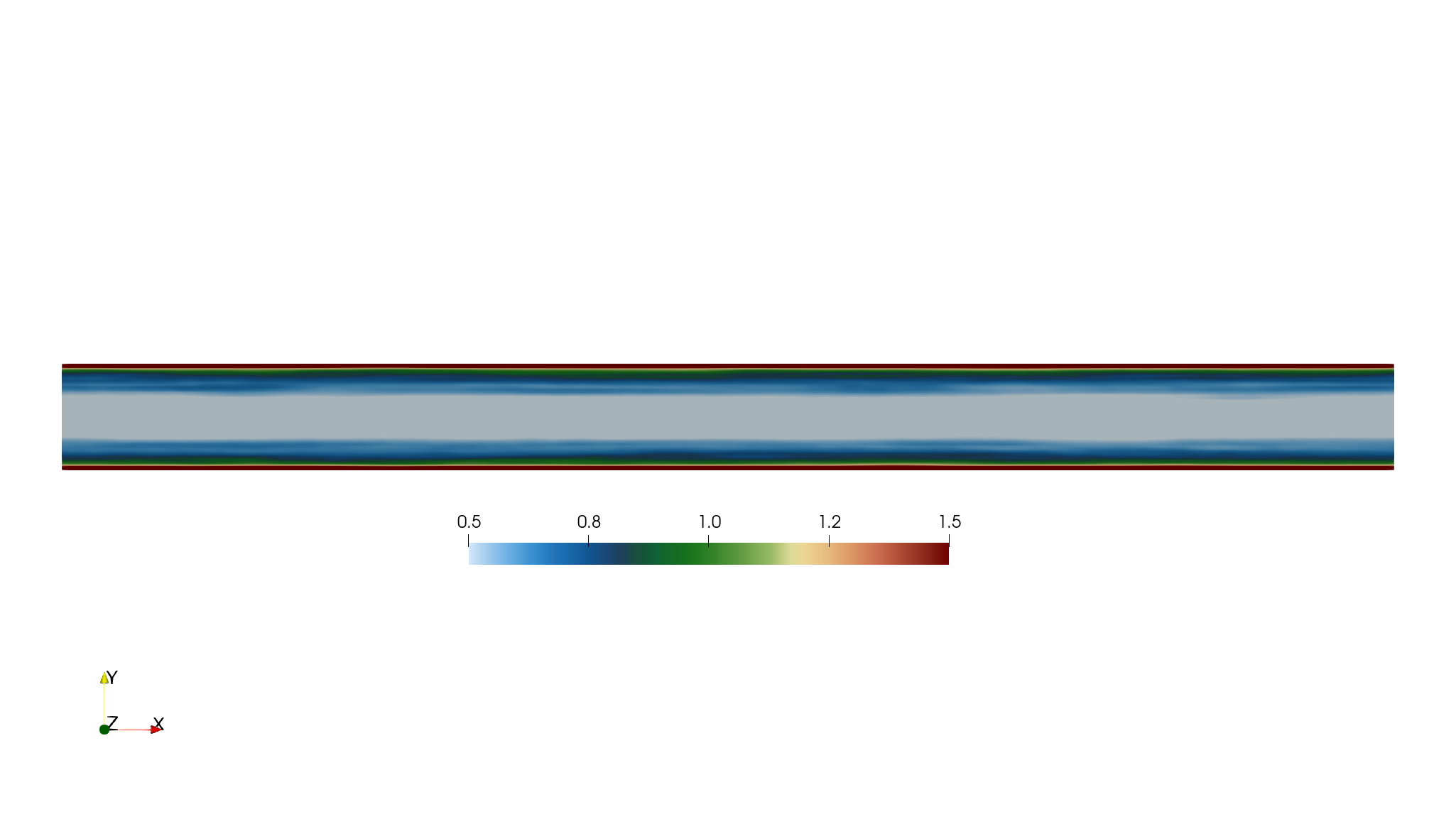}
\put(1.0,12.5){\color{black}\rotatebox{90}{0.4}}
\end{overpic}\hspace{-0.7cm}
\includegraphics[trim={1cm 18cm 55cm 18cm},clip=true,width=0.35\linewidth]{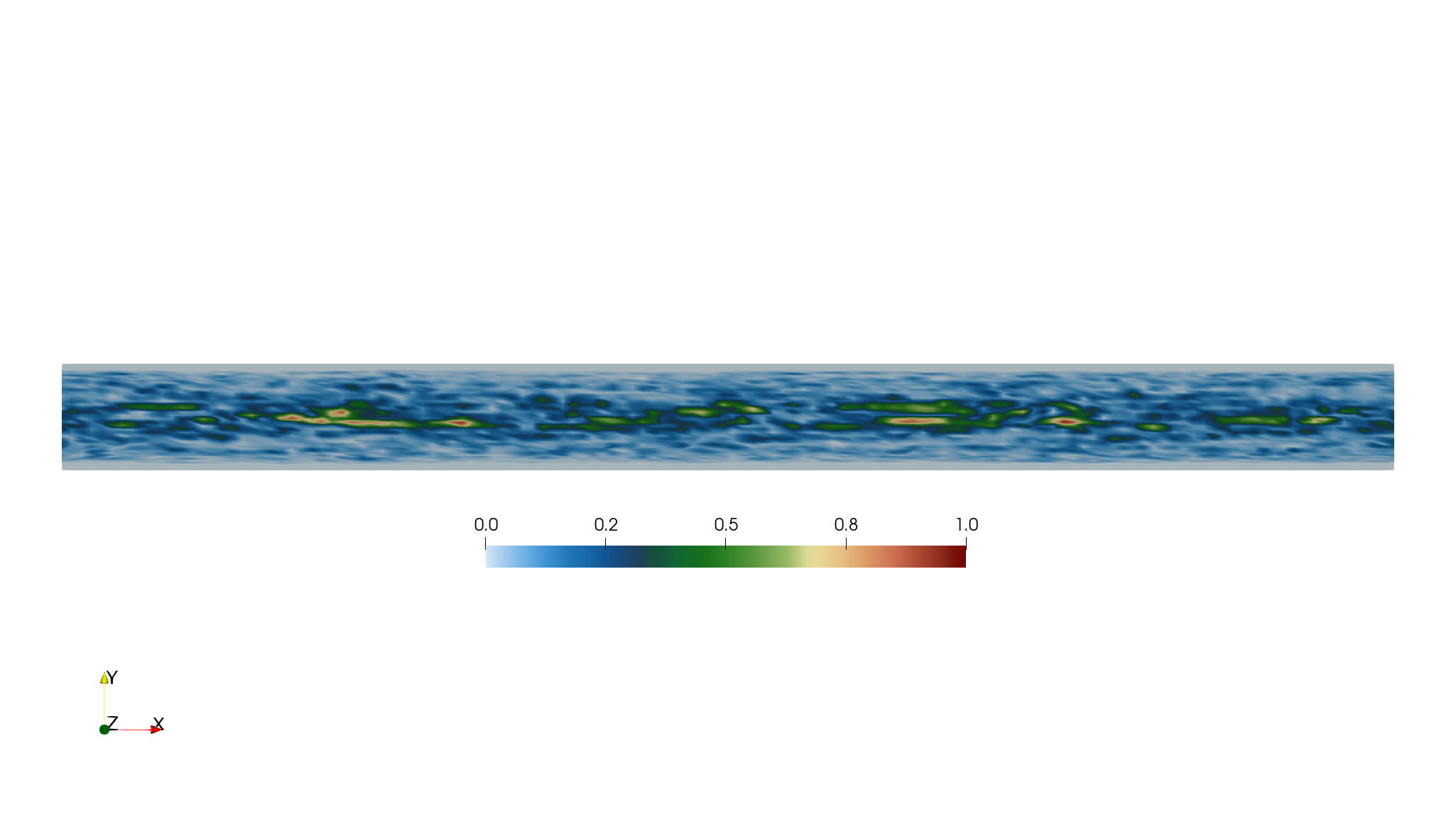}\hspace{-0.7cm}
\includegraphics[trim={1cm 18cm 55cm 18cm},clip=true,width=0.35\linewidth]{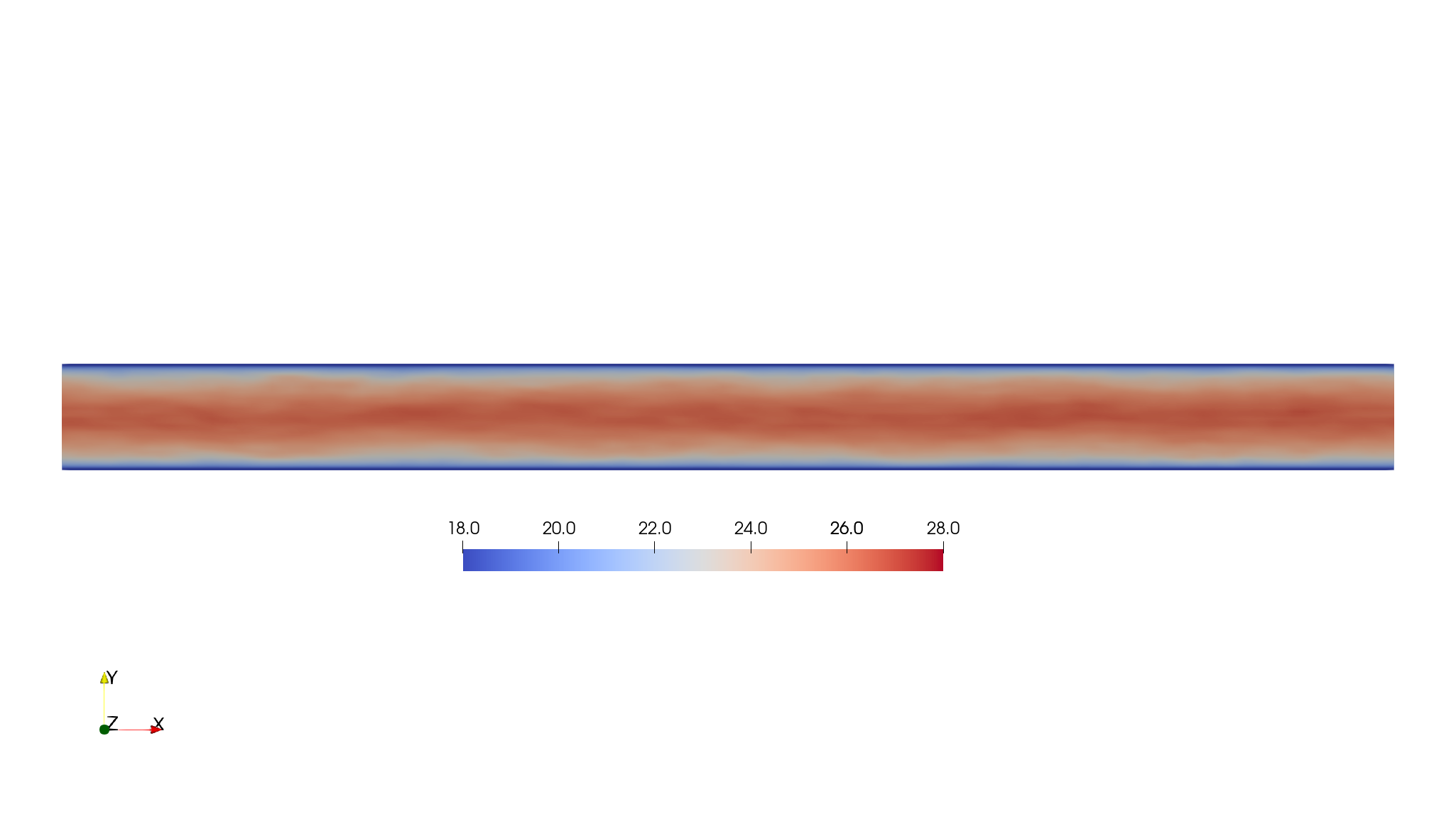}
%
% ROW 3
\begin{overpic}
[trim={1cm 18cm 55cm 18cm},clip=true,width=0.35\linewidth]{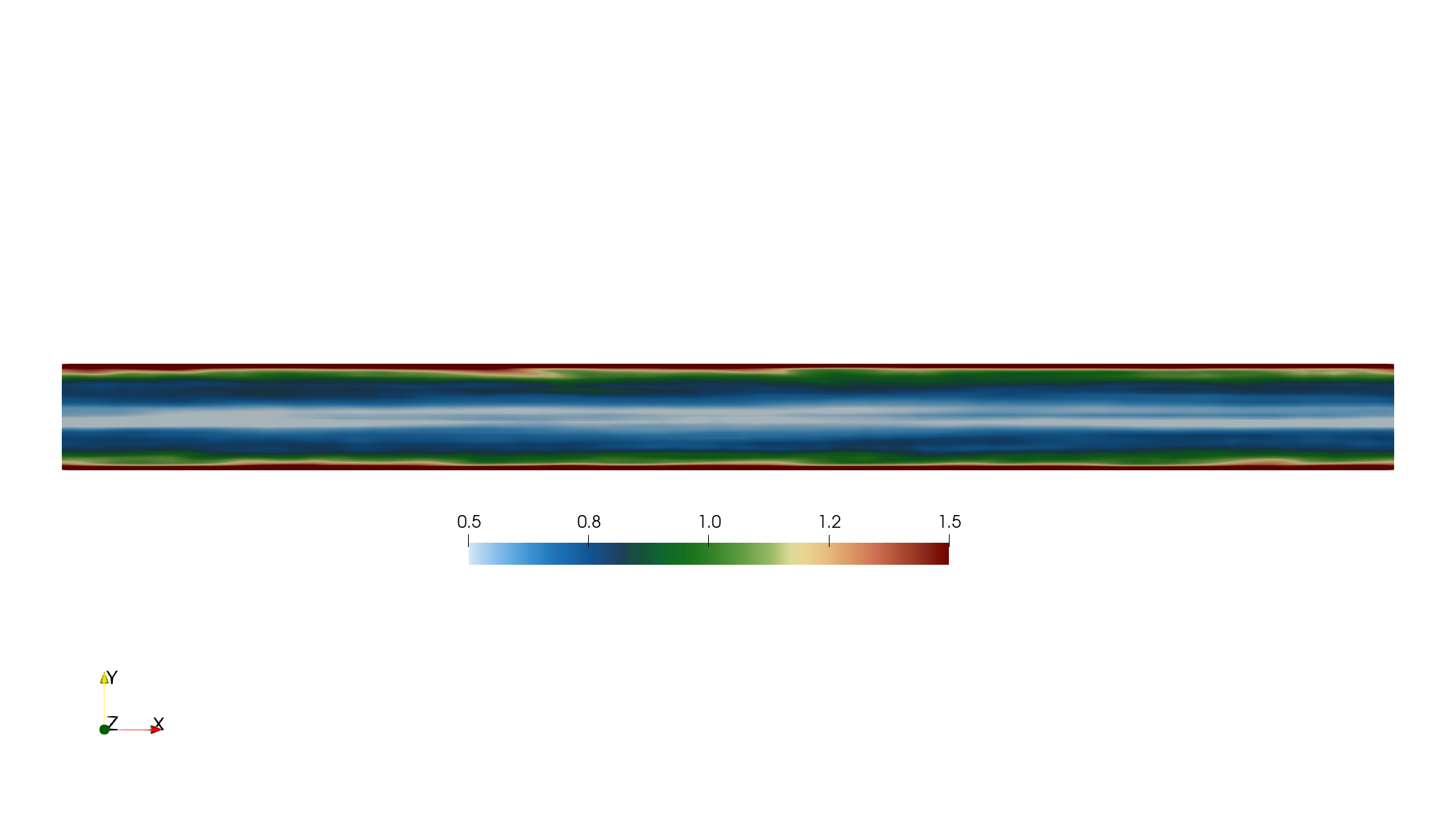}
\put(1.0,12.0){\color{black}\rotatebox{90}{0.8}}
\end{overpic}\hspace{-0.7cm}
\includegraphics[trim={1cm 18cm 55cm 18cm},clip=true,width=0.35\linewidth]{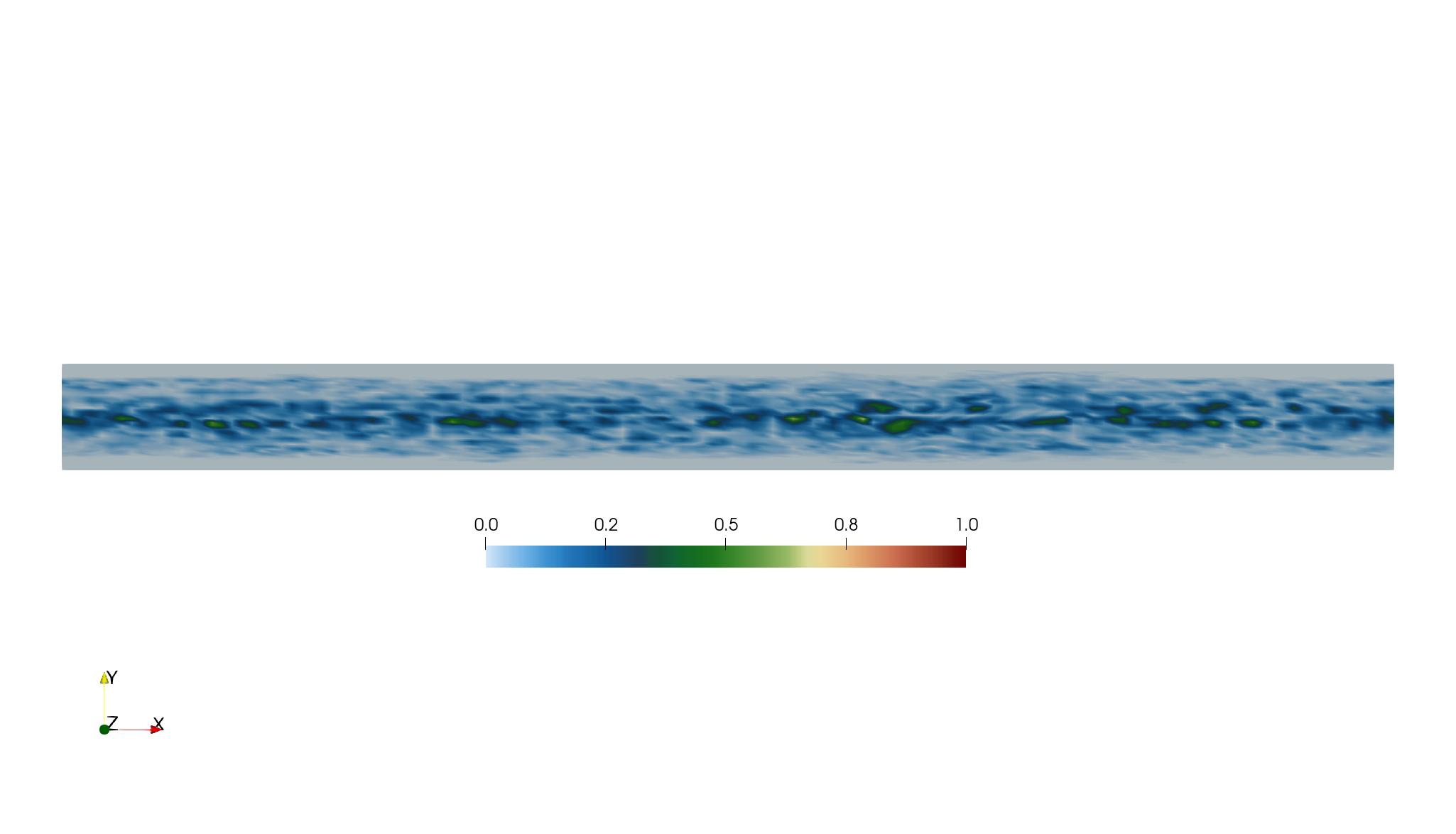}\hspace{-0.7cm}
\includegraphics[trim={1cm 18cm 55cm 18cm},clip=true,width=0.35\linewidth]{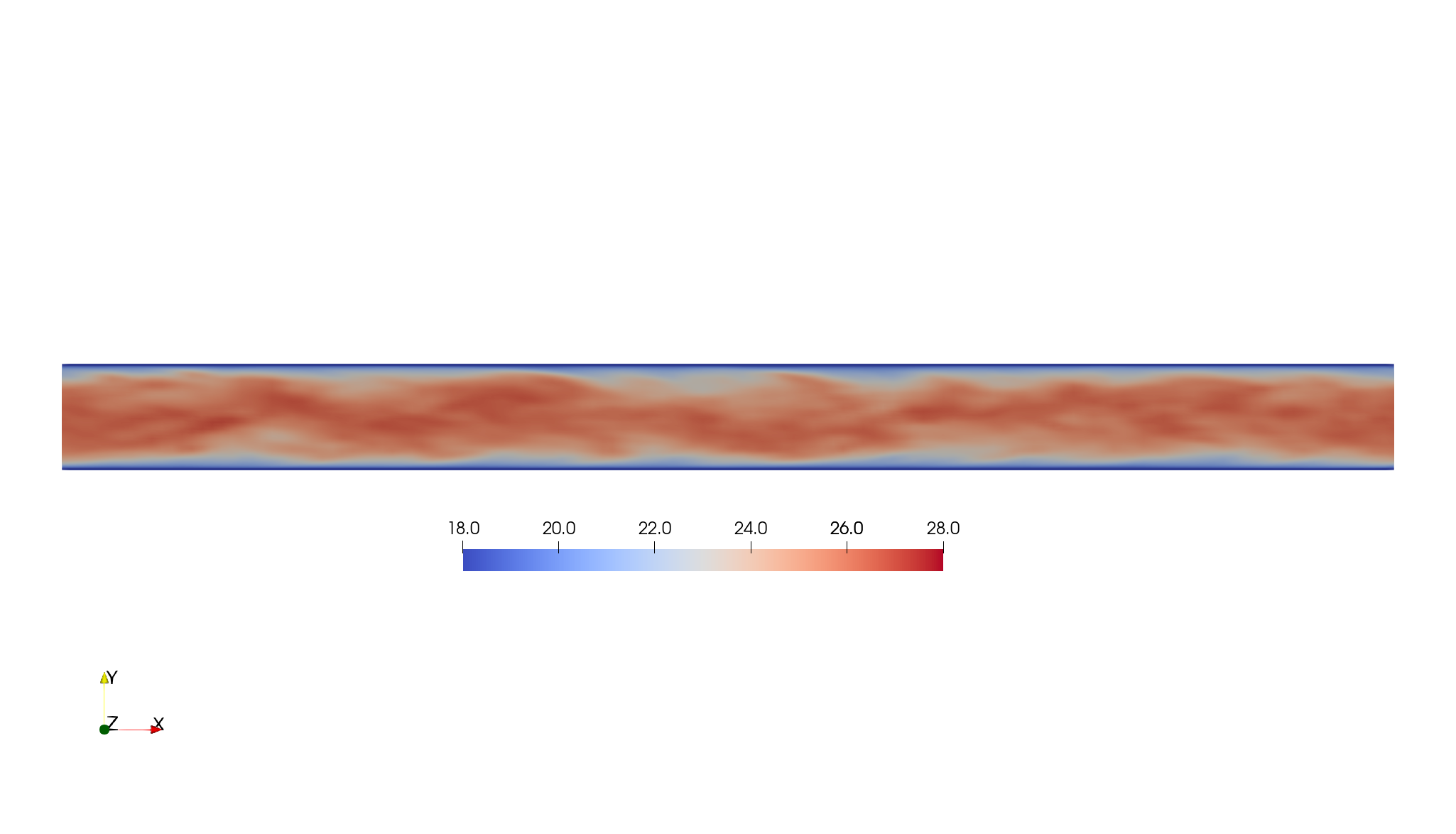}
%
% ROW 4
\begin{overpic}
[trim={1cm 18cm 55cm 18cm},clip=true,width=0.35\linewidth]{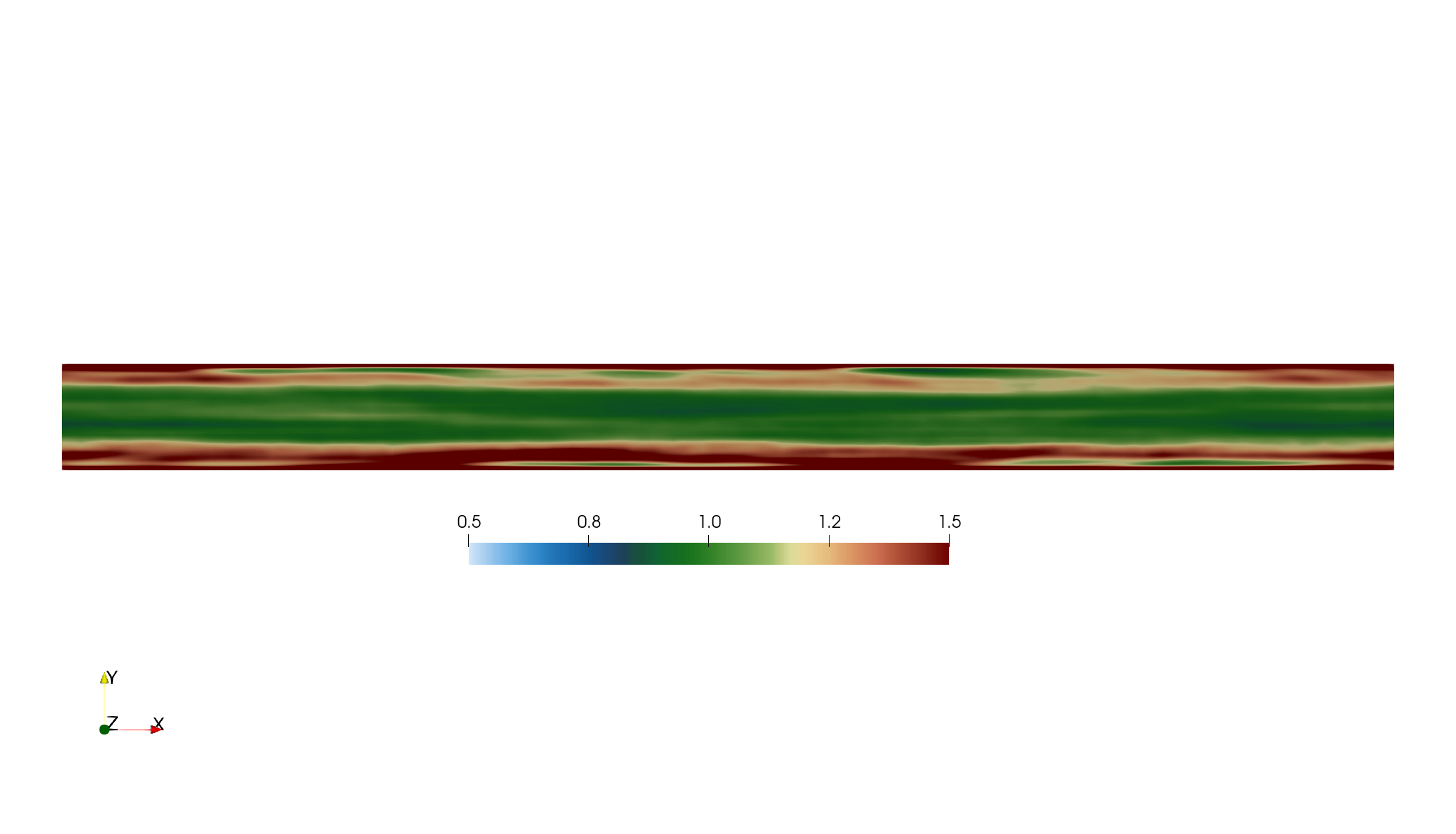}
\put(1.0,14.0){\color{black}\rotatebox{90}{1.6}}
\end{overpic}\hspace{-0.7cm}
\includegraphics[trim={1cm 18cm 55cm 18cm},clip=true,width=0.35\linewidth]{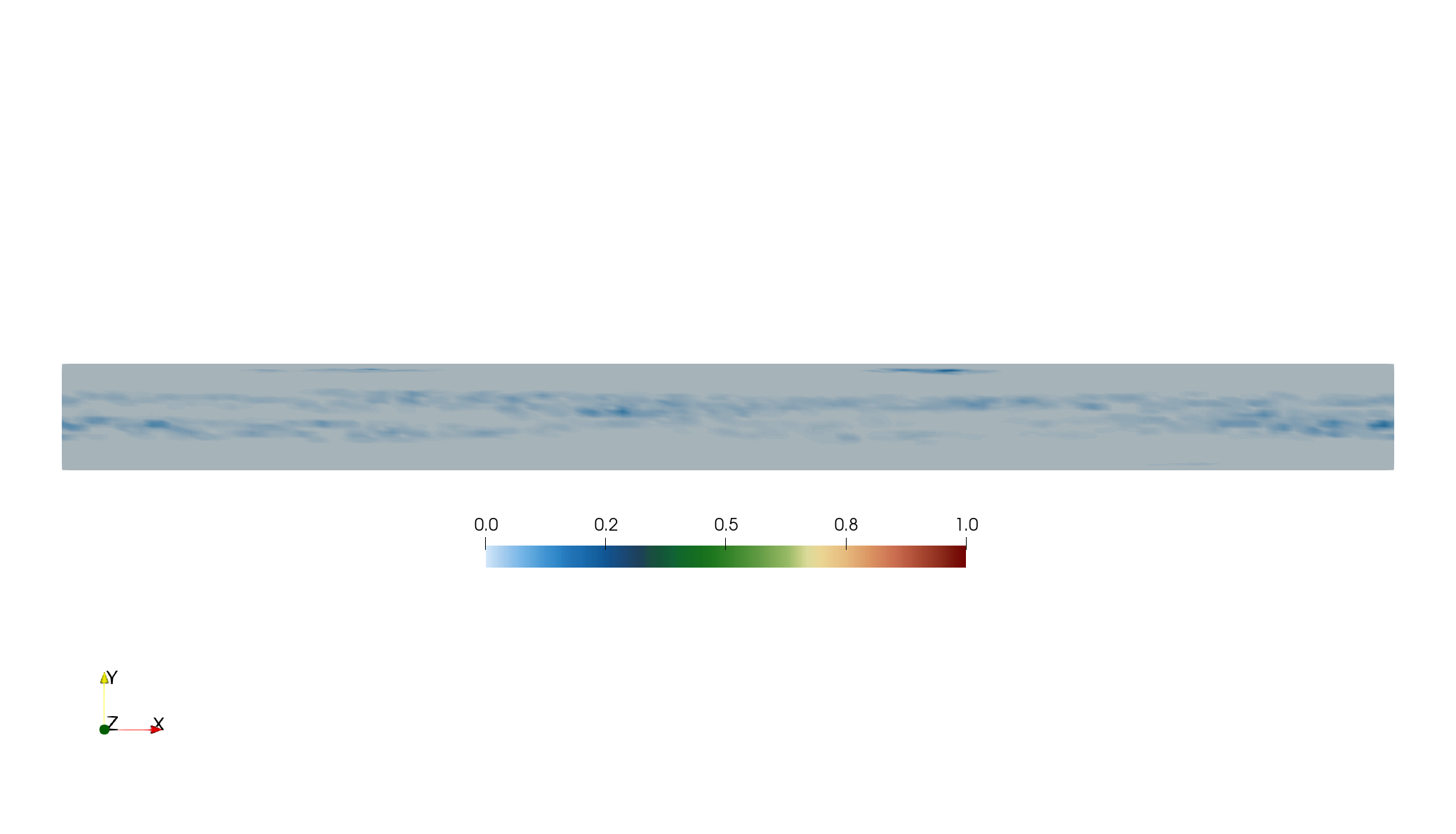}\hspace{-0.7cm}
\includegraphics[trim={1cm 18cm 55cm 18cm},clip=true,width=0.35\linewidth]{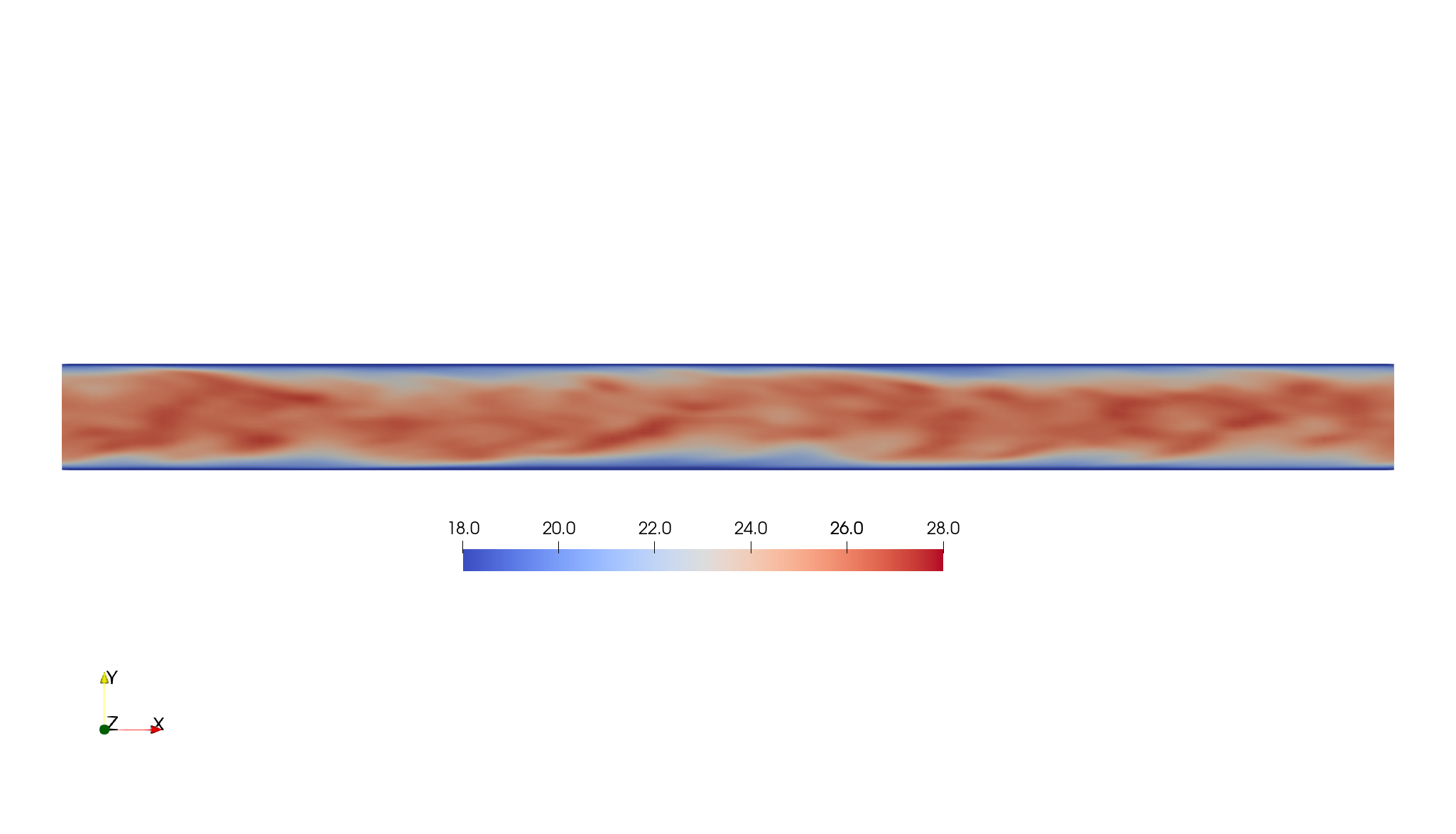}
%
% ROW 5
\begin{overpic}
[trim={1cm 16cm 55cm 18cm},clip=true,width=0.35\linewidth]{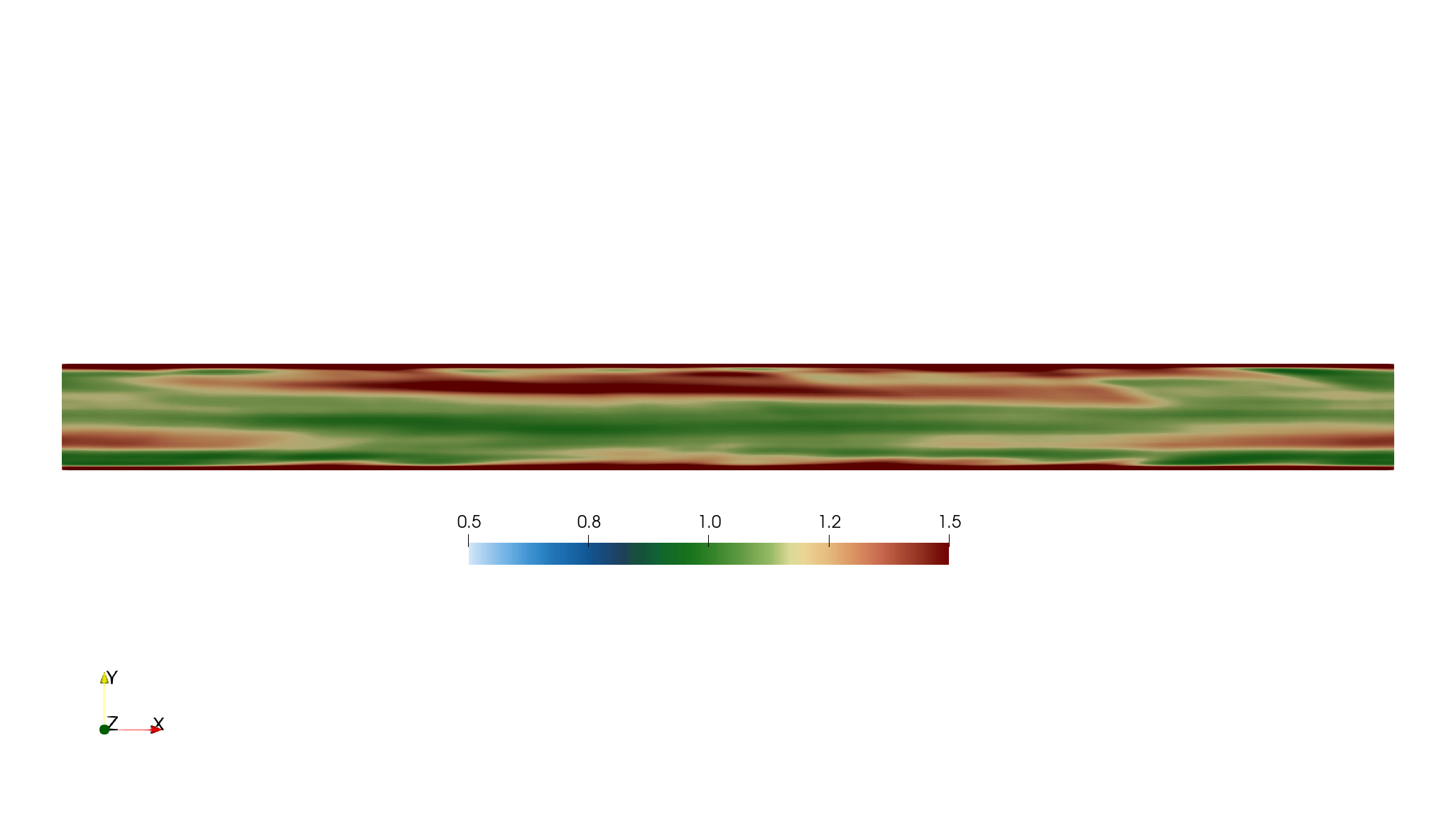}
\put(1.0,25.0){\color{black}\rotatebox{90}{4.0}}
\end{overpic}\hspace{-0.7cm}
\includegraphics[trim={1cm 16cm 55cm 18cm},clip=true,width=0.35\linewidth]{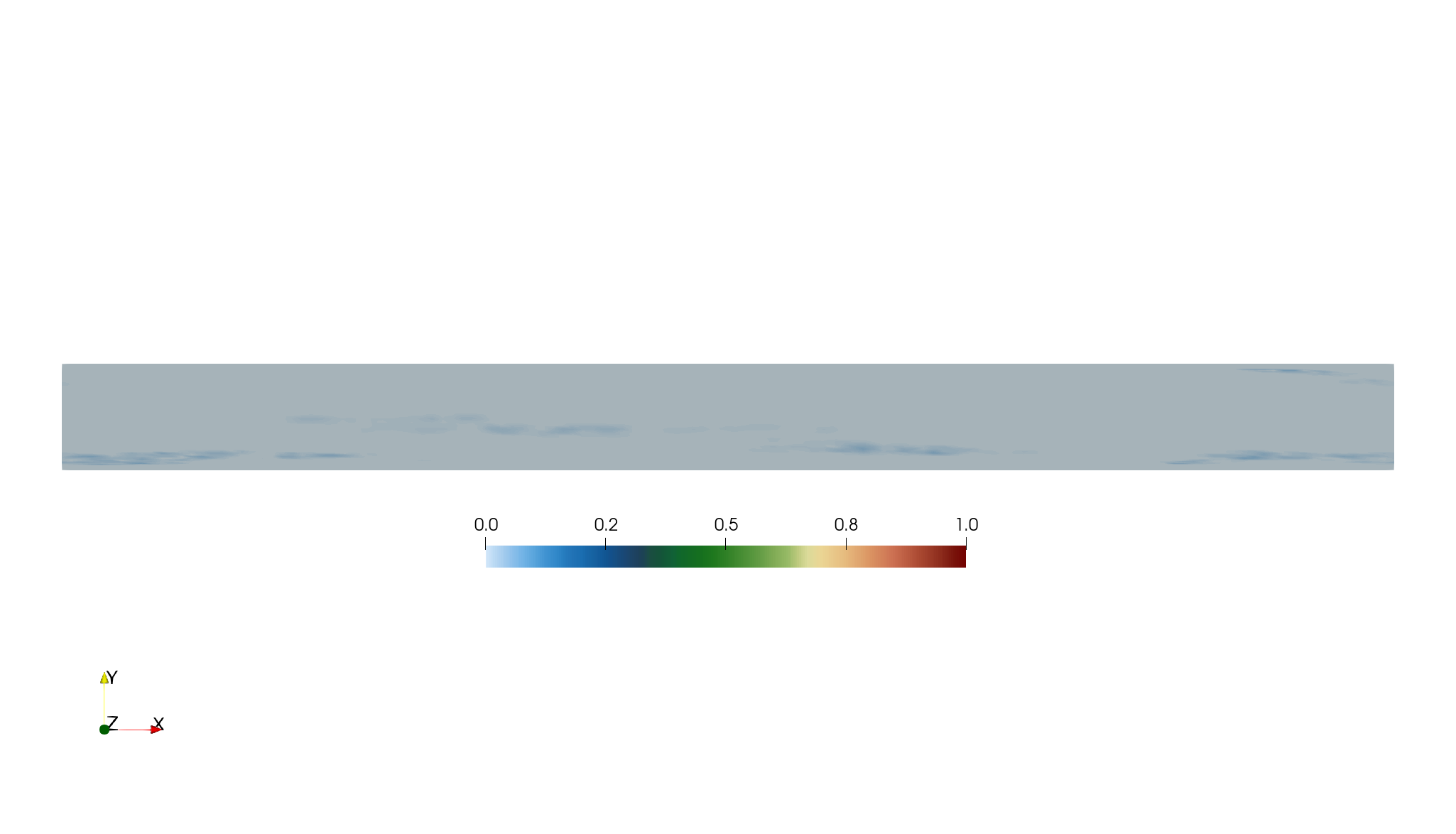}\hspace{-0.7cm}
\includegraphics[trim={1cm 16cm 55cm 18cm},clip=true,width=0.35\linewidth]{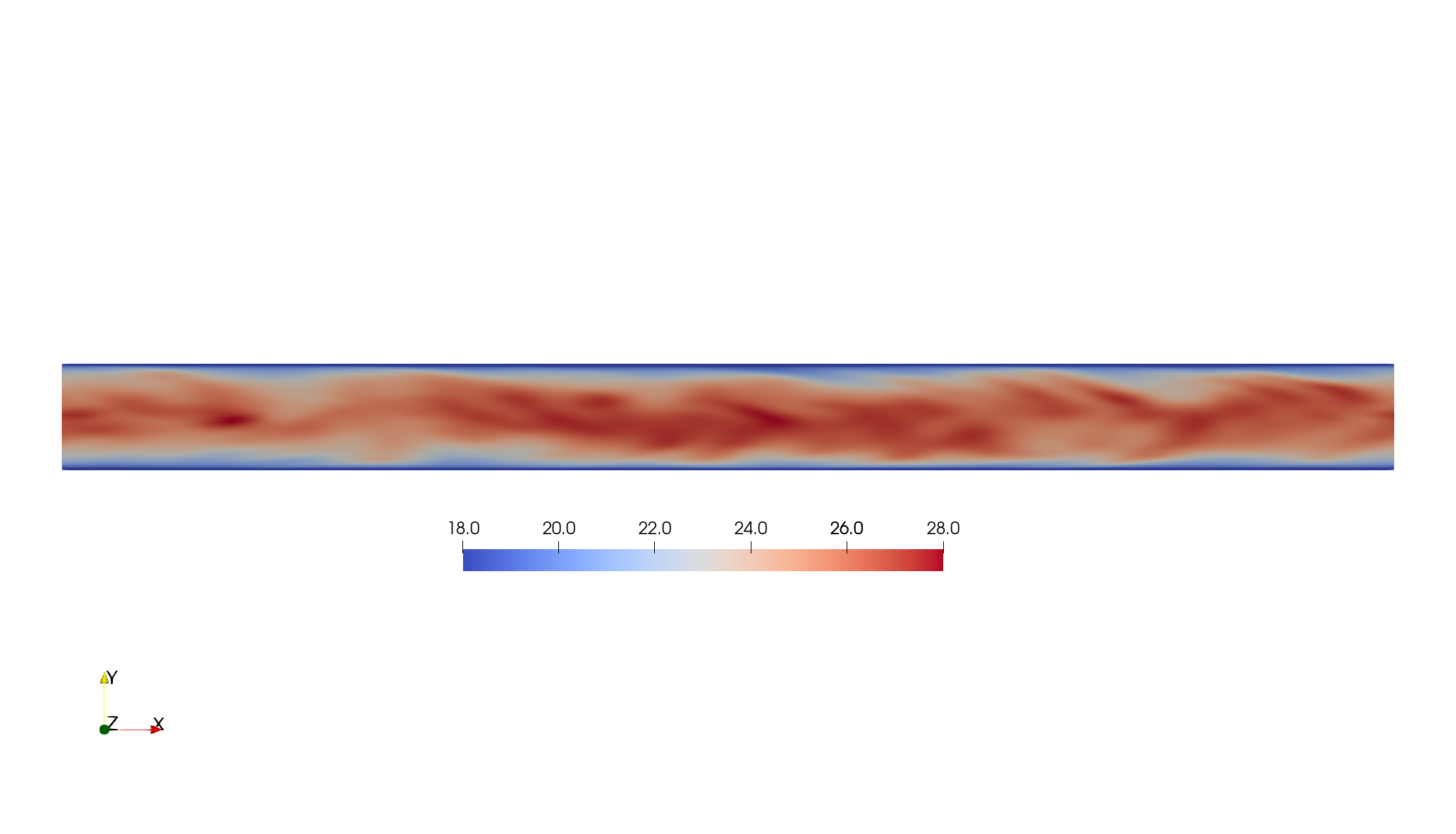}
%[trim={5cm 38.5cm 75cm 38.5cm},clip=true,width=0.45\linewidth]{figures/Fmag_tdev2000.png}
%[trim={5cm 38.5cm 75cm 38.5cm},clip=true,width=0.45\linewidth]{figures/ux_tdev2000.png}
%
%\hspace{-2.0cm}
\begin{overpic}
[trim={18cm 5cm 25cm 26.5cm},clip=true,width=0.32\linewidth]{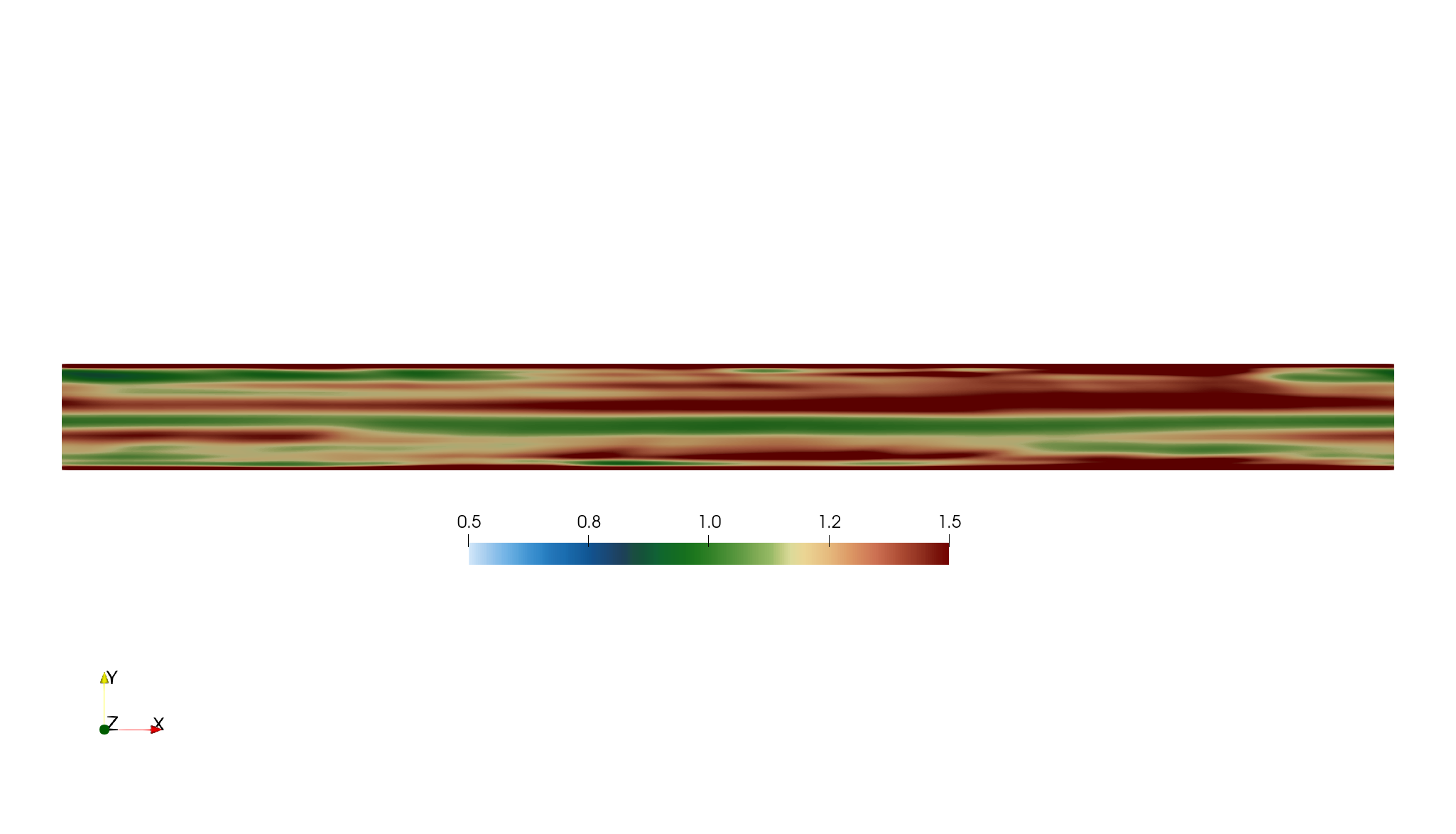}
\put(55.0,15.0){\color{black}\rotatebox{0}{$\la{}r_\mathcal{M}\ra$}}
%\put(11.0,10.0){\crule[white]{8cm}{0.4cm}}
\put(14.5,35.0){\color{black}\rotatebox{0}{0.5}}
\put(55.0,35.0){\color{black}\rotatebox{0}{1.0}}
\put(95.0,35.0){\color{black}\rotatebox{0}{1.5}}
\end{overpic} % 0.5 1.5
\hspace{-0.5cm}
\begin{overpic}
[trim={18cm 5cm 25cm 26.5cm},clip=true,width=0.32\linewidth]{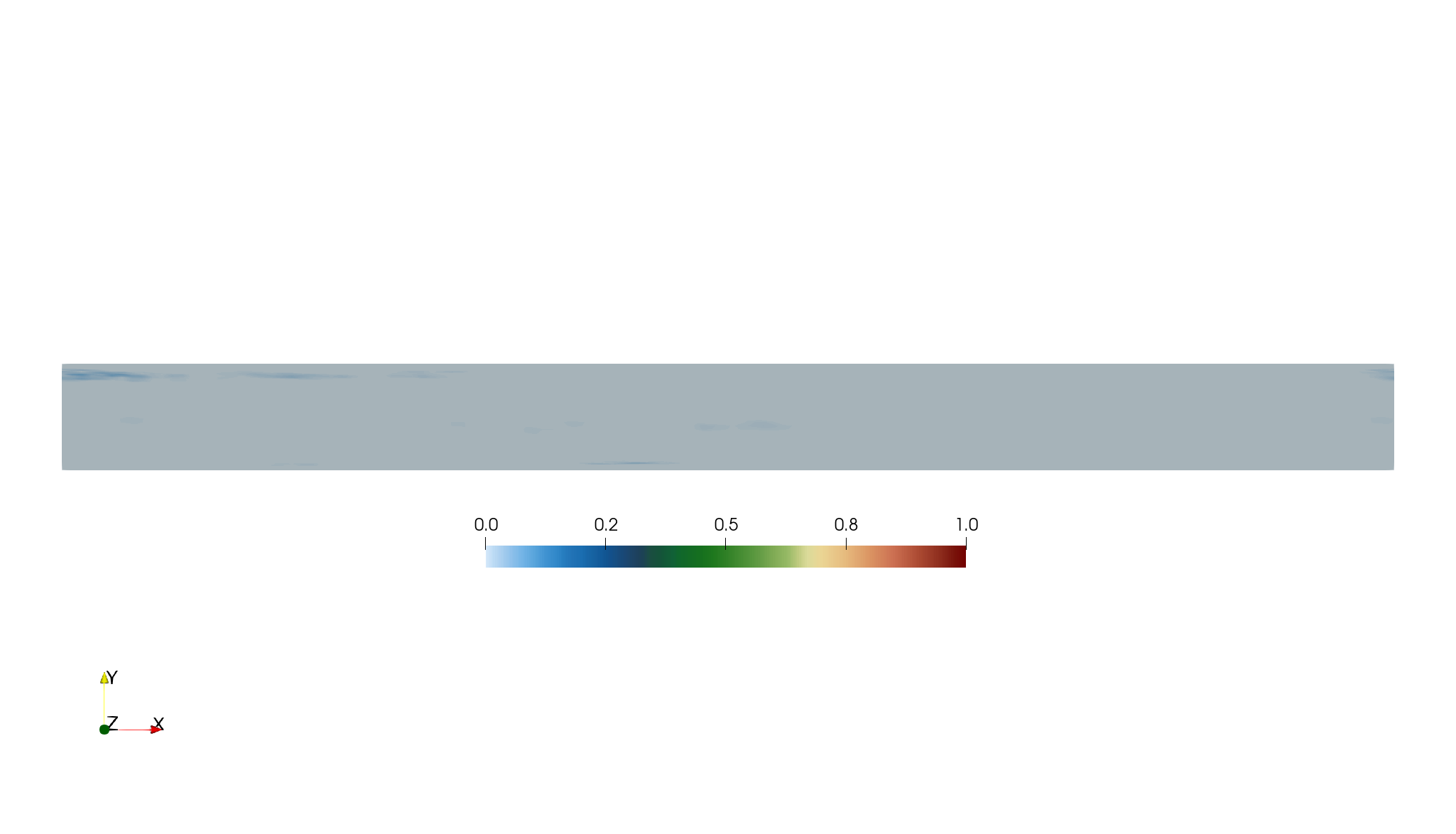}
\put(55.0,15.0){\color{black}\rotatebox{0}{$|\hat{F}|$}}
%\put(11.0,10.0){\crule[white]{8cm}{0.4cm}}
\put(16.5,35.0){\color{black}\rotatebox{0}{0.0}}
\put(57.0,35.0){\color{black}\rotatebox{0}{0.5}}
\put(95.0,35.0){\color{black}\rotatebox{0}{1.0}}
\end{overpic} % 0.5 1.5
%
%\hspace{-0.5cm}
\begin{overpic}
[trim={18cm 5cm 25cm 27cm},clip=true,width=0.32\linewidth]{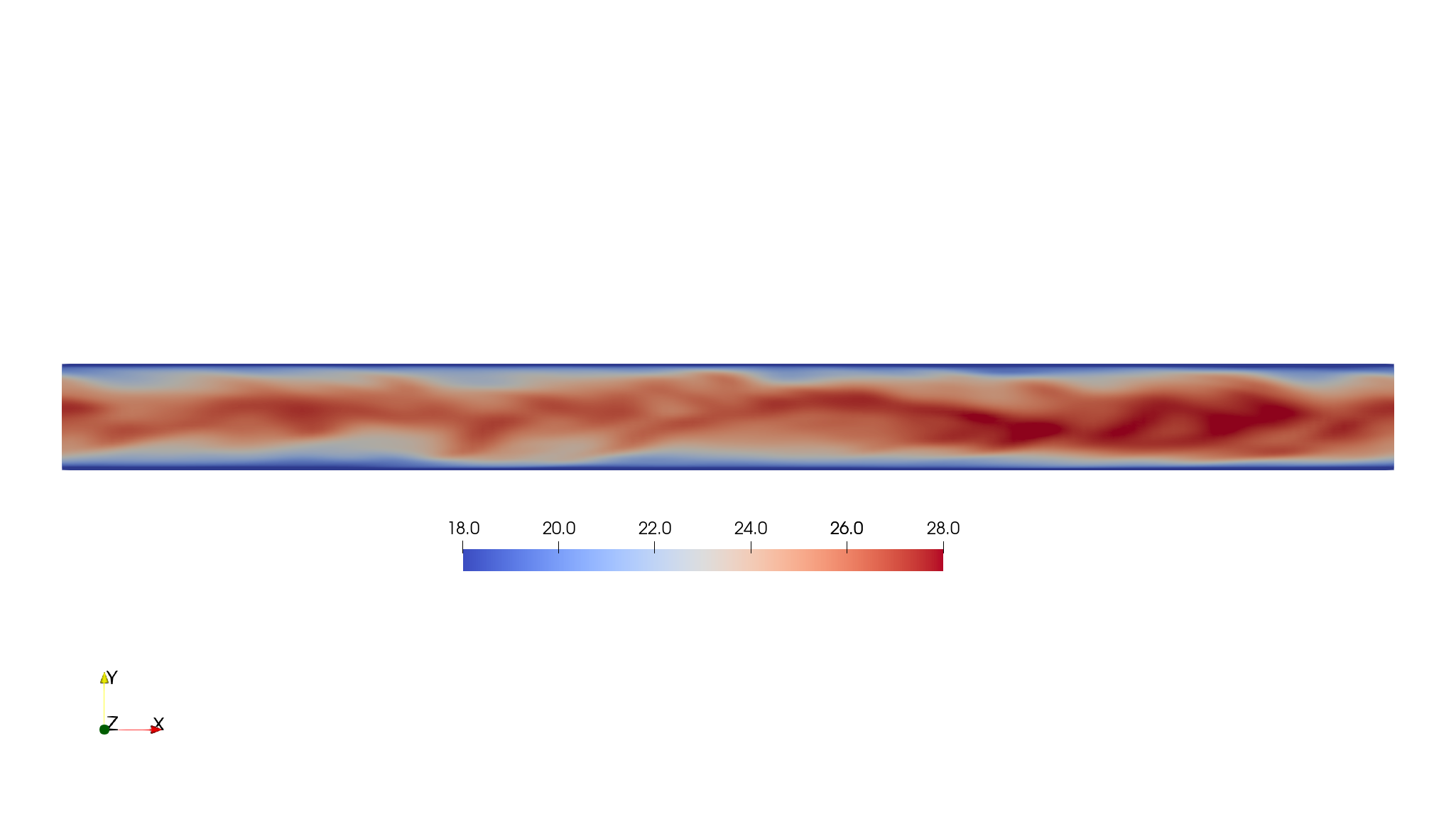}
\put(55.0,15.0){\color{black}\rotatebox{0}{$\overbar{u_x^+}$}}
%\put(11.0,10.0){\crule[white]{8cm}{0.4cm}}
\put(14.0,35.0){\color{black}\rotatebox{0}{18}}
\put(30.5,35.0){\color{black}\rotatebox{0}{20}}
\put(46.0,35.0){\color{black}\rotatebox{0}{22}}
\put(62.3,35.0){\color{black}\rotatebox{0}{24}}
\put(78.5,35.0){\color{black}\rotatebox{0}{26}}
\put(94.75,35.0){\color{black}\rotatebox{0}{28}}
\end{overpic} % 18 to 28
\vspace{-1.0cm}
\end{center}
\caption{Snapshots of (left) mean resolution adequacy parameter, $\la{}r_\mathcal{M}\ra$ (\ref{rM}), (middle) magnitude of scaled forcing vector field, $\hat{F}_i=F_i/F_c$ where $F_c=0.25(\Delta_x\Delta_y\Delta_z)^{-1/3}k^{3/2}_{tot}$ (\ref{Fi_actual}), and  
(right) resolved streamwise velocity in $x$-$y$ planes for the temporally developing
hybrid simulation of channel flow at $Re_\tau\approx 5200$. Note that $k_{tot}$ used to define $F_c$ is a function of wall distance.  Only results for the coarse resolution are shown here .
Time since initiation of forcing $t/t_f$ is shown, where $t_f$ is the
flow-through time.\label{fig:rd_ux}}
%\todo[inline]{RDM: need to redesign this figure and figure 6 to better represent the fording field.} 
\end{figure}

\begin{figure}[th]
\begin{center}
\subfigure[$y=0.8$]{
\includegraphics[trim={6cm 9cm 6cm 9cm},clip=true,width=0.48\linewidth]{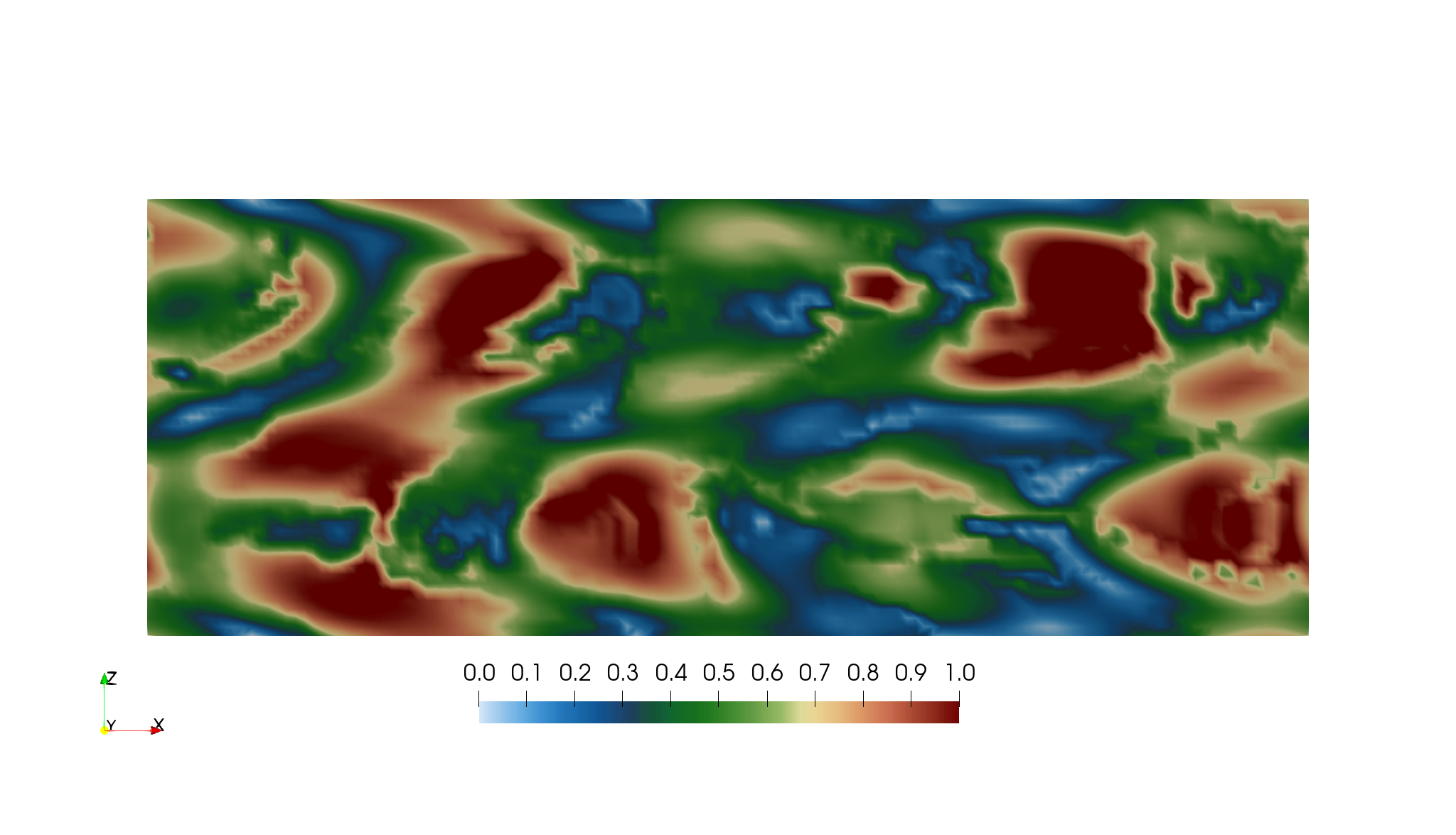}
}
\subfigure[$y=0.6$]{
\includegraphics[trim={6cm 9cm 6cm 9cm},clip=true,width=0.48\linewidth]{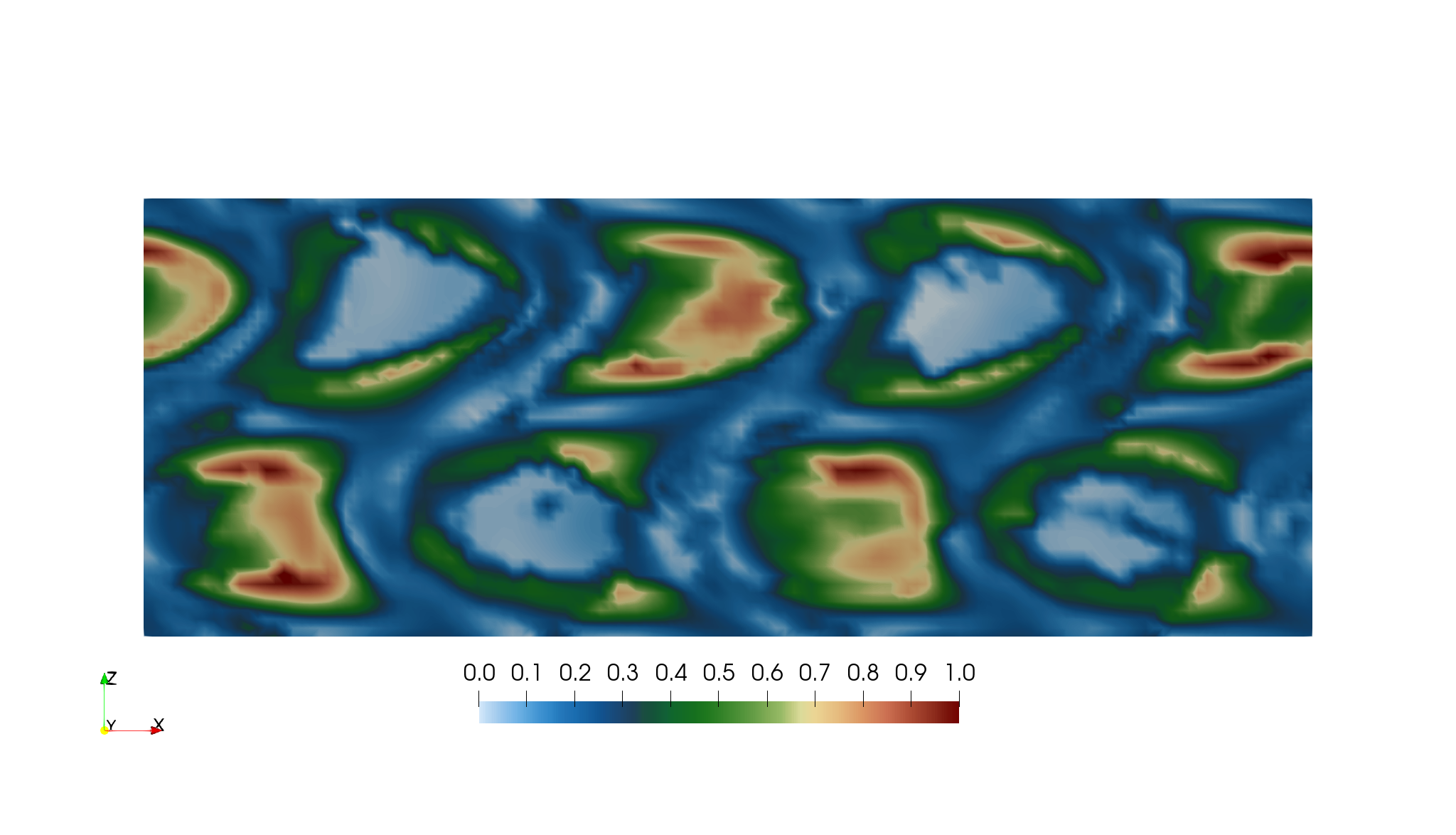}
}\\
\subfigure[$y=0.4$]{
\includegraphics[trim={6cm 9cm 6cm 9cm},clip=true,width=0.48\linewidth]{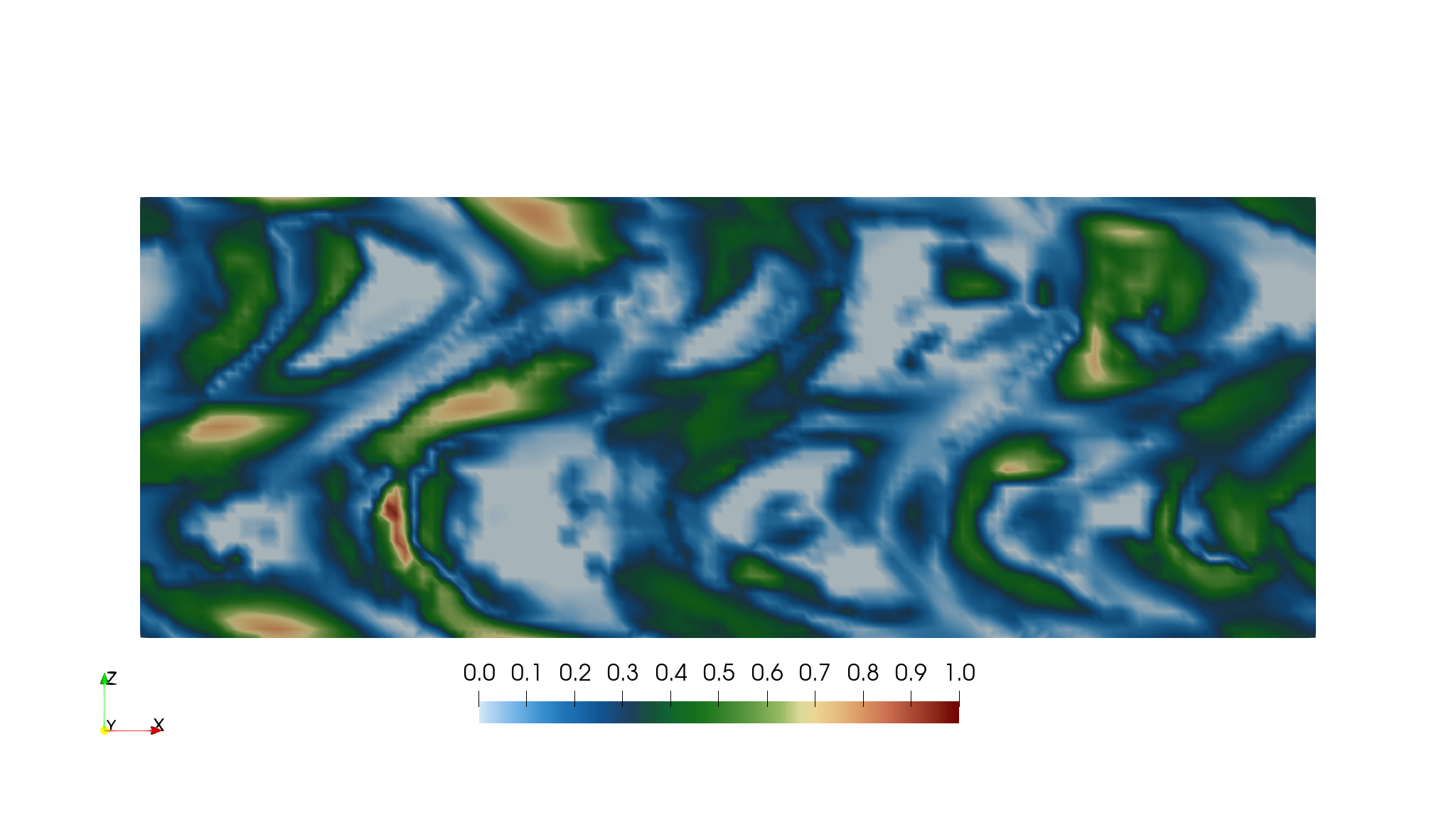}
}
\subfigure[$y=0.2$]{
\includegraphics[trim={6cm 9cm 6cm 9cm},clip=true,width=0.48\linewidth]{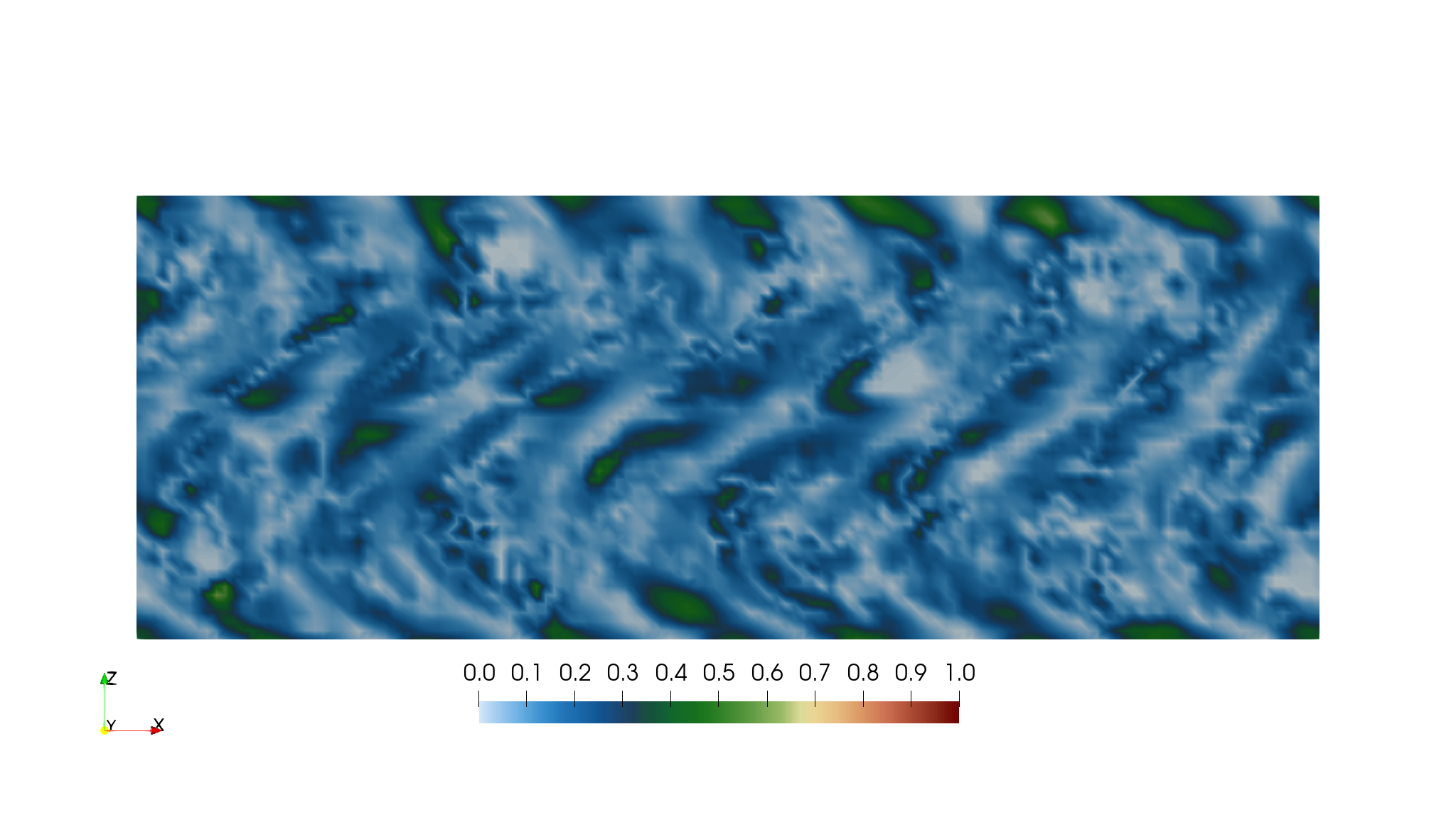}
}\\
\begin{overpic}
[trim={20cm 7cm 20cm 75cm},clip=true,width=0.45\linewidth]{figures/Fmag_0p8h.png}
\put(45.0,-9){\color{black}\rotatebox{0}{$|\hat{F}|$}}
%\put(11.0,5.0){\crule[white]{8cm}{0.4cm}}
\put(9.0,2.0){\color{black}\rotatebox{0}{0.0}}
\put(46.0,2.0){\color{black}\rotatebox{0}{0.5}}
\put(83.0,2.0){\color{black}\rotatebox{0}{1.0}}
\end{overpic}
\end{center}
\caption{\hspace{0.2cm}Magnitude of scaled forcing vector field, $\hat{F}_i=F_i/F_c$ where $F_c=0.25(\Delta_x\Delta_y\Delta_z)^{-1/3}k^{3/2}_{tot}$, at multiple wall-parallel planes 
a $t=0.04$ after initiating forcing from the steady RANS solution.  Entire spanwise 
domain is displayed.  Early forcing experiences no clipping yielding large coherent 
acceleration with the size of the structures decreasing with proximity 
to the wall. Note that $k_{tot}$ is dependent on wall distance.  \label{fig:f_planes}}
%\todo[inline]{Need to redesign this figure and figure 4 to better represent the forcing structure.}
\end{figure}

As previously discussed, the hybridization is driven by forcing.  If no forcing is 
applied, the simulation would remain a RANS simulation.  The process of evaluating the 
resolution and introducing resolved turbulence is summarized through the resolution 
adequacy parameter (\ref{rM}), forcing field (\ref{Fi_actual}), and resolved velocity for the coarse case in 
Fig.~\ref{fig:rd_ux}.  An example of the wall-parallel forcing field is shown in Fig.~\ref{fig:f_planes}.  Note that by construction, a variable length scale and clipped 
Taylor-Green field is not divergence-free.  For an incompressible solver, the divergence 
of the forcing field is projected out in the pressure solution step so overall behavior 
of the hybrid method is not affected by forcing divergence.  
Thus, for incompressible applications, an additional pressure-Poisson-like solve is not necessary however, it may be important 
for compressible solvers.  %For the forcing 
%fields visualized in Fig.\ref{fig:rd_ux}, an additional divergence-free projection has been applied 
%to the raw $F_i$ to show the structure of the forcing that actually
%impacts the resolved field.  
% trim = left bottom right top
%\put(7.0,8.0){\color{white}\rule{15.7pt}{2pt}}
%\put(-1.0,9.0){\color{black}\rotatebox{90}{1.6}}
\begin{figure}[th!]
\begin{center}
\subfigure[Temporally developing]
{\includegraphics[trim={1.7cm 0cm 2cm 0cm},clip=true,width=0.49\linewidth]{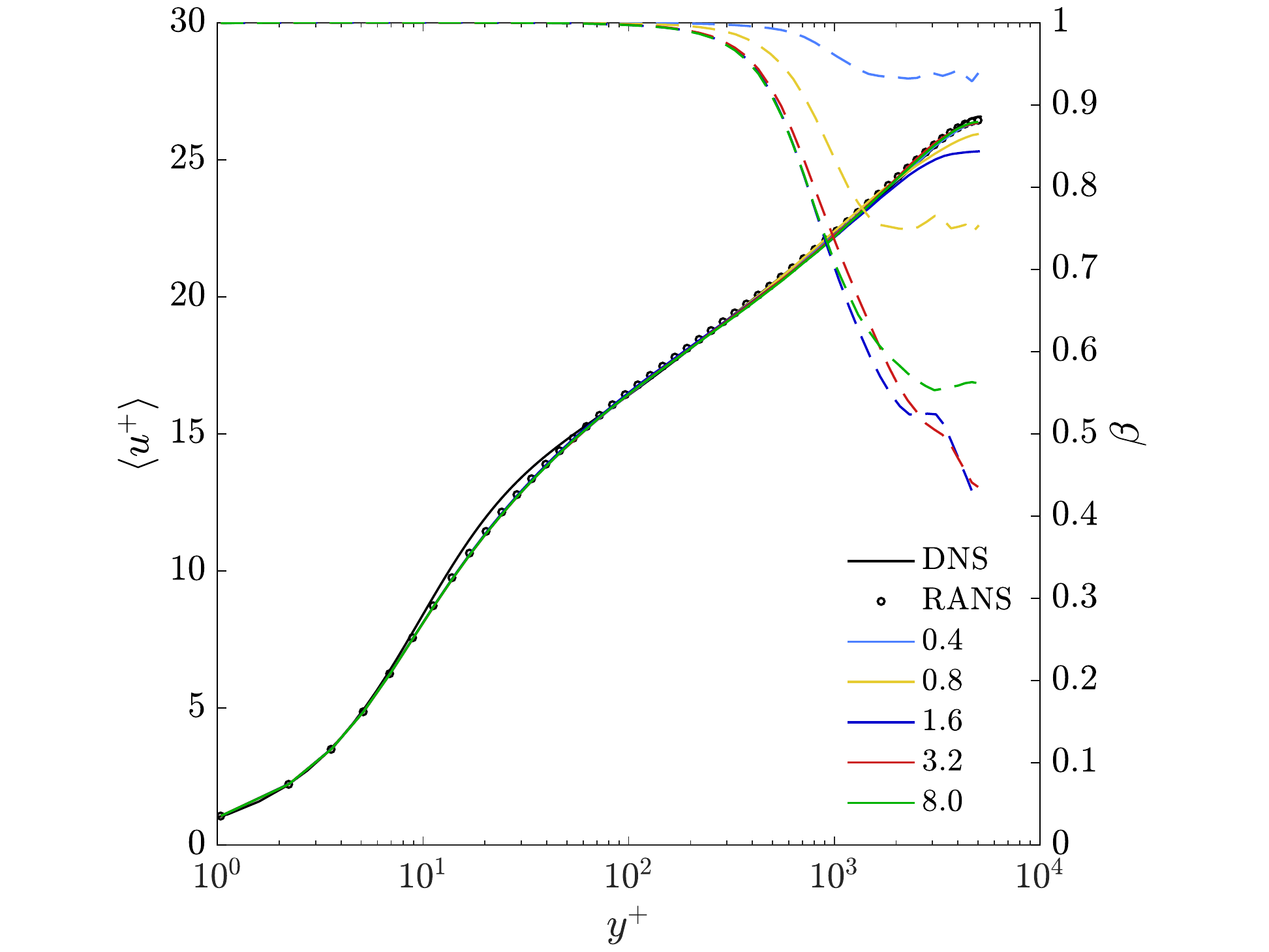}\label{tdev_ux_cf8}}
\subfigure[Spatially developing]
{\includegraphics[trim={1.7cm 0cm 2cm 0cm},clip=true,width=0.49\linewidth]{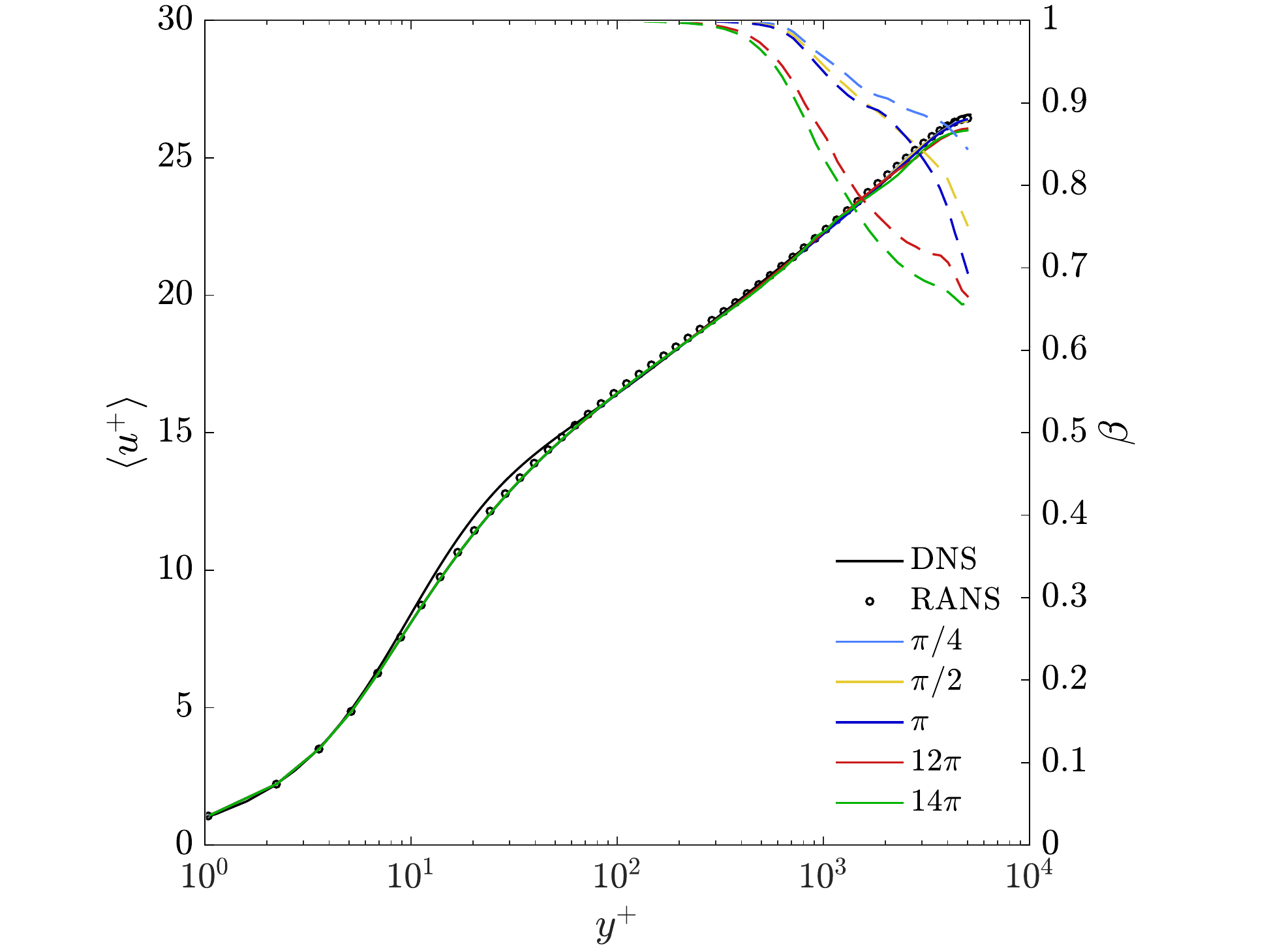}\label{sdev_ux_cf8}}
\end{center}
\caption{
%\todo[inline]{Vertical axis labels: $\la\bar{u}^+_x\ra$ ??}
Mean streamwise velocity for fully developed channel flow
at $Re_\tau\approx5200$ along with the fraction of unresolved
turbulent kinetic energy $\beta$ (dashed). In (a) the simulation
evolves in time from the stationary RANS solution to the stationary
hybrid solution. Numbers in the legend refer to $t/t_f$, the time in
flow through units. In (b), the hybrid simulation evolves in the
streamwise direction from the inlet RANS solution to a hybrid
solution. Numbers in the legend refer to $x/\delta$, the down-stream
distance normalized by channel half-width.\label{fig:poc}}
\end{figure}

At $t_f = 0$, forcing is activated, and the state evolves in time from 
the RANS initial condition to a statistically stationary hybrid state as shown in the 
previous section.   During the first few steps of forcing (Fig.\ref{fig:rd_ux},
$t_f=0.0^+$) a large region with excess resolution is present,
indicated by $\la{}r_\mathcal{M}\ra<1$, and shown as blue in the
figure.  Note that $\la{}r_\mathcal{M}\ra$ is much larger than 1.5
near the wall in regions which are incapable of resolving turbulence
and remain RANS.  Coherent, large fluctuations are excited as the
scale of the forcing is the large scale turbulent lengthscale.
Some clipping of the forcing (\ref{Ptest}) still occurs due to the approximation 
of means through time-averaging.  The local magnitude of the forcing follows the RANS
$k$ profile and the hybrid solution is still nearly identical to the
RANS state at this point.  The forcing field is characterized by very long streamwise, 
but thin wall-normal, structures.  Shortly after ($t_f=0.4$), some
fluctuations become evident and the forcing field is drastically
altered with much shorter streamwise structures.  This is due to reduction of the subgrid turbulence length
scale ($L_{sgs}\sim(\beta{}k_{tot})^{3/2}/\varepsilon$) and
as a result, the forcing lengthscale,
%\todo{RDM: where did this come from? SWH: fixed?}
and to more local clipping where the prescribed Taylor-Green
field would remove energy.  By $t_f=0.8$,
distinct turbulent structures are visible and the over-resolved core
has been reduced and flanked by bands of
$\la{}r_\mathcal{M}\ra\approx1$ extending towards the center of the
channel.

At $t_f=1.6$, we begin to see a shortcoming in the current
formulation.  A grid-resolved LES region extends through the entire
channel (green regions of $\la{}r_\mathcal{M}\ra\approx1$) as
intended, however, there are still small region of slight active forcing and
bands of under-resolved regions (red regions of
$\la{}r_\mathcal{M}\ra>1$).  This is another artifact of time
averaging being used as a surrogate for the true expected value of
$r_\mathcal{M}$.  That is, small fluctuations in the puesdo-mean of
$r_\mathcal{M}$ result in local forcing activation when
$\la{}r_\mathcal{M}\ra$ is only slightly less than unity whereas there
is no counterpart when $\la{}r_\mathcal{M}\ra$ is slightly greater
than one.  The result, on average, is that excess energy is being
continually added to the small scales.  An \emph{ad hoc} modification
to the M43 model coefficient to remove this excess energy is described
in Appendix \ref{app:M43} (see Eq. \ref{M43_hack}), but, it appears to do
so too slowly.  The
effects of this under-resolution may contribute to the temporary disruption of the mean
velocity profiles in Fig. \ref{fig:poc}a.  Keeping in mind that the
steady-state $\beta$ for the coarse mesh only achieves a minimum of
about 0.6 (Fig. \ref{fig:ux_res}a), we see the resolved field
over-shoots this value to 0.4 at $t_f=1.6$.  Thus, as indicated by the
red $\la{}r_\mathcal{M}\ra>1$ bands in Fig.
\ref{fig:rd_ux}, more resolved turbulence has been added than can be
properly resolved.  The blunting of the mean velocity profile in the
center of the channel may be a result.  After this over-shooting, the field gradually
heals with the M43 modification removing excess energy.  By $t_f=8$,
the mean velocity has nearly attained its steady-state \S\ref{sec:stationary}.
% [trim=left bottom right top, clip]

Though we have not performed a rigorous spectral analysis of the resolved 
velocity field, it appears that the largest scales of turbulence are not excited 
until later in the simulation.  For instance, compare the streamwise oscillations 
at $t=0.4$ and $t=8.0$ snapshots of Fig. \ref{fig:rd_ux}.  Contrary to the 
intended forcing behavior, the largest turbulence scale should be 
excited at the start of the simulation, suggesting a potential deficiency in 
the current forcing structure.  This was alluded to in the previous remark about the wall-normal 
forcing structure being too small.  More interestingly, because the forcing is 
apparently not exciting such large scale motions directly, a transfer of energy 
to these large scales of motion is necessary.  While some of the energy may 
be coming from the excited turbulence scales, it may also be transferred 
directly from the mean.  A drain of energy from the mean would also be
consistent with the 
temporary blunting of the velocity profiles at early time.  The mean then recovers
through the action of the applied mean pressure gradient.  The structure of the
forcing at early time is plotted 
in several wall-parallel planes in Fig. \ref{fig:f_planes} in addition to the wall-normal plane shown in  Fig. \ref{fig:poc}.  The prescribed forcing field does result in large structures away from the wall, but only in the wall-parallel planes. Such granularity in the wall-normal direction forcing certainly contributes to the largest turbulence 
structures not being excited.  Correcting the prescribed 
early forcing structure would likely  help reduce the small distortion of the mean during 
the forcing. 

A potential remedy to the discussed under-resolution would be to allow the forcing field to also remove
energy in regions satisfying the conditions $\alpha<1$,
$\la{}r_\mathcal{M}\ra>1$, and $\mathcal{P}^{test}_F<1$
(\ref{Ptest}). Alternatively, and certainly more simply, the
coefficient $C_F$ (\ref{Ftar}) could be reduced so that the forcing is
more gradual and there is no healing time necessary to correct the
overshoot.  Drastic transitions
in time from RANS to grid-resolved LES would not actually occur in a
simulation other than during the initial startup or in some highly unsteady problem.  Therefore, such a
coefficient reduction may not be generally necessary for practical
application of the AMS method.  This viewpoint is supported by the
results of the next section.

\subsubsection{Channel with spatially evolving hybrid-state}
Next we consider the case of a
hybrid-state developing in space.  Unlike the previous case of
transition from RANS to LES in time, transition in space will occur in
many practical HRL simulations.  That is, one would like to construct
a mesh with LES-quality resolution only where needed, specify only
mean inlet velocity and scalar profiles with no resolved turbulence,
and rely on the HRL framework to transition from RANS to LES
without any loss of simulation fidelity.  Poor transitional behavior
may corrupt the entire solution downstream of the transition region.
Therefore, spatial transition is important to the stationary state
and needs to be tested.
%This critical spatial development has been 
%largely overlooked in the literature.  

\begin{figure}[th!]
\begin{center}
% ROW 1
\begin{overpic}
[trim={7.5cm 35.7cm 118cm 41.2cm},clip=true, width=0.45\linewidth]{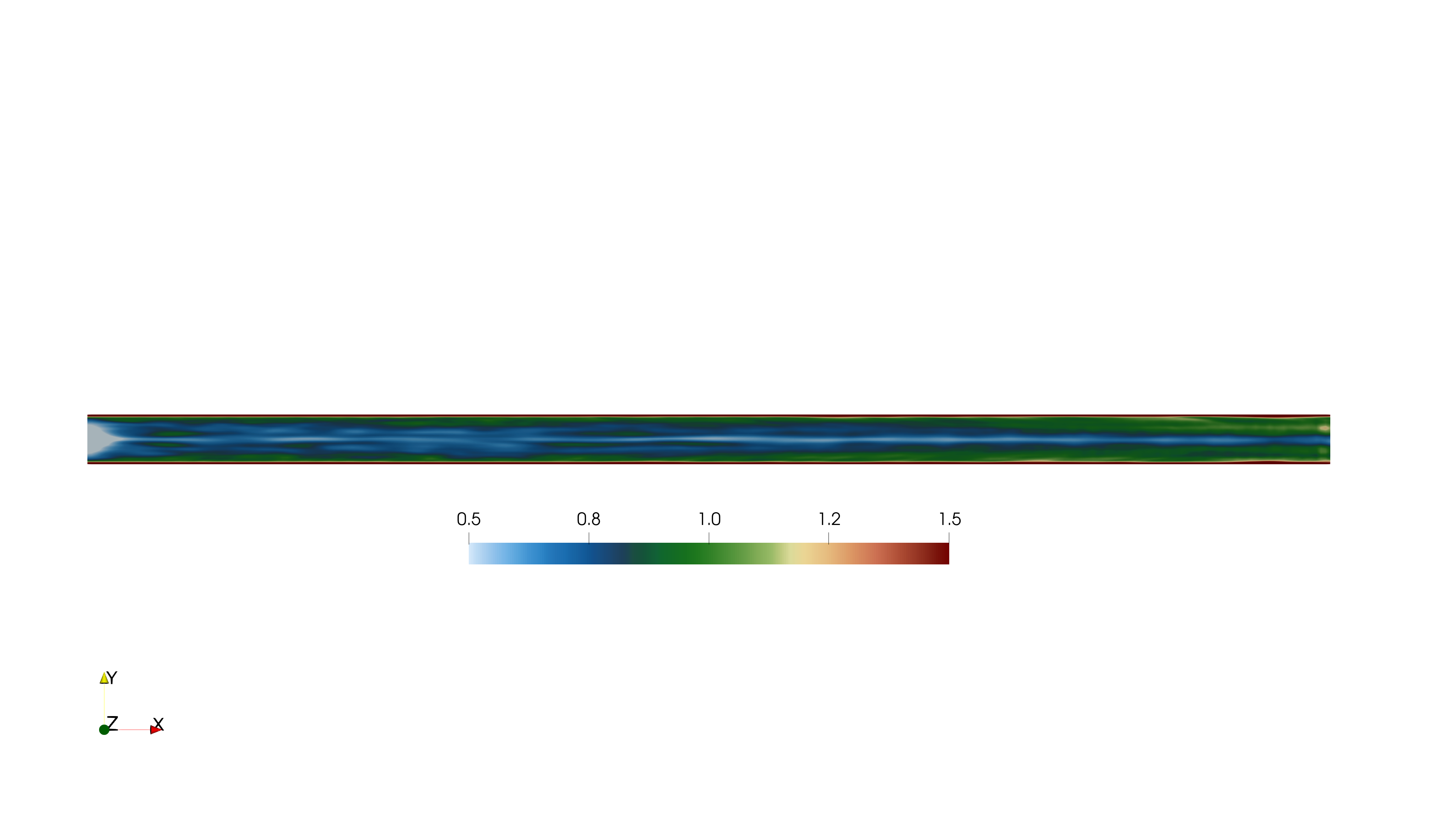}
\end{overpic}\hspace{0.2cm}
\includegraphics[trim={114.5cm 35.7cm 11cm 41.2cm},clip=true,width=0.45\linewidth]{figures/rM_sdev.png}\vspace{-0.3cm}
%
% ROW 2
%\vspace{-0.2cm}
\begin{overpic}
[trim={7.5cm 35.7cm 118cm 41.2cm},clip=true, width=0.45\linewidth]{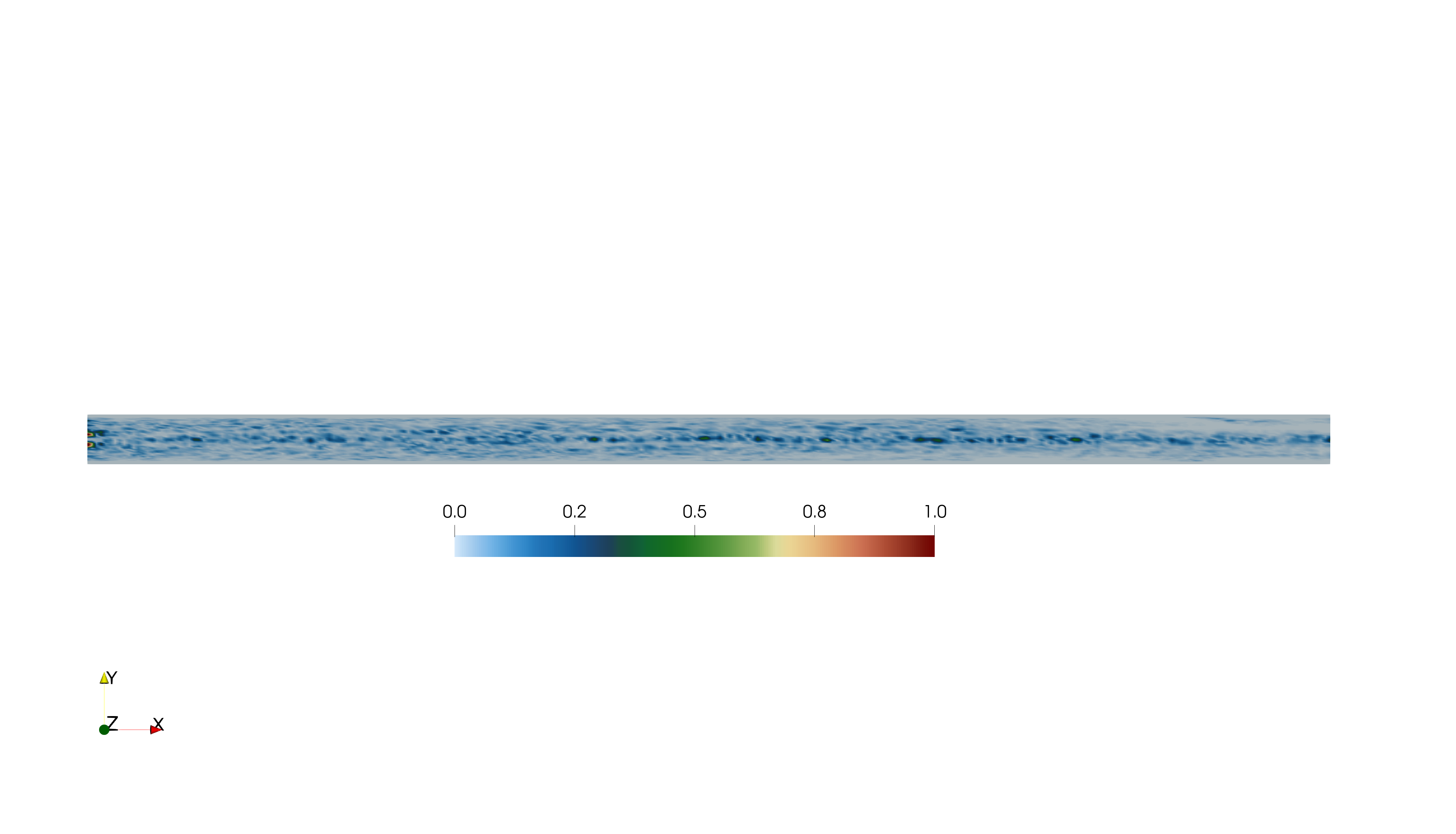}
\put(-5.5,-3.0){\color{black}\rotatebox{90}{\emph{Inflow: {R}{A}{N}{S} Dirichlet}}}
\put(202.5,7.0){\color{black}\rotatebox{90}{\emph{Outlet Face}}}
\end{overpic}\hspace{0.2cm}
\includegraphics[trim={114.5cm 35.7cm 11cm 41.2cm},clip=true,width=0.45\linewidth]{figures/Fmag_sdev.png}\vspace{-0.3cm}
%
% ROW 3
%\vspace{-0.2cm}
\begin{overpic}
[trim={7.5cm 35.7cm 118cm 41.2cm},clip=true, width=0.45\linewidth]{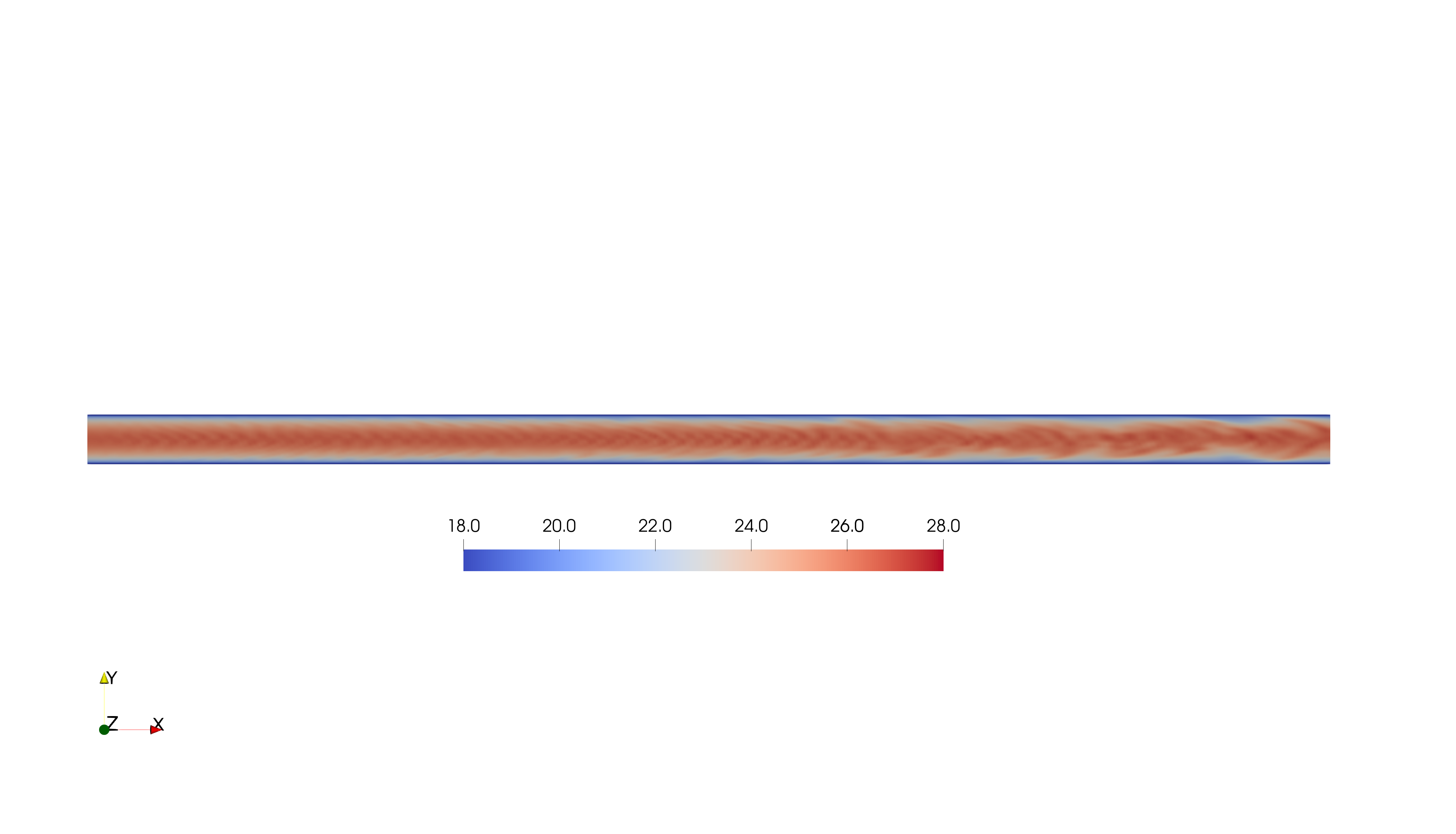}
\end{overpic}\hspace{0.2cm}
\includegraphics[trim={114.5cm 35.7cm 11cm 41.2cm},clip=true,width=0.45\linewidth]{figures/ux_sdev.png}\vspace{+0.2cm}
\hspace{-1cm}\begin{overpic}
[trim={18cm 5cm 25cm 26.5cm},clip=true,width=0.32\linewidth]{figures/rM_tdev2000.png}
\put(55.0,15.0){\color{black}\rotatebox{0}{$\la{}r_\mathcal{M}\ra$}}
%\put(11.0,10.0){\crule[white]{8cm}{0.4cm}}
\put(14.5,35.0){\color{black}\rotatebox{0}{0.5}}
\put(55.0,35.0){\color{black}\rotatebox{0}{1.0}}
\put(95.0,35.0){\color{black}\rotatebox{0}{1.5}}
\end{overpic}\hspace{-0.2cm} % 0.5 1.5
\hspace{-0.5cm}
\begin{overpic}
[trim={18cm 5cm 25cm 26.5cm},clip=true,width=0.32\linewidth]{figures/Fmag_tdev2000.png}
\put(55.0,15.0){\color{black}\rotatebox{0}{$|\hat{F}|$}}
%\put(11.0,10.0){\crule[white]{8cm}{0.4cm}}
\put(16.5,35.0){\color{black}\rotatebox{0}{0.0}}
\put(57.0,35.0){\color{black}\rotatebox{0}{0.5}}
\put(95.0,35.0){\color{black}\rotatebox{0}{1.0}}
\end{overpic}\hspace{-0.2cm} % 0.5 1.5
\hspace{-0.5cm}
\begin{overpic}
[trim={18cm 5cm 25cm 27cm},clip=true,width=0.32\linewidth]{figures/ux_tdev2000.png}
\put(55.0,15.0){\color{black}\rotatebox{0}{$\overbar{u_x^+}$}}
%\put(11.0,10.0){\crule[white]{8cm}{0.4cm}}
\put(14.0,35.0){\color{black}\rotatebox{0}{18}}
\put(30.5,35.0){\color{black}\rotatebox{0}{20}}
\put(46.0,35.0){\color{black}\rotatebox{0}{22}}
\put(62.3,35.0){\color{black}\rotatebox{0}{24}}
\put(78.5,35.0){\color{black}\rotatebox{0}{26}}
\put(94.75,35.0){\color{black}\rotatebox{0}{28}}
\end{overpic}\vspace{-0.9cm} % 18 to 28
\end{center}
\caption{Inlet (left) and outlet (right) sections of the spatially
evolving channel with coarse resolution. Shown are (top) pseudo-mean resolution adequacy parameter $\la{}r_\mathcal{M}\ra$, (middle) magnitude of the scale forcing $|\hat{F}|$, and (bottom) resolved streamwise velocity $\overbar{u_x^+}$. Inflow/outflow streamwise sections of $4h$, out of the total $16\pi$, are shown.}
\label{fig:sdev_cont}
\end{figure}

In this section, rather than using periodicity in the streamwise direction, a RANS 
inlet profile is specified.  Again using the resolution 
of the coarse case, the streamwise length of the domain is extended to $16\pi$ while 
maintaining the same span and height.  The inlet and outlet sections of the extended channel 
are shown in Fig. \ref{fig:sdev_cont} and mean velocity profiles, averaged over specific 
streamwise slices, are presented in Fig. \ref{fig:ux_res}b.

Contrary to the temporally developing case, no $\beta$ overshoot and no 
LES regions of $\la{}r_\mathcal{M}\ra>1$ are present.  By the end of 
the channel, the mean velocity and $\beta$ have nearly reached the 
coarse grid stationary hybrid-state.  There is still a small over-resolved 
core-region at the outlet of the channel. Further, the mid channel velocity 
is slightly lower than the DNS and periodic hybrid simulation.  Forcing is 
most active just at the inlet ($x=0$) but it is sustained at low amplitude
over the entire channel length.  This early region, $x<1$, produces  the largest relative 
drop in $\beta$.  Such a rapid addition in resolved turbulence without 
disruption to the mean is highly desirable for hybrid applications as it 
indicates that only small regions of LES resolution upstream of a flow
feature of interest are necessary.  

Unfortunately, the relatively slow drop in $\beta$ after $x=\pi/2$ indicate 
the current forcing formulation is not able to rapidly fill in all
resolvable scales.  This behavior seems to be particular 
to the spatial development case.  For instance, after $t=1.6$ in the 
time-developing channel, $\beta$ has dropped below $0.5$.  Based on the 
bulk velocity, the fluid has traveled just over $12\pi$ at this time. However, 
after traveling $12\pi$ from the RANS inlet in the spatial developing case, 
$\beta$ has only dropped to just under $0.7$.  Thus, it seems there are 
subtleties to the current forcing structure that relies on periodicity in the 
streamwise direction to be most effective.  Nonetheless, the AMS 
approach is successful in transitioning a hybrid state from RANS to LES 
in space.  This also indicates that the current $C_F$ value is not excessive 
for practical applications of AMS.

\subsubsection{High Reynolds number channel}
\respb{
To conclude our study of basic AMS in channel flow, we consider a much
higher Reynolds number, $Re_\tau=20,000$.  We have also exercised AMS
in lower Reynolds number cases ($Re_\tau=180$ and 
2000, not shown)  with results not substantively different than the $Re_\tau=5200$ cases 
discussed in the previous sections.
%Therefore, these results have been omitted.  
However, at higher Reynolds number, we can expect the outer kinetic energy peak 
to increase due to additional large scale turbulent structures away from the wall 
\cite{lee:2018}.%\todo{RDM: is this really the only or best reference
%for this?} \todo{SWH: was it not you two who showed the "zombie"
%structures are responsible for the outer peak?}
In general, RANS models do not represent this outer peak
well, yet they can still produce excellent mean velocity profiles with
a well-defined log-layer.  This is a result of the previously
mentioned error-cancellation in RANS models.  As indicated in
Fig. \ref{fig:ux_res}, the AMS hybrid formulation improves the
prediction of the outer $k$-peak, and one might suspect that this
improvement could disrupt the error cancellation in the RANS
representation of the subgrid Reynolds stress used in AMS and
thereby affect the mean.  At $Re_\tau=5K$ and the resolutions
considered, this improvement is small and the overall performance of
AMS is still excellent.
}
\begin{figure}[th!]
\begin{center}
\subfigure[Mean $\bar{u}^+_x$ and $\beta$]
{\includegraphics[trim={1.9cm 0cm 2cm 0cm},clip=true,width=0.45\linewidth]{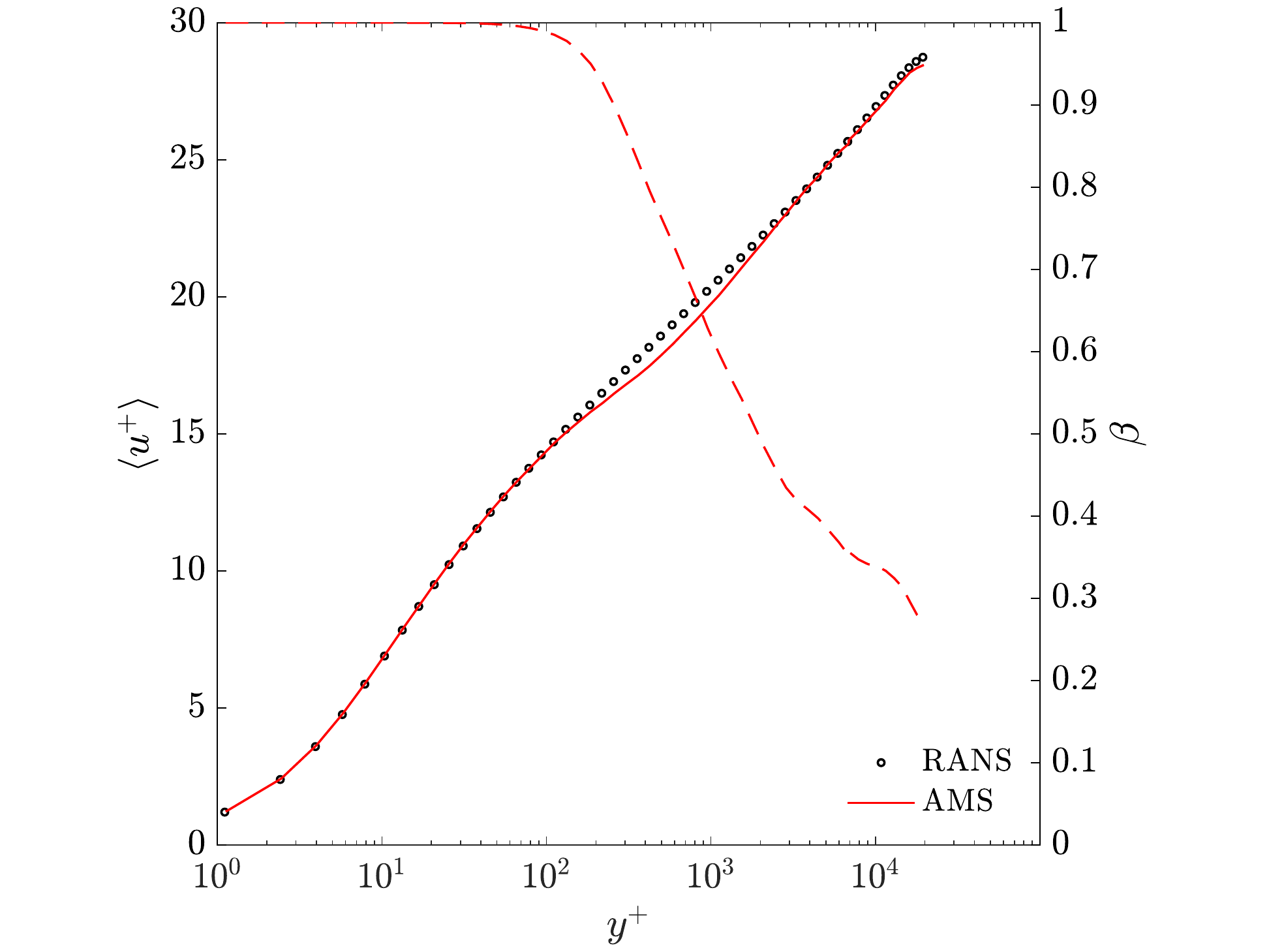}}
\label{muxhre}
\subfigure[Turbulent kinetic energy $k^+$]{
\includegraphics[trim={1.9cm 0cm 2cm 0cm},clip=true,width=0.45\linewidth]{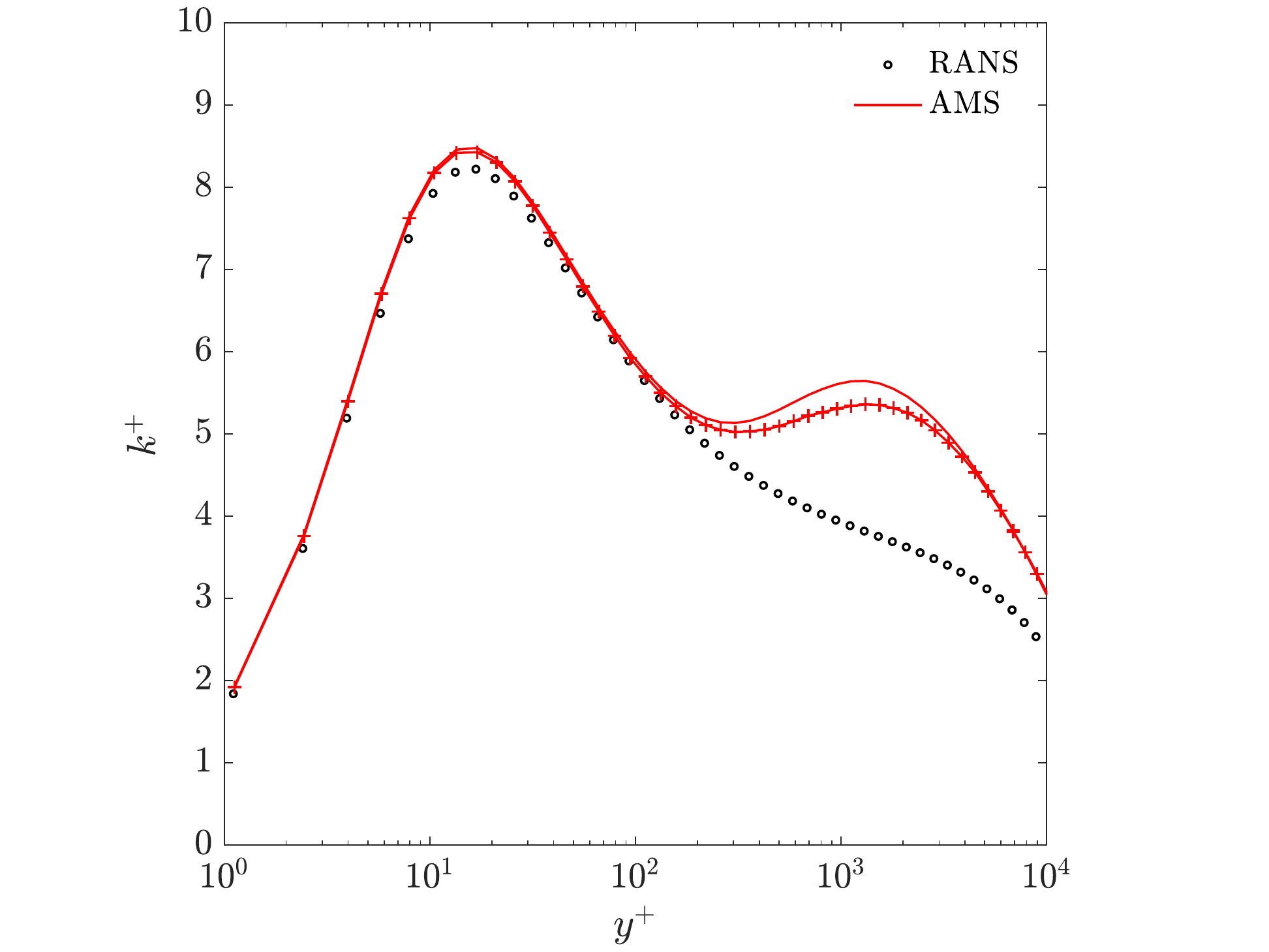}
\label{mtkehre}}
\end{center}
\caption{
%\todo[inline]{Vertical axis labels: $\la\bar{u}^+_x\ra$ ??, $k^+$}
Mean streamwise velocity in wall units (a) for fully
developed channel flow at $Re_\tau=20K$ along with the fraction
of unresolved turbulence $\beta$ (dashed), and (b) turbulent
kinetic energy $k^+$ of the statistically steady solution.  In (b),
lines marked with + symbols
are $k$ obtained directly from the hybrid RANS model while the
unmarked lines are the time-averaged resolved turbulence plus $\alpha$
times the RANS $k^+$.  Simulations have been run for nearly 100
convective times\label{fig:re20k}.}
\end{figure}
\respb{
The largest structures that are responsible for the increase in the
outer kinetic energy peak with increasing with Reynolds number, should
be easily resolved in AMS. So, at higher Reynolds number, AMS will produce
larger improvements in the outer kinetic energy peak, making a higher
Reynolds number channel simulation a simple test of whether improving the
representation of $k$ in AMS can degrade the performance of the RANS-based
subgrid Reynolds stress model.
}
\respb{
To this end, an AMS simulation was performed with the fine resolution of
Table \ref{tab:grids} at $Re_\tau=20K$ but with the first wall-normal grid point shifted to maintain $y^+=1$ for the higher $Re_\tau$.  Therefore, the wall-normal resolution in the center of the channel is more coarse than the $Re_\tau=5200$ fine resolution case.
%\todo{RDM: what about y-resolution, that must be finer near wall, right? Need to report.}
Chien's $k$-$\varepsilon$ model was found to perform poorly at this
Reynolds number in the near-wall to buffer region.  Since we do not
expect AMS to improve the simulation this close to the wall with such a grid, it was
necessary to use a different base RANS model for this AMS test.  While
there are no DNS or detailed experimental results available at $Re_\tau=20K$,
the $v^2$-$f$ model yields expected mean velocity and is used with AMS
here.
}
\respb{
Mean velocity and $k$ profiles from the $Re_\tau=20K$ simulation are
shown in Fig. \ref{fig:re20k}.  There is a depature from the log
profile from $y^+=200$ to $y^+=2000$.  In comparison with
Fig. \ref{fig:ux_res}, $\beta$ is lower in the center of the channel
and values below one extend to lower values of $y$ for the same
resolution.  This is a direct result of large turbulence scales which
are not present at lower Reynolds number and are resolvable at the
current resolution.  The largest deviation from the baseline RANS
profiles is in the outer-layer peak in $k$. This is so for both $k$
obtained directly from the hybrid RANS $k$ equation and for $k$
determined from the resolved fluctuations plus $\alpha$ times the
hybrid RANS $k$. While we cannot directly confirm how well the AMS
outer peak corresponds to the true peak, the existence of the peak is
already an improvement over the basic RANS model. As suspected, the
increase in $k$ relative to RANS affects the RANS model for the
subgrid Reynolds stress, resulting in a distortion of the mean. Note
that the departure starts where $k$ begins to increase relative to baseline RANS
and ends at about the location of the outer peak.  The problem with
the RANS model for the Reynolds stress is that the large scales
responsible for the outer layer peak in $k$ and for the majority of
the Reynolds stress away from the wall also extend closer to the wall,
where they do not carry significant Reynolds
stress \cite{lee:2018}. As a consequence, in this intermediate region
where the AMS mean velocity departs from expectations, $k$ is not a
good predictor of the Reynolds stress. In baseline RANS this is
mitigated by the compensating error that the outer layer peak in $k$
is absent. In AMS, the improved representation of $k$ will have two consequences: the hybrid RANS
estimate of the turbulent viscosity will be inconsistent; and $\beta$
will not adequately represent the contribution of the resolved
turbulence to mean momentum transfer.
}
\respb{
These results highlight the sensitivities of the AMS formulation to
the fidelity of the hybridized RANS model. To improve this fidelity,
it may be fruitful to further modify the hybrid RANS model to make
more use of knowledge of the resolved scales. Another possibility is
to reformulate the hybrid RANS model to represent only the subgrid
turbulence or to enrich the representation of the subgrid Reynolds
stress and its dependence on $\beta$. These are out of scope for the
current paper, but are important directions for future development.
}

\subsection{Periodic Hill}
\label{sec:phill}

In the previous channel cases, RANS models perform well, and the simulations 
demonstrate that the AMS hybrid framework does not disrupt this
performance as  
resolution varies and hybrid states evolve.  In this section, we move to 
a case where RANS is known to fail and evaluate whether the AMS formulation 
can improve over the RANS results.  Flow separation and smooth-wall reattachment are 
of great interest in many engineering applications.  However, RANS models 
tend to incorrectly predict the recirculation region, often delaying 
reattachment. This leads to erroneous prediction of critical quantities of interest 
such as lift and drag.  With localized regions of model deficiency, such flow 
scenarios are precisely where HRL should be of greatest benefit.  The periodic 
hill test case geometry consists of a channel with a 2D-hill which
produces a 
separation region of approximately three hill heights.  Having been studied both 
experimentally~\cite{rapp:2011,breu:2009} and numerically~\cite{breu:2009}, this case offers a rare combination of ``truth'' data as 
well as established resolution requirements for accurate simulation of the flow with 
existing LES models. The periodic hill is therefore an ideal test case
to assess the utility of the AMS framework. In what follows, we 
therefore present periodic hill simulation results using both a baseline RANS
model and an AMS hybrid formulation. %\todo{swh: lead-in needs work}
\begin{table}
\centering
\begin{tabular}{ c c c c c c c }
\hline
$Case$  & $N_x$ & $N_y$ & $N_z$ & $\Delta_x^+$ & $\Delta_z^+$ & $Reduction$ \\
\hline
\hline
Fine & 190 & 120 & 40 & 47 & 112 & 14 \\
Medium & 140 & 120 & 30 & 63 & 150 & 25 \\
Coarse & 95 & 110 & 20 & 95 & 225 &  62 \\
\hline
\end{tabular}
\caption{Resolutions used for periodic hill simulations at $Re_h\approx10K$ 
reported here. For all simulations $\Delta_y^+(\mbox{wall})\approx1$
at the bottom surface and $\Delta_y^+(\mbox{wall})\approx2$ at the
top, based on the wall shear stress at the top of the hill.  Note that
$\Delta_x^+$ is an average value as streamwise clustering in the
recirculation region is used on all grids.  Grid size reductions are
reported in the last column as the ratio of the WRLES \cite{breu:2009}
grid size to that of the AMS simulations.\label{tab:grids2}}
\end{table}

Flow over a periodic hill geometry at $Re_h = 10,600$ (where $h$ is the hill height), 
case UFR~3-30 (available: \url{http://qnet-ercoftac.cfms.org.uk/w/index.php/UFR_3-30_Test_Case})
in the ERCOFTAC database, is simulated.   The computational domain consists of 
a single peak-to-peak period of the periodic hill in the streamwise direction of 
dimensions $9.5h \times 3.035h \times 4.5h$.  Periodic boundary conditions are 
applied in both the streamwise and spanwise directions.  The domain is discretized 
as shown in Table \ref{tab:grids2} with plus units based on the friction velocity at the 
peak of the hill, or at $x=0$.  Simulation grids consist of approximately 0.9M,  0.5M, and 0.2M cells in total. 
The wall-resolved LES used for comparison \cite{breu:2009} uses over 13M cells.  
As in the channel, a streamwise body force (mean pressure gradient) is required
to maintain a constant bulk velocity. In this case it is applied only in the upper 
portion of the domain ($y>1$). This avoids a non-physical interaction of the forcing 
with the inclined hill surfaces.
%
%\todo{RDM: is this what is done in the LES? If so, we should say so.}
% they arent very clear about just how the forcing was done
The base RANS model used for the periodic hill is a slightly modified
version of the ``code-friendly''~\cite{lien:2001} $v^2$-$f$
model \cite{durb:1995} (appendix~\ref{app:rans}, Eq. \ref{Rf}).
Elliptic relaxation RANS models have the advantage of not requiring
wall distance or wall shear stress, making them attractive in
simulations of complex domains with unstructured grids.  The low-$Re$
$k$-$\omega$ model~\cite{wilcox:2006} would also be a potential
candidate for AMS application to complex geometries as it satisfies
the requirements of providing both a turbulent length and time scale
and the model $k$ captures near-wall and log-layer behavior.
% trim={left bottom right top}
\begin{figure}[th]
\begin{center}
\subfigure[RANS]
{\includegraphics[trim={4cm 18.4cm 4cm 19cm}, clip=true, width=0.43\linewidth]{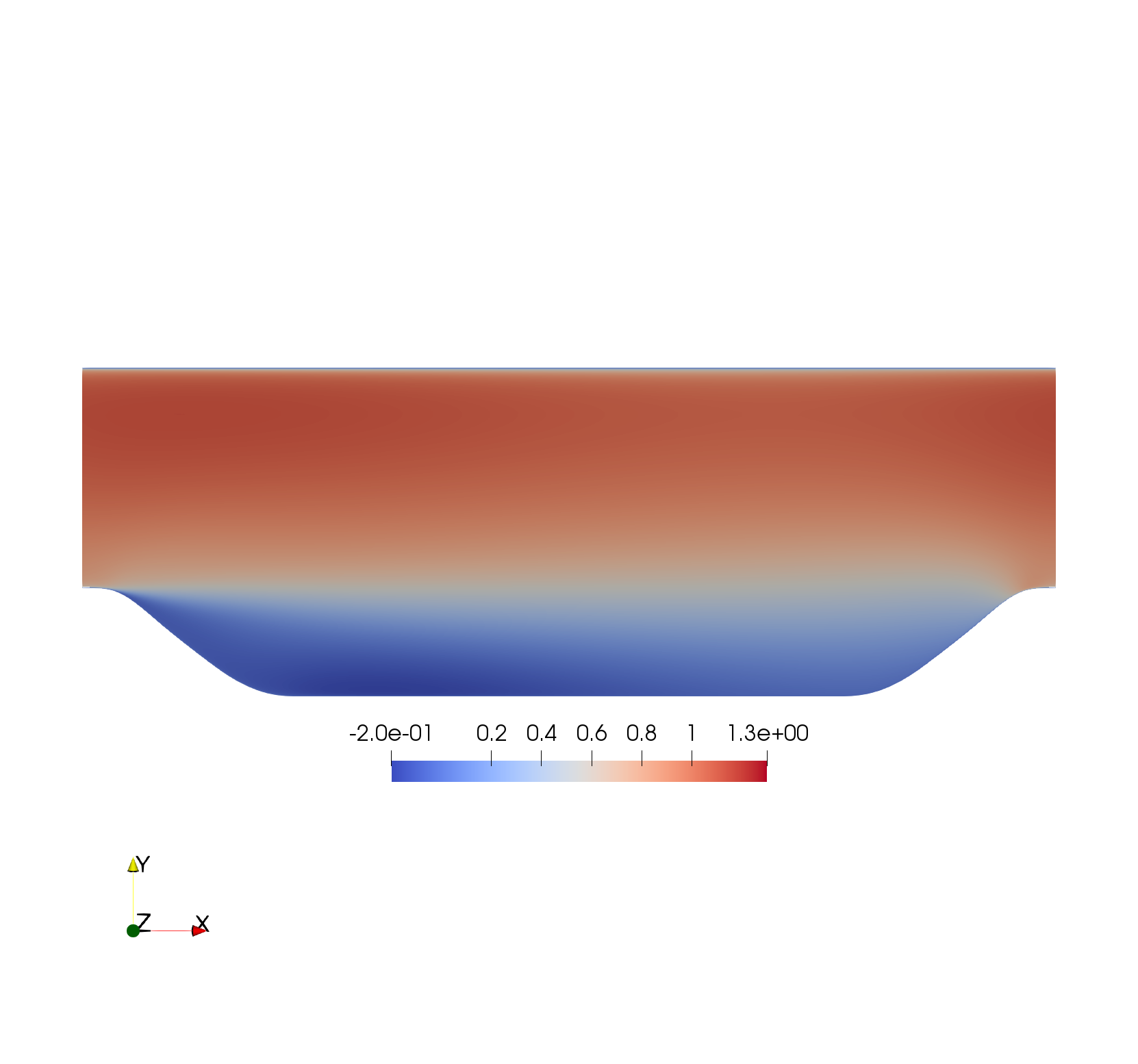}}
\subfigure[Coarse]
{\includegraphics[trim={5.5cm 12cm 5.5cm 12cm}, clip=true, width=0.45\linewidth]{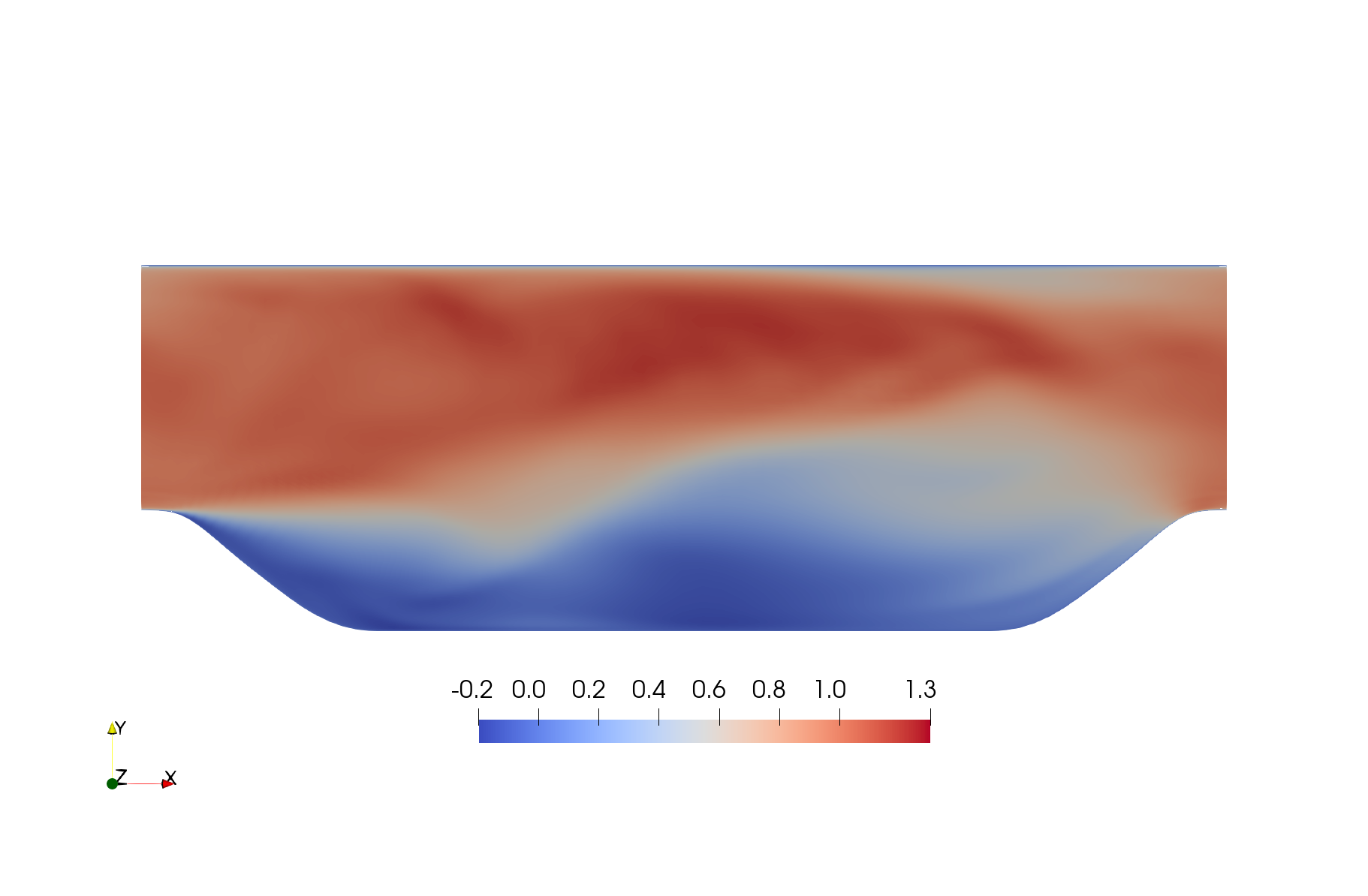}}\\
\subfigure[Medium]
{\includegraphics[trim={5.5cm 12cm 5.5cm 12cm}, clip=true, width=0.45\linewidth]{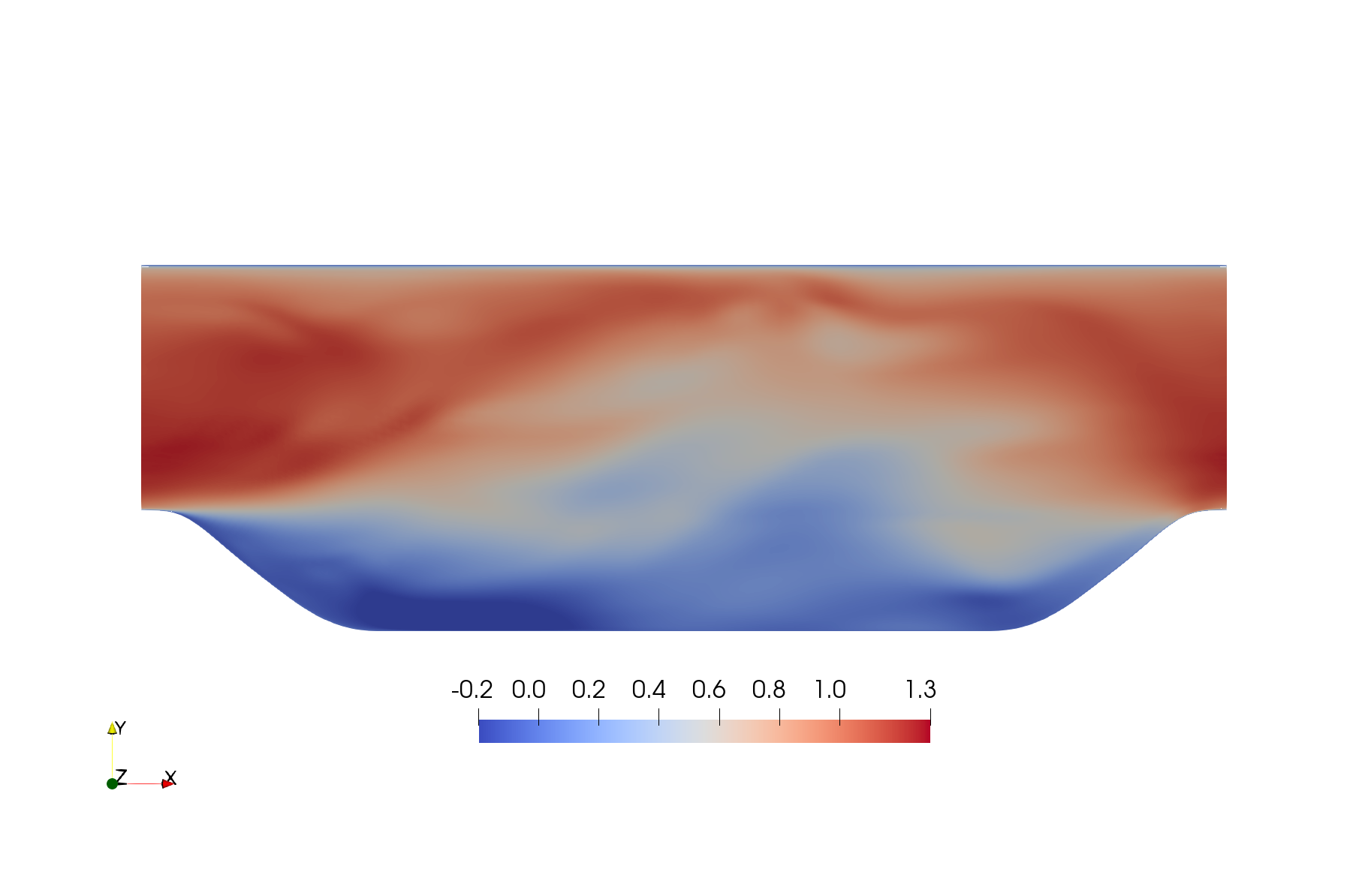}}
\subfigure[Fine]
{\includegraphics[trim={5.5cm 12cm 5.5cm 12cm}, clip=true, width=0.45\linewidth]{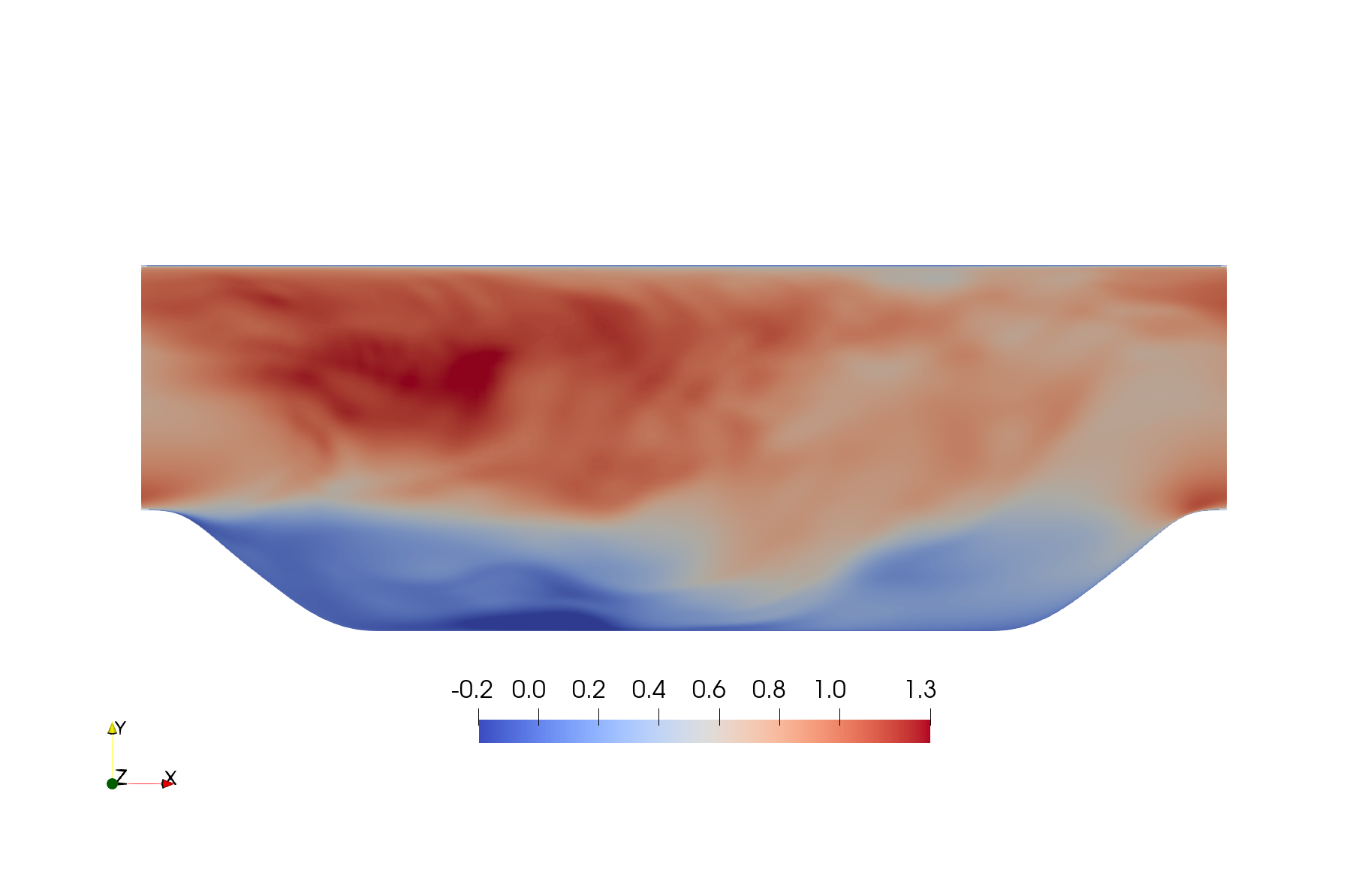}}\\
\begin{overpic} % this is just the colorbar
[trim={0cm 14cm 0cm 36cm}, clip=true, width=0.9\linewidth]{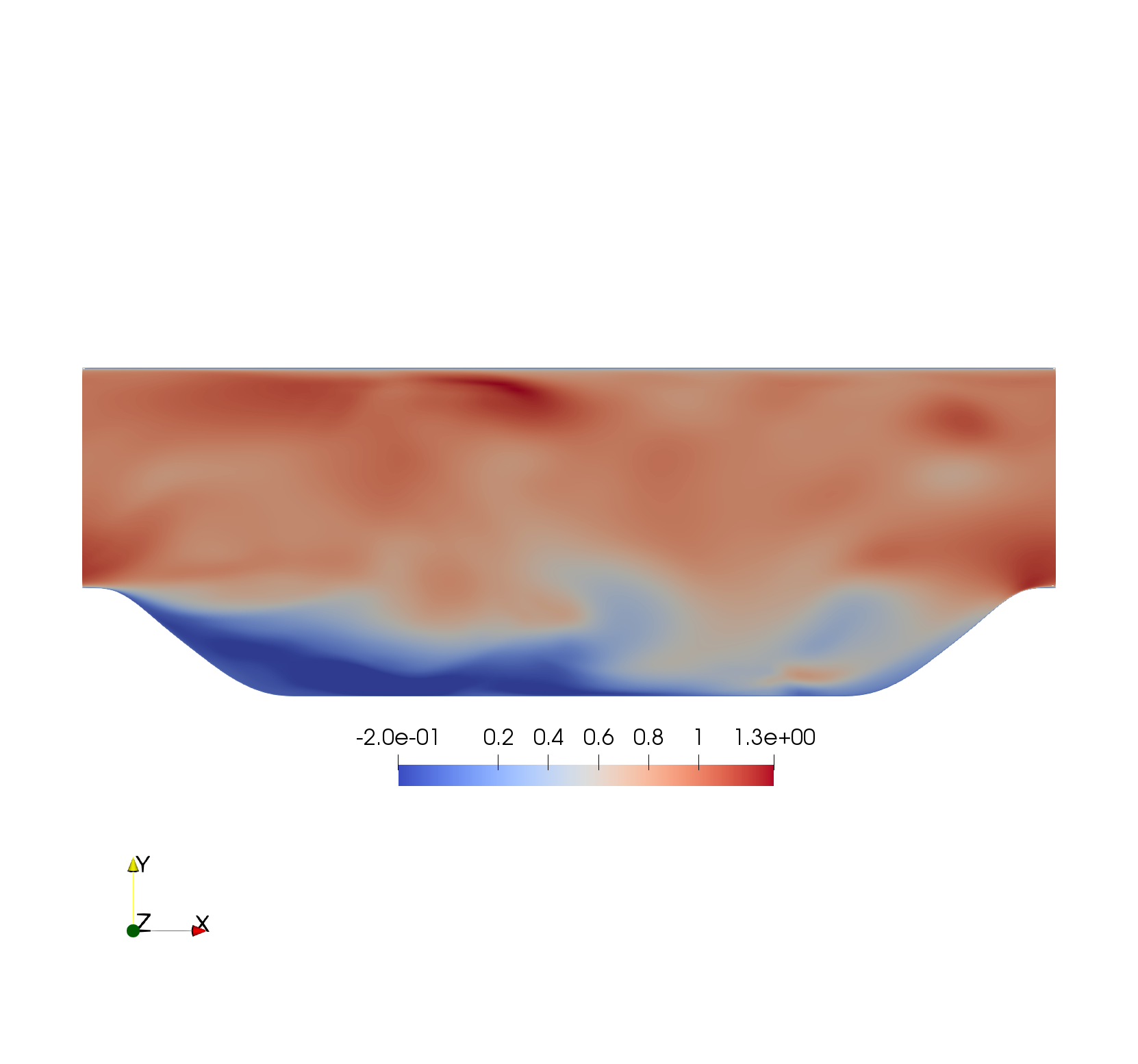} 
\put(28.0,3.5){\crule[white]{8cm}{0.5cm}}
\put(32.5,4.5){\color{black}\rotatebox{0}{-0.2}}
\put(42.0,4.5){\color{black}\rotatebox{0}{0.2}}
\put(51.0,4.5){\color{black}\rotatebox{0}{0.6}}
\put(59.5,4.5){\color{black}\rotatebox{0}{1.0}}
\put(66.1,4.5){\color{black}\rotatebox{0}{1.3}}
\end{overpic}
\end{center}
\caption{Streamwise velocity contours for the periodic hill test case
using (a) $\bar{v^2}$-$f$ RANS and (b-d) AMS using the coars through
fine grids of Table \ref{tab:grids2}.  Instantaneous AMS snapshots are shown
here to illustrate the degree of resolved turbulence, mean velocity
profiles for AMS are shown in Fig. \ref{fig:phill_ux_profsw}.}
\label{fig:phill_ux_cont}
\end{figure}
Contrary to the channel results, where profiles were simply averaged over planes at a 
fixed $y$-location, profiles presented here are obtained by averaging 
quantities (\S\ref{sec:ms}) using 50 samples over $20$ flow-throughs.  The 
pseudo-mean was found to be slowly fluctuating even after the simulation was 
brought to a quasi-steady state after 20 flow-throughs.  Further, the resolved stress 
as defined through the psuedo-mean excludes low-frequency structures (see \S\ref{sec:stress}).  
To enable calculation of the total Reynolds stress to compare with experiments, the 
resolved stress is calculated directly through $\la\bar{u}_i\bar{u}_j\ra-\la\bar{u}_i\ra\la\bar{u}_j\ra$.  Again, the simulation 
was initialized from the RANS solution.
\begin{figure}[h!]
\begin{center}
\includegraphics[trim={0cm 2.5cm 0cm 3.5cm}, clip=true, width=0.98\linewidth]{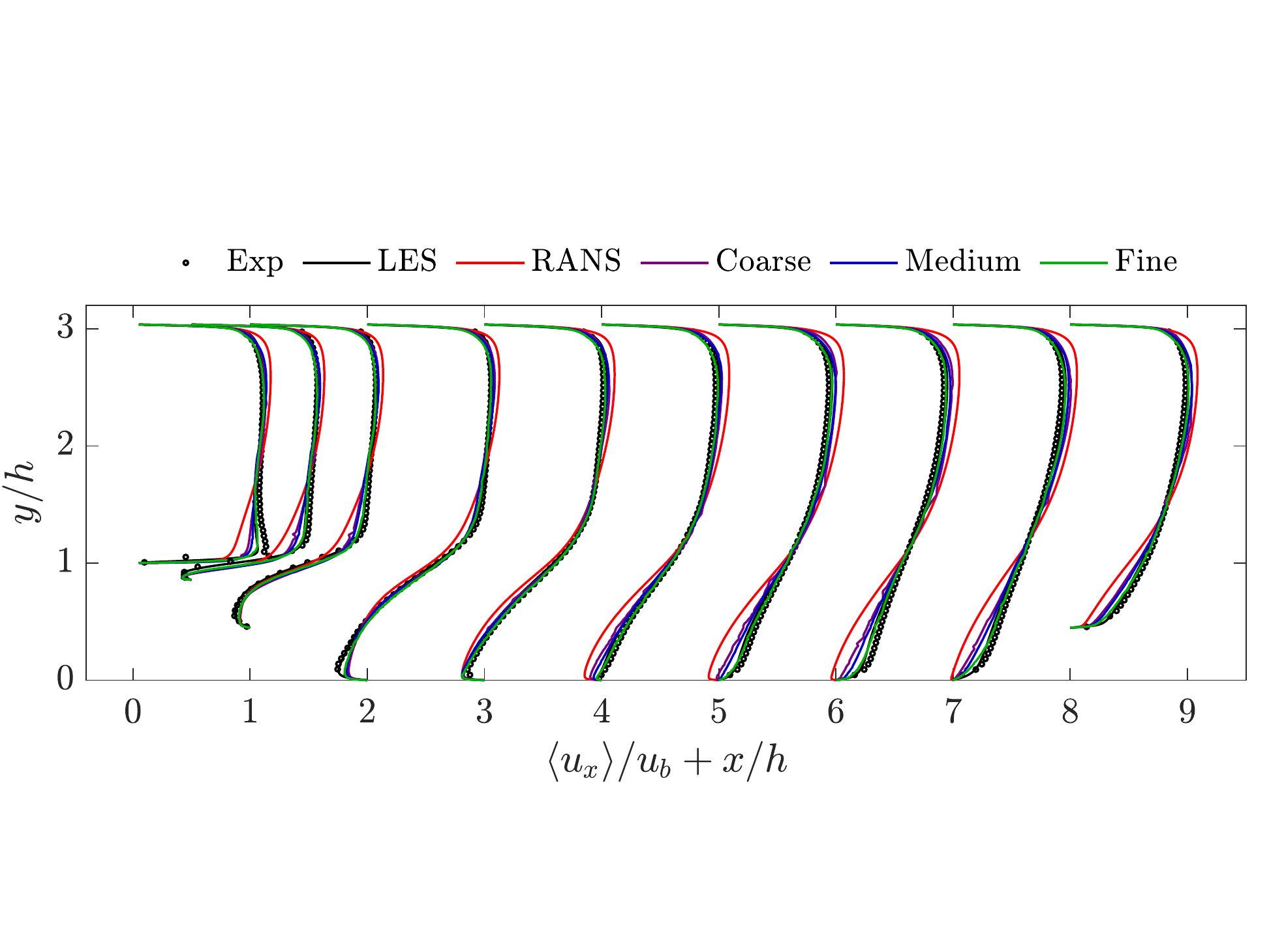}
\includegraphics[trim={0cm 2.5cm 0cm 4.5cm}, clip=true, width=0.98\linewidth]{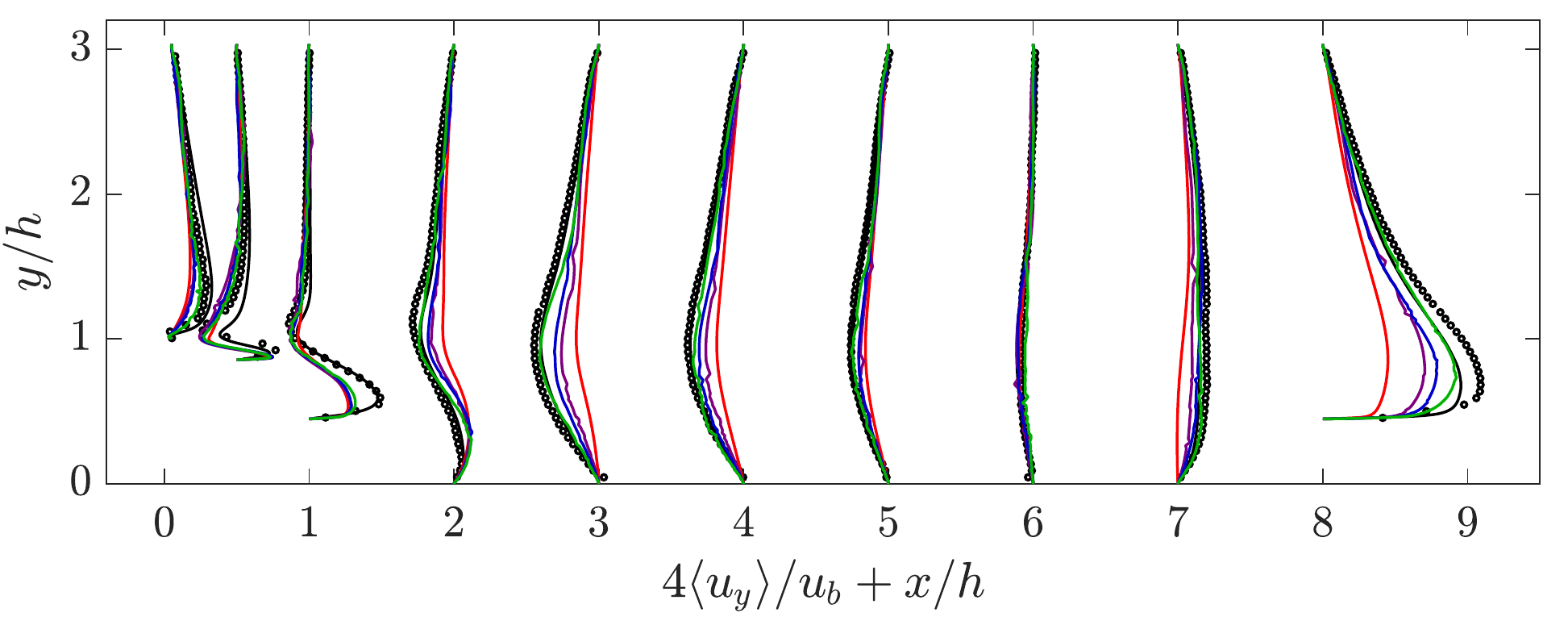}
\end{center}
\caption{
%\todo[inline]{Axis labels need to be explicitly normalized.}
Mean streamwise (top) and vertical (bottom) velocity
profiles normalized by the bulk velocity, $u_b$, offset by streamwise location normalized by the hill height, $h$, for all resolutions in
Table \ref{tab:grids2}, along with data from the experiments
of \cite{rapp:2011} and WRLES of \cite{breu:2009}.
%\todo[inline]{are these the right references?}
Note that the vertical
velocity is multiplied by a factor of four to enhance visibility.}
\label{fig:phill_ux_profsw}
\end{figure}

Figure~\ref{fig:phill_ux_cont} shows streamwise velocity contours for
both the baseline RANS $\bar{v^2}$-$f$ model and an instantaneous
field from hybrid simulations for all grids, illustrating the
scales of resolved structures in the hybrid simulation.  Comparisons
of experimental, LES, RANS, and AMS mean velocity profiles at various
streamwise locations are displayed in Fig.~\ref{fig:phill_ux_profsw}.
Clearly, AMS improves RANS predictions, matching experiments and WRLES
quite well for all resolutions considered.  Of primary importance to
aerodynamic applications is the reattachment point at $x \approx4$.  A
gradual improvement of the reattachment length is obtained with
increasing resolution for AMS while the RANS reattachment is delayed
downstream to approximately $x=7$.  There is little difference in
reattachment length between the medium and fine resolutions,
indicating \respa{consistency of the hybrid
solutions once sufficiently high resolution is achieved.}  In addition to the reattachment,
the hybrid simulations also improves the streamwise velocity profile
nearly everywhere in the domain. In particular, a drastic improvement
is seen in the near-wall region at the top of the hill and everywhere
after the reattachment point on the bottom wall.  The fine resolution
considered here yields further improvements near the wall 
and just past the reattachment point.  Similar improvements to the mean vertical
velocity profiles with increasing resolution are also observed.  Though the vertical velocity is
improved over nearly all the streamwise locations, at $x=0.5$ and
$1.0$, the hybrid results are nearly identical to RANS.  The likely
reason for this is apparent from examination of the hybrid state in
this region.
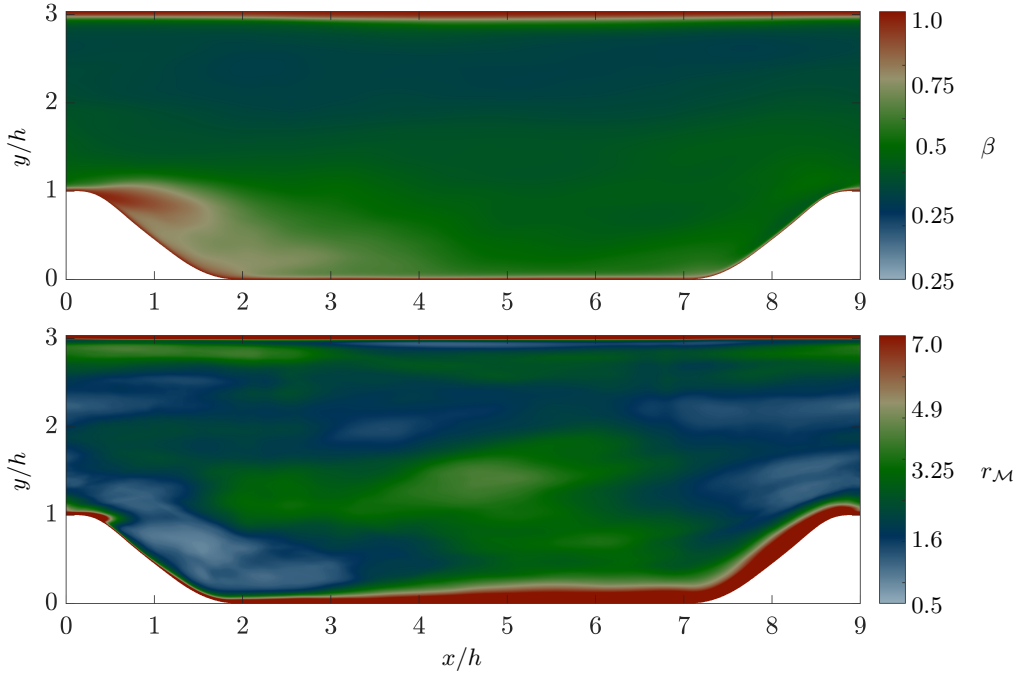
\begin{figure}[th!]
\begin{center}
\begin{overpic}
[trim={0cm 3.7cm 0.7cm 4cm}, clip=true, width=0.70\linewidth]{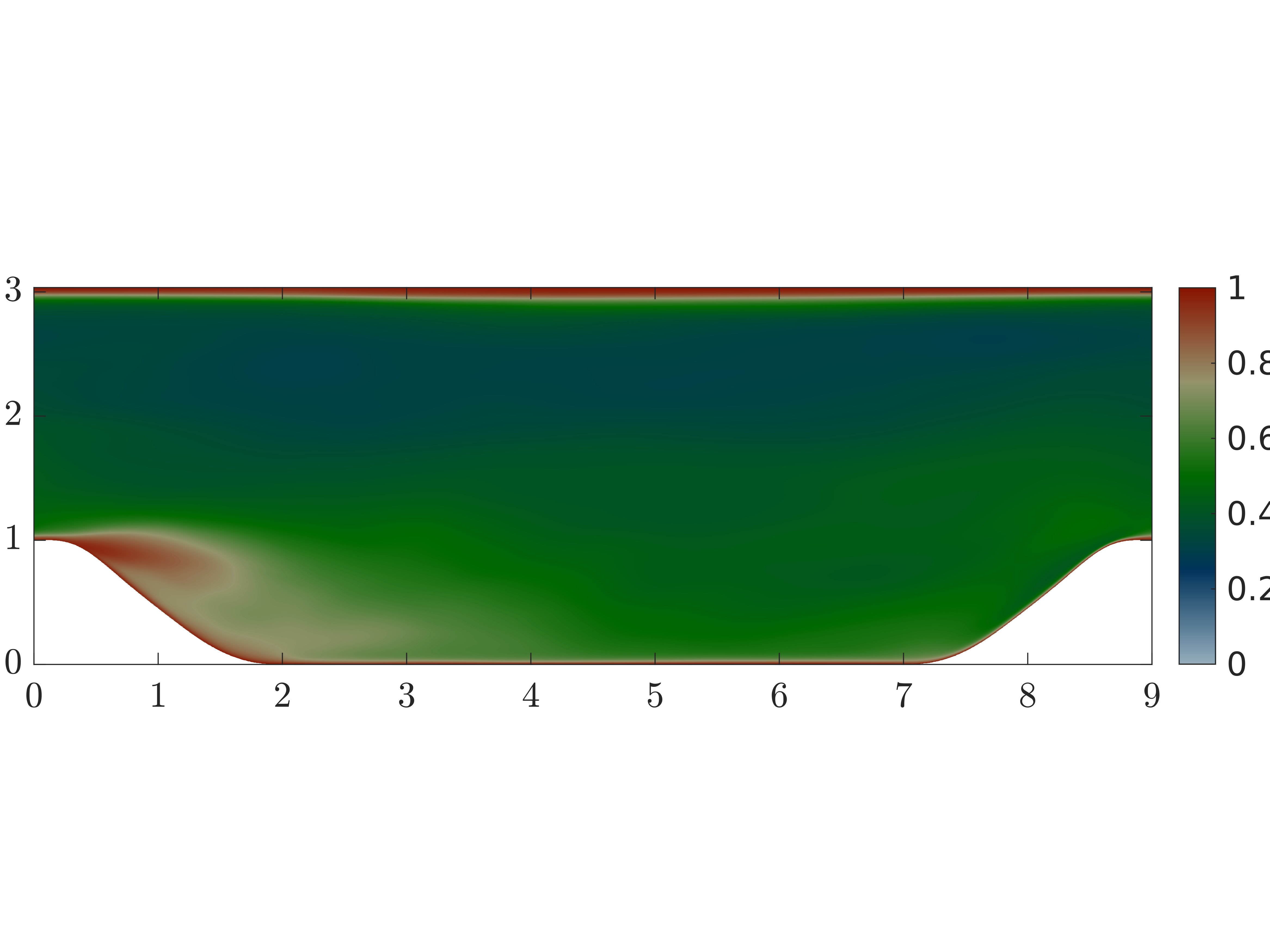}
\put(-4.0,17.0){\color{black}\rotatebox{90}{$y/h$}}
\put(108.0,18.5){\color{black}\rotatebox{0}{$\beta$}}
\put(100.0, 33.0){\color{black}\rotatebox{0}{$1.0$}}
\put(100.0,25.5){\color{black}\rotatebox{0}{$0.75$}}
\put(100.5,18.5){\color{black}\rotatebox{0}{$0.5$}}
\put(100.0,10.5){\color{black}\rotatebox{0}{$0.25$}}
\put(100.0,3.0){\color{black}\rotatebox{0}{$0.25$}}
\end{overpic}
\begin{overpic}
[trim={0cm 3.7cm 0.7cm 4cm}, clip=true, width=0.70\linewidth]{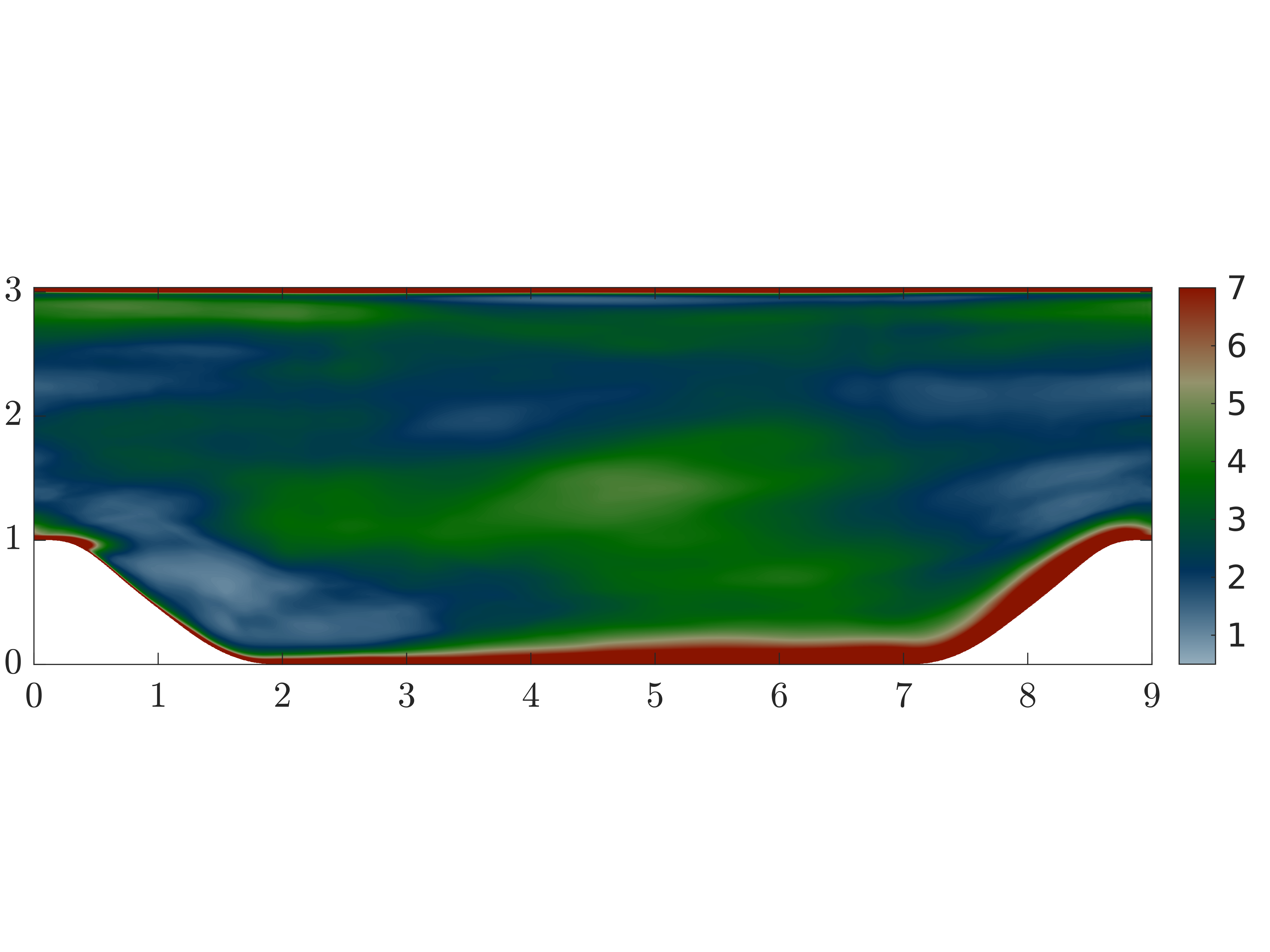}
\put(-4.0,17.0){\color{black}\rotatebox{90}{$y/h$}}
\put(108.0,18.5){\color{black}\rotatebox{0}{$r_\mathcal{M}$}}
\put(100.0, 33.0){\color{black}\rotatebox{0}{$7.0$}}
\put(100.0,25.5){\color{black}\rotatebox{0}{$4.9$}}
\put(100.0,18.5){\color{black}\rotatebox{0}{$3.25$}}
\put(100.0,10.5){\color{black}\rotatebox{0}{$1.6$}}
\put(100.0,3.0){\color{black}\rotatebox{0}{$0.5$}}
\put(46.0,-3.0){\color{black}\rotatebox{0}{$x/h$}}
\end{overpic}
\end{center}
\caption{Mean contours of the unresolved turbulence energy fraction $\beta$ (top)
and resolution adequacy $r_\mathcal{M}$ (bottom) for the fine resolution simulation.}
\label{fig:phill_beta}
\end{figure}  

Spatial distributions of the fraction of unresolved turbulence energy
and the resolution adequacy parameter are shown in
Fig.~\ref{fig:phill_beta} for the fine grid case.  As expected,
$\beta=1.0$ in the RANS regions near the walls.  In the large hybrid
regions throughout most of the domain $\beta\approx0.5$ with a minimum
of about 0.3 near the top of the channel.  In the recirculation
region, $\beta$ increases to about 0.7 and very nearly unity in the
actual separation shear layer.  This is a clear indication that the
fine resolution used here is not sufficient to resolve the fine
near-wall structures that propagate into the shear layer.  This is the
likely cause of the poor vertical velocity predictions in this region.
However, the resolution used here is sufficient to improve
representation of all other flow
characteristic, in particular the separation, indicating resolving
fairly large structures is all that is generally necessary for good
prediction of this flow.  That is, representing the mixing of high-momentum upper
channel fluid with the recirculation region requires only coarse
resolutions.

As expected, $r_\mathcal{M}$ is very high in RANS regions near the wall.  
Interestingly, $r_\mathcal{M}$ varies significantly in hybrid regions from just below 
unity in the recirculation regions, to approximately unity above the hill, to well above 
unity ($r_\mathcal{M}\approx4$) in the center of the channel.  As discussed in 
\S\ref{sec:forcing}, hybrid regions of $r_\mathcal{M}>1$ indicate locally 
under-resolved turbulent structures.  Perhaps a combination of the
resolved structures in this region being more three-dimensional than
those around the hill, coupled with the fact that the spanwise grid
spacing is coarser than the streamwise and wall-normal resolution,
lead to this region of under-resolution.  However, this does indicate
that the modification to the M43 model (see Appendix \ref{app:M43},
Eq. \ref{M43_hack}) is not generally sufficient to remove resolved
turbulence as it is convected into under-resolved regions.  This
suggests that introducing additional terms to the resolution adequacy
parameter that are sensitive to the convective gradient of $\beta$
or $\la{}r_\mathcal{M}\ra$, may be necessary.  Fortunately, the mean flow
seems to be tolerant of these under-resolved regions, as we see
no local disagreement between the hybrid results and experiments in
the mean velocity profiles. Perhaps, this insensitivity is indicative of
$C_r$ in (\ref{rM}) being overly restrictive. We will evaluate this
hypothesis in the next section.  The local over-resolution in
the recirculation region leads to continuous local forcing of
resolved structures.  The observed distribution of
$r_\mathcal{M}$ indicates that, for complex geometries and flows, the
convection of resolved turbulence leads to a \emph{continuous} need to
locally remove excess resolved turbulence in some regions while also
having to add resolved fluctuations in others.
 %
 %RDM removing Reynolds stress plots
%\begin{comment} trim=left bottom right top
\begin{figure}[H]%h!
\begin{center}
\includegraphics[trim={0cm 2.5cm 0cm 3.5cm}, clip=true, width=0.98\linewidth]{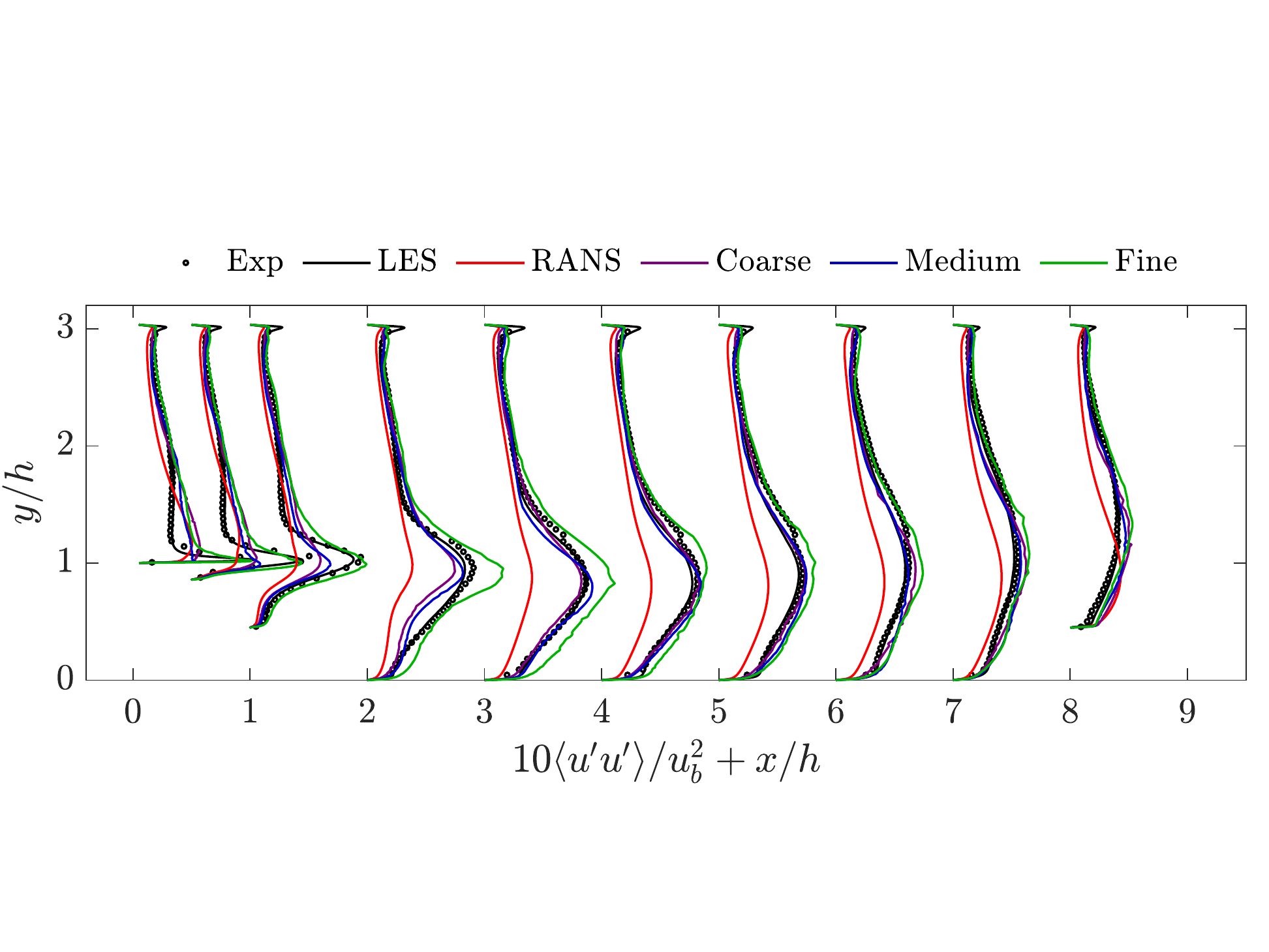}
\includegraphics[trim={0cm 2.5cm 0cm 4.5cm}, clip=true, width=0.98\linewidth]{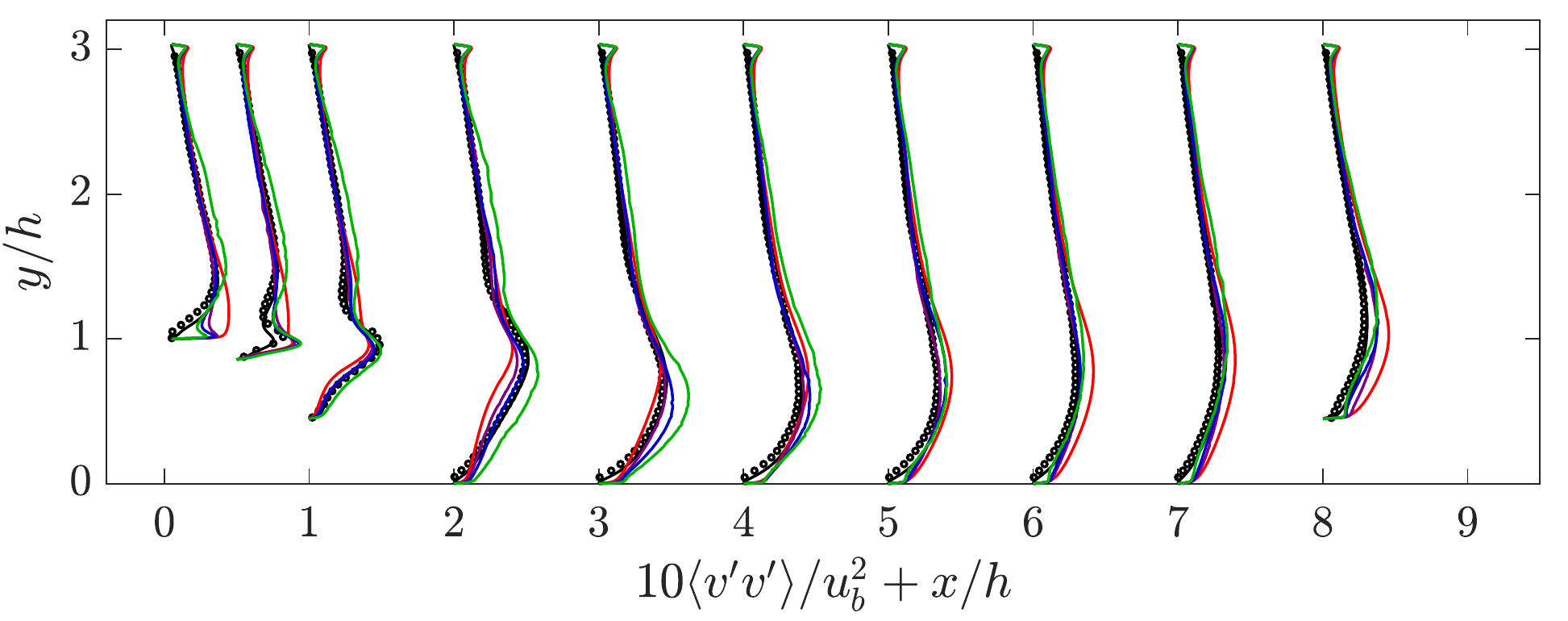}
\includegraphics[trim={0cm 2.5cm 0cm 4.5cm}, clip=true, width=0.98\linewidth]{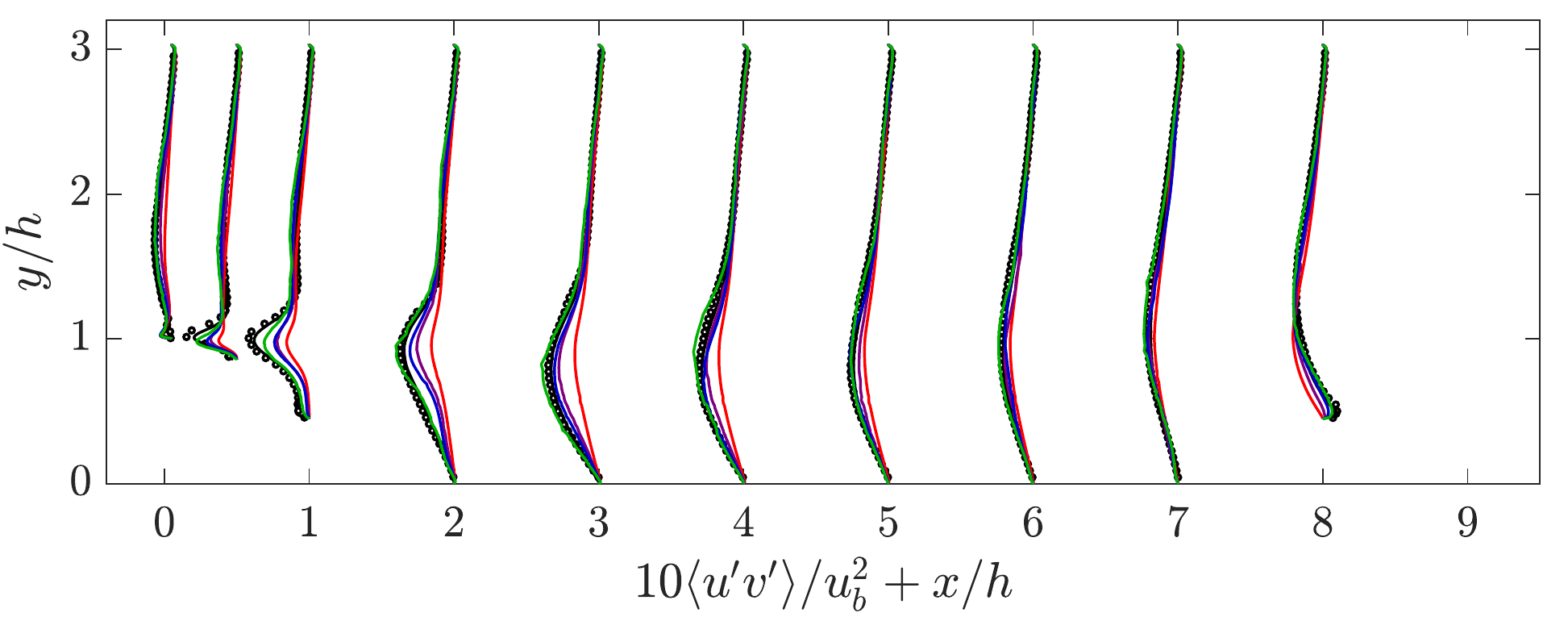}
\end{center}
\caption{
%\todo[inline]{Axis labels need to be explicitly normalize}
Profiles of the mean total (model plus resolved) Reynolds stress 
components normalized by the bulk velocity, $u_b$, offset by streamwise location normalized by the hill height, $h$, for all resolutions in
Table \ref{tab:grids2}, along with data from the experiments
of \cite{rapp:2011} and WRLES of \cite{breu:2009}.
%\todo[inline]{are these the right references?} Shown are the streamwise velocity 
variance (top) vertical velocity variance (middle) and Reynolds shear
stress (bottom).  The prime here indicates the total fluctuations,
that is, using fluctuations relative to the true mean.  Note
amplification by a factor of 10 for visibility.}
\label{fig:phill_uv_profs}
\end{figure}
%\end{comment}

\begin{comment}
An exception to the general improvements in the mean velocity is
visible in the vertical mean velocity profiles at the 
$x=0.5$ and $x=1.0$.  It is 
curious that the streamwise velocity profiles would be so much better
than RANS in 
these regions while the vertical velocity is worse than for basic RANS.  
From the $\alpha$ plot, it is evident that this region is the least resolved 
hybrid state with $\alpha\approx0.7$.  The resolution adequacy in this 
region of the separated shear layer is larger than unity (not shown)
\todo{swh: should I make this plot? i should probably make this
plot... RDM: yes, please make this plot} indicating 
locally under-resolved turbulence.  This is a consequence of well-resolved 
turbulence at the top of the hill being convected into the under-resolved 
shear-layer region.  It is a more extreme, but similar problem to 
the overshoot seen with the temporally evolving hybrid-state in channel 
flow (\S\ref{fig:rd_ux}).  Given that this overshoot is driven by convection of turbulence, 
the modification to the M43 model (\ref{M43_hack}) cannot act rapidly enough to remove the 
excess turbulence\todo{update}. Thus, as previously suggested, it may be beneficial to 
have the forcing term also active in such areas to remove resolved
turbulence energy.  However, such errors suggest introducing additional terms to 
the resolution adequacy parameter, which are sensitive to the convective 
gradient of $\alpha$ or $\la{}r_\mathcal{M}\ra$, may be necessary.  Such additions will be the subject of 
future work.
\end{comment}

Despite the minor discrepancies described above, the AMS formulation
yields nearly identical results to the wall-resolved LES
of \cite{breu:2009} at a resolution reduction of over an
order-of-magnitude and computational costs reduced by a factor of 30
or more due to larger timesteps.  These improvements are a direct
result of the model-split formulation being capable of providing
improved modeled stress predictions for arbitrary levels of resolved
turbulence.
%RDM  Remove Reynolds stress plots
% SWH Adding back with new results
%\begin{comment}
This claim is supported by examining components of the Reynolds stress
tensor in Fig. \ref{fig:phill_uv_profs}.  AMS stress profiles
generally show large improvements over RANS.  However, some hybrid
stress components become excessive.  For instance, the fine resolution
results in excesses of $\la{}u^\prime{}u^\prime\ra$ and
$\la{}v^\prime{}v^\prime\ra$ in the recirculation region.  The
aforementioned sustained local addition of artificial fluctuations in
the recirculation region may contribute to these
disagreements.  \respb{Another possibility is the improved production
used in the RANS model (\ref{Pk}) which enhances production of TKE in
the separated shear layer from $x=0$ to $x=1$. While correct, in the
context of the RANS system and its numerous modeling limitations this may
cause an excess in TKE and the observed excesses
in both normal stress components in the vicinity of the shear layer at
$x=2$, which then diffuses downstream.
%RDM this seems overly speculative, and is not necessary.
%Such an issue would suggest we
%should sensitize the dissipation equation to $\alpha$ in addition to
%the production, which is not surprising.
This is another example of the limitations imposed by the RANS model on the
fidelity of the TKE.}  Shear stress,
$\la{}u^\prime{}v^\prime\ra$, is uniformly improved with the fine
resolution only deviating from the experimental values in the
separated shear layer just after separation at $x=1$.

Of course, the periodic hill has been simulated successfully by
several other hybrid and bridging frameworks. Successful simulations
have been performed using DDES\cite{sari:2007}, PANS\cite{razi:2017},
and PITM \cite{chao:2010} all at approximately 1M cell counts using
codes with second-order numerics.  \respb{AMS and these other
hybrid methods are clearly more cost-effective than WRLES
(Table~\ref{tab:grids2}). Note that at worst only a factor of 2 more
equations are solved in AMS, due to the inclusion of RANS, than in the dynamic Smagorinsky WRLES
of \cite{breu:2009}, while obviating the dynamic procedure, making AMS on even the finest grid at least an order of
magnitude less costly. 
But, it is appropriate to ask how the cost of AMS would compare to
a WMLES of the same problem.  Assuming that the wall-model is valid in
the separation, recirculation, and reattachment regions and that it
does not require a separate highly-refined grid~\cite{park:2014}, we
can estimate the cost for a WMLES based on resolution requirements for
boundary layers \cite{choi:2012,lars:2016} in which the streamwise, spanwise
and wall normal grid sizes should be no larger than 8\%, 2\% and 5\% of the
boundary layer thickness, respectively. 
%$(\Delta_x/\delta, \Delta_y(wall)/\delta, \Delta_z/\delta)\approx(0.08,0.02,0.05)$
%where $\delta$ is the local boundary layer thickness \cite{choi:2012,
%lars:2016}.
% and by assuming the wall-model both does not require a separate highly-refined grid \cite{park:2014} and that the wall-model is valid at the separation, recirculation, and reattachment regions.
The most restrictive point is the top of the hill, where
$\delta_{99}\approx0.08h$, and for a structured grid, this will set
the number of grid cells in the wall-normal and spanwise directions
throughout the domain. Streamwise spacing, however, can adjust to the
local boundary layer thickness, which is so small for only about
$0.5h$ in the streamwise direction. Assuming that $\delta_{99}\approx
1h$ in the remainder of the domain yields a total required grid of
approximately 12M cells, which is only slightly smaller than that
required in WRLES \cite{breu:2009}. The primary driver of this large
cell count is the spanwise resolution requirement at the top of the
hill which, due to the structured grid, is used throughout the
domain. If one could use a grid in which the spanwise and wall-normal
resolution varies in the streamwise direction, then the WMLES grid
could be reduced to about 1.6M cells, which is still 
larger than the fine AMS grid in Table~\ref{tab:grids2}. However, this
analysis greatly understates the cost reductions of HRL relative to
WMLES for the types of external flows in which HRL are likely to be
applied (e.g. an airfoil or even aircraft). When applied to such
flows, a HRL can represent the turbulent boundary layer over most of
the body with RANS, requiring very coarse LES resolution only in
critical regions such as near separation. In a WMLES, however, LES
resolution is needed for all turbulent boundary layers, which will be
orders of magnitude more expensive.}

\begin{comment}
However, this is only appropriate over about $0.5h$ of the streamwise
direction.  For the remainder of the domain, we use
$\delta_{99}\approx1h$.  With a structured grid, the spanwise and
wall-normal spacing is fixed by the most restrictive point while the
streamwise spacing would be free to coarsen away from the top of the
hill.  Thus, we estimate
$(\Delta_x(hill)/\Delta_x, \Delta_y(wall), \Delta_z) \approx(0.0064/0.08,
0.0016, 0.004)$, with $5\%$ of the grid using $(\Delta_x(hill)$ and
the remainder using $\Delta_x$, which leads to $(N_x, N_y, N_z) =
(190, 60, 1125)$ or $N_{total}\approx{}12M$ cells. Interestingly, this
estimate is only slightly below the WRLES \cite{breu:2009} and is due
to the very restrictive spawise grid spacing imposed on the entire
domain by the top of the hill.  If we assume an overset grid is being
used, the spanwise spacing can additionally be restricted to only
being used around the hill and top wall to arrive at a much smaller
grid spacing of $N_{total}\approx{}1.6M$ cells.  Thus, while the
wall-normal grid spacing is indeed less expensive, the spanwise and
streamwise grid spacing cause WMLES to be significantly more expensive
than HRL for this problem, unless overset grids and codes are used.
Similar comments are likely to hold for other problems with a
two-dimensional mean.}\todo{TAO: Did edit to last sentence expand your
meaning too far Sigfried?}
\end{comment}

Caveat the slightly excessive normal stress components just after the the separation, the reduced grid sizes 
necessary with AMS are encouraging.  In addition to revealing an avenue for 
formulation improvement, this test case has shown the ability of AMS models to 
accurately predict complex flow features at reasonable 
computational costs. The tolerance of under-resolution (Fig. \ref{fig:phill_beta}) indicates that perhaps 
even more coarse resolutions could be used by relaxing the $r_\mathcal{M}$ 
requirements.  We examine this possibility next.

\begin{comment}

ALL RELATIVE TO TOP OF HILL
dy_wall = 1e-3
dy_wall+ = 1
nu = 9.438414e-5
delta_99 = 0.08
delta_h = 1

u_tau = (y+/y) * nu = 0.09438414
tau_wall = u_tau^2 * rho = 0.00890836588

LARSSON RECS
(?x/?, ?yw/?, ?z/?) ? (0.08, 0.02, 0.05)

based on d99
(?x, ?yw, ?z) ? (0.0064, 0.0016, 0.004)
(Nx, Nyw, Nz) = (1484, 60*, 1125) (*approx from grid growth)
100M

based on dhalf
(?x, ?yw, ?z) ? (0.08, 0.02, 0.05)
(Nx, Nyw, Nz) = (118, 50, 90) (*approx from grid growth)
500K

\end{comment}

%Similar to the channel, at the top wall, the hybrid simulation 
%remains locally RANS ($\alpha=1$).  Though it is difficult to see without a log-scale, 
%a much thinner region of RANS is also maintained along the bottom wall.  

%\subsection{Sensitivity to $C_r$}
\subsection{Sensitivity to resolution parameter}
\label{sec:cr_sens}

The assumption that $C_r=1$ in the evaluation of the available
resolution in Eq. \ref{rM} has not been strongly justified to this
point.  This coefficient essentially indicates how many grid lengths
are necessary to resolve some convecting turbulent structure.  Clearly
this coefficient should change depending on the characteristics of the
spatial discretization, including the order of accuracy and dispersion
relation~\cite{yalla2020effects}.  For increasing order of accuracy
and decreasing dispersion errors, small structures approaching the
grid scale are effectively better resolved, and hence a lower $C_r$
should be possible.
%More precisely, it is determined by the 
%dispersion relation of the numerical method which is determined by its spatial 
%order-of-accuracy \cite{yalla2020effects}.  The higher the order fo the method, 
%the smaller the dispersion errors for small structures near the grid scale and 
%a lower $C_r$ should be possible.
Determining exact values of $C_r$ for different order methods is
beyond the scope this paper.  Motivated by the observed tolerance in
the 
under-resolved regions of periodic hill problem, we
now examine the effects of reducing $C_r$ with the second-order finite
volume numerics of CDP.
\begin{figure}[th!]
\begin{center}
\subfigure[Mean $\bar{u}^+_x$ and $\beta$]
{\includegraphics[trim={1.9cm 0cm 2cm 0cm},clip=true,width=0.45\linewidth]{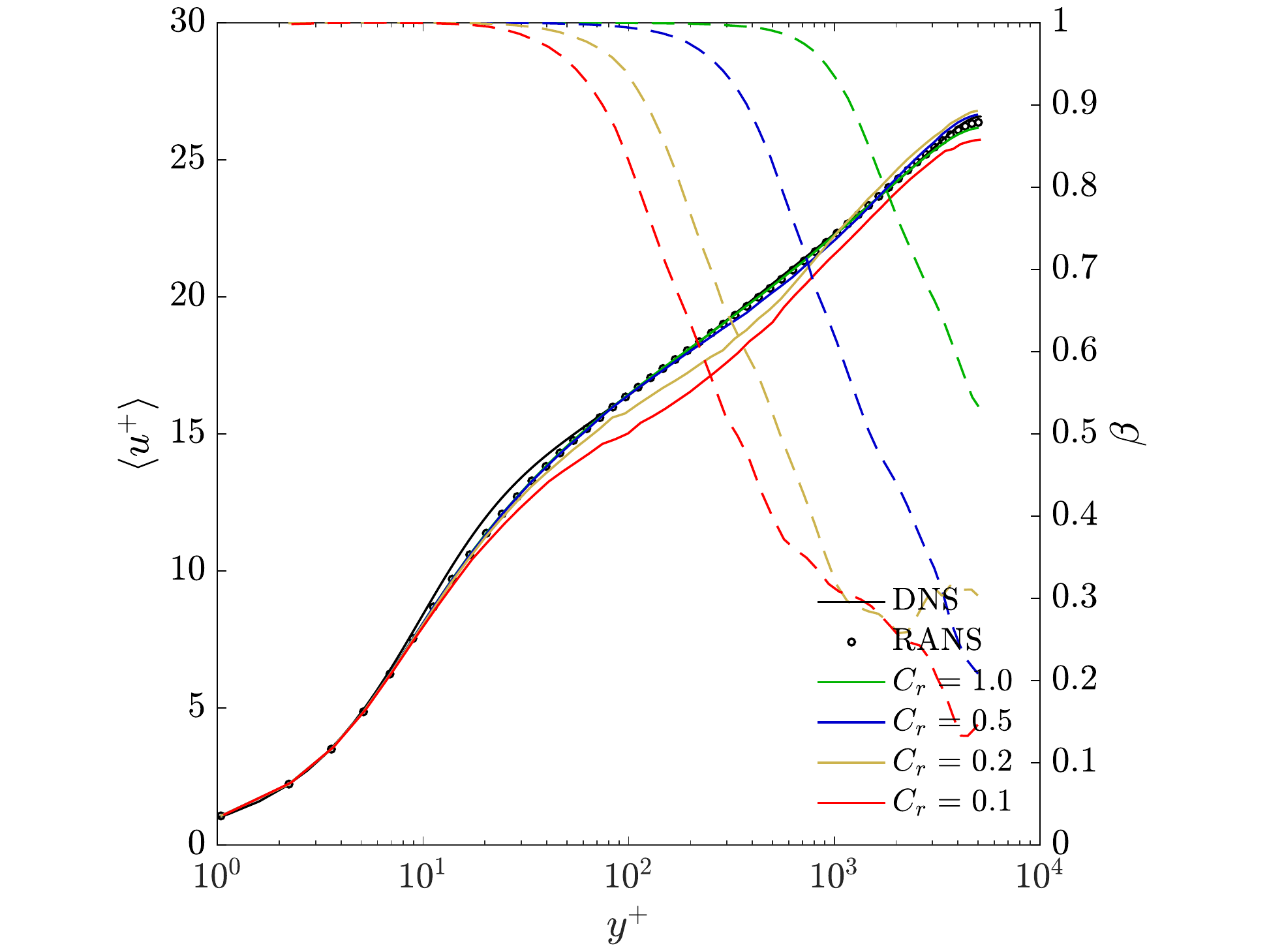}}
%\label{mux}
\subfigure[Turbulent kinetic energy $k^+$]{
\includegraphics[trim={1.9cm 0cm 2cm 0cm},clip=true,width=0.45\linewidth]{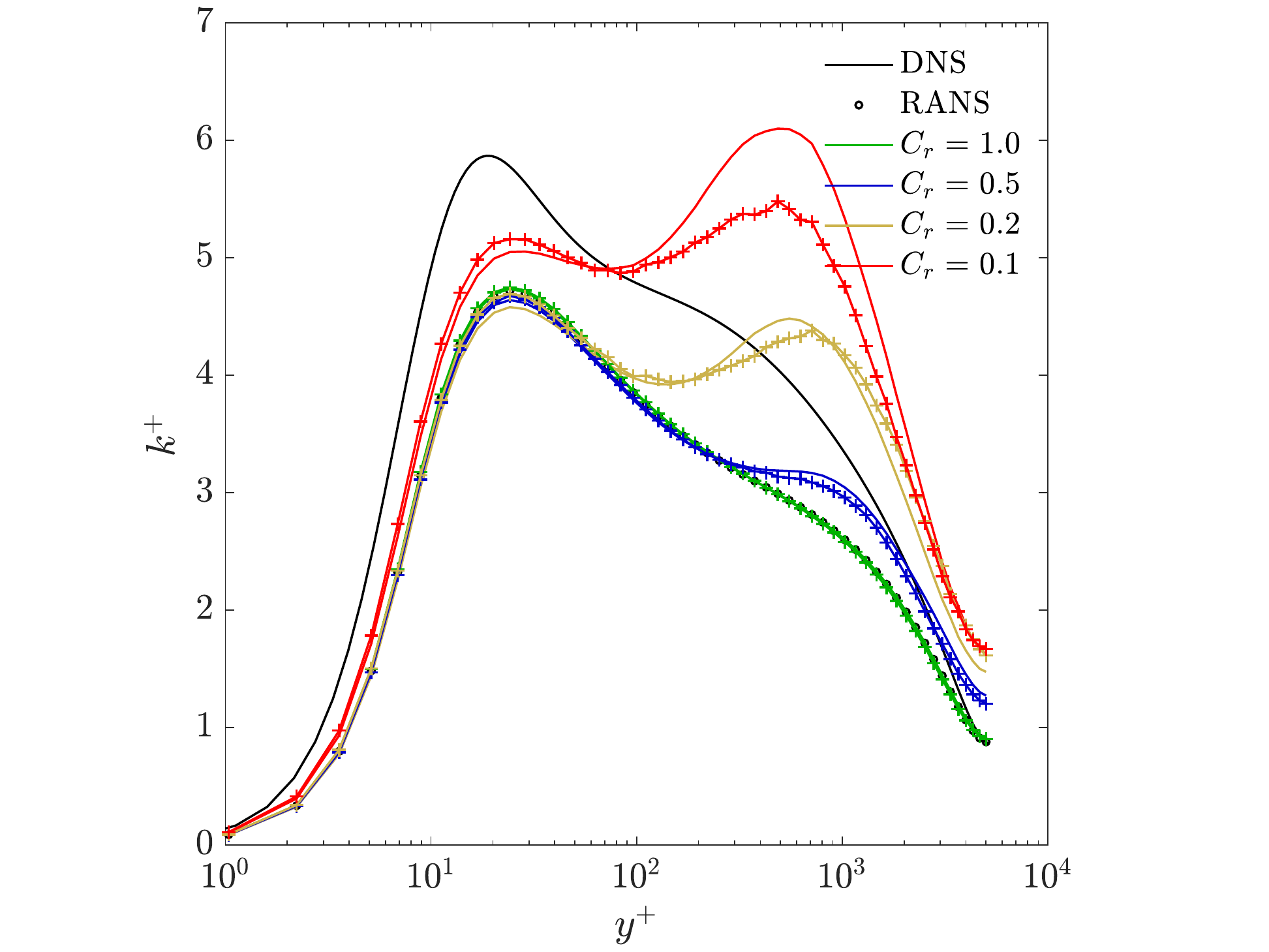}}
%\label{mtke}}
\end{center}
\caption{
%\todo[inline]{Vertical axis labels: $\la\bar{u}^+_x\ra$ ??, $k^+$}
Mean streamwise velocity in wall units (a) for fully
developed channel flow at $Re_\tau\approx5200$ along with the fraction
of unresolved turbulence $\beta$ (dashed), and (b) turbulent kinetic
energy $k^+$ of the statistically steady solution highlighting the
sensitivity to the parameter $C_r$.  In (b), the lines marked with +
symbols are $k^+$ obtained directly from the RANS model while the
unmarked lines are the time-averaged resolved turbulence plus $\beta$
times the RANS $k^+$.  Simulations have been run for approximately 80
flow-throughs.\label{fig:ux_cr}}
\end{figure}

First, consider the coarse resolution channel with $C_r=0.5$, $0.2$, 
and $0.1$ (Fig. \ref{fig:ux_cr}).  Lowering $C_r$ from unity to $0.5$ results in 
a drop in the center-channel $\beta$ from just above $0.5$ to $0.2$ with resolved 
turbulence moving towards the wall from $y^+\approx500$ to $100$.  Only a 
slight change in the log-layer slope and middle channel mean velocity are 
observed.  Both measures of the total TKE (modeled and resolved plus 
model times $\beta$) move toward the DNS value with a small
excess TKE at the center of the channel.   However, reducing $C_r$ 
below 0.5 results in both distortions to the mean velocity and large errors 
in TKE.  Errors introduced into the TKE are so pronounced that the outer 
peak is predicted to be larger than the inner with $C_r=0.1$.  Thus, it seems 
the small errors incurred with $C_r=0.5$ may be a tolerable exchange 
for the increase in resolved turbulence on a given grid, but moving 
below $C_r=0.5$ does not appear to be acceptable.
\begin{figure}[h!]
\begin{center}
\includegraphics[trim={0cm 2.5cm 0cm 3.5cm}, clip=true, width=0.98\linewidth]{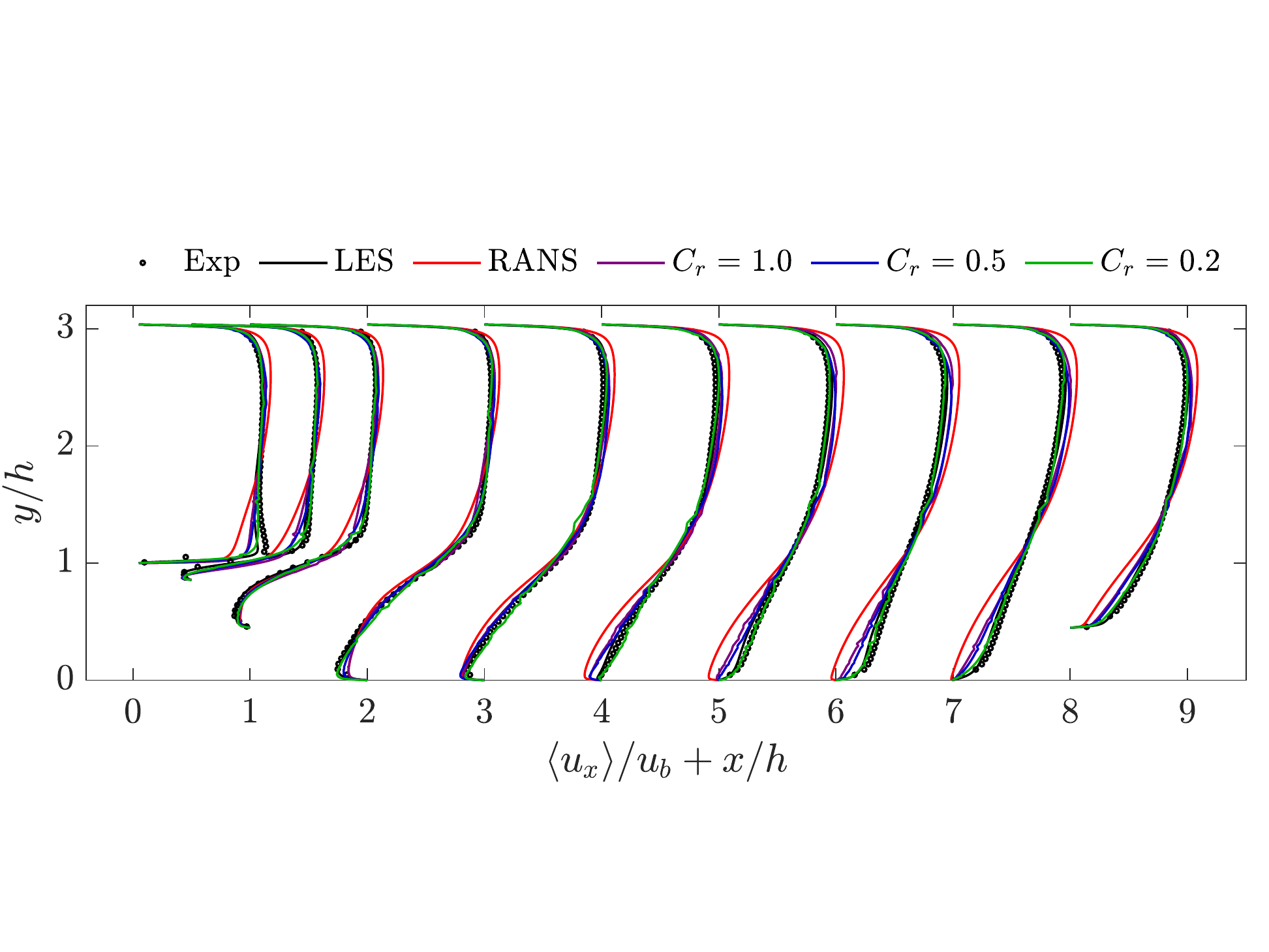}
\includegraphics[trim={0cm 2.5cm 0cm 4.5cm}, clip=true, width=0.98\linewidth]{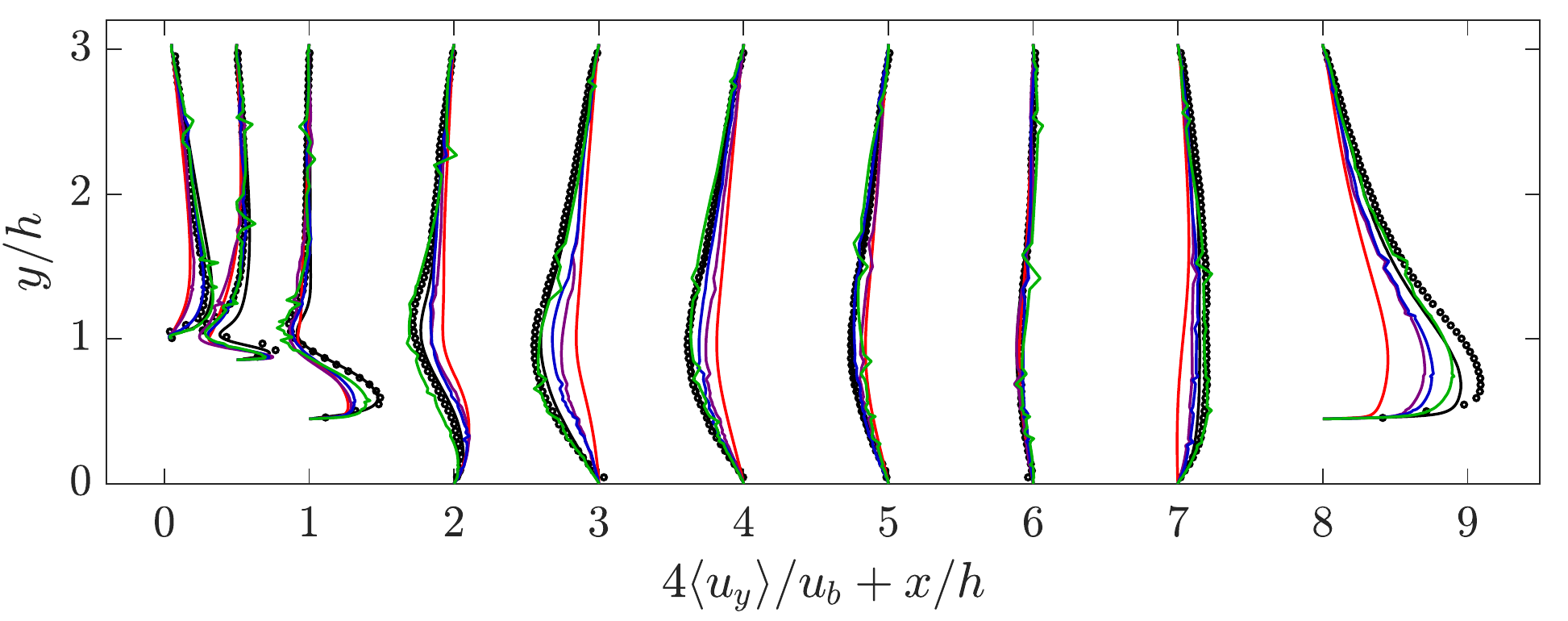}
\end{center}
\caption{
%\todo[inline]{Axis labels need to be explicitly normalized}
Mean streamwise (top) and vertical (bottom) velocity profiles normalized by the bulk velocity, $u_b$, offset 
by streamwise location normalized by the hill height, $h$, for only the coarse grid in Table \ref{tab:grids2} using 
different $C_r$ values, along with data from the experiments
of \cite{rapp:2011} and WRLES of \cite{breu:2009}.
%\todo[inline]{are these the right references?} Note that the vertical velocity is multiplied by a factor 
of four to enhance visibility.}
\label{fig:phill_ux_cr}
\end{figure}
%
%SWH: adding new results, remove if desired
%\begin{comment}
\begin{figure}[H]
\begin{center}
\includegraphics[trim={0cm 2.5cm 0cm 3.5cm}, clip=true, width=0.98\linewidth]{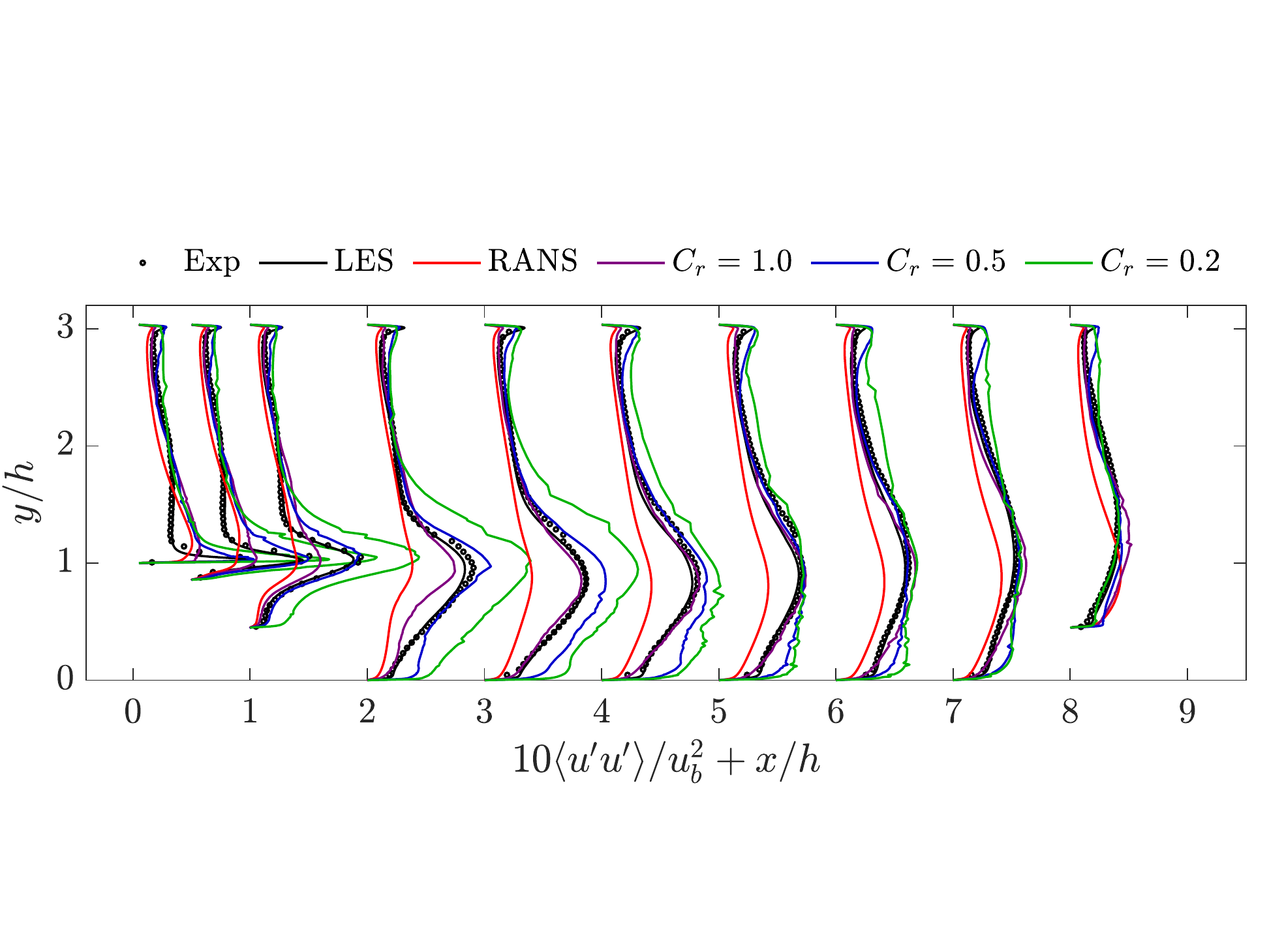}
\includegraphics[trim={0cm 2.5cm 0cm 4.5cm}, clip=true, width=0.98\linewidth]{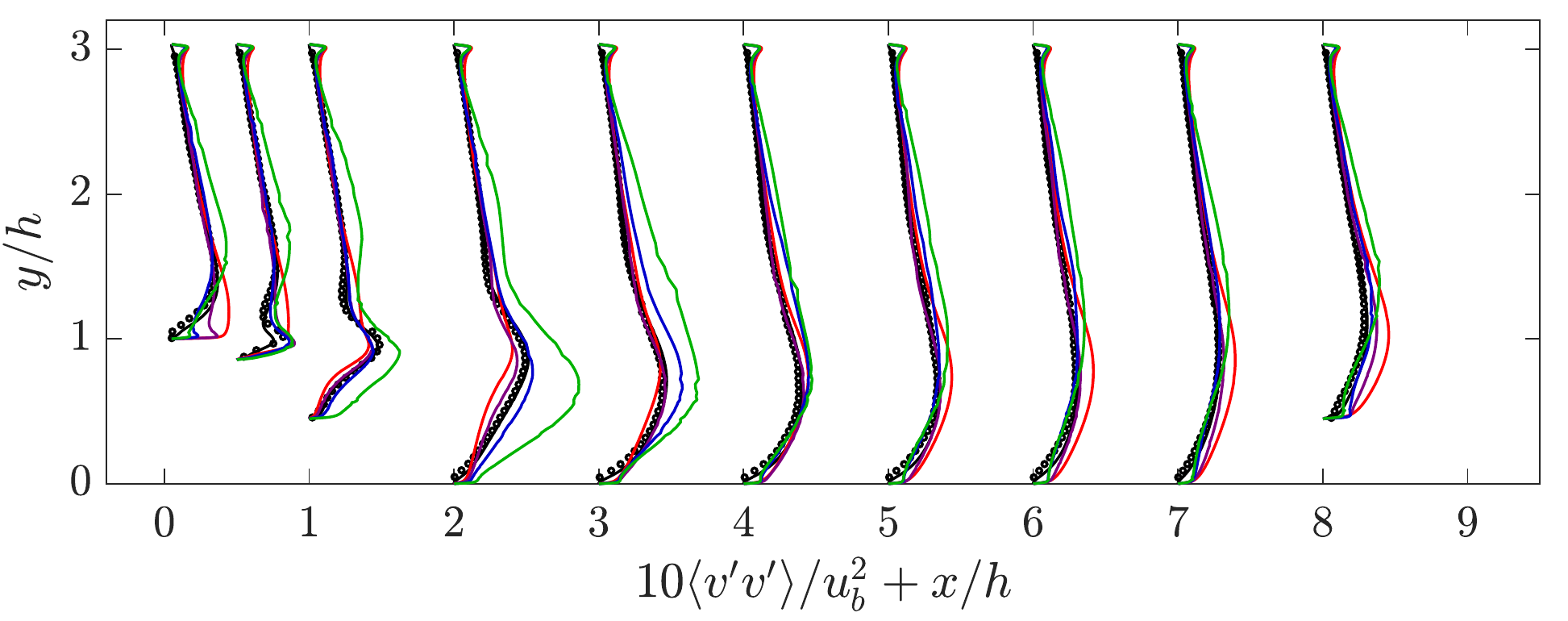}
\includegraphics[trim={0cm 2.5cm 0cm 4.5cm}, clip=true, width=0.98\linewidth]{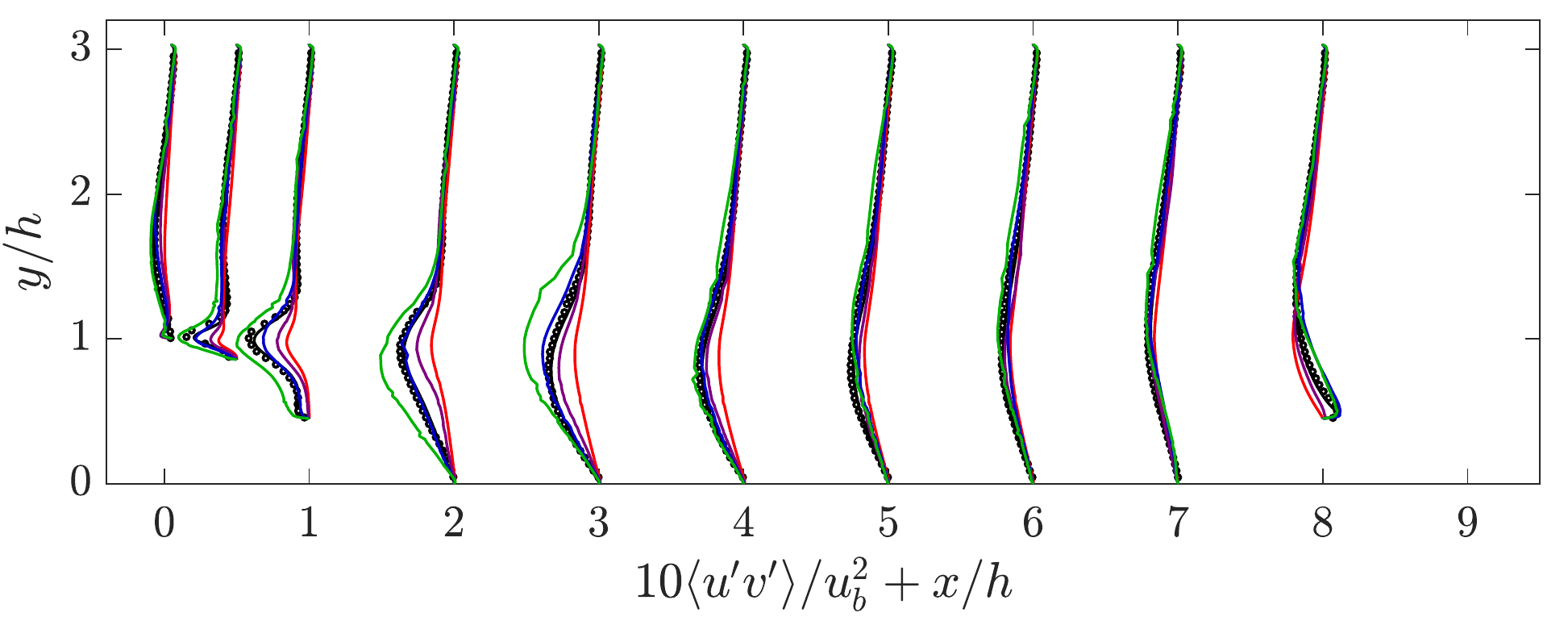}
\end{center}
\caption{
%\todo[inline]{Axis labels need to be explicitly normalized}
Profiles of the mean total (model plus resovlved) Reynolds stress
components normalized by the bulk velocity, $u_b$, offset by streamwise location normalized by the hill height, $h$, for only the coarse grid in
Table \ref{tab:grids2} using different $C_r$ values, along with data
from the experiments of \cite{rapp:2011} and WRLES
of \cite{breu:2009}.
%\todo[inline]{are these the right references?}
Shown are the streamwise velocity variance (top) vertical velocity
variance (middle) and Reynolds shear stress (bottom).  Note
amplification by a factor of 10 for visibility.}
\label{fig:phill_uv_profs_cr}
\end{figure}
%\end{comment}
%FIXED: \todo{RDM: $C_r=1$ and 0.5 are missing from the vertical velocity profiles.}

Next, we examine the effects of reduced $C_r$ in the more complex case 
of the periodic hill.  Figure \ref{fig:phill_ux_cr}
shows how reducing $C_r$ has the same basic effect as increasing the
resolution with $C_r=1$.  For $C_r=0.5$, the reattachment location is
well predicted though profiles at $x=6$ and $7$ still deviate slightly
from the experimental data. Contrary to the channel, $C_r=0.2$
actually gives the best prediction of the mean streamwise velocity
with values nearly identical to the experiment and the fine resolution
hybrid case with $C_r=1$. 
%RDM: removing Reynolds stress plots for now
%SWH: adding new results, remove if desired
%\begin{comment}
However, the Reynolds stress components (Fig. \ref{fig:phill_uv_profs_cr}) 
are over-predicted as observed 
in the channel.  This is most pronounced in the prediction of the streamwise 
variance where the hybrid profiles are significantly higher than the experimental 
value at $x=2$, $3$, and $4$ in the recirculation region.  The good mean 
velocity results for $C_r=0.2$ appear to be a result of error canceling.  Again 
consistent with the channel results, lowering $C_r$ to $0.5$ appears to
introduce only small errors in exchange for the most resolved 
turbulence possible for a given resolution.
%\end{comment}
So, it would seem that $C_r=1$ is too
conservative and the value of 0.5 may be more
appropriate for the numerical schemes used here. 

% SWH: I'm not sure about this line
%The lower $C_r$
%is particularly valuable in the periodic hill case, presumably because
%RANS does poorly so the accurate predictions require sufficient
%resolved fluctuations.%\todo{RDM: do you buy this?}

\subsection{Alternative energy transfer models}
\label{sec:aetm}

To this point, we have used the tensor-diffusivity M43 model for the
energy transfer portion of the model-split formulation.  Though the
model has been shown to perform well, many existing RANS-based codes
do not currently support a tensor diffusivity, making
implementation and adoption of the AMS method more difficult. It is
therefore useful to investigate the performance of AMS using
more easily implemented models for $\tau^e_{ij}$.
The three additional models discussed here are formulated in terms of
expected values so that they do not contribute to mean stress portion
of AMS.
\begin{figure}[th!]
\begin{center}
\subfigure[Mean $\bar{u}^+_x$ and $\beta$]
{\includegraphics[trim={1.9cm 0cm 2cm 0cm},clip=true,width=0.45\linewidth]{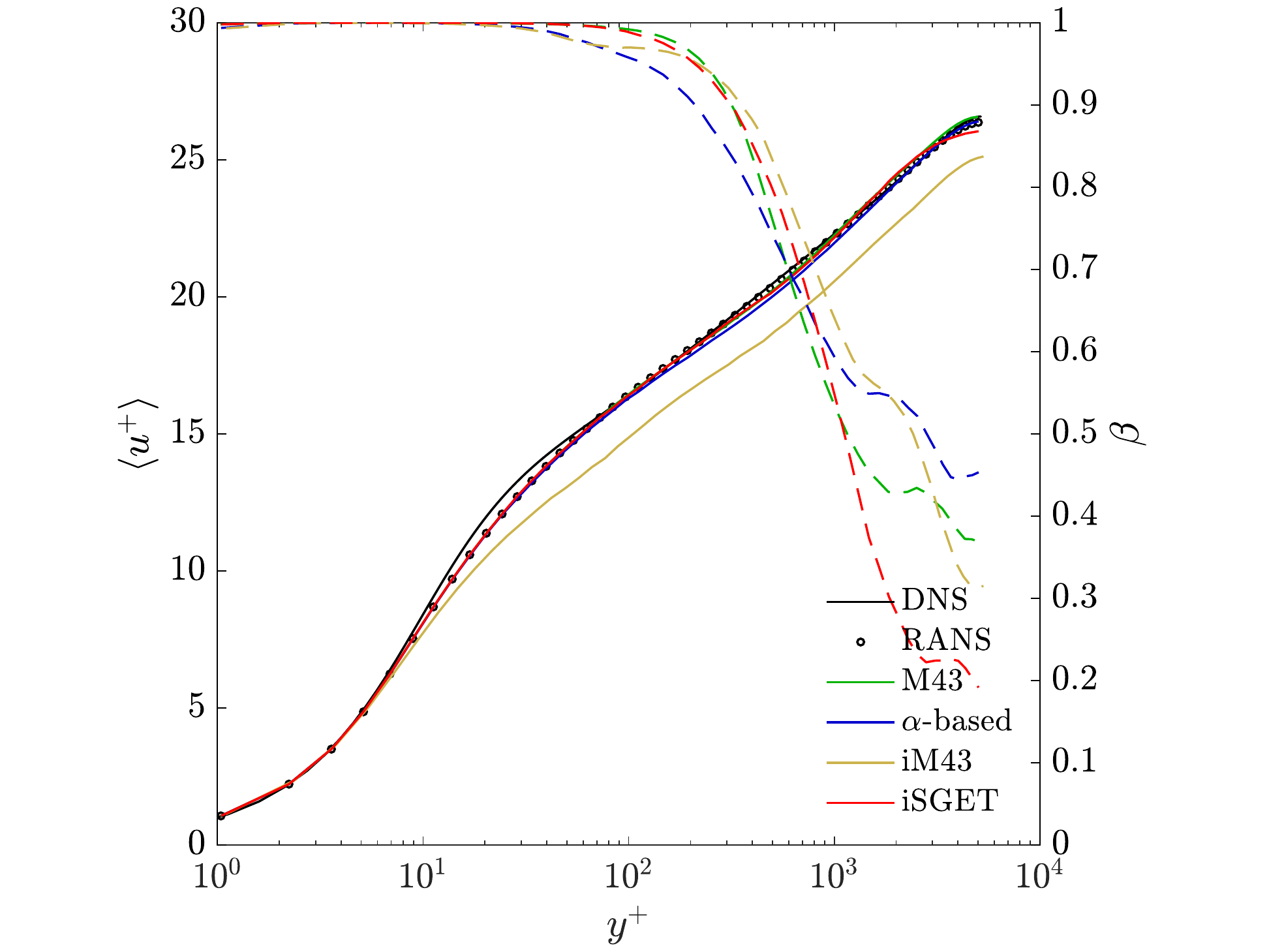}}
%\label{mux}
\subfigure[Turbulent kinetic energy]{
\includegraphics[trim={1.9cm 0cm 2cm 0cm},clip=true,width=0.45\linewidth]{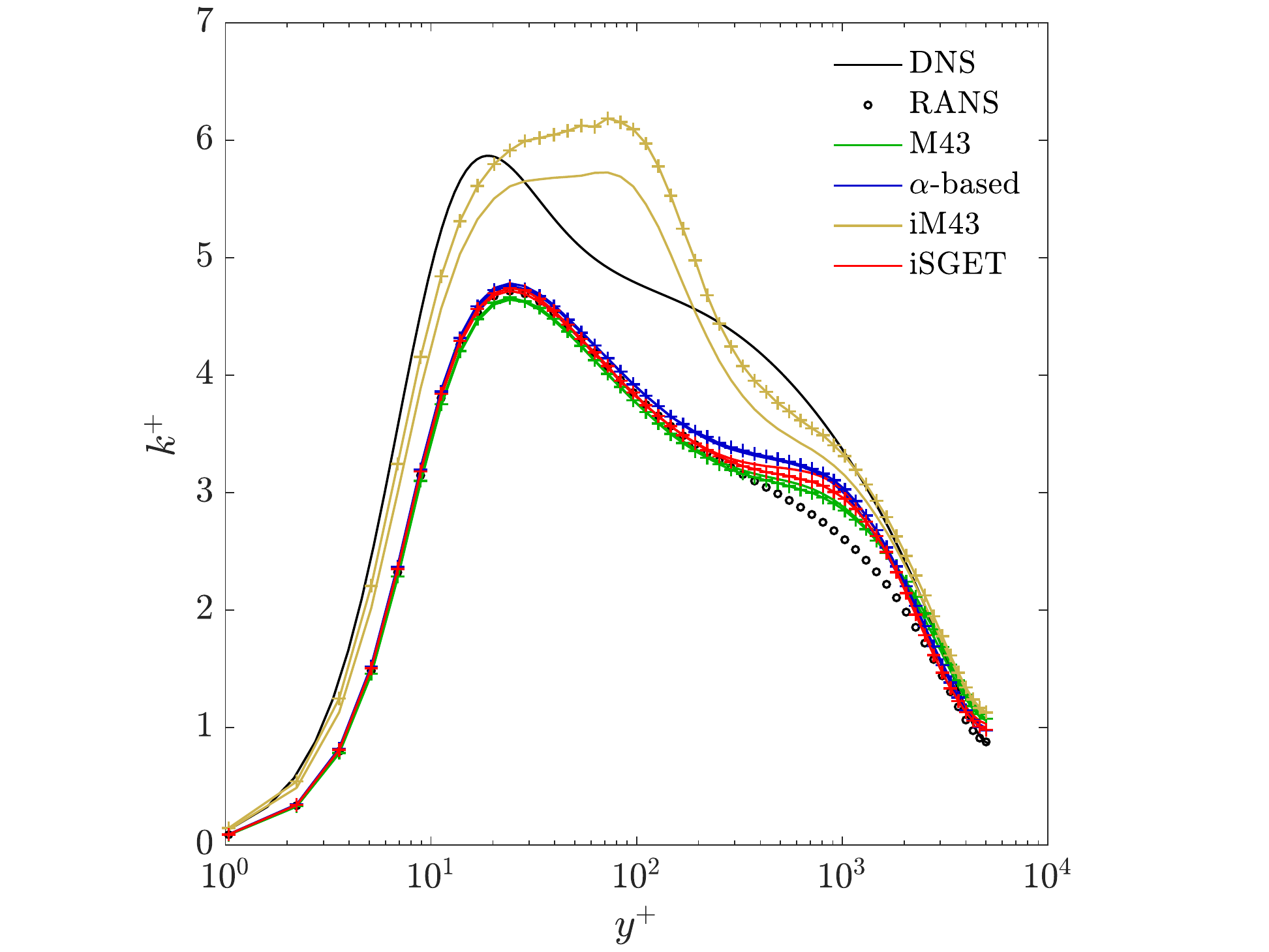}}
%\label{mtke}}
\end{center}
\caption{
%\todo[inline]{Vertical axis labels: $\la\bar{u}^+_x\ra$ ??, $k^+$}
Mean streamwise velocity in wall units (a) for fully
developed channel flow at $Re_\tau\approx5200$ along with the fraction
of unresolved of resolved turbulence $\beta$ (dashed), and (b) turbulent
kinetic energy $k^+$ of the statistically steady solution using
different energy transfer models and the fine resolution of
Table \ref{tab:grids}.  In (b), the lines marked with + symbols
are $k^+$ obtained directly from the RANS model while the
unmarked lines are the time-averaged resolved turbulence plus $\beta$
times the RANS $k^+$.  Simulations have been run for approximately 80
flow-throughs.\label{fig:ux_sget}}
\end{figure}

The first such model (labeled $\alpha$-based) was introduced in \S\ref{sec:genLES}.  Based on
expanding the total sugbrid term in light of arguments leading to eddy
viscosity formulations, it was suggested that the fluctuating gradient
contribution to the model should scale with $\alpha$ as
%\todo{SWH: probably too strong of a word here}
$\alpha(1-\alpha)\nu_{tot}{}S^>_{ij}$ (see Eq.~\ref{tau_ur2}), where
$\nu_{tot}$ is the RANS eddy viscosity.  This $\alpha$ dependence has
the correct RANS and DNS limits, since it vanishes where there are no
resolved fluctuations ($\alpha=1$) and where all turbulence is
resolved ($\alpha=0$).  The second simple energy transfer model (iM43
for isotropic M43) is an isotropic version of the M43 model, with the
length scale defined as the cell-diagonal, \emph{i.e.}
$\nu_e=C(\mathcal{M})\varepsilon^{1/3}\delta_{diag}^{4/3}$ where
$\delta_{diag}=(\mathcal{M}_{ij}\mathcal{M}_{ji})^{1/2}$.  Finally,
the third model (iSGET for implicit subgrid energy transfer) is not
formulated in terms of an explicit eddy viscosity.  It is well-known
that first-order upwinding of the convection term results in numerical
dissipation.  Motivated by the streamwise-upwind Petrov-Galerkin
(SUPG) method
\cite{broo:1982}, we introduce upwinding depending on a cell-Reynolds
number, but in this case the Reynolds number is based on the subgrid turbulence 
intensity, as determined from $\beta{}k_{tot}$ (see
Appendix~\ref{app:iSGET} for details).

Results for these alternative models are presented in
Fig. \ref{fig:ux_sget} for the fully-developed channel.  Here, the
fine resolution defined in Table \ref{tab:grids} is used because the
energy transfer models are most important with higher levels of
resolved turbulence.  The previously presented M43 results are also
included for comparison.  While the anisotropic M43 model performs
the best, both the $\alpha$-based and iSGET models also perform well.  The
$\alpha$-based model produces only a slight deviation from the log law
centered around $y^+\approx{}300$, while iSGET slightly distorts the
mean velocity profile above $y^+\approx{}2000$.  Interestingly, the
$\alpha$-based model results in the least resolved turbulence and
iSGET resolves significantly more turbulence above $y^+\approx{}2000$.  The
likely reason for the latter is that we have not counted the numerical
dissipation as part of the production of subgrid energy in the
definition of $r_\mathcal{M}$ in Eq. \ref{rM}.
%\todo{RDM: in light if
%this, i.e. that the iSGET model was not completely integrated into
%AMS, I'm inclined to eliminate it here. My following explanation for
%why we're still reporting this is tenuous.}
This is an inconsistency in the implementation of iSGET into AMS,
which would need to be corrected by calculating the mean rate of
numerical dissipation introduced by the upwinding. Many CFD codes,
including CDP used here, are not instrumented for this. The relatively
good results obtained using this incomplete integration of iSGET into
AMS suggest that implementing the numerical dissipation diagnostics
required to complete the AMS integration may be worthwhile. 
Perhaps correcting this
would also improve the slight mean velocity distortion. The good
results with iSGET also confirm that the AMS formulation can be used
even with implicit LES models for the energy transfer.
However, the
iM43 model performs poorly, with a significant shift in the
log-layer due to over-prediction of the wall shear stress.  The reason
for this is clear from the turbulent kinetic energy.  While the other
models produce similar $k_{tot}$ profiles, the iM43 results in
$k_{tot}$ values that are far too high for $y^+\lessapprox{}1000$.
The higher turbulence levels enhance momentum transfer towards the
wall, increasing the wall shear stress.  Thus, it appears the M43
model does require the anisotropic form to perform well.

From the perspective of both flexibility in model selection and
implementation of AMS, these results are encouraging.  The
$\alpha$-based model piggy-backs on the mean stress portion of the
split-model while iSGET simply exploits commonly used numerical
approximations.  Neither model requires a tensor-diffusivity.  As
mentioned above, both these models are formulated in terms of expected
values of eddy viscosity or cell Reynolds number so they do not
contribute to the mean stress. In AMS, using energy transfer models
with fluctuating eddy viscosities, \emph{e.g.} Smagorinsky, Vreman,
WALE, \emph{etc.}, or a fluctuating cell $Re$ in implicit methods,
would require that the contribution of the energy transfer term to the
mean stress be computed, and the mean stress model adjusted
accordingly.

%\subsection{Wall-Mounted Hump}
%\label{sec:wmh}
%\input{wmh}

\section{Conclusions}
\label{sec:conclusion}
%\todo{SWH: this conclusion feels a bit lean for the heft of the paper.
%Please add if you think of anything.}

The deficiencies of RANS models in representing complex flow features
like separation, and the expense of using LES to represent, for
example, broad expanses of a boundary layer, make a hybridization of
these two model paradigms (HRL) compelling.  Unfortunately, active
development of such HRL for more than 20 years has not resulted in
widely applicable robust and accurate HRL models. As detailed in
Sec.~\ref{sec:motivation}, there are four interrelated deficiencies common to
most HRL formulations that appear responsible for this state of affairs:
first is the use of a single eddy viscosity to represent both the mean
subgrid stress and the transfer of energy to the small scales; second is
the blending of RANS and LES models to hybridize them;  third is the application of RANS-based
transport models to fluctuating quantities; and last is reliance on
passive self-generation of resolved turbulence fluctuations when such
fluctuations need to be introduced.  These deficiencies lead to common
HRL failure modes, such as log-layer miss-match and model stress
depletion.

%\todo{RDM: added blending, is that OK?}

The Active Model Split (AMS) hybrid formulation described in
Section~\ref{sec:ms} was developed specifically to eliminate these
deficiencies. It does so by ``splitting'' the model into separate
models to represent the mean subgrid stress and the transfer of energy
to the unresolved scales, and by actively stirring where necessary to
produce resolved fluctuations. The model splitting 
eliminates the overloading of a single eddy viscosity with two roles,
since separate models are used for mean stress and energy transfer.
It also eliminates blending because the model is essentially LES
everywhere, with a consistent model for the mean subgrid stress. 
%\todo{swh:  minor point, prefer not to refer to it as a 'RANS model
%active everywhere'.  While true, stating it this way may confuse
%readers.  It is more LES everywhere with a subgrid model that
%actually models the stress}.
This allows LES to
be used with much coarser resolution than would otherwise be required,
since it is not necessary for the subgrid contribution to the mean stress
to be negligible, which is important since in HRL there will generally
be regions of very coarse LES.  Further, model splitting allows RANS
models to be used as designed, only for the mean stress and acting
only on average quantities. Finally, active stirring explicitly
eliminates the need for fluctuations to develop due to natural
instabilities of the mean.

The model split formulation does not just address challenges in hybrid
RANS LES modeling, it also addresses similar challenges in LES. In
particular, by eliminating the need for the resolved fluctuations to
carry the majority of the Reynolds stress, model splitting reduces the
resolution requirements for LES of inhomogeneous turbulent
flows. Further, by reverting to RANS if the LES resolution is not
sufficient to represent near-wall fluctuations, the model splitting
formulation provides a natural wall representation for LES, as
demonstrated in the channel results reported in Section~\ref{sec:fdc}.
%\todo{swh: Splitting hairs, I don't believe we should call this
%`WMLES'.  As we mentioned in the intro, WMLES should be used where
%LES is active over the entire domain.}
One can argue then that model splitting like that described here is
how LES should generally be performed.

The model tests in channel flow and the periodic hill indicate that
the AMS formulation has indeed addressed the primary shortcomings
of HRL models. In the channel flow, it produces generally consistent results for
horizontal resolution ranging over a factor of at least 2.2, showing
both resolution independence and a lack of log-layer
miss-match. Further, the AMS solution remains consistent as the model
transitions from RANS to LES in either time or space. In the more
complex periodic hill case, in which RANS performs poorly due to flow
separation, the AMS formulation produces mean velocity in good
agreement with both experiments and wall-resolved LES, with a grid
that is up to 60 times smaller than the LES. These test results suggest
that the AMS approach is a solution to the common hybrid modeling
issues identified above and successfully eliminates hybridization
artifacts such as modeled-stress depletion. Evaluations of the AMS
formulation in more complex flow scenarios are clearly warranted.  These
should include common aerodynamic test cases with varying degrees of smooth
wall separation and reattachment, and flows that are unsteady in the
mean.

While the model test results are quite good, the AMS solutions are not
perfect. Further, there are several details that must be specified to
complete the definition of a particular AMS implementation. For the
model and tests presented here, these details were described in
Section~\ref{sec:ms} and are summarized in Appendix~\ref{app:rans}
and \ref{app:M43}, but
further investigation and refinements would clearly be useful. The most
significant of these are listed here.
%\todo{RDM: is this list complete? Does something not belong on it?}
\begin{enumerate}
\item \textbf{Forcing Formulation:} The forcing formulation described
in Section~\ref{sec:forcing} is an ad hoc place holder and has shortcomings. A formulation
that is more realistic in structure, does not need to be clipped, has
a more controllable energy injection rate and can extract resolved
energy when needed would be a great improvement. Such improvements
could allow more rapid transitions from RANS to LES.
\item \textbf{Resolution Inhomogeneity:} As is well known, when LES
resolution is inhomogeneous, the filter operator that defines the
resolved scales and the spatial derivative operator do not
commute. When there is a mean flow through the inhomogeneous grid, the
commutator represents the transfer of energy between resolved and
unresolved fluctuations, as needed as the resolution changes. This
generally goes unmodeled, and its effects are particularly acute with
HRL because hybrid simulations are commonly done with strongly
inhomogeous grids, indeed that is the objective. Thus, the commutator needs
to be modeled, and the forcing formulating may be useful in this regard.
\item \textbf{Energy Transfer Model:} As discussed in
section~\ref{sec:energyXfer}, the M43 models has been used here to represent
energy transfer to the unresolved scales. It accounts for the effects
of resolution anisotropy, but not turbulence anisotropy. In complex
turbulent flows typical of HRL applications, turbulence is expected to
be highly anisotropic and LES resolution is generally coarse, so the
subgrid turbulence is anisotropic too. Energy transfer models that
account for turbulence and grid anisotropy are thus needed.
\item \textbf{Pseudo-mean Definition:} The pseudo-mean is defined in
Section~\ref{sec:pseudoMean} as a causal temporal filter with a time constant
determined by the turbulence time scale $k/\epsilon$. However, neither
the filter definition or the averaging time scale have been carefully
investigated, so refinements are likely to be appropriate. For
example, in the presence of a mean velocity, the convective
turbulent time scale $k^{3/2}/(\varepsilon|\la{}u\ra|)$ is also relevant. The
appropriate averaging time scale is of some importance because, as
discussed in Section~\ref{sec:stress}, the interactions of the largest
turbulent scales with the subgrid are poorly described by gradient
transport models, so these largest scale fluctuations are best
included in the pseudo-mean.  Finally, the interaction of the
pseudo-mean time averaging with mean unsteadiness needs to be
investigated.
\end{enumerate}
These opportunities for further refinement are essentially to improve
the LES component of the AMS formulation. Despite the need for some
refinements, by addressing the primary shortcomings of most HRL
formulations, AMS provides a framework for robust, predictive, and
cost-effective simulation complex turbulent flows.

\section*{Acknowledgements}
The authors acknowledge generous financial support provided primarily
by the National Aeronautics and Space Administration (cooperative
agreement number NNX15AU40A).  Additional funding was provided by the
Air Force Office of Scientific Research (grant FA9550-11-1-007), the
Exascale Computing Project (17-SC-20-SC), a collaborative effort of
two U.S. Department of Energy (DOE) organizations (Office of Science
and the National Nuclear Security Administration), the DOE Energy
Efficiency and Renewable Energy (EERE) (contract DE-AC02-06CH11357)
and the National Science Foundation (CBET-1904826).  Thanks are also
due the Texas Advanced Computing Center at The University of Texas at
Austin for providing HPC resources that have contributed to the
research results reported here (\url{http://www.tacc.utexas.edu}).

Finally, the authors would like to thank the Center for Predictive Engineering and Computational Sciences (PECOS) research group, including M. Lee, G. Yalla, J. Melvin, and C. Pederson, for support and lively discussions.

\appendix
%\section*{Appendix}
%\label{sec:appendix}

%\noindent\textbf{{I. RANS model details}}\\
\section{RANS model details} \label{app:rans}
For completeness, the models used in the presented results are defined here.  
In \S\ref{sec:fdc}, Chien's $k$-$\varepsilon$ model is used for basic channel flow.  A 
modified version of the ``code-friendly'' variant~\cite{lien:2001} of Durbin's  
$\overbar{v^2}$-$f$ model~\cite{durb:1995} is used in \S\ref{sec:phill} for the periodic 
hill case.  Note that we have only considered incompressible flow here and the 
density is assumed to be unity everywhere.  Further, the mean strain magnitude, 
$\la{}S\ra=(\la{S}_{ij}\ra\la{S}_{ij}\ra)^{1/2}$, is used and not the fluctuating strain 
magnitude.  The associated equations for turbulent kinetic energy, $k$, turbulent 
dissipation rate, $\varepsilon$, minimum turbulent stress component, 
$\overbar{v^2}$, and the redistribution rate, $f$, can be expressed with the 
generic form
\begin{equation}
\pd_t\phi +\la{}u_i\ra\pd_i\phi= \mathcal{P}_\phi-\mathcal{D}_\phi+\partial_k\big{(}\partial_k\kappa_\phi\phi\big{)},
\label{eqn:gen}
\end{equation}
with each term provided in Tables \ref{tab:ke} and \ref{tab:v2f}.  The
Chien wall treatment is in terms of $\delta^+$, which is defined in
terms of the distance to the wall $\delta_w$ and the mean wall shear stress
at the nearest wall:
\begin{align}
\delta^+ &= \delta_w\frac{u_\tau} {\nu}\nonumber\\
u_\tau &= \sqrt{\tau_w}\nonumber\\
\tau_w & = \nu \pd{}_n{}\la{}u\ra_x\nonumber
\end{align}
however, the channel flows presented here have been normalized so that $u_\tau=1$.
Eddy viscosities for each model can be expressed as $\nu_t=C_\mu\zeta{}kT$ with 
terms as shown in Table \ref{tab:ev}.
\begin{table}
\centering
\begin{tabular}{ c c c }
\hline
$\phi$  & $k$ & $\varepsilon$ \\
\hline
\hline
$\mathcal{P}_\phi$ & Eq. \ref{Pk} & $C_{\varepsilon{}1}f_1\frac{\mathcal{P}_k}{T}$ \\
$\mathcal{D}_\phi$ & $\varepsilon+2\nu\tfrac{k}{\delta_w^2}$ &  $C_{\varepsilon{}2}f_2\tfrac{\varepsilon}{T}+2\nu\tfrac{\varepsilon}{\delta_w^2}e^{-0.5\delta^+}$\\
$\kappa_{\phi}$ & $\nu+\nu_t$ & $\nu+\tfrac{\nu_t}{1.3}$   \\
$\phi_{wall}$ &  0 & 0 \\
\hline
\end{tabular}
\caption{Terms for generic RANS transport model (\ref{eqn:gen}) for Chien $k$ and $\varepsilon$ RANS model}
\label{tab:ke}
\end{table}
\begin{table}
\centering
\begin{tabular}{ c c c c c }
\hline
$\phi$  & $k$ & $\varepsilon$ & ${v^2}$ & $f^*$ \\
\hline
\hline
$\mathcal{P}_\phi$ & Eq. \ref{Pk} & $C_{\varepsilon{}1}\frac{\mathcal{P}_k}{T}$ & $kf$ & $-\frac{1}{(C_LL)^2}\bigg{(}R_f\Big{(}\frac{\overbar{v^2}}{k}(C_1-6)-\frac{2}{3}(C_1-1)\Big{)}-C_2\frac{\mathcal{P}_k}{k}\bigg{)}$  \\
$\mathcal{D}_\phi$ & $\varepsilon$  & $C_{\varepsilon{}2}\frac{\varepsilon}{T}$ & $6\frac{\overbar{v^2}}{k}\varepsilon$ & $\frac{f}{(C_LL)^2}$  \\
$\kappa_{\phi}$ & $\nu+\nu_t$ & $\nu+\tfrac{\nu_t}{1.3}$  & $\nu+\nu_t$ & 1  \\
$\phi_{wall}$ &  0 & $2\nu\big{(}\frac{k}{\delta^2_{w}}\big{)}_1$ & 0 & 0 \\
\hline
\end{tabular}
\caption{Terms for generic RANS transport model (\ref{eqn:gen}) for $\overbar{v^2}$-$f$ RANS model.  The $1$ subscript for the $\varepsilon_{wall}$ boundary condition indicates the values at the first wall-normal grid point. $^*$Note there is no unsteady or convective term in the elliptic $f$-equation.}
\label{tab:v2f}
\end{table}
\begin{table}
\centering
\begin{tabular}{ c c c c }
\hline
Model & $C_\mu$ & $\zeta$ & $T$ \\
\hline
\hline
Chien & 0.09 & $f_\mu{}$ & $\max\Big{(}\frac{k}{\varepsilon},6\frac{\sqrt{\nu}}{\varepsilon} \Big{)}$ \\
${v^2}$-$f$ & 0.2  & $\tfrac{v^2}{k}$ & $\min\bigg{(} \max\Big{(}\frac{k}{\varepsilon},6\frac{\sqrt{\nu}}{\varepsilon} \Big{)}, \frac{0.6k}{\sqrt{6}C_\mu\overbar{v^2}\la{}S\ra}\bigg{)}$  \\
\hline
\end{tabular}
\caption{Eddy viscosity terms for Chien and $\overbar{v^2}$-$f$ RANS model.}
\label{tab:ev}
\end{table}
Additional Chien wall functions are 
\begin{align}
f_\mu &=1-e^{-0.0115\delta^+}\nonumber\\
f_1 &=1\nonumber\\
f_2 &=1-\frac{0.4}{1.8}e^{-\frac{Re^2_T}{36}}\nonumber
\end{align}
where the turbulent Reynolds number is $Re_T=k^2/(\nu\varepsilon)$.   Finally,
the following coefficients close Chien's model
\begin{align*}
C_{\varepsilon{}1} &= 1.35, {  } C_{\varepsilon{}2} = 1.8.
\end{align*}
The length scale used in the $f$-equation for $\overbar{v^2}$-$f$ is given by
\begin{equation}
L= \max\bigg{(} \min\Big{(}\frac{k^{3/2}}{\varepsilon},   \frac{k^{3/2}}{\sqrt{6}C_\mu\overbar{v^2}\la{}S\ra} \Big{)},  C_\eta\frac{\nu^{3/4}}{\varepsilon^{1/4}} \bigg{)}
\end{equation}
and an additional time scale modification is made here with 
\begin{equation}
R_f=\min\Big{(} \frac{1}{T}, \frac{\la{}S\ra}{3}\Big{)}
\label{Rf}
\end{equation}
which was observed to not affect basic RANS behavior but to be
necessary for use with AMS.  Without this modification,
$\overbar{v^2}$ becomes excessive in hybrid simulations.  The general
applicability of this modification deserves further study. The model
is complete with the following coefficients
\begin{align*}
C_\eta &= 70, { } C_1 = 1.4, { } C_2 = 0.3, C_L = 0.23,\\
C_{\varepsilon{}1} &= 1.4\Big{(}1+0.005(k/\overbar{v^2})^{1/2} \Big{)}, { } C_{\varepsilon{}2} = 1.9.
\end{align*}

%Due to how $r_\mathcal{M}$ is formulated 
%based on $\overbar{v^2}$, this error resulted in forcing being activated too near the wall.  Therefore, when not using 
%the modification in (\ref{Rf}), additional near-wall protection is necessary.   An additional scaling-factor of 
%\begin{equation}
%\eta_{wall}=\frac{1}{2}\Big{(}\tanh\big{(} C_\zeta(\zeta-\zeta_c)\big{)} +1 \Big{)}
%\end{equation}
%applied to (\ref{Fi_actual}), with $\zeta=1.5\overbar{v^2}/k$, was found to be sufficient. 

%\textcolor{red}{SST model if results also shown}\\*
%\noindent\textbf{\emph{SST:}}
%This model was the earliest version

%\noindent\textbf{{II. M43 model details}}\\
\section{M43 model details} \label{app:M43}
The M43 model is discussed in detail elsewhere \cite{haer:2019b}.  \respa{In short, a tensor eddy viscosity is used to describe the dissipation anisotropy resulting from anisotropic filtering implied by the resolution.  In this way, unrealistic spectral energy pile-ups in coarse grid directions are avoided and the resolved stress is not corrupted by commonly used stretched grids.  The model assumes the unresolved turbulence is in the Kolmogorov inertial range.}  Here, we make 
the addition of a resolution adequacy modifier and present the model form and 
provide coefficients appropriate for the second-order finite volume numerics used 
in this work.  The purpose of the modifier is to sensitize the model to potential local 
under-resolution, \emph{i.e.} $r_\mathcal{M}>1$.  The basic M43 model, with the 
resolution adequacy modifier, is
\begin{equation}
\nu^E_{ij}=f(r_\mathcal{M})C(\cM)\varepsilon^{1/3}\mathcal{M}^{4/3}_{ij}
\label{M43}
\end{equation}
where the resolution tensor, $\mathcal{M}_{ij}$ has been discussed in
\S\ref{sec:ms},
\respa{$\mathcal{M}^{4/3}$ is determined by raising the eigenvalues of
  $\mathcal{M}$ to the 4/3 power},
%  the power is understood as being applied to the eigenvalues of
%  $\mathcal{M}_{ij}$},
and 
the dissipation is taken directly from the RANS transport models.  Use of the total 
$\varepsilon$ is not strictly correct, as it should be the subgrid dissipation, \emph{i.e.} the 
total less $2\nu{}\la{}\pd_j{}u^>_i\pd_j{}u^>_i\ra$.  However, for the coarse LES 
considered here, the resolved dissipation is negligible.  With an 
anisotropic diffusivity, the general expression for the energy transfer part 
of the subgrid stress in the model-split formulation takes the form 
\begin{equation}
\tau^e_{ij}=\nu^E_{ik}\pd_k{}u^>_{j} + \nu^E_{jk}\pd_k{}u^>_{i} - \frac{2}{3}\nu^E_{mn}S^>_{mn}\delta_{ij}.
\end{equation}
Let $\lambda^\cM_3$ be the smallest eigenvalue of $\cM$ and $\lambda^\cM_1$ 
the largest. The coefficient $C$ is a function of the eigenvalues of ${\cM}$ as
\begin{equation}
C(\cM) = C^\circ_\cM\sum_{i=0}^4\sum_{j=0}^{{4-i}} c_{ij}x^iy^j,
\label{fits}
\end{equation}
where $x=\ln(r)$, $y=\ln(\sin(2\theta))$, $r^2=(\lambda_1^{\hat\cM})^2 + (\lambda_2^{\hat\cM})^2$ and 
$\theta=\cos^{-1}(\lambda_1^{\hat\cM}/r)$.  Coefficients appropriate for second-order 
finite volume numerics are provided in Table \ref{tab:coefs}.
\begin{table}
\centering
\begin{tabular}{ c c }
\hline
\hline
$C^\circ_\cM$ & 0.11\\
$c_{00}$ & 0.9719\\   %1
$c_{10}$ & 0.06559\\   %2                            
$c_{01}$ & 0.07110\\     %3                         
$c_{20}$ & 0.04992\\       %4                        
$c_{11}$ & -0.05690\\        %5                      
$c_{02}$ & 0.09797\\           %6                  
$c_{30}$ & -0.01559\\  %7
$c_{21}$ & 0.002004\\   %8                                                  
$c_{12}$ & 0.002177\\       %10                      
$c_{03}$ & 0.03423\\             %11                  
$c_{40}$ & 0.001219\\ %12
$c_{31}$ & 0.0004179\\  %13      
$c_{22}$ & 0.0004211\\      %14                   
$c_{13}$ & 0.001224\\            %15                
$c_{04}$ & 0.003695\\ %16
\hline
\end{tabular}
\caption{Values of the fitting coefficients in (\ref{fits}) based on
the fitting method outined in \cite{haer:2019b} as applied to second-order 
finite volume numerics.}
\label{tab:coefs}
\end{table}
Finally, the overall scaling is modified as a function of the
resolution adequacy as
 \begin{equation}
 f(r_\mathcal{M})=\max{}\big{(}\min{}(\la{}r_\mathcal{M}\ra^2,30),1\big{)}.
 \label{M43_hack}
 \end{equation}
This modification is motivated by the fact that $r_\mathcal{M}$ is a length
scale ratio.  When $r_\mathcal{M}>1$ while resolved fluctuations are
not zero, the simulation is locally under-resolved by an amount
indicated by $r_\mathcal{M}$.  Thus, the length scales in the M43 eddy
viscosity are modified by this length scale ratio.  However, it is
essentially \emph{ad-hoc}. A more principled formulation is needed.\\

%\vspace{1cm}

\section{Upwinding-based energy transfer model} \label{app:iSGET}
The upwinding performed in \S\ref{sec:aetm} for the iSGET model is based 
on SUPG \cite{broo:1982} with an upwinding weight of the form
\begin{equation}
w = \frac{1}{\tanh(Re_f)}-\frac{1}{Re_f}
\end{equation}
which is used to determine face-values for the finite-volume 
approximation of convective term gradient.  When $w$ is 0, a central-difference
approximation is recovered, whereas $w=1$ results in first-order upwinding.  
However, instead of using a cell Reynolds number based on convection velocity, here 
we define the cell Reynolds number for each grid face as
\begin{equation}
Re_f = \frac{u^<_{rms}|s|}{\nu}
\end{equation}
where $u^<_{rms}=(2/3\beta{}k_{tot})^{1/2}$ is the subgrid turbulence intensity, and 
$|s|$ is the distance between face-sharing cell centers.
%\todo{SWH: is this enough?  Don't really want to get into basic of FV method...}

%the Air Force Office of Scientific Research (grant FA9550-11-1-007), 

%\section{References}
%\bibliographystyle{aiaa}    
\bibliography{tams}

\end{document}